\documentclass[letterpaper,11pt]{yalephd}
\usepackage{geometry} % you need this for yalephd.cls to work.
\usepackage{graphicx} % you probably want the rest of these.
\usepackage{dcolumn}
\usepackage{bm}
\usepackage{bbm} % for identity operator
\usepackage{amsmath}
\usepackage{amsfonts}
\usepackage{amssymb}
\usepackage{appendix}
\usepackage{comment}
\usepackage{cite}
\usepackage{notoccite}
\usepackage{lmodern}
\usepackage{subfig}
\usepackage{slashed} % for feynman slash notation
\usepackage{simplewick} % for wick contractions

\newcommand{\identity}{\mathbbm{1}}
\newcommand{\SU}{\text{SU}}
\newcommand{\U}{\text{U}}
\newcommand{\vev}[1]{\ensuremath{\left\langle #1 \right\rangle} }

\newlength\longest

\begin{document}

% Need to define title before the abstract.
%\title{Nearly Conformal Dynamics as a Mechanism for Higgs Compositeness (working title)}
%\title{Strongly Coupled Systems with Nearly Conformal Dynamics}
\title{Studies of Conformal Behavior \\ in Strongly Interacting Quantum Field Theories}
\author{Andrew David Gasbarro}
\advisor{Thomas Appelquist and George Tamminga Fleming}
\date{December, 2018} % usually not \today.

% All the stuff at the front of your thesis.
\frontmatter

\begin{abstract}

In this dissertation, we present work towards characterizing various conformal and nearly conformal quantum field theories nonperturbatively using a combination of numerical and analytical techniques. 
A key area of interest is the conformal window of four dimensional gauge theories with Dirac fermions and its potential applicability to beyond the standard model physics.
%We are interested in gauge theories near and inside the conformal window for their possible applicability to beyond the standard model physics.

In the first chapter, we review some of the history of models of composite Higgs scenarios in order to motivate the study of gauge theories near the conformal window.
In the second chapter we review lattice studies of a specific theory, $\SU(3)$ gauge theory with eight flavors of Dirac fermions in the fundamental representation of the gauge group.  
We place a particular emphasis on the light flavor-singlet scalar state appearing in the spectrum of this model and its possible role as a composite Higgs boson.
We advocate an approach to characterizing nearly conformal gauge theories in which lattice calculations are used to identify the best low energy effective field theory (EFT) description of such nearly conformal gauge theories, and the lattice and EFT are then used as complementary tools to classify the generic features of the low energy physics in these theories.
 %and to use a combined lattice and EFT approach to classify the generic features of the low energy physics of these theories.
%Progress on characterizing nearly conformal gauge theories would be greatly improved with a valid effective field theory description of nearly conformal gauge theories to be used in conjunction with lattice studies.
We present new results for maximal isospin $\pi\pi \rightarrow \pi\pi$ scattering on the lattice computed using L{\"u}scher's finite volume method.
This scattering study is intended to provide further data for constraining the possible EFT descriptions of nearly conformal gauge theory.
%We demonstrate that the maximal isospin scattering length is in good agreement with leading order chiral perturbation theory ($\chi$PT) when expanded in physical quantities, but in poor agreement with chiral perturbation theory when expanded in bare quantities, meriting further analysis with other effective field theory methods.
In Chapter 3, we review the historical development of chiral effective theory from current algebra methods up through the chiral Lagrangian and modern effective field theory techniques.
We present a new EFT framework based on the linear sigma model for describing the low lying states of nearly conformal gauge theories.
We place a particular emphasis on the chiral breaking potential and the power counting of the spurion field.

In Chapter 4, we report on a new formulation of lattice quantum field theory suited for studying conformal field theories (CFTs) nonperturbatively in radial quantization.
We demonstrate that this method is not only applicable to CFTs, but more generally to formulating a lattice regularization for quantum field theory on an arbitrary smooth Riemann manifold.
The general procedure, which we refer to as \emph{quantum finite elements} (QFE), is reviewed for scalar fields.
Chapter 5 details explicit examples of numerical studies of lattice quantum field theories on curved Riemann manifolds using the QFE method. 
We discuss the spectral properties of the finite element Laplacian on the 2-sphere. 
Then we present a study of interacting scalar field theory on the 2-sphere and show that at criticality it is in close agreement with the exact $c=1/2$ minimal Ising CFT to high precision.
We also investigate interacting scalar field theory on $\mathbb{R} \times \mathbb{S}^2$, and we report significant progress towards studying the 3D Ising conformal fixed point in radial quantization with the QFE method.
In the near future, we hope for the QFE method to be used to characterize the four dimensional conformal fixed points considered in the first half of this dissertation.
\end{abstract}

\maketitle
\makecopyright{2018} % change as needed.
\tableofcontents
\listoffigures % remove this if you have no figures.
\listoftables % remove this if you have no tables.

\chapter{Acknowledgments} % this needs to be before \mainmatter.
I would like to thank my advisor, Doctor George Fleming, for introducing me to the fascinating physics of nonperturbative quantum field theory and lattice gauge theory, for the many years of guidance and academic advice, for acquainting me with the lattice gauge theory community and introducing me to all the right people, and for serving as a role model for how to navigate life inside of academia; 
my ``advisor away from home,'' Professor Richard Brower, for inspiring me with his enthusiasm and creativity, and for motivating me to be bold in my work and to question conventional wisdom;
Professor Thomas Appelquist for standing in as my formal advisor, for always exemplifying careful and detailed thinking, and for teaching me to communicate my ideas clearly and with conviction;
Doctor Evan Weinberg, for showing me the ropes in lattice gauge theory, for helping me greatly as I struggled through some of my first projects as a graduate student, and in general for playing the role of the wise older graduate student (even though we are nearly the same age);
James Ingoldby, for being a great friend during my entire time in graduate school, for the years of bouncing ideas around and talking shop, and for more recently being a great collaborator and office mate;
Doctor Pavlos Vranas and Doctor Arjun Gambhir for hosting me at Lawrence Livermore National Laboratory in the Fall of 2017;
Professor Andre Walker-Loud for hosting me at Lawrence Berkeley National Laboratory in the Fall of 2017; 
the Fermilab theory group for hosting me during the Summers of 2014, 2015, and 2016;
my collaborators in the Lattice Strong Dynamics (LSD) collaboration and the Quantum Finite Elements (QFE) collaboration;
my friends in the Yale GSAS and working in particle physics around the work, for making my time as a graduate student such an enjoyable and enriching experience, and for continuing to inspire me with the passion and conviction that they bring to their work.
You have all helped me to become the scientist that I am today, and this dissertation would not have been possible without you.

Finally, I would not be the man that I am today without my family.  You have always been the biggest supporters of my work and my dreams, from grade school to graduate school.  Thank you for keeping me grounded through the doubts and confusions of graduate student life.  A physicist couldn't ask for better parents.  This work is dedicated to you, Lisa and David Gasbarro.

%A lot of people are awesome. Probably your family, friends, 
%advisor, and that one super special high school teacher who
%believed in you.

\clearpage
\thispagestyle{empty}
\null\vfill
\settowidth\longest{\huge\itshape Make it something that it's not and measure it.}
{
\centering
\parbox{\longest}{%
  \raggedright{\huge\itshape%
   Make it something that it's not and measure it. \par\bigskip
  }   
  \raggedleft\Large\MakeUppercase{Prestia Family Motto}\par%
}

\vfill\vfill
}
\clearpage

% Starts proper Arabic numbering of pages and chapters.
\mainmatter

% Add additional \chapter{}s as necessary.

% use \cite{} to cite a reference in your bibliography file.
% use \ref{} to reference a \label{} from an equation, figure, or table.

% for sets of equations use align or gather:
%\begin{align}
%\end{align}

% for long equations, use multline.

% for figures:
%\begin{figure}[ht]
%\centering
%\includegraphics[width=.45\textwidth]{name_of_figure.eps}
%\caption{A caption! \label{a_figure}}
%\end{figure}

% for tables:
%\begin{table}
%\begin{tabular}{c|c|c}
% 1 & 2 & 3 \\
%\hline
%\end{tabular}
%\caption{Another caption! \label{a_table}}
%\end{table}

\chapter{Background and Motivation \label{chapter:background}} 
The standard model has been confirmed in nearly all facets by high energy and precision experiments over the last several decades.  
It consists of two seemingly disparate sectors: the strong nuclear sector and the electroweak sector.  
The discovery of a light Higgs boson at the LHC in 2012 \cite{Aad:2012tfa,Chatrchyan:2012xdj} raises questions about the completeness of the standard model electroweak sector, which as it stands appears fine tuned.
The characteristic scale of the electroweak sector, $\Lambda_{EW} \approx 246\,\text{GeV}$, enters by tuning the quadratic term in the Higgs potential, which is the only relevant operator in the standard model.
On the other hand, the strong nuclear sector is asymptotically complete in the UV, and the characteristic scale of the strong interactions, $\Lambda_{QCD} \approx 300 \,\text{MeV}$, arises naturally in the infrared from the strong dynamics.  
%The flavor hierarchy of the quark and lepton masses, though unexplained, is still technically natural.  

This work is motivated by the possibility that the electroweak sector may also reveal itself to be an asymptotically free gauge theory whose low energy scales are born out of the dynamics of underlying strong dynamics in a natural way.  
This premise is not new.  
In 1979, Weinberg \cite{Weinberg:1975gm} and Susskind \cite{Susskind:1978ms} introduced the idea that new Yang-Mills gauge dynamics may be responsible for driving electroweak symmetry breaking, now known as \emph{technicolor}.  
The technicolor gauge dynamics were originally assumed to be QCD-like, but precision electroweak experiments seem to disfavor QCD-like gauge dynamics for dynamical electroweak symmetry breaking (DEWSB).  
Furthermore, because QCD does not produce a light scalar in its spectrum, these theories do not yield a viable light Higgs candidate.  
However, recent lattice studies of Yang Mills gauge theories with different quark contents have shown novel infrared behavior around the \emph{conformal window}.  
The novel gauge dynamics may be capable of ameliorating many of the problems of QCD-like technicolor.  
DEWSB with nearly conformal gauge dynamics stands as a serious candidate to address the Higgs hierarchy problem and to guide experimental searches for beyond standard model physics.  

\section{Lessons from QCD}
To set the storyline for how the electroweak sector may be UV completed by new strong dynamics, let us briefly review the history of the development of the strong nuclear sector. 
%In the technicolor framework, many aspects of this history are expected to repeat themselves in the electroweak sector.  
Attempts were made to model the strong interaction between protons and neutrons as early as 1935.  
At the time a major goal was to understand how the protons and neutrons were bound in the atomic nucleus. 
Yukawa \cite{Yukawa:1935xg} put forward a theory of an $\SU(2)$ doublet of nucleons $\psi$ that interact via the Yukawa interaction, $\vec{\pi} \cdot \bar{\psi} \gamma_5 \vec{\sigma}\psi$, with a force mediated by an $\SU(2)$ triplet of $\pi$ mesons.  
The Yukawa theory consists of fundamental scalars, and when viewed as an effective field theory it faces similar fine tuning problems to the Higgs sector in the Standard Model.  
%A log divergence in the pion four-point amplitude leads to the generation of a quartic self interaction.  
%This interaction will generate additive corrections to the pion mass, which are power law divergent just as in the renormalization of the standard model Higgs mass.  
%Since the physical pion masses are an order of magnitude smaller than the nucleon masses and the cutoff must be larger than the nucleon masses, this theory has a hierarchy problem.  

Before notions of effective field theory and fine tuning were considered, other mesons and baryons were discovered which suggested a more fundamental description of the strong nuclear sector than Yukawa's theory.  
The hadrons in the ``particle zoo'' were categorized using group theoretic methods by Gell-Mann which led to the notion of quarks and quark flavor in the eightfold way \cite{GellMann:1962xb}.  
The Pauli exclusion principle for quarks inside the hadrons suggested that there should be some additional quantum number besides spin and isospin.  
Han, Nambu, and Greenberg \cite{Greenberg:1964pe,Han:1965pf} posited that quarks possess an additional $\SU(3)$ gauge degree of freedom -- the last major missing ingredient of what is now known as QCD.  
Thusly, the low energy theory written down by Yukawa was replaced by a more fundamental description without any fine tuning problems whose low energy scales emerge from strong dynamics.  
Gross, Wilczek, and Politzer \cite{Gross:1973id,Politzer:1973fx} demonstrated that QCD was asymptotically free.  
This allowed high energy scattering to be studied in perturbation theory, which provided many of the initial confirmations of QCD.  
Furthermore, asymptotic freedom allows one to remove the cutoff from QCD using perturbation theory such that the theory is ultraviolet complete, or valid up to arbitrarily high energy scales. 

In perturbative QCD, one can also show by computing the two loop beta function that the theory becomes strongly coupled in the infrared at a scale, $\Lambda_{QCD}$, the perturbative confinement scale.  
This low energy scale is generated in a natural way and is not due to any dynamics beyond QCD itself.  
Wilson provided evidence of quark confinement using the nonperturbative lattice regulator and the strong coupling expansion \cite{Wilson:1974sk}, and lattice Monte Carlo computations pioneered by Creutz \cite{Creutz:1979kf,Creutz:1979zg,Creutz:1980zw} have explored the QCD spectrum in great detail.   

\section{Technicolor and Its Shortcomings}
%
%The electroweak sector may eventually be demystified in much the same way as the strong sector.  
The standard model electroweak sector is analogous to Yukawa's description of the strong nuclear sector in 1935.
Assuming that beyond the standard model physics exists, the standard model electroweak sector is a fine tuned low energy effective description of yet undiscovered high energy dynamics.  
Seeking to mimic the success of QCD, Weinberg and Susskind \cite{Weinberg:1975gm,Susskind:1978ms} suggested that the standard model electroweak sector may be UV completed by a new strongly interacting gauge sector built on an asymptotically free Yang-Mills theory.  
%Though the standard electroweak theory is in close agreement with all present experiments, it is possible that the true low energy effective theory of the electroweak sector has small augmentations from the standard model values which have so far been undetectable by experiments.  
%In particular, it is extremely difficult to probe the coupling in the Higgs potential independently at the LHC.
%There is not enough experimental evidence at this time to conclude that the linear sigma model is the unequivocal description of the Higgs potential.
%For this reason, we consider the standard electroweak theory more analogous to Yukawa's theory of the strong sector than of chiral perturbation theory.  
%I will revisit this point in more detail when I discuss possible EFT descriptions of walking technicolor.  
%

In technicolor, a new set of quarks -- the techniquarks -- have an $\SU_L(N_f)\times \SU_R(N_f)$ flavor symmetry which is broken down to $\SU_V(N_F)$ by the chiral condensate $\langle \bar{q}q\rangle$.  
One or more $\SU_{I,L}(2) \times \SU_{I,R}(2)$ ``isospin'' subgroups of the techniquark flavor group are gauged under electroweak symmetry, $\SU_L(2)\times \U(1)$, which also breaks when the techniquark condensate is formed.  
An isotriplet of massless $0^{-}$ states forms when the isospin symmetry is spontaneously broken, which are the Nambu-Goldstone bosons of the spontaneously broken $\SU_{I,L}(2) \times \SU_{I,R}(2) \rightarrow \SU_{I,V}(2)$ flavor subgroup. 
These are the states which are responsible for giving mass to the weak gauge bosons in the Higgs mechanism.

Let us review some of the issues that arise when trying to construct a model of dynamical electroweak symmetry breaking in the standard technicolor picture.  
In the section that follows, we will explain how some of these issues may be mitigated by nearly conformal gauge dynamics which motivates further study of theories near the conformal window.  
The first is the problem of how a light Higgs boson ($M_{\text{Higgs}} / \Lambda_{EW} \approx 1/2$) arises from a strong technicolor gauge sector.  
During most of the development of technicolor theories the Higgs mass was not known, and so this is a more recent phenomenological consideration for technicolor-like models.
In one scenario, the Higgs arises as the lightest composite state with $0^{++}$ quantum numbers. 
In QCD, the lightest $0^{++}$ state, known as the $f_0(500)$ or simply the $\sigma$, has a mass of roughly 500MeV, whereas the pion decay constant -- which can be used as a characteristic scale for the chiral symmetry breaking -- is roughly $F_\pi \approx 90 \text{MeV}$.  
The characteristic scale of the electroweak symmetry breaking is 246 GeV; therefore a simple rescaling of QCD will not produce a light enough Higgs candidate.
%In many cases, the flavor group is rather large and many more light PNGB's will form.  
%Lifting the other PNGBs is one of the model building freedoms and challenges of technicolor models.  
In a different composite Higgs scenario, the Higgs is taken to be a PNGB of the spontaneously broken chiral symmetry in order to explain its light mass \cite{Dugan:1984hq}.
We will not discuss the composite PNGB Higgs scenario in this work, but we remark that it is an active area of investigation by members of the lattice BSM community (c.f. \cite{Brower:2015owo}).
In Chapter~\ref{chapter:Lattice}, we will discuss the appearance of light $0^{++}$ states in lattice studies of nearly conformal gauge theories and their possible realization as a light Higgs candidate.
%; in others, the lightness of the Higgs arises from a spontaneous breaking of conformal symmetry.  
%This work focuses on the latter scenario.  
%The theoretical framework behind spontaneous breaking of conformal symmetry is not completely understood, and one purpose of this work is to shed light on this mechanism.  

A second issue arises when one attempts to incorporate fermion mass generation into a technicolor scenario.  
While we will not address fermion mass generation in this work, in principle this must be accommodated for in any UV completion of the electroweak sector. 
The most common way to incorporate standard model fermion mass generation into the technicolor framework is to imagine that standard model fermions and techniquarks are both charged under a larger gauge group known as the extended technicolor group, $G_{ETC} \supset G_{TC}$.  
At some large scale, $\Lambda_{ETC}$, extended technicolor breaks down to $G_{TC}$. 
The standard model fermions are singlets of the remaining technicolor subgroup, and four Fermi operators are generated in the broken theory,
\begin{equation}
\frac{\bar{\psi}\psi \bar{\Psi}\Psi}{\Lambda_{ETC}^2},\, \frac{\bar{\Psi}\Psi\bar{\Psi}\Psi}{\Lambda_{ETC}^2},\, \frac{\bar{\psi}\psi\bar{\psi}\psi}{\Lambda_{ETC}^2}
\end{equation}
where $\Psi$ denote techniquarks and $\psi$ denote standard model fermions.  
These operators encode the effect of the coupling of standard model fermions to techniquarks by ETC gauge boson exchange.  
When the techniquarks condense, the first operator above gives mass to the standard model fermions, and the second operator can give mass to the technipions.  
However, the third operator leads to flavor changing neutral currents.  
Experimental measurements of kaon mixing as well as other rare processes require $\Lambda_{ETC} \gtrsim 1000\text{TeV}$ \cite{Lane:1993wz} for simple extended technicolor models, though this picture can be more complicated in theories with multiple ETC scales \cite{Appelquist:2003hn,Appelquist:2004ai}.  
While increasing $\Lambda_{ETC}$ suppresses FCNC, it also suppresses the SM fermion mass terms to such a degree that if the techniquark condensate is similar in magnitude to the condensate in QCD, $\langle \bar{\Psi}\Psi\rangle / \nu^3 \approx 25$, where $\nu = 2^{-1/4} G_F^{-1/2} = 246 \text{GeV}$, the strange quark mass would be much too light, $m_s \approx 0.4 \text{MeV}$ \cite{Lane:1993wz,Fleming:2008gy}.  
However, if non-QCD-like dynamics lead to a significant increase in the size of the chiral condensate, then physical quark masses may be achievable even with a large extended technicolor scale.  
We will see that nearly conformal dynamics may produce exactly this effect.

In early models of technicolor, it was assumed by analogy with QCD that all techniquarks were paired to form electroweak doublets.  
Peskin and Takeuchi devised a set of parameters, S T and U, which quantify the vacuum polarization corrections to four fermion scattering processes compared to the standard model prediction \cite{Peskin:1991sw}.  
They are referred to as ``oblique corrections'' because they only affect the mixing and propagation of gauge bosons and do not depend on the fermions in the initial and final states.
The S parameter is proportional to the derivative of the left-right current correlator and is related to the number of chirally gauged fermion species.  
Using the identity $J^{\mu}_{L,R} = J^\mu_V \pm J^\mu_A$, it can be written in terms of the vector - vector and axial vector - axial vector current correlators as \cite{Peskin:1991sw}
\begin{equation}
S = 4 \pi \frac{N_f}{2} \frac{d}{d q^2} \left.\left( \Pi_{VV}(q^2) - \Pi_{AA}(q^2) \right)\right|_{q^2 = 0} \label{eq:Sparam}
\end{equation}
For naive technicolor with QCD-like dynamics and all techniquarks carrying electroweak charge, the S parameter in technicolor is estimated to have the lower bound
\begin{equation}
S \gtrsim \frac{1}{6\pi}\left[\frac{N_c N_f}{2} \right]
\end{equation}
where $N_f$ is the number of techniquark flavors charged under electroweak and $\SU(N_c)$ is the technicolor gauge group.  
This coarse estimate is in significant tension with phenomenological constraints \cite{ALEPH:2005ab} for the value of S, which seems to rule out simple technicolor models with many electroweak doublets and QCD like dynamics for $N_{TC},N_{TF} \gtrsim 4$.  
Technicolor models with $N_{TC},N_{TF} \lesssim 4$ may still be consistent, albeit somewhat disfavored by electroweak precision experiments \cite{Shintani:2008qe,Fleming:2008gy}. 
These estimates rely on two basic assumptions about the technicolor sector.  
First, it is not necessary for all technicolor flavors to form electroweak doublets.  
If only one pair of flavors is given electroweak charges (minimal technicolor), electroweak gauge interactions will break the TC flavor symmetry $\SU_L(N_f)\times \SU_R(N_f) \rightarrow \SU_{I,L}(2)\times \SU_{I_R}(2) + \SU_{F,L}(N_f-2) \times \SU_{F,R}(N_f-2)$, but this does not cause any problems.  
It is usually imagined that $N_f-2$ of the flavors will be lifted by explicit mass terms or by four fermi interactions arising from ETC in order to get rid of the extra PNGBs.  
Therefore, there need not be an excessive number of BSM electroweak doublets contributing to S.  
Even small numbers of doublets may be in tension with precision measurements, but this tension is based on the assumption of QCD-like dynamics and the validity of chiral perturbation theory.  
For non-QCD-like gauge dynamics, this estimate may break down. 

\section{Nearly Conformal Gauge Dynamics}
Next we discuss how nearly conformal dynamics may decrease the S parameter, increase the size of the chiral condensate, and reduce the Higgs mass, thus addressing many of the issues of QCD-like technicolor.  
First, we will briefly review how IR fixed points may form in the running of the Yang-Mills coupling.  
Such theories are said to be inside the \emph{conformal window}.  
We will then explain how the dynamics of theories inside and near the conformal window may mitigate many of the issues in standard technicolor.

\subsection{Infrared Fixed Points in Yang-Mills and the Conformal Window}
The perturbative two loop beta function for the gauge coupling in $\SU(N_c)$ gauge theory with $N_f$ flavors of Dirac fermions in the fundamental representation of the gauge group takes the form
\begin{equation}
\mu \frac{\partial}{\partial \mu} g(\mu) = \beta(g) = -b_0 g^3 + b_1 g^5 + \mathcal{O}(g^7)
\end{equation}
where the first two coefficients in the expansion are known to be independent of renormalization scheme.
\begin{equation}
b_0 = \frac{1}{(4\pi)^2} \left(\frac{11}{3}N_c - \frac{2}{3}N_f\right), \: b_1 = \frac{1}{(4\pi)^4}\left(\frac{34}{3} N_c^2 - \left(\frac{13}{3} N_c - \frac{1}{N_c} \right)N_f\right)
\end{equation}
For asymptotic freedom, one must require $b_0 > 0 \rightarrow N_f < 11/2 N_c$.  
For $N_f$ just below this bound, fixed point solutions exist for which the fixed point coupling is perturbative and therefore the perturbative analysis is self consistent. 
These fixed points are known as Caswell-Banks-Zaks fixed points \cite{Caswell:1974gg,Banks:1981nn}. 
The asymptotic freedom boundary $N_f = 11/2 N_c$ constitutes the top of the conformal window. 

As $N_f$ is decreased further, the fixed point coupling value increases monotonically until eventually perturbative control is lost.  
It is assumed that the fixed point continues to exist for $N_f$ below the perturbative region.  
The fixed point would then be strongly coupled.  
At sufficiently small $N_f = N_f^c$, the coupling runs to a strong enough value to confine before reaching any fixed point; the fixed point disappears and the theory is chirally broken in the IR.  
This constitutes the bottom of the conformal window. 
For $N_f$ just below $N_f^c$, the theory confines in the infrared, but there is a large range of scales over which the running coupling evolves slowly (or \emph{walks}).  
A confining gauge theory in this region of parameter space is referred to by various equivalent terms: slowly running, walking, or nearly conformal.
This slowly running coupling can significantly alter the low energy physics as we will discuss in Section~\ref{sec:DEWSBwalk}.  
An approximate phase diagram for asymptotically free gauge theories with $N_c$ colors and $N_f$ flavors in various representations of the gauge group was presented by Dietrich and Sannino \cite{Dietrich:2006cm} in which the lower boundary of the conformal window is computed by estimating the onset of spontaneous chiral symmetry breaking using the ladder approximation \cite{Appelquist:1988yc,Cohen:1988sq}.
We will discuss recent studies of conformal window gauge theories on the lattice in Section~\ref{sec:latticewindow} and in Chapter~\ref{chapter:Lattice}.

A more physical picture of the conformal window comes from considering the competing screening and antiscreening effects of quarks and gluons in the vacuum of Yang-Mills theory.  
In pure Yang-Mills, antiscreening by gluons alone pushes the coupling very rapidly into the confined phase.  
In QCD (only considering the light flavors), $N_f=2$ and the story is not changed much by the small amount of screening by the quarks; the theory still rapidly confines.  
As $N_f$ increases, the screening effect of the many quark flavors becomes more and more prominent until the effect of the quarks and gluons balance.  
This results in infrared conformality.  
%One can estimate perturbatively that this should occur around $N_F = 3 N_C$, so it is no surprise that $N_F = 8, N_C = 3$ is an interesting theory to study nonperturbatively.

\subsection{DEWSB with Nearly Conformal Gauge Dynamics \label{sec:DEWSBwalk}}
With an understanding of the conformal window established, let us now consider how conformal or near conformal behavior in the gauge theory may provide for a better mechanism on which to build a model of DEWSB.  
As we have discussed, one shortcoming of QCD-like dynamics is that the generation of fermion masses by extended technicolor will also lead to FCNC which exceed experimental bounds from measurements of rare mixing processes.  
In 1986, Appelquist et. al. \cite{Appelquist:1986an} studied the fermion masses arising from extended technicolor with a slowly running gauge coupling.  
They estimated the fermion self energy function $\Sigma(p)$ by approximately solving the perturbative gap equation.  
The SM fermion masses are given by the equation
\begin{equation}
m_f =  \frac{g_{ETC}^2}{\Lambda_{ETC}^2}\langle 0 | \bar{\Psi}\Psi | 0 \rangle_{ETC} \approx \frac{g_{ETC}^2}{4\pi^2}\frac{N_{ETC,F}}{\Lambda_{ETC}^2} \int^{\Lambda_{ETC}} p dp \Sigma(p)
\end{equation}
$\Sigma(p=0)$ will typically take some nonzero value on the order of the confinement scale.  
For large $p$, $\Sigma(p)$ eventually damps very rapidly, but there could be a substantial range over which $\Sigma(p)$ falls slowly before rapidly damping.  
It was found that a slowly running coupling increases the range over which the fermion self energy is slowly falling, and thus enhances the techniquark condensate and the resultant SM fermion masses.  
Numerical estimates found that for a cutoff large enough to suppress FCNCs, first and second generation quark masses were obtainable from ETC.  

We have discussed that even a minimal number of techniquarks charged under electroweak (a single pair of techniquarks forming one doublet), may still be in tension with experimental bounds on the S parameter.  
%S is related to the derivative of the left-right current correlator in momentum space \cite{Fleming:2008gy}.
%\begin{equation}
%\Pi_{LR}^{\mu\nu}(q) = ig^{\mu\nu}\Pi_{LR}(q^2) + (q^{\mu}q^{\nu} \text{terms}) =  \int d^4 x e^{-iqx} \langle J^{\mu}_L(x) J^{\nu}_R(0) \rangle
%\end{equation}
%\begin{equation}
%\Pi'_{LR}(0) = \frac{d}{dq^2}\Pi_{LR}(q^2)|_{q^2 = 0}
%\end{equation}
The standard model contribution is always subtracted off in the definition of the S parameter so that nonzero values of S only come from BSM particles in the current-current correlator loops.  
Say we have an $N_f$ flavor technicolor theory with 2 flavors gauged in a doublet under electroweak.  
Clearly the three ``pions''  (NGBs of $SU_{L}(2) \times SU_{R}(2) \rightarrow SU_{V}(2)$ subgroup) and the one ``$\sigma$'' that form after confinement and chiral symmetry breaking will contribute to current-current correlators at loop level, but these are exactly the particles that will play the role of the electroweak gauge boson longitudinal components and the Higgs boson.  
They do not contribute BSM signals to the S parameter.  
On the other hand, it is also possible for the techniquarks charged under electroweak to form mesons with the electroweak neutral techniquarks.  
These would be kaon-like mesons in the sense that they mix techniquark generations.  
In addition, higher spin mesons such as the techni-rho will also form.  These mesons will also contribute to the current-current correlator loops giving nonzero contributions to S.

In nearly conformal theories, large contributions to the S parameter may be avoidable.
As we have shown in Eq.~\ref{eq:Sparam}, the S parameter is proportional to the difference of the vector and axial vector current correlators.
%By virtue of the identity $J^\mu_{L,R} = J^\mu_V \pm J^\mu_A$,
%\begin{equation}
%\Pi'_{LR}(0)  = \Pi'_{VV}(0) - \Pi'_{AA}(0)
%\end{equation}
The S parameter will remain small if there is a cancellation between the vector-vector and axial-axial current correlators arising from a degeneracy or near degeneracy of even and odd parity mesons.  
%That is, if there is a degeneracy between even and odd parity mesons running around the loops, the S parameter will remain small.  
In QCD, the splitting between the vector meson ($\rho$) and the axial vector meson ($a_1$) is a result of chiral symmetry breaking.  
In a chirally symmetric theory, the $\rho$ and the $a_1$ should be degenerate.  
It is plausible that the splitting between even and odd parity partner states will be smaller in a nearly conformal theory and as such the S parameter may be reduced.
An argument for a reduced vector - axial vector mass splitting in a theory with a slowly running coupling based on dispersion relations was given in \cite{Appelquist:1998xf}.
An estimation of the S parameter on the lattice with $N_F=2$ and $N_F = 6$ quarks in the fundamental representation of $\SU(3)$ was performed by the LSD collaboration using domain wall fermions \cite{Appelquist:2010xv}.  
It was found that as $N_F$ was increased, the spectrum becomes more parity doubled and the S parameter per electroweak doublet decreases.
%If in the IR, rather than going to a chirally broken phase, the theory runs to a conformal phase which is chirally symmetric then one would expect an exact degeneracy between parity partners -- so called ``parity doubling.''  
%In this scenario, the BSM loop contributions to the S parameter would all cancel and the S parameter would be exactly zero.  
%One could go as far as to think of the S parameter as an order parameter for the phase transition between the conformal and chirally broken IR phases as one tunes $N_{TF}$. 
%For theories near but outside the conformal window, one expects the $\rho$ and $a_1$ to be very close in mass in the zero quark mass limit and thus that the S parameter would be small but nonzero.  
%All of this is only heuristic.  

Finally we consider the obstacle of producing a light composite Higgs boson.
At the time of the development of early technicolor, this was not an issue because the Higgs had not been discovered.  
In fact, many theories conjectured that the Higgs was very heavy and could be integrated out of the low energy effective theory.  
We now know that a light Higgs boson exists whose mass is about half the characteristic scale of the electroweak symmetry breaking ($M_{\text{Higgs}} / \Lambda_{EW} = 125/246 \approx 1/2$).
% possible for such a particle to be born out of a technicolor model?  
One idea is that the lightness of the $0^{++}$ resonance in the technicolor sector could be born out of the approximate scale invariance of a nearly conformal theory \cite{Yamawaki:1985zg,Bando:1986bg}.  
Such a particle is referred to as a techni-dilaton.  
The word dilaton signifies that the particle is the Nambu-Goldstone boson of spontaneously broken dilatation symmetry.  
In the case of nearly conformal theories, there is a small explicit breaking of the scale symmetry by the slow running of the gauge coupling. 

%This story is somewhat complicated by the conformal anomaly inherent in all massless Yang-Mills theories. 
In a typical scenario of spontaneous symmetry breaking, the classical potential of the field has a degenerate ground state; a particular vacuum is chosen which breaks the symmetry of the classical action and the fields are expanded about this vacuum and quantized.  
In Yang-Mills theories with massless fermions, the theory is classically scale invariant, but it is not the classical dilatation symmetry which is spontaneously (or explicitly) broken.  
When the theory is quantized, the scale symmetry is broken by the running of the gauge coupling.  
This is the conformal or trace anomaly.  
The theory develops an effective potential at the quantum level which explicitly breaks the scale symmetry.  
But scale invariance may reappear in the effective potential at a particular value of the gauge coupling if the coupling runs to an infrared fixed point.
%This is the fixed point at which the the conformal symmetry is recovered at the quantum level.  
This picture is similar to the Coleman-Weinberg mechanism in which spontaneous symmetry breaking arises from the effective potential in scalar QED \cite{Coleman:1973jx}.  
In a nearly conformal gauge theory, the effective potential only has an approximate scale invariance in a range of gauge couplings.
%In this region, one can imagine the analog of a Mexican hat potential with a slight tilt that explicitly breaks the symmetry.  
%This is a schematic picture of how a light scalar may emerge in a nearly conformal gauge theory.

In Section~\ref{sec:latticewindow} and Chapter~\ref{chapter:Lattice} we will discuss the appearance of light $0^{++}$ scalar states in recent lattice studies of conformal and nearly conformal gauge theories.  
The dilaton idea is one possible explanation of the appearance of a new light state in these spectra, but some other mechanism may be responsible (e.g. \cite{Eroncel:2018dkg}).
The origin of these light scalars is an area of active investigation.

\section{Lattice Studies of Conformal Window Gauge Theories \label{sec:latticewindow}}

The challenge of classifying gauge theories near and inside the conformal window and of characterizing the low energy physics of these theories has been taken up by lattice theorists over the past decade.
Some studies have been aimed specifically at assessing the walking technicolor scenario.
In one study, the Lattice Strong Dynamics (LSD) collaboration examined the phenomenon of condensate enhancement near the conformal window.  They reported an enhancement in the chiral condensate in $\SU(3)$ gauge theory with six flavors of fermions in the fundamental representation compared to two flavors in the fundamental representation \cite{Appelquist:2009ka}.
For the same two theories, the LSD collaboration also studied the S-parameter and the phenomenon of parity doubling, reporting that the six flavor theory has a smaller S-parameter per electroweak doublet and is more parity doubled compared to the two flavor theory \cite{Appelquist:2010xv}.

More recently, the behavior of the flavor-singlet scalar (composite Higgs boson candidate) near the conformal window has been studied by several collaborations in a variety of theories.
The latKMI collaboration first reported a low mass scalar in $\SU(3)$ gauge theory with eight flavors of fermions in the fundamental representation \cite{Aoki:2014oha}.
The eight flavor theory has been subsequently studied by the LSD and latKMI collaborations in greater detail \cite{Appelquist:2016viq,Aoki:2016wnc,Appelquist:2018prep} confirming the existence of this light state.
Light scalar states have also been reported in $\SU(3)$ gauge theory with two flavors in the symmetric (sextet) representation of the gauge group \cite{Fodor:2015vwa,Fodor:2016pls}, $\SU(3)$ gauge theory with four light flavors and eight heavy flavors in the fundamental representation of the gauge group \cite{Brower:2015owo}, $\SU(3)$ gauge theory with twelve degenerate flavors in the fundamental representation of the gauge group \cite{Aoki:2013zsa}, and $\SU(2)$ gauge theory with two flavors in the adjoint representation of the gauge group \cite{DelDebbio:2010hx}.
We will discuss the phenomenon of light scalar states in more detail in Chapter~\ref{chapter:Lattice}.

%Outside of the walking technicolor scenario, asymptotically free gauge theories near and inside the conformal window remain of great theoretical interest and have a variety of possible applications to beyond the standard model physics.
The characterization of the conformal window is a worthwhile theoretical exercise in its own right even outside the context of a particular phenomenological application.
One challenge is to identify the lower boundary of the conformal window at the critical number of flavors $N_f = N_f^c$.
In supersymmetric QCD, the lower boundary of the conformal window is known from Seiberg duality \cite{Seiberg:1994pq}, but in nonsupersymmetric theories the extent of the conformal window remains a difficult nonperturbative question.
On the lattice, this question can be investigated by simulating particular gauge theories and attempting to map out the phase diagram point-by-point in theory space by assessing whether each individual gauge theory exhibits infrared conformality or not.
Early studies were carried out on $\SU(2)$ gauge theory with two flavors in the symmetric representation of the gauge group \cite{Catterall:2007yx} which has been reported to be inside the conformal window \cite{Catterall:2008qk,Hietanen:2009az}.
The LSD collaboration performed early work on $\SU(3)$ gauge theory with fundamental fermions and reported that $N_f = 12$ is IR conformal -- and so $16 \geq N_f \geq 12$ are within the conformal window -- while $N_F=8$ was determined to be chirally broken \cite{Appelquist:2007hu}.
The existence of an infrared fixed point in twelve flavor $\SU(3)$ gauge theory has been investigated by many groups and continues to be debated \cite{Hasenfratz:2016dou,Fodor:2016zil,Lin:2015zpa}.
Another widely studied theory is $\SU(3)$ gauge theory with two flavors of fermions in the symmetric (sextet) representation \cite{Fodor:2016pls,Fodor:2015vwa} which is another model in which the existence of an infrared conformal fixed point has been debated \cite{Hasenfratz:2015ssa,Fodor:2015zna}.
A comprehensive review from 2012 by Neil details the wide range of gauge theories that had been investigated up to that time and gives one a sense of the extent of the lattice BSM effort and the broad range of theories considered \cite{Neil:2012cb}.
More recent reviews which cover the issue of conformality in the twelve flavor theory and the light scalar in the eight flavor theory amongst other things are found in Refs.~\cite{Nogradi:2016qek,Svetitsky:2017xqk}.

The characterization of conformal and nearly conformal gauge theories also has implications for model building scenarios besides the standard walking technicolor picture.
One example is the composite pseudo-Nambu-Goldstone-Higgs (two Higgs doublet) scenario \cite{Ma:2015gra} which corresponds to a gauge theory with four light flavors such as the theory studied in Ref.~\cite{Brower:2015owo}.
Another example is the mechanism of partial compositeness \cite{Kaplan:1991dc} in which standard model fermion masses arise by the linear mixing of standard model fermions with (possibly composite) heavy fermions.
A UV completion of a phenomenologically viable model of partial compositeness requires a large anomalous dimension for the baryon operator, which may arise near the lower boundary of the conformal window.
Other nonperturbative mechanisms that are not necessarily tied to the conformal window have also been studied on the lattice.
The partial compositeness scenario has been investigated recently for UV completions with fermions in two distinct representations of the gauge group \cite{Ayyar:2017qdf,Ayyar:2018zuk}.
Phenomenologically viable models of composite dark matter have been proposed in recent years and studied on the lattice \cite{Appelquist:2015zfa,Appelquist:2015yfa}.
In summary, the characterization of the conformal window and other novel gauge dynamics which arise at strong coupling is a rich field for future phenomenology, and the lattice is a powerful tool for making progress in this area.

\section{Organization of this Work}
In Chapter~\ref{chapter:Lattice}, we will review aspects of lattice studies of nearly conformal gauge theories.
We will focus on the appearance of light scalar states in the spectra and the assessment of these states as possible light Higgs boson candidates.
As a particular example, we will review the study of $N_f =8$ QCD by the Lattice Strong Dynamics collaboration.
We will discuss the challenges of studying nearly conformal theories using lattice methods that are not present in traditional lattice QCD calculations, such as the approach to the chiral limit and the interpretation of different scale setting schemes.
From the collected evidence of lattice studies of many different gauge theories near and inside the conformal window, we will argue that the appearance of light scalar states in such theories may be a generic phenomenon.
The common features of the low energy physics suggest that it may be possible to develop a low energy EFT description for nearly conformal gauge theories which will help to unify the various lattice calculations and to guide future studies.
Because the scalar is similar in mass to the pions and well separated from the heavier states in the theory, the low energy EFT should contain the light scalar state as a dynamical degree of freedom along with the PNGBs.
As a step toward constraining the possible forms of this EFT, we will present a new lattice study of maximal isospin $\pi\pi$ scattering in $N_f=8$ QCD.

In Chapter~\ref{chapter:EFT}, we  discuss chiral effective theory starting from the current algebra efforts in the 1960s up through the modern picture of the chiral Lagrangian and more general effective field theory methods.  
After this review, we present a new effective field theory framework for describing nearly conformal gauge theory that is based on the linear sigma model.
The linear sigma EFT framework incorporates scalar states along with the PNGBs, one of which is the $0^{++}$ or $\sigma$.
We will focus on the role of chiral breaking terms in the EFT and the possibly large quark mass effects.
This will lead us to consider a more general power counting for the spurion field than is typically used in chiral perturbation theory.
Chapters~\ref{chapter:Lattice} and~\ref{chapter:EFT} together detail an effort to combine traditional lattice methods and effective field theory methods to develop a generic unified picture of the low energy dynamics of nearly conformal Yang Mills gauge theories.

In the second half of this manuscript, we discuss a separate but closely related effort to reformulate lattice gauge theory on curved manifolds.
The original motivation for this effort was to develop a lattice formulation of \emph{radial quantization} for studying conformal field theories, and a key future goal of this effort is a characterization of the conformal window of four dimensional gauge theories.
Because radial quantization is naturally formulated on $\mathbb{R}\times \mathbb{S}^{d-1}$ which is a curved geometry, the development of the methodology naturally led us to a more general development for lattice field theory on arbitrary smooth Riemann manifolds.

In Chapter~\ref{chapter:QFE1}, we explain the general method for formulating lattice quantum field theory on an arbitrary curved Riemann manifold.
In Chapter~\ref{chapter:QFE2}, we present explicit lattice calculations for scalar field theory on the manifolds $\mathbb{S}^2$ and $\mathbb{R}\times \mathbb{S}^{2}$.  
The former is equivalent to the minimal 2D Ising CFT at criticality.
We confirm this by explicitly comparing the lattice calculation on the curved manifold to the exact solution, which is a first confirmation of the viability of the method.
The latter should be equivalent to the 3D Ising CFT at criticality studied in radial quantization.
We present early results that the method seems to be converging to the critical point.

\chapter{Lattice Results for a Nearly Conformal Gauge Theory \label{chapter:Lattice}}

%\section{Key Concepts in Lattice Gauge Theory}
%\subsection{Euclidean Correlators and Spectroscopy}
%\subsection{Staggered Fermions}

We have discussed in Chapter~\ref{chapter:background} that a realistic scenario of dynamical electroweak symmetry breaking by a new strong force requires confining gauge dynamics which differ significantly from QCD in certain respects. 
In particular, a reduced electroweak S-parameter, an enhanced chiral condensate, and a light flavor-singlet scalar meson are favorable features of the gauge dynamics.
We have discussed how these features may arise near the lower boundary of the conformal window.  
While general approximate results may be computed analytically using the ladder approximation and other such techniques, the lattice is the best tool for studying details of nonperturbative QFTs.
In the past decade, lattice theorists have begun to explore gauge theories near and inside the conformal window and to assess their viability as foundations for models of dynamical electroweak symmetry breaking.

A disadvantage of the lattice approach is that one must choose a specific Lagrangian with fixed number of flavors $N_f$ and number of colors $N_c$ to study.
A particular quantity in a specific field theory may take months or years to calculate to high accuracy.
In this work, we attempt to mitigate this problem by advocating an approach that combines lattice calculations carried out for a particular Lagrangian with effective field theory analyses which should be generally applicable to any nearly conformal gauge theory.  
In this Chapter, we present numerical lattice calculations and discuss their implication for determining and constraining the correct EFT description of nearly conformal gauge theories.

\section{$N_f=8$ QCD on the Lattice \label{sec:nf8}}
For our lattice studies, we have chosen to investigate SU$(3)$ gauge theory with $N_f = 8$ flavors of Dirac fermions in the fundamental representation. 
The continuum Lagrangian in Minkowski metric reads
\begin{equation}
\mathcal{L} = -\frac{1}{4}F_{\mu\nu}^a F^{a \mu \nu} + \sum_{f=1}^8 \bar{\psi}_f \left ( i \slashed{D} - m_f \right) \psi_f
\end{equation}
%In previous sections, I have used capital $\Psi$ and lowercase $\psi$ to distinguish techniquarks and standard model quarks, but here we have only techniquarks which I now denote with lowercase $\psi$.  
In the lattice studies of the $N_f=8$ theory discussed here, we will always take the flavors to be mass degenerate, $m_f = m_q \; \forall f$. 
We are often chiefly interested in the chiral limit $m_q \rightarrow 0$ in which the theory will have exact Nambu-Goldstone bosons. 
%For thoughts on model building applications, I may pontificate about splitting the masses, but all lattice calculations will be done with degenerate fermions.  
In the context of a dynamical electroweak symmetry breaking scenario, we imagine that the $N_f=8$ theory or a similar nearly conformal gauge theory will serve as the new strong interaction to complete the electroweak sector of the standard model.
But, in the present work we consider the $N_f = 8$ theory in isolation.  
We are interested in studying the low energy behavior of the nearly conformal gauge sector without complications from standard model couplings.  
As in the case of QCD and chiral perturbation theory, we imagine that once the low energy effective theory of the gauge theory sector in isolation is understood, standard model effects can be added into the EFT as perturbations.  

One complication that would arise from including electroweak charges that couple to standard model fermions is top quark loops.  
If the top mass were to arise from an ETC scenario, one would have to include a four-fermi operator in the theory whose coefficient is large.   
It is an open question how the coupling to the top quark will affect the vacuum structure of the gauge sector.  
Other model building frameworks for generating fermion masses such as partial compositeness \cite{Kaplan:1991dc} may be more innocuous.  
We do not attempt to address such issues in this work.  
In general, these questions are difficult to address on the lattice because four-fermi operators usually introduce a sign problem into the action, which greatly impedes the ability to perform numerical calculations.  

We are interested in the nonperturbative regime of this theory, and observables will be computed using Monte Carlo techniques.  
The theory must first be regulated by moving from the continuum to the lattice.  
The pure gauge action may be formulated with exact gauge invariance on the lattice using Wilson's plaquette action \cite{Wilson:1974sk}.  
For the quark fields, there are several choices for lattice fermion formulations, each of which has its own costs and rewards.  
Kogut-Susskind staggered fermions retain an exact $U(1)_V \times U(1)_A$ lattice chiral symmetry, which is enough to derive Noether currents.  
One staggered flavor produces four flavors of continuum fermions -- a remnant of \emph{fermion doubling} --  but because the number of flavors in this theory is a multiple of four, this is advantageous (no rooting is required).  
For the lattice action there will be two staggered flavors, which also gives us an exact lattice $SU_L(2) \times SU_R(2)$ flavor symmetry.  
For the lattice fermion fields, we adopt here the standard notation for staggered fermions, $\chi(x)$.  
The lattice action is given by
\begin{equation}
S = a^4 \sum_x \sum_{f=1}^2 \bar{\chi}_f (x) \sum_\mu \frac{\eta_\mu(x)}{2a} \left( U_\mu(x) \chi_f(x+\mu) - U_\mu^\dag (x-\mu) \chi_f(x-\mu) \right) + S_\text{Gauge} \left [ U_\mu(x) \right ]
\end{equation}
where $a$ is the lattice spacing and $S_\text{Gauge}$ is Wilson's lattice action for the gauge fields, $U_\mu (x)$.  
$\eta_\mu(x) = (-1)^{x_1+x_2+x_3+x_4 - x_\mu}$ is the staggered phase which plays the role of the Dirac gamma matrix for staggered fermions.
In the staggered formulation, the fermionic fields have only one component; there is no Dirac index.  
The four spin and four flavor components are spread out across the sixteen corners of each hypercube.  
One combines the appropriate spin and flavor components into mesonic operators by including \emph{staggered phases} and by applying gauge invariant shifts.  
Further details on the formulation and uses of staggered fermions can be found in standard lattice field theory texts (c.f. \cite{Gattringer:2010zz}).
In practice, the theory is not simulated directly with the Grassmann-valued fermion fields.  
Because the action is quadratic in the fermion fields, one can integrate them out and arrive at an effective action for the gauge fields.  
Many improved gauge actions have been developed which are designed to reduce lattice discretization errors.  
The work that we will discuss has been carried out on ensembles generated using an nHYP smeared staggered action \cite{Hasenfratz:2001hp,Hasenfratz:2007rf}.

Much work has been carried out on the $N_f = 8$ theory already. 
A key point of investigation by the lattice community has been whether this theory is conformal or confining in the infrared. 
Recent studies of the running coupling found that the running was slow and that there was no evidence of an IR fixed point \cite{Hasenfratz:2014rna,Fodor:2015baa}. 
Others maintain that evidence points to conformal behavior in the IR \cite{Ishikawa:2015iwa,daSilva:2015vna,Noaki:2015xpx}.  
Nonetheless, the most common stance is that the theory is chirally broken \cite{Hasenfratz:2014rna,Fodor:2015baa,Aoki:2013xza,Appelquist:2014zsa,Appelquist:2016viq,DeGrand:2015zxa}.  
In this work, we will not focus on answering the question of whether the theory is conformal or confining.
Where necessary, such as in the EFT analysis of Section~\ref{sec:EFTFits}, we will assume that the theory is confining.
We remark that a study by the LSD collaboration also found evidence for a reduced S parameter \cite{Appelquist:2010xv,Appelquist:2014zsa}, which is one of the expected favorable features of a nearly conformal gauge theory.

%The LSD collaboration \cite{8fLSD1}\cite{8fLSD2} and the latKMI collaboration \cite{8fKMI1} \cite{8fKMI2} \cite{8fKMI3} have both computed parts of the low lying spectrum in detail.  
%There is evidence for the existence of a light $0^{++}$ state which could serve as a dilatonic Higgs, but the case is far from closed on this matter \cite{degrandBSM}.  
%The recent paper by the LSD collaboration \cite{8fLSD2} explores the spectrum on ensembles of the lightest quark masses computed to date for eight flavor QCD.  
%They find that the $0^{++}$ state is degenerate with the pions and well separated from the $\rho$.  
%A plot from \cite{8fLSD2} demonstrating this phenomenon is reproduced here in Fig.~\ref{fig:sigmaplot}.  
%If the theory is indeed chirally broken in the IR, one would expect the $\sigma$ to peel off from the pions and arrive at some nonzero mass in the chiral limit.  
%Much work needs to be done to understand the nature of this light resonance.  
%One of the major objectives of the present study is to shed light onto the mechanism through which the $\sigma$ becomes light and to understand how much like the standard model Higgs this particle behaves.
Next let us review the recent lattice study by the LSD collaboration of $N_f=8$ QCD at small bare quark masses with staggered fermions.
We emphasize that the present author, though a member of the LSD collaboration, was not a key contributor to these spectral analyses.
The spectral results have been presented in the references \cite{Appelquist:2016viq,Gasbarro:2017fmi}, and are reproduced here to set the stage for the $\pi\pi$ scattering study of the $N_f = 8$ QCD theory presented in Section~\ref{sec:scattering} and the EFT analysis of the $N_f = 8$ QCD theory presented in Section~\ref{sec:EFTFits}.
We will also replot the spectral data in various ways in order to emphasize the aspects of the data set that are most important for these discussions.

\begin{figure}[t]
	\includegraphics[width=0.75\textwidth]{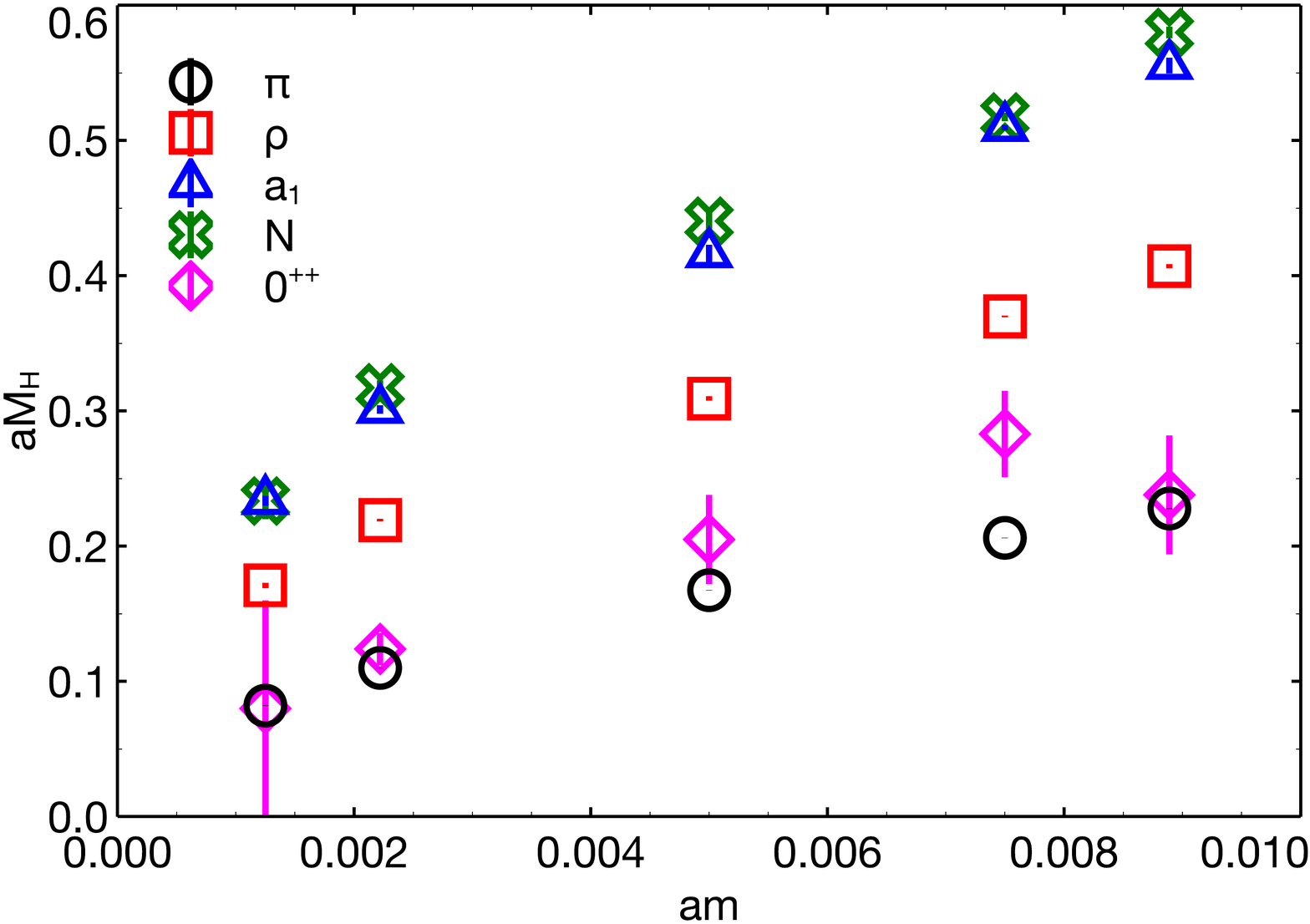}
	\centering
	\caption{The spectrum of eight flavor QCD computed by the LSD collaboration and presented in Refs.~\cite{Appelquist:2016viq,Gasbarro:2017fmi}.  The spectrum is computed using nHYP smeared staggered fermions.  Quantities are plotted in bare lattice units. \label{fig:spectrum1}}
\end{figure}
The spectrum of $N_f=8$ QCD computed by the LSD collaboration and presented in references \cite{Appelquist:2016viq,Gasbarro:2017fmi} is shown in Fig.~\ref{fig:spectrum1} plotted in bare lattice units.
For our purposes, the most interesting state in the spectrum is the lightest flavor-singlet scalar, also referred to as the $0^{++}$ or $\sigma$ state.
In QCD, the $\sigma$ is significantly heavier than the pions and unstable at a comparable distance from the chiral limit \cite{Briceno:2016mjc,Guo:2018zss}.
In Fig.~\ref{fig:spectrum1}, we see that the $\sigma$ state is similar in mass to the pions and well separated from the $\rho$ and other heavy resonances over a wide range of bare quark masses.
%To demonstrate this gap in the spectrum more clearly, we replot only the masses $M_\pi$, $M_\sigma$, and for comparison $M_\rho$ in bare lattice units on the left panel of Fig.~\ref{fig:latticeunits}.
The light $\sigma$ state in $N_f = 8$ QCD was first discovered by the latKMI collaboration \cite{Aoki:2014oha}.
Since then, the spectrum has been studied in more detail both by latKMI \cite{Aoki:2016wnc} and by the LSD collaboration \cite{Appelquist:2016viq,Appelquist:2018prep} including the present data set.
%To demonstrate this gap in the spectrum more clearly, we plot only the masses $M_\pi$, $M_\sigma$, and for comparison $M_\rho$ in bare lattice units on the left panel of Fig.~\ref{fig:latticeunits}.
Light scalar states have also been observed in $\SU(3)$ gauge theory with two flavors in the symmetric (sextet) representation of the gauge group \cite{Fodor:2015vwa,Fodor:2016pls}, $\SU(3)$ gauge theory with four light flavors and eight heavy flavors in the fundamental representation of the gauge group \cite{Brower:2015owo}, $\SU(3)$ gauge theory with twelve degenerate flavors in the fundamental representation of the gauge group \cite{Aoki:2013zsa}, and $\SU(2)$ gauge theory with one \cite{Athenodorou:2018fia} and two \cite{DelDebbio:2010hx} flavors in the adjoint representation of the gauge group.
The collected evidence suggests that light scalar resonances may be a generic feature of gauge theories which are nearly conformal in the infrared.
%A light sigma state with a mass well separated from the rho and similar to the pseudo-Nambu-Goldstone bosons over a wide range of bare quark masses appearing in a variety of nearly conformal theories suggests that it may be beneficial or even necessary to include the $\sigma$ state as a dynamical degree of freedom in a low energy EFT description of nearly conformal theories.
In Chapter~\ref{chapter:EFT}, we will consider how one might build an EFT description of nearly conformal gauge theories which includes this light flavor singlet scalar.

\begin{figure}[t]
	\includegraphics[width=0.45\textwidth]{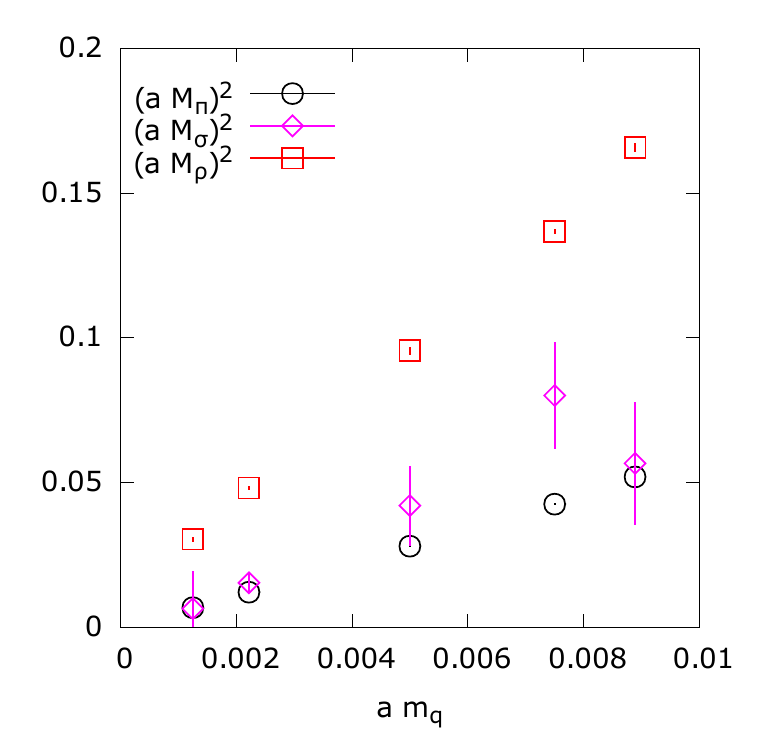}
	\includegraphics[width=0.45\textwidth]{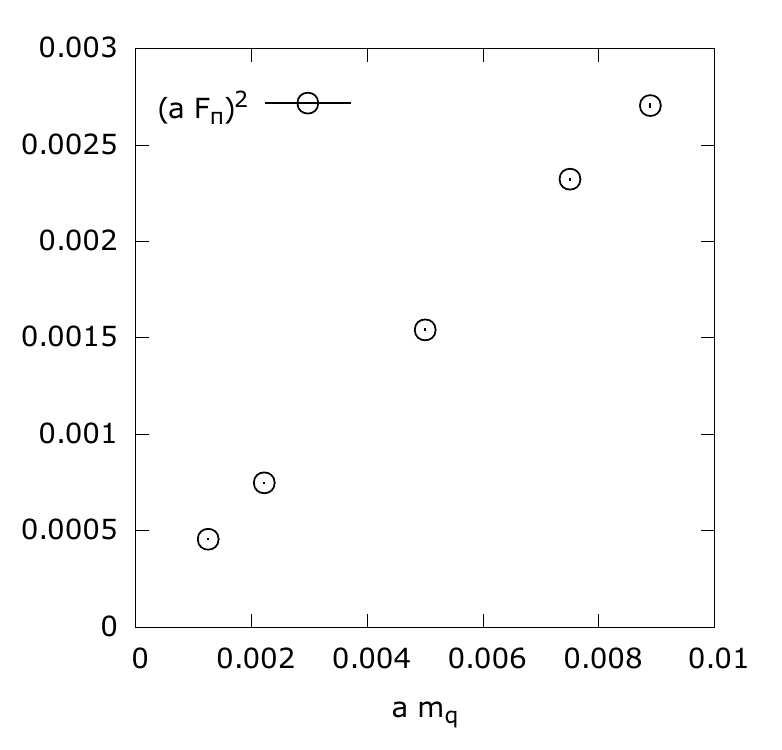}
	\centering
	\caption{The pion, scalar, and rho masses (left) and the pion decay constant (right) plotting in bare lattice units.  This data originates from Refs.~\cite{Appelquist:2016viq,Gasbarro:2017fmi}.  The spectrum is computed using nHYP smeared staggered fermions.  \label{fig:latticeunits}}
\end{figure}
The interpretation of the spectrum requires some special considerations for a theory that is nearly conformal compared to the more familiar case of lattice QCD.
The spectrum plotted in bare lattice units in Figs.~\ref{fig:spectrum1} can be misleading in certain ways.
In QCD, the heavy states that are tied to the confinement scale such as the nucleon mass and the $\rho$ mass do not vary much as one changes the quark mass.
This allows one to easily \emph{set the scale} of the lattice calculation by comparing the dimensionless mass computed on the lattice to the measured mass of the true physical particle and setting the lattice spacing so that they match: $a = M_{\text{lat}}/M_{\text{phys}}$.
One criterion for a good scale is that it has a weak dependence on the quark masses.
Common scales include the proton mass, the omega mass, the pion decay constant, the Sommer scale, the string tension, and scales that can be defined via Wilson flow \cite{Sommer:2014mea}.
However, in Fig.~\ref{fig:spectrum1} we see a strong quark mass dependence in the masses of all states.
How can we set the lattice scale consistently from mass point to mass point if none of the dimensionful quantities are independent of the quark mass?

\begin{figure}[t]
	\includegraphics[width=0.5\textwidth]{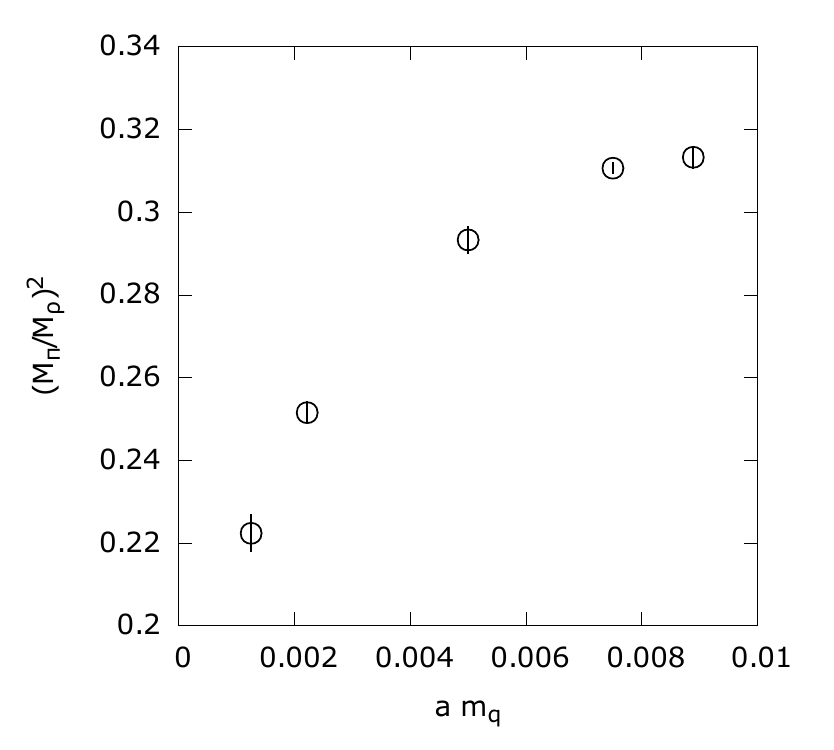}
	\centering
	\caption{Squared pion mass divied by squared rho mass vs bare quark mass in lattice units for $N_f=8$ QCD.  Quantities computed by the LSD collaboration and presented in Refs.~\cite{Appelquist:2016viq,Gasbarro:2017fmi}.  The spectrum is computed using nHYP smeared staggered fermions. \label{fig:MpioMrho}}
\end{figure}
While the masses and decay constants are varying substantially with the quark mass, they are not doing so independently of one another.
In Fig.~\ref{fig:latticeunits}, we plot the squared masses $(a M_\pi)^2$, $(a M_\sigma^2)$, and $(a M_\rho)^2$ and the squared decay constant $(a F_\pi)^2$ in lattice units against the quark mass in lattice units $a m_q$.
One sees that all of the squared dimensionful quantities have an approximately linear dependence on the quark mass.
In ratios, the large approximately linear dependence of the dimensionful quantities on the quark mass cancels and a small nonlinear behavior is left behind.
\begin{figure}[t]
	\includegraphics[width=0.45\textwidth]{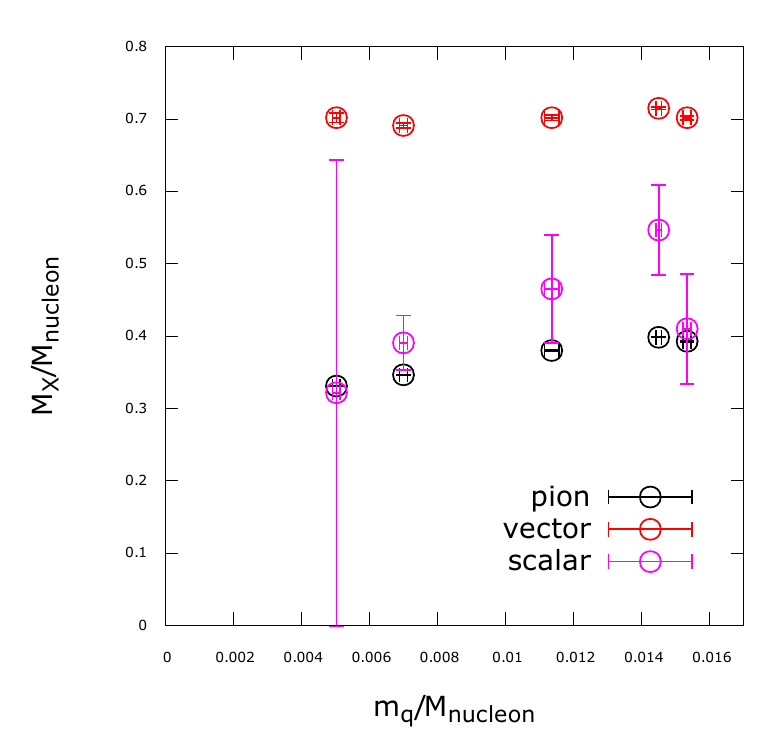}
	\includegraphics[width=0.45\textwidth]{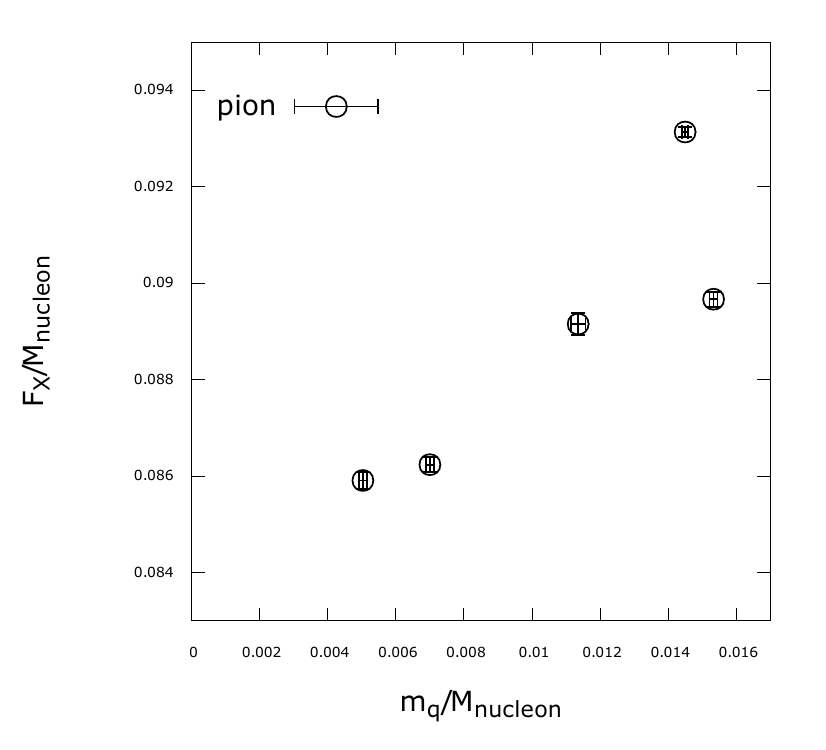}
	\centering
	\caption{The pion, scalar, and rho masses (left) and the pion decay constant (right) plotted in units of the nucleon mass.  This data originates from Refs.~\cite{Appelquist:2016viq,Gasbarro:2017fmi}.  The spectrum is computed using nHYP smeared staggered fermions.  \label{fig:nucleonunits}}
\end{figure}
In Fig.~\ref{fig:MpioMrho}, we show that the ratio of the pion to the rho mass (squared) is slowly varying but that it does change by about thirty percent over the range of quark masses studied.
In Fig.~\ref{fig:nucleonunits}, we plot $M_\pi$,$M_\rho$, and $M_\sigma$ as well as $F_\pi$ in units of the mass of the nucleon.  
Again, these ratios vary slowly with the quark mass after the dominant linear behavior has been canceled by taking dimensionless ratios.

The shared, dominant, linear behavior of the dimensionful quantities in the $N_f=8$ theory may be interpreted in two ways.
One possibility is that the confinement scale has a strong dependence on the quark mass, and all quantities tied to the confinement scale -- the nucleon mass, the rho mass, etc -- are varying along with it.
The other interpretation is that the confinement scale is relatively insensitive to the quark mass, but the lattice spacing is varying with the quark mass.
In a lattice computation, these two scenarios are somewhat a matter of perspective since the lattice only tells us about ratios of scales, not absolute scales.  
All that we can say for certain is that $(a \Lambda_{\text{conf}})^2$ depends significantly and approximately linearly on the bare quark mass $a m_q$.

The choice of units used to express the data is reflective of whether one interprets the confinement scale to be fixed and the lattice spacing to be varying or vice versa.
In the context of an effective field theory analysis, we consider it most sensible to consider the confinement scale to be relatively insensitive to the quark mass so that the EFT has a well defined cutoff that doesn't change much as one varies the quark mass.
We can use the nucleon mass (as a proxy for the confinement scale) as a unit against which to measure other dimensionful quantities as in Fig.~\ref{fig:nucleonunits}.
The cutoff of an EFT for the pions and the sigma, which we take to be roughly the mass of the lightest excluded state, i.e. the rho mass, is approximately independent of the quark mass.  
On the other hand, in lattice units for which we consider the confinement scale to be varying and the lattice spacing to be fixed, we see in Fig.~\ref{fig:spectrum1} and Fig.~\ref{fig:latticeunits} that the rho mass is varying and that it is less clear how to identify the cutoff for the EFT.

%If one plots dimensionless ratios of dimensionful quantities in the $N_f=8$ theory, the ratios depend very weakly on the bare quark mass.
%In Fig.~\ref{fig:nucleonunits}, we demonstrate this by plotting $M_\pi$, $M_\sigma$, $M_\rho$, (left) and $F_\pi$ (right) against the quark mass $m_q$ where all quantities are expressed in ratios with (or in units of) the nucleon mass.
%While all dimensionful quantities depend very strongly on the quark mass in bare lattice units, they depend very weakly on the bare quark mass when plotted in physical units, in this case units of the nucleon mass.
%Another way to say it is that the masses and decay constants of the hadrons change 
%There are two possible interpretations of this behavior.
%One possibility is that the confinement scale has a strong dependence on the quark mass, and all quantities tied to the confinement scale -- the nucleon mass, the rho mass, etc -- are varying along with it.
%The other interpretation is that the confinement scale is relatively insensitive to the quark mass, but the lattice spacing is varying with the quark mass.
%However, in a lattice computation, these two scenarios really are nothing more than interpretation.
%On the lattice, there are no absolute units, there are only relative scales.
%All that we can say for certain is that the quantity $a\Lambda_{\text{conf}}$ depends strongly on $a m_q$.

%
\begin{figure}[t]
	\includegraphics[width=0.45\textwidth]{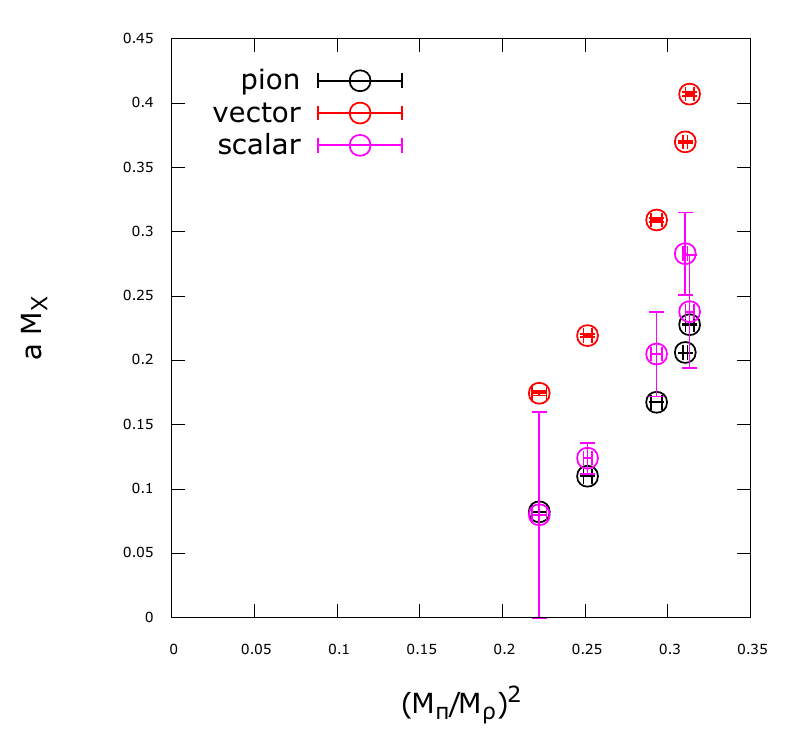}
	\includegraphics[width=0.45\textwidth]{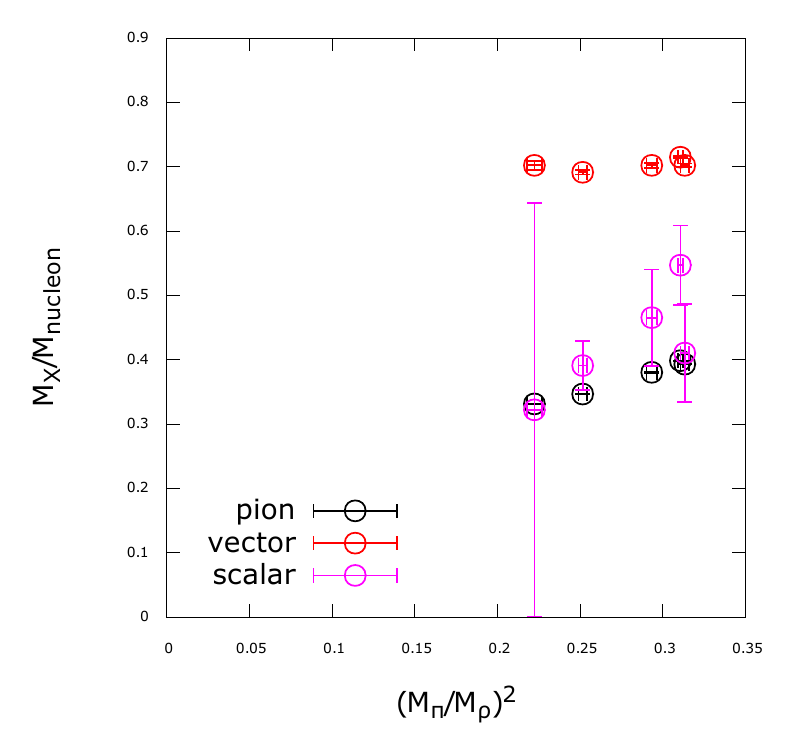}
	\centering
	\caption{The pion, scalar, and rho masses in bare lattice units (left) and in nucleon units (right) plotted against $(M_\pi / M_\rho)^2$.  This data originates from Refs.~\cite{Appelquist:2016viq,Gasbarro:2017fmi}.  The spectrum is computed using nHYP smeared staggered fermions.  \label{fig:againstmpiomrho}}
\end{figure}
Another detail worth considering is the distance from the chiral limit.
It is tempting to suppose that as the bare quark mass is tuned towards zero, one approaches the chiral limit in a linear fashion.
However, for the reason mentioned above, much of the effect of changing the bare quark mass goes into the variation of the quantity $a\Lambda_{\text{conf}}$, and ratios of hadron masses and decay constants are relatively insensitive to the quark mass.
This leads to a great difficultly in approaching the chiral limit.
One can measure the distance from the chiral limit in a more physical way by considering the ratio $M_\pi^2 / M_\rho^2$.
In QCD, the pion mass squared is linear in the quark mass and the rho mass is relatively insensitive to the quark mass, so this quantity linearly approaches zero with the quark mass.
In Fig.~\ref{fig:MpioMrho} we see that the ratio is not linearly approaching zero but decreasing much more slowly with the quark mass.
We see that while the bare quark mass is changing by nearly an order of magnitude, $M_\pi^2 / M_\rho^2$ only changes by about thirty percent.

We visualize the approach of $M_\pi$, $M_\rho$, and $M_\sigma$ to the chiral limit in a more physical way by plotting $M_\pi^2 / M_\rho^2$ on the x-axis in Fig.~\ref{fig:againstmpiomrho}.
Again we see that despite the wide range of bare quark masses considered, there is only a modest movement towards the chiral limit.
Thus, in an effective field theory analysis, it should be kept in mind that the data only represents a limited range of distances from the chiral limit.

\section{$\pi\pi$ Scattering in $N_f = 8$ QCD \label{sec:scattering}}
In this section, we present new lattice calculations in $N_f=8$ QCD intended to further constrain the possible landscape of effective field theory descriptions of nearly conformal gauge theories.
We will study s-wave $\pi\pi$ scattering in the maximal isospin channel.  
This scattering channel is the simplest to study numerically because it does not contain any disconnected diagrams.
It is an interesting channel for distinguishing between chiral perturbation theory and a low energy EFT which includes a light flavor singlet scalar.
In the chiral Lagrangian, the only diagram contributing to $\pi\pi$ scattering at tree level is the four pion vertex.
In a theory including a light scalar, this scattering amplitude includes contributions from t-channel and u-channel exchange of the scalar along with the four-pion contact interaction.

In the gauge theory, we compute the scatting phase shift nonperturbatively on the lattice using L{\"u}scher's finite volume formalism.
We will first review the key ideas of this formalism in Section~\ref{sec:Luscher}.
Then we will discuss the specific case of maximal isospin scatting in the $N_f=8$ theory and present the new lattice results in Section~\ref{sec:latscat}.

\subsection{L{\"u}scher's Method \label{sec:Luscher}}
L{\"u}scher's finite volume approach relates the energy levels of multi-particle states on the lattice to the phase shift of a corresponding scattering process.
The physical picture for the case of $2\rightarrow 2$ scattering is that when two particles are placed in a finite box, their wavefunctions will overlap by an amount depending on the relative size of the Compton wavelengths of the individual particles and the size of the box, and they will interact in some way that depends on the range of the interaction.
As the particles are squeezed together, the energy of the two particle state will increase due to the overlap of the wavefunctions and the nontrivial scattering between the particles.
The resulting energy level can then be related to the scatting phase shift at a particular momentum.

The formalism was originally developed for the simplified case of two dimensional quantum field theories \cite{Luscher:1990ck}.
When the Compton wavelengths and the range of the interaction are both small compared to the finite spatial extension of the system $L$, then the wave function outside of the interaction region is a plane wave and the effect of the interaction manifests itself in a momentum dependent phase shift $\delta(k)$.
The scattering state must satisfy the periodicity condition on a finite, periodic, spatial volume (a circle).
\begin{equation}
e^{2i\delta(k) + i k L} = e^{i k 0} = 1
\end{equation}
Then the quantization condition for the scattering momentum $k_n$ is given by
\begin{equation}
k_n L + 2 \delta(k_n) = 2 \pi n \quad,\quad n \in \mathbb{Z}   \label{eq:scatquant2d}
\end{equation}
Through this quantization condition, one can relate a given scattering momentum to the phase shift evaluated at that momentum.
The scattering momenta are acquired from the two particle energy levels by the relativistic dispersion relation,
\begin{equation}
E_n = 2 \sqrt{k_n^2 + m^2}  \label{eq:scatdispersion2d}
\end{equation}
for the scatting of identical particles of mass $m$.
In two dimensions, for $2\rightarrow 2$ scattering of scalar particles of equal mass, these are all the relations that are needed to extract the phase shift.
Procedurally, one computes the two particle energy on the lattice, uses the relativistic dispersion relation Eq.~\ref{eq:scatdispersion2d} to compute the scattering momentum for each two particle energy level, and finally one computes the phase shift at each scattering momentum through the plane wave quantization condition on a periodic ring Eq.~\ref{eq:scatquant2d}.

In practice, one will typically desire to compute the phase shift at many different scattering momenta.
For example, in the case of scattering in a resonance channel, one must compute the phase shift at many scattering momenta in order to map out the change in the phase shift through the resonance.
In a purely elastic channel for scattering below threshold, it is still desirable to have a range of scattering momenta in order to extract the scattering length and effective range in the small momentum effective range expansion.
\begin{equation}
k\cot(\delta(k))=\frac{1}{A} + r \frac{m^2}{2}\left(\frac{k^2}{m^2}\right) + \mathcal{O}\left(\left(\frac{k}{m}\right)^4\right)
\end{equation}
We will not consider resonance scatting in our study, instead only focusing on the latter case of extracting the scattering length and effective range.
One way to extract many energy levels and therefore many scatting momenta is by simply computing a large number of excited state energies.
This can be facilitated greatly by computing the energy using a large basis of interpolating operators as was demonstrated in a recent study of I=0 scatting to extract the $\sigma$ pole in QCD \cite{Briceno:2016mjc}.
Another simple way to get different scatting momenta is by changing the spatial box size.
Finally, one can ``fake'' a different spatial box size by computing the two particle energy in a moving (boosted) frame.  
In a moving frame, the box size is relativistically length contracted in the direction of movement leading to a smaller effective box size and a larger scattering momentum.

So far we have discussed the L{\"u}scher formalism for the simplified case of two dimensional quantum field theories.
In four spacetime dimensions, the quantization condition that relates the scattering phase shift to the scattering momentum is substantially more complicated.
The generalization of the scattering formalism to four dimensions was derived by L{\"u}scher in \cite{Luscher:1990ux}.
In the continuum, one could simply move to an angular momentum basis in the spatial directions and define the scatting phase shift for each angular momentum quantum number $\delta_l$.
However, the lattice explicitly breaks the rotational symmetry down to the cubic subgroup, preventing a straightforward generalization.
For s-wave scattering ($l=0$) the energy level $E_n$ is allowed by the cubic symmetry on the finite 3-torus if and only if the associated scattering momentum $k_n$ satisfies
\begin{equation}
e^{2i\delta_0(k)} = \frac{\mathcal{Z}_{00}(1;q^2) + i \pi^{3/2}q}{\mathcal{Z}_{00}(1;q^2) - i \pi^{3/2}q}
\end{equation}
where $q = kL/(2\pi)$ and the Riemann zeta function is given by
\begin{equation}
\mathcal{Z}_{00}(s;q^2) = \frac{1}{\sqrt{4\pi}} \sum_{\vec{n} \in \mathbb{Z}^3}(\vec{n}^2 -q^2))^{-s} 
\end{equation}
initially defined for $\text{Re}(s) > 3/2$ and then through analytic continuation.
The quantization condition can equivalently be expressed for the quantity $k\cot\delta_0(k)$ as
\begin{equation}
k \cot{\delta_0(k)} = \frac{2\pi}{L} \pi^{-3/2} \mathcal{Z}_{00}(1;q^2)
\end{equation}
We refer the reader to \cite{Luscher:1990ux} for further details of the derivation of the four dimensional quantization condition.

\subsection{New Results for Maximal Isospin $\pi\pi$ Scattering \label{sec:latscat}}
Now we turn to our study of interest, maximal isospin $\pi\pi$ scattering in the $N_f=8$ model.
First let us explain the group theoretic setup of the scattering problem and our use of the term ``isospin'' in a theory with eight flavors.
We have explained that the lattice theory is formulated in terms of the staggered fermion discretization.
For this lattice action, each species of lattice fermions becomes four species of continuum Dirac fermions in the continuum limit.
At finite lattice spacing, these four \emph{tastes} (which is the term used for flavor in this context) do not have a continuous $\SU(4)\times \SU(4)$ flavor symmetry.
The flavor symmetry is broken at finite lattice spacing to a discrete subgroup known as the taste group \cite{Kilcup:1986dg}.
For the $N_f=8$ theory, we work with two flavors of lattice staggered fermions which become eight flavors of continuum Dirac fermions.
At finite lattice spacing, we do have an exact $\text{U}(2)\times \text{U}(2)$ global flavor symmetry for the two staggered flavors at the classical level; the $\text{U}_A(1)$ is broken at the quantum level by the anomaly in the usual way.
Thus we can organize our calculation around this exactly realized $\SU(2)\times \SU(2)$ ``isospin'' subgroup of the full $\SU(8)\times \SU(8)$ flavor group when we work at fixed taste.

Let us denote the two staggered flavors by $(\chi_1,\chi_2)$.
The interpolating operator for the distance zero Goldstone pion is given by
\begin{equation}
\pi^{+}(\vec{x},t) = \bar{\chi}_2(\vec{x},t)\epsilon(\vec{x},t)\chi_1(\vec{x},t)
\end{equation}
where $\epsilon(\vec{x},t) = (-1)^{x+y+z+t}$ is the staggered phase corresponding to the spin-taste structure $\gamma_5 \times \xi_5$ \cite{Kilcup:1986dg,Daniel:1987aa}.
When we discuss maximal isospin scattering in the context of $N_f=8$, we are really considering the scattering of the highest weight state in the $\mathbf{63}$ representation of $\SU_V(8)$.  
The ``orientation'' of the highest weight state within the adjoint multiplet is a matter of convention, so we may take the distance zero $\pi^+$ -- which is already the highest weight state within the isospin subgroup -- to also be the highest weight state within the larger adjoint multiplet.

In order to compute the s-wave phase shift, $\delta_{l=0}(k)$, it is sufficient to consider zero momentum scattering. 
The pion operator projected onto zero spatial momentum is given by
\begin{equation}
\pi^{+}(t) = \sum_{\vec{x}} \pi^+(\vec{x},t)
\end{equation}
We construct the simplest two body operator which sources the maximal isospin two pion state.
\begin{equation}
\mathcal{O}_{I=2}(t) = \pi^+(t)\pi^+(t+1) \label{eq:OIeq2}
\end{equation}
We have chosen the pions to be separated by one time slice in order to avoid projection onto unwanted states through Fierz rearrangement identities \cite{Fu:2013ffa}.
The energy of the  $\pi^+\pi^+ \rightarrow \pi^+\pi^+$ scattering state is computed from the two point function of $\mathcal{O}_{I=2}(t)$ at well separated time slices.
\begin{figure}[t!]
	\begin{center}
		\includegraphics[width=0.75\textwidth]{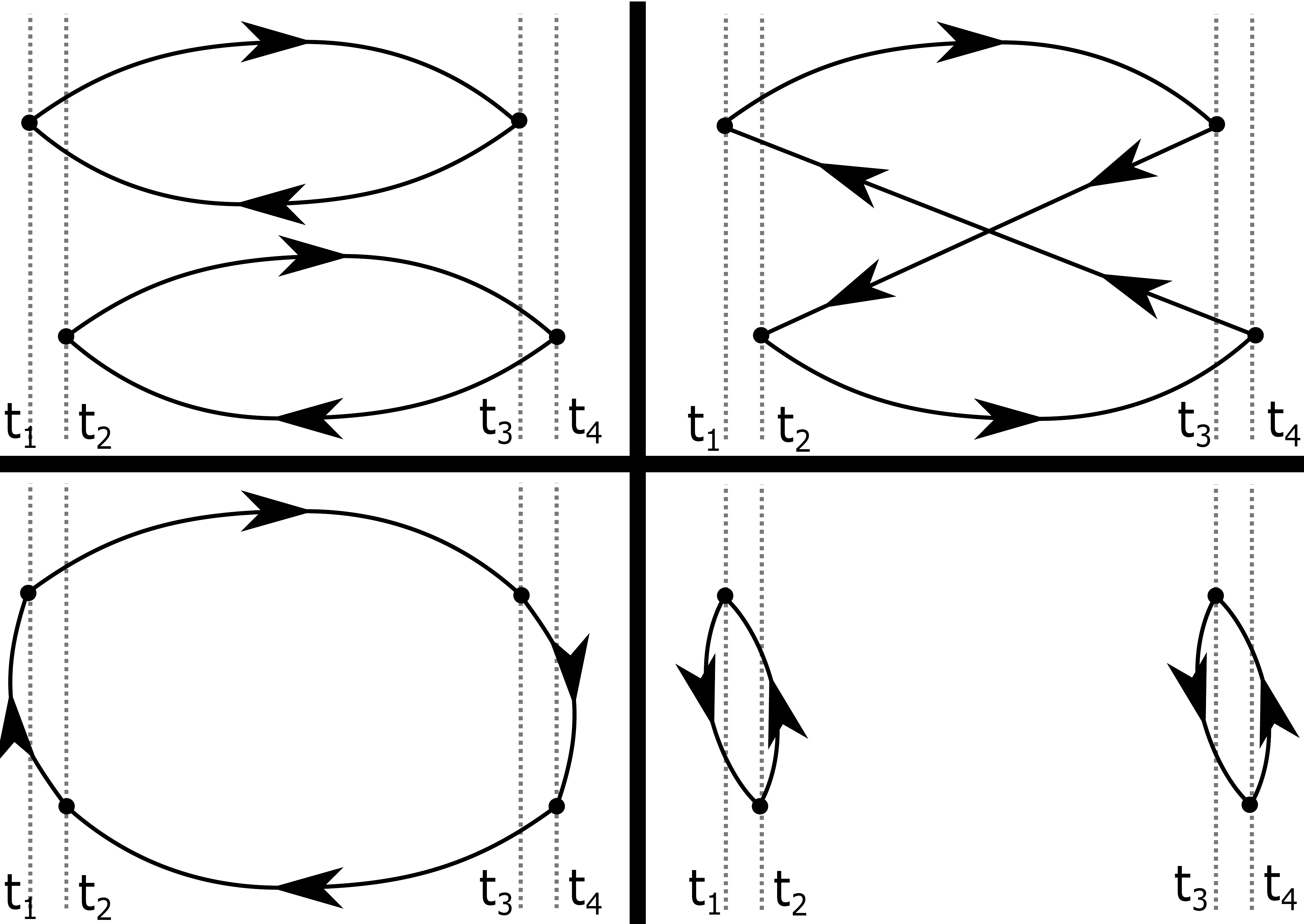}
	\end{center}
	\caption{Valence quark diagrams appearing in $2\rightarrow 2$ scattering processes.  The top left cell is the ``direct'' (D) channel, the top right is the ``crossed'' (C) channel, the bottom left is the ``rectangle'' (R) channel, and the bottom right is the ``vacuum'' (V) channel.  In all diagrams, time flows from left to right.}
	\label{fig:scatt_all}
\end{figure}
%The four pion amplitude corresponding to the interpolating operator Eq.~\ref{eq:OIeq2} is of the form
\begin{eqnarray}
&&\langle \mathcal{O}_{I=2}^\dagger(t_1)  \mathcal{O}_{I=2}(t_3) \rangle = \langle {\pi^+}(t_1)^\dagger {\pi^+}(t_2)^\dagger  {\pi^+}(t_3)  {\pi^+}(t_4) \rangle \\
&&= \sum_{\vec{x_1}\vec{x_2}\vec{x_3}\vec{x_4}} \epsilon(x_1) \epsilon(x_2) \epsilon(x_3) \epsilon(x_4) \langle \bar{\chi}_1(x_1) \chi_2(x_1) \bar{\chi}_1(x_2) \chi_2(x_2) \bar{\chi}_2(x_3) \chi_1(x_3) \bar{\chi}_2(x_4) \chi_1(x_4) \rangle  \nonumber
\end{eqnarray}
where $t_2 = t_1 + 1$ and $t_4 = t_3 + 1$ or in general $t_3,t_4 \gg t_1,t_2$.  
The possible Wick contractions are
\begin{align*}
\langle \bar{\chi}_1(x_1) \chi_2(x_1) \bar{\chi}_1(x_2) \chi_2(x_2) \bar{\chi}_2(x_3) \chi_1(x_3) \bar{\chi}_2(x_4) \chi_1(x_4) \rangle \nonumber \\ 
\contraction[2ex]{  = \langle }{ \bar{\chi}_1}{(x_1) \chi_2(x_1) \bar{\chi}_1(x_2) \chi_2(x_2) \bar{\chi}_2(x_3) }{ \chi_1}
\contraction{= \langle \bar{\chi}_1(x_1) }{ \chi_2}{(x_1) \bar{\chi}_1(x_2) \chi_2(x_2)}{ \bar{\chi}_2}
\bcontraction[2ex]{ = \langle \bar{\chi}_1(x_1) \chi_2(x_1)}{ \bar{\chi}_1}{(x_2) \chi_2(x_2) \bar{\chi}_2(x_3) \chi_1(x_3) \bar{\chi}_2(x_4)}{ \chi_1}
\bcontraction{= \langle \bar{\chi}_1(x_1) \chi_2(x_1) \bar{\chi}_1(x_2) }{) \chi_2}{(x_2) \bar{\chi}_2(x_3) \chi_1(x_3)}{ \bar{\chi}_2}
= \langle \bar{\chi}_1(x_1) \chi_2(x_1) \bar{\chi}_1(x_2) \chi_2(x_2) \bar{\chi}_2(x_3) \chi_1(x_3) \bar{\chi}_2(x_4) \chi_1(x_4) \rangle \nonumber \\
\contraction[2ex]{+ \langle}{ \bar{\chi}_1}{(x_1) \chi_2(x_1) \bar{\chi}_1(x_2) \chi_2(x_2) \bar{\chi}_2(x_3) \chi_1(x_3) \bar{\chi}_2(x_4) }{\chi_1}
\contraction{+ \langle \bar{\chi}_1(x_1)}{\chi_2}{(x_1) \bar{\chi}_1(x_2) \chi_2(x_2) \bar{\chi}_2(x_3) \chi_1(x_3) }{ \bar{\chi}_2}
\bcontraction[2ex]{+ \langle \bar{\chi}_1(x_1) \chi_2(x_1)}{\bar{\chi}_1}{(x_2) \chi_2(x_2) \bar{\chi}_2(x_3) }{ \chi_1}
\bcontraction{+ \langle \bar{\chi}_1(x_1) \chi_2(x_1) \bar{\chi}_1(x_2)}{\chi_2}{(x_2) }{\bar{\chi}_2}
+ \langle \bar{\chi}_1(x_1) \chi_2(x_1) \bar{\chi}_1(x_2) \chi_2(x_2) \bar{\chi}_2(x_3) \chi_1(x_3) \bar{\chi}_2(x_4) \chi_1(x_4) \rangle \nonumber \\
\contraction{+ \langle \bar{\chi}_1(x_1) \chi_2(x_1) \bar{\chi}_1(x_2)}{ \chi_2}{(x_2)}{\bar{\chi}_2}
\contraction[2ex]{+ \langle}{\bar{\chi}_1}{(x_1) \chi_2(x_1) \bar{\chi}_1(x_2) \chi_2(x_2) \bar{\chi}_2(x_3)}{ \chi_1}
\bcontraction{+ \langle \bar{\chi}_1(x_1)}{\chi_2}{(x_1) \bar{\chi}_1(x_2) \chi_2(x_2) \bar{\chi}_2(x_3) \chi_1(x_3)}{\bar{\chi}_2}
\bcontraction[2ex]{+ \langle \bar{\chi}_1(x_1) \chi_2(x_1)}{\bar{\chi}_1}{(x_2) \chi_2(x_2) \bar{\chi}_2(x_3) \chi_1(x_3) \bar{\chi}_2(x_4)}{ \chi_1}
+ \langle \bar{\chi}_1(x_1) \chi_2(x_1) \bar{\chi}_1(x_2) \chi_2(x_2) \bar{\chi}_2(x_3) \chi_1(x_3) \bar{\chi}_2(x_4) \chi_1(x_4) \rangle \nonumber  \\
\contraction[2ex]{+ \langle}{\bar{\chi}_1}{(x_1) \chi_2(x_1) \bar{\chi}_1(x_2) \chi_2(x_2) \bar{\chi}_2(x_3) \chi_1(x_3) \bar{\chi}_2(x_4)}{ \chi_1}
\contraction{+ \langle \bar{\chi}_1(x_1) \chi_2(x_1) \bar{\chi}_1(x_2) }{\chi_2}{(x_2) \bar{\chi}_2(x_3) \chi_1(x_3)}{\bar{\chi}_2}
\bcontraction{+ \langle \bar{\chi}_1(x_1) \chi_2(x_1)}{\bar{\chi}_1}{(x_2) \chi_2(x_2) \bar{\chi}_2(x_3)}{ \chi_1}
\bcontraction[2ex]{+ \langle \bar{\chi}_1(x_1)}{ \chi_2}{(x_1) \bar{\chi}_1(x_2) \chi_2(x_2)}{ \bar{\chi}_2}
+ \langle \bar{\chi}_1(x_1) \chi_2(x_1) \bar{\chi}_1(x_2) \chi_2(x_2) \bar{\chi}_2(x_3) \chi_1(x_3) \bar{\chi}_2(x_4) \chi_1(x_4) \rangle \nonumber 
\end{align*}
The first two terms are referred to as the ``direct'' channel, and the second two terms are the ``crossed'' channel.  Employing $\gamma_5$ hermiticity and taking into account the anticommutative properties of the Grassman valued fields, one arrives at
\begin{equation}
\langle {\pi^+}(t_1)^\dagger {\pi^+}(t_2)^\dagger  {\pi^+}(t_3)  {\pi^+}(t_4) \rangle = C_D(13;24) + C_D(14;23) - C_C(1324) - C_C(1423)
\end{equation}
with
\begin{eqnarray}
C_D(i k;j l) = \text{Tr}(G_{x_i x_k}^\dagger G_{x_i x_k} )\text{Tr}_C(G_{x_j x_l}^\dagger G_{x_j x_l} ) \\ 
C_C(i j k l) = \text{Tr}(G_{x_i x_k}G_{x_j x_k}^\dagger G_{x_j x_l} G_{x_i x_l}^\dagger )
\end{eqnarray}
where Tr(...) denotes a color trace as well as a spatial sum over time sheets and $G_{x y}$ is a quark propagator from $x$ to $y$.

Valence quark diagrams in Fig.~\ref{fig:scatt_all} help to visualize the different scattering channels.  
The ``rectangle'' and ``vacuum'' diagrams only contribute to pion scattering with nonmaximal isospin, and they tend to be noisier and computationally more expensive \cite{Fu:2013ffa}.  
The absence of these diagrams makes maximal isospin scattering a good first channel to investigate.  

\begin{figure}[t]
	\includegraphics[width=0.45\textwidth]{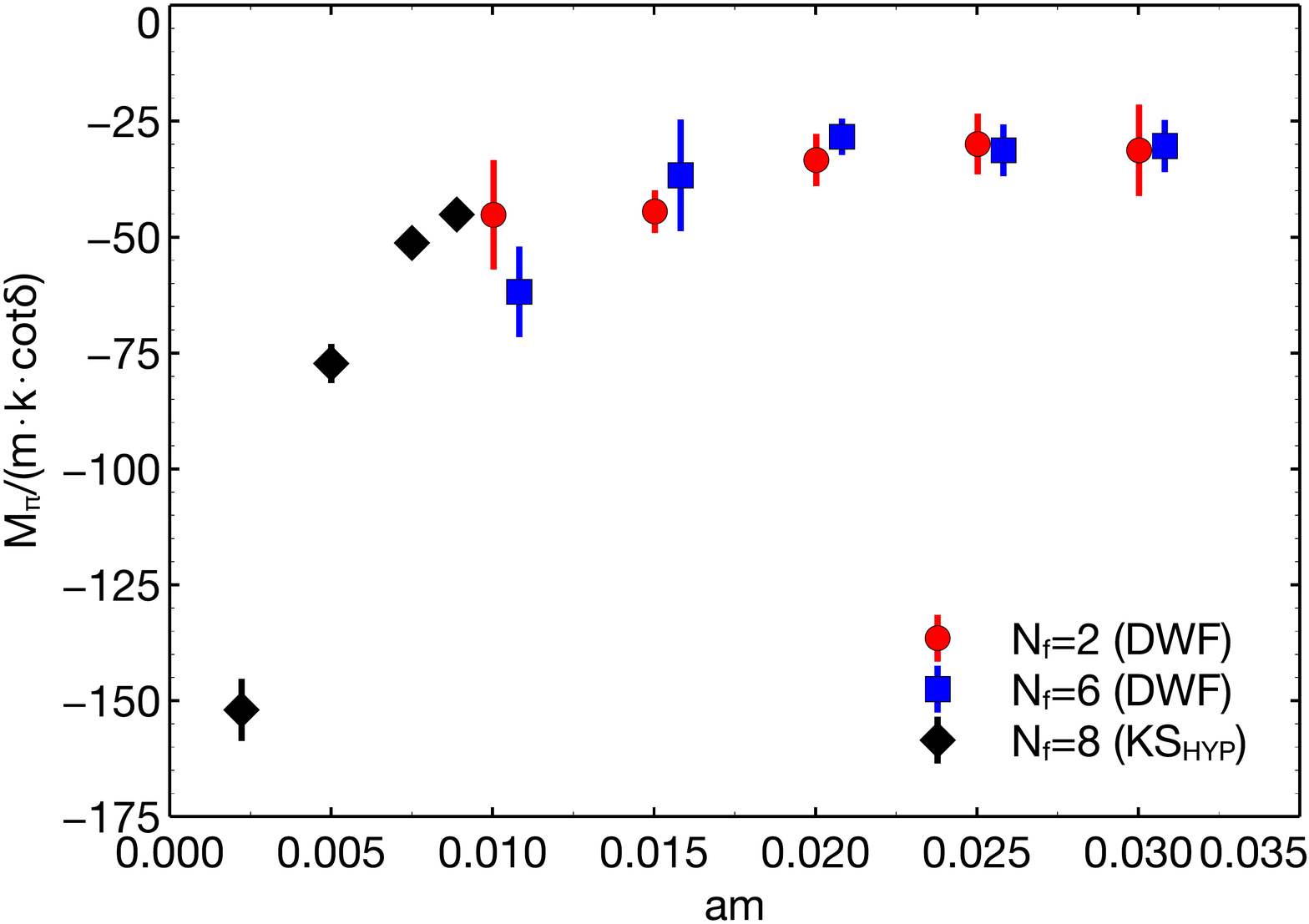}
	\includegraphics[width=0.45\textwidth]{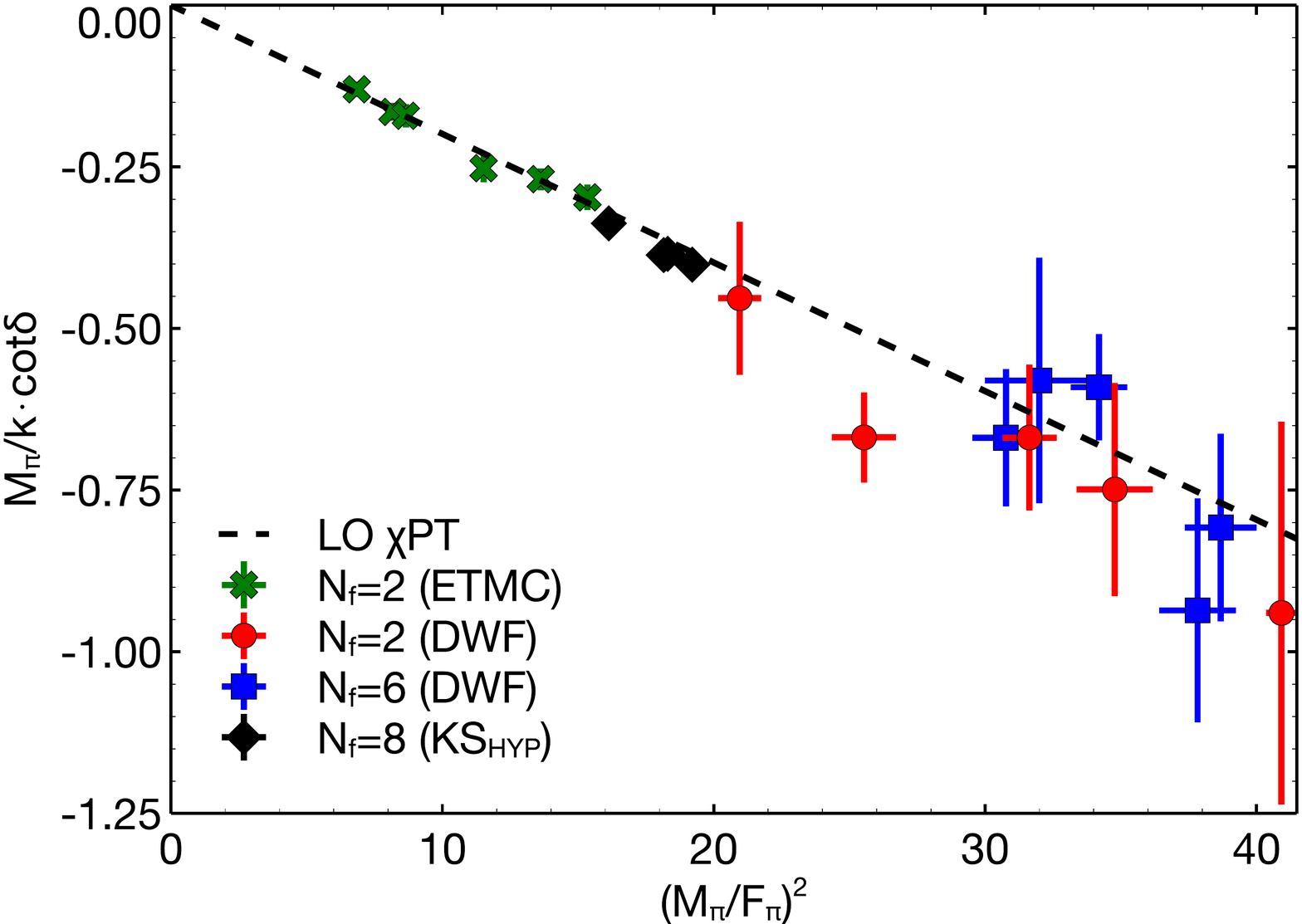}
	\centering
	\caption{The maximal isospin $\pi\pi$ scattering length.  Left: $M_\pi/(m_q k \cot(\delta)) \approx M_\pi A / m_q$ is plotted against the bare quark mass.  The data for eight flavor QCD with Kogut-Susskind staggered fermions is shown in black.  This data was originally presented in Ref.~\cite{Gasbarro:2017fmi}.  For comparison, we also include data for $N_f = 2$ and $N_f=6$ QCD with domain wall fermions from Ref.~\cite{Appelquist:2012sm} in red and blue respectively.  Right: The dimensionless ratio $M_\pi/(k \cot(\delta)) \approx M_\pi A$ is plotted against the dimensionless ratio $(M_\pi / F_\pi)^2$.  Again the data for $N_f=2$ and $N_f=6$ from Ref.~\cite{Appelquist:2012sm} is included.  We also include data for $N_f=2$ at light quark masses from Ref.~\cite{Feng:2009ij}.   \label{fig:scattering1}}
\end{figure}
For this first study of pion scattering in the $N_f=8$ theory, we have only considered a single interpolating operator and we have only extracted a single two particle energy and a single scattering phase shift.
We have found that the extracted scattering momentum is very small such that the effective range term is negligible and $k\cot(\delta_0(k)) \approx 1/A$.
In a future study, one may consider multiple interpolating operators, multiple volumes, and boosted frames in order to extract energy levels corresponding to larger scattering momenta and to extract the effective range also from the phase shift.

The results of the L{\"u}scher analysis of the maximal isospin correlator are presented in Fig.~\ref{fig:scattering1}.
We also include data from $N_f=2$ QCD from Refs.~\cite{Appelquist:2012sm,Feng:2009ij} and for $N_f=6$ QCD from Ref.~\cite{Appelquist:2012sm} for comparison.
In the left plot, we show $M_\pi / (m_q k \cot(\delta)) \approx M_\pi A / m_q$ vs the bare quark mass $m_q$.
The quantity on the y-axis has dimensions of inverse mass and is plotted in units of the lattice spacing, and the x-axis has dimensions of mass and is plotted in units of the inverse lattice spacing.
The data for $N_f=2$ is nearly flat as a function of the bare quark mass and is statistically consistent with a constant.
The data for $N_f=6$ shows some small amount of curvature at the lightest mass point, but is still relatively flat.
The new results for $N_f = 8$ show a marked difference from the other two data sets, with the data demonstrating a high degree of curvature.
Some of this effect may be due to the variation in the lattice spacing from mass point to mass point as we discussed in Section~\ref{sec:nf8}.
We also remark that on the left plot of Fig.~\ref{fig:scattering1}, the comparison between the various calculations for $N_f = 2$, $6$, and $8$ is only qualitative because the dimensionful quantities are plotted in lattice units, and the lattice scale varies between the different studies.

To get some bearing for the expected behavior of the phase shift, let us examine the expression for the phase shift in chiral perturbation theory.  The scattering length in $\chi$PT at next-to-leading order is given by \cite{Bijnens:2009qm,Bijnens:2011fm}
\begin{flalign}
M_\pi A = &\frac{-M^2}{16 \pi F^2} \left[1 + \frac{N_f M^2}{(4\pi F)^2}\left(-256\pi^2\left(\left(1-2/N_f\right)\left(L_4^r-L_6^r\right)+\frac{1}{N_f}\left(L_0^r+2L_1^r+2L_2^r+L_3^r\right) \right)\right.\right. \nonumber \\
& \left.\left.-2\frac{N_f - 1}{N_f^3} + \frac{2-N_f + 2N_f^2+N_f^3}{N_f^3}\log\left(M^2/\mu^2\right)\right)\right]
\end{flalign}
where $M^2 = 2Bm_q$. $B$ and $F$ are the usual low energy constants for the condensate and decay constant in the leading order chiral Lagrangian, and $L_i^r$ are the renormalized Gasser Leutweyler NLO low energy constants.
The leading order prediction is that $M_\pi A/m_q$ is a constant.
We see on the left panel of Fig.~\ref{fig:scattering1} that the $N_f = 2$ calculation agrees well with this leading order prediction, being relatively constant as a function of the quark mass.  
The $N_f = 6$ data starts to show a small amount of curvature but is roughly consistent with a constant.
The new data points that we have computed for $N_f = 8$ show a large amount of curvature over a comparatively small range of bare quark masses and are inconsistent with the leading order $\chi$PT prediction.

The expression for the scattering length in $\chi$PT may be re-expanded in the physical quantities $M_\pi^2$ and $F_\pi^2$ replacing the bare quantities $M^2$ and $F^2$ \cite{Appelquist:2012sm}.
\begin{flalign}
M_\pi A = &\frac{-M_\pi^2}{16 \pi F_\pi^2} \left[1 + \frac{M^2}{(4\pi F_\pi)^2}\left(-256\pi^2\left(L_0^r +2L_1^r + 2L_2^r + L_3^r -2L_4^r-L_5^r+2L_6^r+L_8^r\right)\right.\right. \nonumber \\
& \left.\left.-2\frac{N_f - 1}{N_f^2} + \frac{2(1-N_f +N_f^2)}{N_f^2}\log\left(M^2/\mu^2\right)\right)\right]
\end{flalign}
To examine this data in light of this alternate expansion, we plot in the right panel of Fig.~\ref{fig:scattering1} the dimensionless quantity $M_\pi A$ verses $(M_\pi / F_\pi)^2$.
In terms of the new expansion parameter, the leading order $\chi$PT prediction is that $M_\pi A$ is linear in $M_\pi^2 / F_\pi^2$ with the slope predicted to be $-1/(16\pi)$.
We see that there is good agreement between the leading order $\chi$PT prediction and the data for $N_f = 2$, $6$, and $8$.
It is puzzling that the agreement should be quite poor when plotting the data against the bare quark mass and in good agreement when plotted against $M_\pi^2 / F_\pi^2$.  
It is possible that we are not yet in a regime in which the chiral expansion in terms of bare quantities is converging, but after resumming the expansion in terms of the physical quantities $M_\pi$ and $F_\pi$ the expansion is convergent.
However, it is also possible that this is a peculiarity of this particular observable in this expansion, just as $M_\pi^2$ is coincidentally well described by leading order $\chi$PT when expanded in the bare quark mass.

In an upcoming work, we will consider a simultaneous analysis of the maximal isospin pion scattering length along with the spectral data including $M_\pi$, $F_\pi$, $M_\sigma$, and $\langle\bar{\psi}\psi \rangle$ of the $N_f=8$ model.
We plan to analyze the data with a variety of possible EFT frameworks including the general $N_f$ chiral Lagrangian \cite{Bijnens:2009qm,Bijnens:2011fm} and the linear sigma EFT framework \cite{Gasbarro:2017ccf,Gasbarro:2018prep}.
In Chapter~\ref{chapter:EFT} we review effective field theory approaches to describing strongly coupled gauge theories with chiral symmetry breaking at low energies.  
We will discuss a new EFT framework based on the linear sigma model and argue that it may provide an improved description of the low energy properties of nearly conformal gauge theories away from the chiral limit \cite{Gasbarro:2017ccf,Gasbarro:2018prep}.

%\section{Vector Form Factor}
%\section{Scalar Form Factor}
%\section{EFT Fits to Lattice Data}
%

%%%%%%%%%%%%%%%%%%%%%%%%%%%%%%
%%%%%%%%%%%%%%%%%%%%%%%%%%%%%%
\chapter{Chiral Effective Theory \label{chapter:EFT}}
We have shown in Chapter~\ref{chapter:Lattice} that there are limitations to applying traditional lattice methods to the problem of nearly conformal gauge theories.
Nearly conformal dynamics lead to an enhancement in the separation between the confinement scale $\Lambda_{\text{conf}}$ and the lattice cutoff $1/a$; the theory must run over a larger range of scales before reaching confinement.
%The problem is made more difficult by the introduction of the quark mass.
This ratio of scales $a \Lambda_{\text{conf}}$ becomes increasingly sensitive to the bare quark mass as one approaches the bottom edge of the conformal window because the quark masses explicitly break the scale symmetry leading to a faster onset of confinement.
In a mass independent scale setting scheme in which $\Lambda_{\text{conf}}$ is fixed by some physical observable like the nucleon mass $M_N$, the lattice spacing varies significantly with the quark mass.
In the regime that is currently accessible, most of the effect of changing the bare quark mass is to change the lattice spacing.  
One may see this by studying dimensionless ratios of hadron masses and decay constants as we have discussed in Chapter~\ref{chapter:Lattice}.  
These ratios evolve very slowly with the bare quark mass compared to QCD.
As a result, it becomes increasingly difficult to reach the chiral limit in a lattice calculation as one approaches the bottom of the conformal window.

Effective field theory (EFT) provides a systematic approach to study the low energy properties of quantum fields theories, whether the UV physics be unknown as in the standard model or incalculable as in the case of strongly interacting gauge theories.
In QCD, chiral perturbation theory ($\chi$PT) is an extremely useful tool for describing the dynamics of pseudo-Nambu-Goldstone bosons away from the chiral limit.
One use of the $\chi$PT framework in the context of lattice theory is the extrapolation of data to the chiral limit.
$\chi$PT is also used to model lattice artifacts and for otherwise gaining intuition about the low energy dynamics of hadrons which can motivate and guide lattice calculations.

In this chapter, we review progress made towards developing an EFT framework that is applicable to nearly conformal gauge theories.
We will begin by reviewing the historical development of EFT methods in the context of hadron physics starting with current algebra.  
While current algebra is a separate topic from effective field theory, it does lead to many general results for gauge theories with global chiral symmetries that EFT descriptions will reproduce.  
Furthermore, effective field theory methods were born out of current algebra methods and S-matrix methods through the work of Weinberg who sought a systematic field theoretical framework for deriving the results of current algebra \cite{PhysRev.166.1568,Schwinger:1967tc,Weinberg:1978kz} such as the soft pion theorems \cite{Weinberg:1966kf}.
Indeed, Weinberg later stated that effective field theory is S-matrix theory made practical \cite{Weinberg:1996kw}.
We will cover the principles of effective field theory and chiral dynamics focusing first on the chiral Lagrangian and its possible applicability to nearly conformal gauge theories.
Finally we will review new work on an EFT framework based on the linear sigma model, which seeks to provide a better description of current lattice data for nearly conformal gauge theories.
%Goldstone.
%ChiPT
%CCWZ
%Linear Sigma

\section{Historical Considerations: Current Algebra and the Linear Sigma Model \label{sec:currentalgebra}}
%Intro
The modern perspective on chiral effective theory finds its historical origins in a body of work by Gell-Mann, Weinberg, and many others developed throughout the 1960s and 1970s attempting to understand the low energy properties of QCD.
In Weinberg's own words, ``It [EFT] all started with current algebra'' \cite{Weinberg:2009bg}.
Gell-Mann introduced the framework of current algebra \cite{PhysRev.125.1067} wherein properties of hadronic matrix elements are derived by considering the algebraic properties of the U$(3)\times$U$(3)$ Noether currents.  
Many results from the current algebra approach can be reproduced with chiral perturbation theory ($\chi$PT), but it is interesting to review the current algebra approach here both for the historical purpose of understanding how and why chiral effective theory was developed and also to highlight the extent to which it was possible to make progress without the tools of effective Lagrangians.
Indeed, current algebra and related topics were part of a large effort to develop a non-Lagrangian approach to particle physics based on the analyticity properties of the S-matrix \cite{Eden:98637,Pietschmann:2011rh}.

Aspects of the current algebra approach are reviewed by Scherer and Schindler \cite{Scherer:2005ri}.
Chiral currents couple to hadronic states with the appropriate quantum numbers.
For example, the axial vector current couples to the pion state with an amplitude $F_\pi$ defined through the matrix element
\begin{equation}
\langle 0| A^\mu_i(x) |\pi_j(\vec{p}) \rangle = i \delta_{ij}\sqrt{2} F_\pi p^\mu e^{-ipx} \label{eq:Fpi}
\end{equation}
$F_\pi$ is the pion decay constant also appearing in the coupling of the pions to the weak gauge bosons.
Matrix elements of chiral currents with other local operators are related by generalized Ward Identities \cite{Peskin:1995ev}.
\begin{flalign}
\partial_{\mu}^x & \langle 0 | T j^{\mu}(x) O_1(x_1)...O_n(x_n) | 0 \rangle = \nonumber \\
& \sum_{m=1}^n \delta(x^0 - x_m^0) \langle 0 |T O_1(x_1)...O_{m-1}(x_{m-1}) [j^0(x),O_m(x_m)] O_{m+1}(x_{m+1}) ... O_n(x_n) | 0 \rangle = \nonumber \\
-i& \sum_{m=1}^n \delta^{(4)}(x-x_m) \langle 0 | T O_1(x_1)...O_{m-1}(x_{m-1}) (\delta O_m(x_m)) O_{m+1}(x_{m+1}) ... O_n(x_n) | 0 \rangle \label{eq:wardid}
\end{flalign}
which are themselves related to S-matrix elements through the LSZ reduction formula.
Eq.~\ref{eq:wardid} is the Schwinger-Dyson equation corresponding to the classical current conservation equation $\partial_{\mu} j^{\mu} = 0$.
In the case $\partial_{\mu} j^{\mu} \neq 0$, an additional term $\langle 0 | T (\partial_\mu j^\mu) O_1 ... O_n | 0 \rangle$ appears on the right hand side.
The contact terms on the right hand side of Eq.~\ref{eq:wardid} contain the operators $\delta O_m$ which are the infinitesimal changes in the operators $O_m$ under the symmetry transformation corresponding to the current $j^\mu$.
\begin{equation}
[Q,O_m(x)] = i \delta O_m(x) \quad , \quad Q = \int d^3\vec{x} j^0(x)
\end{equation}
The algebraic properties of the chiral symmetry currents enter through these commutators into the chiral Ward identities.

Let us review one result from current algebra that will be important in the discussion that follows, the Gell-Mann-Oakes-Renner (GMOR) relation \cite{PhysRev.175.2195}.
In this example as with many of the current algebra results, some assumption has to be made on top of the algebraic properties of the currents.  
For GMOR, one takes as an ansatz the partially conserved axial current (PCAC) relationship, originally introduced by Goldberger and Treiman \cite{PhysRev.110.1178}.
\begin{equation}
\partial_{\mu}A^{\mu}_i(x) =  (m_u + m_d) P_i(x) \label{eq:PCACsu2}
\end{equation}
where $P_i$ is the pseudoscalar density.
It is interesting to note that in some cases the justification for the assumptions made in current algebra calculations -- such as the PCAC assumption that feeds into the GMOR relation -- were justified at the time by their exemplification in popular models such as the linear and nonlinear sigma models \cite{GellMann:1960np}.
We will come back to discuss some of these specific models later in the chapter.

With knowledge of the underlying gauge theory (or, more simply, the corresponding free quark theory), the PCAC relation may be derived at the quark level through a Noether analysis.  
For a general mass matrix in an $N_f$ flavor gauge theory, the generalized PCAC relation becomes
\begin{equation}
\partial_{\mu} A^{\mu}_i = i \bar{\psi} \gamma_5 \{T_i,\mathcal{M}\} \psi = 2 m_0 P_i +  m_j \left(\frac{2}{N_f} \delta_{ij} P^0 + d_{ijk} P_k \right) \label{eq:PCAC}
\end{equation}
where in the second equality we have decomposed the mass matrix as $\mathcal{M} = m_0 \identity + m_i T_i$; $T_i$ are the generators of $SU(N_f)$ normalized such that Tr$(T_i T_j) = \delta_{ij}$ whose symmetric structure constants are normalized such that $\{T_i,T_j\} = 2/N_f \delta_{ij} \identity + d_{ijk} T_k$.
For convenience, a collection of Lie algebra identities for $su(N)$ are given in Appendix~\ref{appendix:lie}.
The expressions for the axial current and pseudoscalar density at the quark level are 
\begin{equation}
A^{\mu}_i(x) = \bar{\psi}(x) \gamma^\mu \gamma_5 T_i \psi(x) \quad , \quad P_i(x) = i \bar{\psi}(x) \gamma_5 T_i \psi(x)
\end{equation}
In what follows, we will specialize to the case of degenerate quarks $\mathcal{M} = m_q \identity$.

We have already defined the amplitude $F_\pi$ through the matrix element in Eq.~\ref{eq:Fpi}.
Analogously, we define $G_\pi$ through the matrix element of the pseudoscalar density between the vacuum and the one pion state.
\begin{equation}
\langle 0|  P_i(x) |\pi_j(\vec{p}) \rangle = \delta_{ij} \sqrt{2} G_\pi e^{-ipx} \label{eq:Gpi}
\end{equation}
It follows from Eqs.~\ref{eq:Fpi},~\ref{eq:PCAC}, and~\ref{eq:Gpi} that
\begin{equation}
\partial_\mu \langle 0 | A^{\mu}_i (0) | \pi_j(\vec{p})\rangle =  \delta_{ij}\sqrt{2} M_{\pi}^2 F_\pi = \delta_{ij} 2\sqrt{2} m_q G_\pi
\end{equation}
and therefore
\begin{equation}
M_\pi^2 = 2m_q \frac{G_\pi}{F_\pi}
\end{equation}
This expression is exact for all $m_q$, following only from the PCAC assumption.

The amplitude $G_\pi$ is related to the quark condensate through the chiral algebra.  
Consider the matrix element 
\begin{equation}
\langle 0 | [A^{\mu=0}_i(\vec{x},t),P_j(\vec{y},t)] | 0 \rangle 
\end{equation}
which appears on the RHS of the axial current Ward identity.
We may insert a complete set of single particle states (assuming pion pole dominance in the pseudoscalar channel)
\begin{equation}
\langle 0 | [A^{0}_i(\vec{x},t),P_j(\vec{y},t)] | 0 \rangle = \sum_l \int \frac{d^3 \vec{k}}{(2\pi)^3 2 |k^0|} \langle 0 | A^{0}_i(\vec{x},t) | \pi_l(\vec{k}) \rangle \langle \pi_l(\vec{k}) | P_j(\vec{y},t) | 0 \rangle - \text{h.c.}
\end{equation}
Inserting the expressions Eq.~\ref{eq:Fpi} and Eq.~\ref{eq:Gpi} and performing the integration over momenta one finds
\begin{equation}
\langle 0 | [A^{0}_i(\vec{x},t),P_j(\vec{y},t)] | 0 \rangle = \delta_{ij} 2 i F_\pi G_\pi \delta^{(3)}(\vec{x} - \vec{y}) \label{eq:commutator1}
\end{equation}
On the other hand, one may compute the commutator directly using the chiral algebra.
\begin{equation}
[A^{0}_i(\vec{x},t),P_j(\vec{y},t)] = -i \delta^{(3)}(\vec{x} - \vec{y}) \bar{\psi}(x) \{T_i,T_j\} \psi(x) = -i\delta^{(3)}\delta(\vec{x} - \vec{y})\left( \frac{2}{N_f} \delta_{ij} S^0 + d_{ijk} S_k \right)
\end{equation}
We have defined the scalar densities $S_0 = \bar{\psi}\identity \psi$ and  $S_i = \bar{\psi}T_i \psi$.
The v.e.v. of each quark flavor is equal at $m_q = 0$: $\langle \bar{\psi}_f \psi_f \rangle = -\nu + \mathcal{O}(\mathcal{M}) \quad \forall f$. 
So, $\langle S^0 \rangle = - N_f \nu$ and $\langle S_i\rangle = 0$ up to $\mathcal{O}(\mathcal{M})$ corrections.
\begin{equation}
[A^{0}_i(\vec{x},t),P_j(\vec{y},t)] = 2 i \nu \delta_{ij} \delta^{(3)}(\vec{x}-\vec{y}) + \mathcal{O}(\mathcal{M}) \label{eq:commutator2}
\end{equation}
Comparing Eq.~\ref{eq:commutator1} and Eq.~\ref{eq:commutator2} we find
\begin{equation}
G_\pi F_\pi = \nu + \mathcal{O}(\mathcal{M}) \label{eq:FpiGpi}
\end{equation}
which gives the GMOR relation for the case of $N_f$ degenerate quarks.
\begin{equation}
M_\pi^2 F_\pi^2 = 2 m_q \nu + \mathcal{O}(\mathcal{M}^2) \label{eq:GMOR}
\end{equation}

Having shown one explicit example, let us quickly review some of the other important results from current algebra without proof in order to demonstrate the scope of the technique.  Carrying out the above derivation for nondegenerate quarks, other interesting relationships between the Goldstone masses may be calculated.  For example, in the case of SU$(3)$ in the isospin limit, $m_u = m_d \neq m_s$, one may derive the Gell-Mann Okubo relationship \cite{osti_4008239,doi:10.1143/PTP.27.949,doi:10.1143/PTP.28.24} between the masses of the mesons in the pseudoscalar octet. 
\begin{equation}
4 M_K^2 = 3 M_\eta^2 + M_\pi^2
\end{equation}
Weinberg showed that one may derive matrix elements for the scattering of soft pions off of other particles and the matrix elements for pion-pion scattering using only PCAC and current algebra relations \cite{PhysRevLett.17.616}.
Using a combination of dispersion relations, PCAC, and current algebra, Alder \cite{Adler:1965ka} and Weisberger \cite{Weisberger:1965hp} based on the work of Fubini \cite{Fubini:1964boa} computed the ratio of the axial vector current coupling to the nucleon to the vector current coupling to the nucleon which is a crucial quantity in beta decay \cite{Pietschmann:2011rh}.
Another example is the  Kawarabayashi Suzuki Riazuddin Fayyazuddin (KSRF) relations \cite{Kawarabayashi:1966kd,Riazuddin:1966sw}
\begin{equation}
F_\rho = \sqrt{2}F_\pi \quad, \quad g_{\rho\pi\pi} = \frac{M_\rho}{\sqrt{2}F_\pi}
\end{equation}
which may be derived using current algebra, PCAC, and vector meson dominance in the p-wave $\pi$-$\pi$ scattering channel \cite{JOHNSON1970247}.

In our recapitulation of the derivation of the GMOR formula, we have demonstrated that calculations in the current algebra framework are somewhat labor intensive and typically require a sequence of assumptions and approximations that are not always obvious or straightforward.
Effective field theory techniques provide a streamlined approach to deriving certain current algebra relationships.
However, the relationships that are derivable in the context of an EFT are typically limited to relationships between states appearing as dynamical degrees of freedom in the EFT.
The EFT makes no predictions about heavy resonances omitted from the construction.
As such, the current algebra techniques are still useful, as exemplified by the Adler-Weisberger sum rule and the KSRF relationships that we have briefly discussed above.

%One may then consider the Axial current Ward identity at zero quark mass with an insertion of a pseudoscalar density.
%\begin{equation}
%\partial_{\mu} \langle 0 | A^\mu_a(x) P_b(y) | 0 \rangle = -i \delta_{ab}\delta(x-y) \langle 0 | \bar{\psi}(x)\psi(x) | 0 \rangle 
%\end{equation}
%This follows from the fact that $A^\mu_a$ is conserved at $m_q = 0$ and the current algebra relation $[A^0_a,P_b] = i\delta_{ab}S$ where $S=\bar{\psi}\psi$ is the scalar density.
%GMOR
%Gell-Mann Okubo
%Goldberger Treiman
%Pion Scattering Length
%Weinberg Sum Rules?
The ansatz of PCAC was motivated by its exemplification in popular models at the time \cite{GellMann:1960np} including the linear and nonlinear sigma models.  
The linear sigma model was introduced originally by Schwinger \cite{Schwinger:1957em}.
%It is an interesting twist of fate that these same theories which motivated the assumptions of current algebra, which in turn motivated the development of effective field theory, may themselves be promoted to effective field theories.
As an introduction to models of chiral symmetry breaking, let us briefly review the simple case of the SU$(2)\times$SU$(2)/$SU$(2)$ (or equivalently $O(4)/O(3)$; we will demonstrate this equivalence later) linear sigma model which is the prototypical example of a theory with spontaneous symmetry breaking \cite{Peskin:1995ev,Scherer:2005ri,Weinberg:1996kr}.
The linear sigma model fields transform in a (bi)linear $(2,\bar{2})$ representation of the full group $\SU_L(2)\times\SU_R(2)$.
\begin{equation}
M_a^{\bar{b}} \rightarrow L_a^c M_c^{\bar{d}} (R^\dagger)_{\bar{d}}^{\bar{b}}
\end{equation}
where $L \in \SU_L(2)$ and $R \in \SU_R(2)$.  
We will sometimes use the notation $(L,R)\circ M = L M R^\dagger$ for the action of the group.  
Rather than independent left and right transformations, we may equivalently consider vector transformations $(T,T) \in \SU_V(2)$ and axial transformations $(T,T^\dagger) \in \SU_A(2)$, where $T \in \SU(2)$.
Indices will be suppressed in the remainder of the discussion wherever possible; when explicit indices on $M(x)$ are warranted, we will use the letters $a,b,c,...$ for fundamental indices to distinguish from $i,j,k,...$ adjoint indices.   
The renormalizable linear sigma Lagrangian containing all operators invariant under the global group up to dimension four is
\begin{equation}
\mathcal{L} = \frac{1}{2}\langle \partial_{\mu} M^\dagger \partial^{\mu}M \rangle - \frac{\mu^2}{2}\langle M^\dagger M \rangle - \frac{\lambda_1}{4} \langle M^\dagger M \rangle^2 - \frac{N_f \lambda_2}{4}\left\langle \left(M^\dagger M\right)^2\right\rangle
\end{equation}
where $\langle ... \rangle$ denotes the trace.
%Choosing to study the real linear representation of $\SU_L(2)\times\SU_R(2)$, we may express the four real degrees of freedom by decomposing the matrix field in terms of Hermitian generators.

We may express the field degrees of freedom in a conventional form by expanding in a basis of Hermitian matrices
\begin{equation}
M(x) = \frac{\sigma(x)}{\sqrt{N_f}} + i \pi_i T_i \label{eq:SU2linear}
\end{equation}
where $T_i = (1/\sqrt{2})\sigma_i$ are the generators of $\SU(2)$ normalized such that $\langle T_i T_j \rangle = \delta_{ij}$.
For the case of $\SU(2)\times\SU(2)$, we may take $\sigma,\pi_i \in \mathbb{R}$ and the representation is closed under the full group.
This follows from the special property of Pauli matrices that they have a simple anticommutator, or equivalently that the symmetric structure constants are all zero for $\SU(2)$.
For $N_f > 2$ we must take $\sigma,\pi_i \in \mathbb{C}$ for the representation to close.
To see this, consider the infinitesimal vector and axial transformations of the fields parametrized as in Eq.~\ref{eq:SU2linear}.
Under a vector transformation, $(T,T) \in \SU_V(N_f)$ with $T = \exp(i \theta_i T_i)$,
\begin{equation}
\delta_V M(x) = i \theta_i [T_i,M(x)] = - \theta_i \pi_j [T_i,T_j] = -i \theta_i \pi_j f_{ijk} T_k
\end{equation}
In terms of the component fields, 
\begin{equation}
\delta_V \sigma = 0 \quad , \quad \delta_V \pi_k = -\theta_i\pi_j f_{ijk}
\end{equation}
The variation in the pions is real because the structure constants are real.
So, the real valued pion and sigma fields form a closed representation of $\SU_V(N_f)$ for any $N_f$.
The sigma transforms as a singlet and the pions transform as an adjoint.
Now consider the axial transformations
\begin{equation}
\delta_A M(x) = i \theta_i \{T_i,M(x)\} = 2 i \frac{\sigma}{\sqrt{N_f}}T_i - \frac{2}{N_f}\theta_i \pi_i \identity - \theta_i \pi_j d_{ijk} T_k
\end{equation}
In terms of the component fields
\begin{equation}
\delta_A \sigma = - \frac{2}{\sqrt{N_f}}\theta_i \pi_i \quad , \quad \delta_A \pi_k \frac{2 \sigma}{\sqrt{N_f}} \theta_k + i \theta_i \pi_j d_{ijk}
\end{equation}
The variation of sigma is real, however for nonzero symmetric structure constants $d_{ijk}$, the pions necessarily pick up a complex contribution from the $\SU_A(N_f)$ rotation.
%implying that the representation will always close for real sigma.
%Indeed, it is only the $\text{U}_A(1)$ transformations that necessitate the inclusion of the complex component of $\sigma$, which is the $\eta'$ degree of freedom.
Therefore, for $N_f=2$ we may take the sigma and pion components to be real, but for $N_f > 2$ we must take the sigma and pion fields to be complex for the representation to close.
In Section~\ref{sec:LSM}, we will discuss the case for general $N_f$ in detail.
We refer to the field basis in Eq.~\ref{eq:SU2linear} as the \emph{linear field basis}.
It is analogous to choosing Cartesian coordinates for field space.
Later we will discuss a \emph{nonlinear basis} for the fields, which is analogous to describing field space in polar coordinates.
%Of course, the field $M(x)$ in Eq.~\ref{eq:SU2linear} is complex valued, but this is a mathematical convenience which allows us to simplify calculations by making use of the Pauli matrix algebra.  

Let us continue the discussion for the case of the real linear representation of $\SU(2)\times\SU(2)$.
Following from the special property of $\SU(2)$ that the generators satisfy $\{T_i,T_j\} = \delta_{ij}\identity$, one may show that the single and double trace quartic operators are not independent: $\left\langle \left(M^\dagger M\right)^2\right\rangle = (1/2)\langle M^\dagger M \rangle^2$.
Redefining the quartic coupling $\lambda = \lambda_1 + \lambda_2$ the Lagrangian may be written purely in terms of bilinear traces.
\begin{equation}
\mathcal{L} = \frac{1}{2}\langle \partial_{\mu} M^\dagger \partial^{\mu}M \rangle - \frac{\mu^2}{2}\langle M^\dagger M \rangle - \frac{\lambda}{4} \langle M^\dagger M \rangle^2 \label{eq:LSMSU2}
\end{equation}
This is much more reminiscent of the $O(4)$ linear sigma model, and for good reason: this action is equivalent to the $O(4)$ linear sigma model.  
Defining the components of an $O(4)$ multiplet as $\vec{\phi} = (\pi_1,\pi_2,\pi_3,\sigma)$, it follows immediately that $\langle M^\dagger M \rangle = \vec{\phi} \cdot \vec{\phi} = \sigma^2 + \vec{\pi}^2$.
The equivalence of the two actions follows from the local isometry between the groups $O(4) \rightarrow O(3)$ and $\SU(2)\times\SU(2) \rightarrow \SU(2)$.
Indeed, it is exactly this isometry that leads to the existence of a real linear representation of $\SU(2)\times\SU(2)$.  
The global chiral symmetry for a larger number of flavors $N_f > 2$ will not in general be isomorphic to an orthogonal group and will only admit complex linear representations.
We will study the general $\SU(N_f) \times \SU(N_f)$ theory in detail in Section~\ref{sec:LSM}.

Let us continue to study Eq.~\ref{eq:LSMSU2} in more detail.
The extrema of the potential are given by 
\begin{equation}
\frac{\partial V}{\partial M^\dagger_{ab}} = \frac{1}{2}\left( \mu^2 + \lambda \langle M^\dagger M \rangle \right) M_{ba} = 0
\end{equation}
For a negative mass term $\mu^2 < 0$ the stable minima are displaced from the origin of field space, and the vacuum becomes nontrivial.
The minimum of the potential is given by
\begin{equation}
\langle M^\dagger M \rangle = \frac{-\mu^2}{\lambda} \equiv f^2
\end{equation}
In this phase, the vacuum breaks the chiral symmetry down to the $\SU_V(2)$ subgroup.
Vector transformations preserve the vacuum while the axial transformations rotate the vacuum.
Let us take the vacuum to be oriented in the direction of the trace (the $\sigma$ direction):  $\langle 0 | M | 0 \rangle = (f/\sqrt{2})\identity \equiv M_0$.
This vacuum is invariant under a vector transformation, but transforms nontrivially under the axial transformations, $(T,T^\dagger) \in \SU_A(2)$ where $T = \exp(i\theta_i T_i) \in \SU(2)$
\begin{equation}
(T,T^\dagger)\circ M_0 = T M_0 T = e^{2 i \theta_i T_i} \frac{f}{\sqrt{2}} \identity \approx \frac{f}{\sqrt{2}}\left( \identity + 2 i \theta_i T_i \right) = M_0 + \delta_A M_0
\end{equation}
The variation in the v.e.v.  under axial transformations has components in the pion directions.
\begin{equation}
\delta_A M_0 = i \sqrt{2} f \theta_i T_i \label{eq:SU2infaxial}
\end{equation}
%Written in components,
%\begin{equation}
%\delta_A \sigma_0 = 0 \quad,\quad \delta_A (\pi_i)_0 = \sqrt{2} f \theta_i
%\end{equation}

Consider the variation in the potential at the minimum $M_0$ due to an infinitesimal axial rotation of the v.e.v.
%\begin{flalign}
%V(M_0 + \delta_A M_0) &= V(M_0) + \sum_{ij} \left.\frac{\partial V}{\partial M_{ij}}\right|_{M_0} (\delta_A M_0)_{ij} \nonumber\\
%& \quad + \frac{1}{2!}\sum_{ijkl} \left.\frac{\partial^2 V}{\partial M_{ij}\partial M_{kl}}\right|_{M_0} (\delta_A M_0)_{ij}(\delta_A M_0)_{kl} + \mathcal{O}(\delta_A M_0^3) %\nonumber\\
%&= V(M_0) + \frac{1}{2} \sum_{ijkl}\mathcal{M}^2_{ij,kl}  (\delta_A M_0)_{ij}(\delta_A M_0)_{kl} + \mathcal{O}(\delta_A M_0^3)
%\end{flalign} 
\begin{flalign}
V(M_0 + \delta_A M_0) &= V(M_0) + \sum_{ab} \left.\frac{\partial V}{\partial M_{ab}}\right|_{M_0} (\delta_A M_0)_{ab} \nonumber\\
& \quad + \frac{1}{2!}\sum_{abcd} \left.\frac{\partial^2 V}{\partial M_{ab}\partial M_{cd}}\right|_{M_0} (\delta_A M_0)_{ab}(\delta_A M_0)_{cd} + \mathcal{O}(\delta_A M_0^3) \nonumber\\
&= V(M_0) + \frac{1}{2} \sum_{abcd}\mathcal{M}^2_{ab,cd}  (\delta_A M_0)_{ab}(\delta_A M_0)_{cd} + \mathcal{O}(\delta_A M_0^3)
\end{flalign} 
The linear term vanishes because $M_0$ is an extremum of the potential, and we have defined the mass matrix $\mathcal{M}^2_{ab,cd} $ as the second derivative of the potential evaluated at the minimum.
The symmetry of the potential under the full group implies that $V(M_0 + \delta_A M_0) = V(M_0)$, and therefore
\begin{equation}
 \sum_{abcd}\mathcal{M}^2_{ab,cd}  (\delta_A M_0)_{ab}(\delta_A M_0)_{cd} = 0 \label{eq:pregoldstone}
\end{equation}
Or equivalently, the mass matrix must satisfy
\begin{equation}
\mathcal{M}^2\delta_A M_0 = 0 \label{eq:goldstone}
\end{equation}
where the indices have been suppressed.
This is Goldstone's theorem  \cite{Goldstone:1961eq,Nambu:1960tm,Peskin:1995ev,Scherer:2005ri,Weinberg:1996kr}. 
Each nonzero component of $\delta_A M_0$ arises from a generator of a symmetry under which the vacuum is not invariant.
Eq.~\ref{eq:goldstone} states that each independent broken symmetry generator is in 1:1 correspondence with a zero eigenvalue of the mass matrix in the broken phase.
In our example, the three nonzero components of $\delta_A M_0$ correspond to the generators of $\SU_A(2)$, and necessitate three massless eigenvalues of the mass matrix.
We confirm this by writing out the Lagrangian in the broken phase, $M \rightarrow M_0 + M$.
\begin{equation}
\mathcal{L} = \frac{1}{2} (\partial \sigma)^2 + \frac{1}{2} (\partial \pi_i)^2 - \frac{m_\sigma^2}{2}\sigma^2 - \frac{m_\sigma^2}{8f^2}(\sigma^2 + \pi_i^2)^2 - \frac{m_\sigma^2}{2 f}\sigma(\sigma^2 + \pi_i^2)
\end{equation}
We have defined the scalar mass in the broken phase $m_\sigma^2 = -2 \mu^2$.  
Indeed, we have an adjoint multiplet of three Goldstone pions corresponding to the three spontaneously broken axial generators.
Expressing Eq.~\ref{eq:goldstone} explicitly in the basis $(\sigma,\vec{\pi})$, the Eigenvalue equation which determines the mass of NGBs is
\begin{equation}
\left( \begin{array}{cccc} 
m_\sigma^2 & 0 & 0 & 0 \\ 
0 & 0 & 0 & 0 \\ 
0 & 0 & 0 & 0 \\ 
0 & 0 & 0 & 0 
\end{array} \right) \left(\begin{array}{c} 0 \\ \sqrt{2} f \theta_1 \\ \sqrt{2}f\theta_2 \\ \sqrt{2}f\theta_3 \end{array} \right) = \left(\begin{array}{c} 0 \\ 0 \\ 0 \\ 0 \end{array} \right)
\end{equation}
and holds for all $\theta_1$, $\theta_2$, and $\theta_3$ as it must.  
This discussion has been carried out for the classical potential, but it may be straightforwardly generalized to the quantum case by replacing the classical potential with the quantum effective potential \cite{Weinberg:1996kr}, so long as the effective potential is invariant under the same symmetry.

To make the connection to our earlier discussion of PCAC and the GMOR formula, we may introduce a term into the linear sigma model potential which explicitly breaks the $\SU_A(2)$ symmetry which will give a small mass to the pions.
For the purposes of this discussion, we introduce the simple breaking potential
\begin{equation}
V \rightarrow V + V_{\text{SB}} =  V - b\langle M + M^\dagger \rangle \label{eq:SU2breaking}
\end{equation}
which is symmetric under $\SU_V(2)$ but breaks $\SU_A(2)$ explicitly.
We will discuss breaking potentials in more detail in Sections~\ref{sec:XPT} and~\ref{sec:LSM}. 
In the presence of the breaking potential, the equation for the extrema of the potential is
\begin{equation}
\frac{\partial V}{\partial M^\dagger_{ab}} = \frac{1}{2}\left( \mu^2 + \lambda \langle M^\dagger M \rangle \right) M_{ba} - b \delta_{ab} = 0
\end{equation}
We may again choose the v.e.v. to be oriented along the $\sigma$ direction, $M_0(b) = F(b)/\sqrt{2} \identity$, where $F(b)$ satisfies
\begin{equation}
\frac{F}{2\sqrt{2}} \left (\mu^2 + \lambda F^2 \right) - b = 0 
\end{equation}
While this equation is exactly solvable, it is sufficient for our discussion to work to first order in $b$.  
Then the solution for the v.e.v. in the presence of the breaking potential is $F(b) \approx f + 2\sqrt{2} b / m_\sigma^2$.

To examine the PCAC relation Eq.~\ref{eq:PCACsu2}, we first write down the axial vector current derived through the Gell-Mann-Levy method \cite{GellMann:1960np} (or the usual Noether procedure \cite{Peskin:1995ev}).
\begin{equation}
A^{\mu}_i = \frac{i}{2}\left\langle (\partial^{\mu} M^\dagger) \{T_i,M\} - \{T_i,M^\dagger\} \partial^\mu M \right\rangle \approx \sqrt{2} F \partial^\mu \pi_i
\end{equation}
In the second approximate equality, we have expanded the expression for the current to first order in the fields.
For the action defined in Eq.~\ref{eq:LSMSU2} without a breaking term, the axial current is exactly conserved $\partial_\mu A^{\mu}_i = 0$.
In the presence of the breaking potential Eq.~\ref{eq:SU2breaking}, the axial vector current is no longer conserved.
\begin{equation}
\partial_{\mu} A^{\mu}_i = 4 b  \pi_i = 4 b P_i
\end{equation}
where $P_i = \pi_i$ is the pseudoscalar density in the linear sigma model.
The divergence of the axial vector current is proportional to the pseudoscalar density.
This is the realization of the PCAC relation in the linear sigma model in Eq.~\ref{eq:PCACsu2}.
Computing the matrix elements $F_\pi$ and $G_\pi$ defined in Eq.~\ref{eq:Fpi} and Eq.~\ref{eq:Gpi}, we find $F_\pi = F$ and $G_\pi = 1/\sqrt{2}$ respectively.
From Eq.~\ref{eq:FpiGpi}, we can deduce the expression for the chiral condensate in the linear sigma model, $G_\pi F_\pi = F/\sqrt{2} = \nu$.
Finally, the GMOR relation dictates that the pion mass should be given by $M_\pi^2 = 2\sqrt{2} b / f + \mathcal{O}(b^2)$.

Now let us check the GMOR prediction by explicitly studying how the Goldstone bosons acquire a mass in the presence of the breaking potential.
Once the symmetry has been explicitly broken, it is no longer the case that $V(M + \delta_A M) = V(M)$.
Eq.~\ref{eq:pregoldstone} becomes
\begin{flalign}
\frac{1}{2} \sum_{abcd}\mathcal{M}_{ab,cd}  (\delta_A M_0)_{ab}(\delta_A M_0)_{cd} &= V_{\text{SB}}(M_0 + \delta_A M_0) - V_{\text{SB}}(M_0) \nonumber \\
&\approx \sum_{ab} \left.\frac{\partial V_{\text{SB}}}{\partial M_{ab}}\right|_{M_0} (\delta_A M_0)_{ab}
\end{flalign}
Differentiating both sides with respect to the fluctuation, we find the matrix equation for pseudo-Nambu-Goldstone bosons (PNGBs) corresponding to Eq.~\ref{eq:goldstone}.
\begin{equation}
\mathcal{M}^2 \delta_A M_0 = \left.\frac{\partial V_{\text{SB}}}{\partial M}\right|_{M_0} \label{eq:pseudogoldstone}
\end{equation}
It's important to notice that $M_0$ has itself been affected by the presence of the breaking potential, $M_0 \rightarrow M_0(b)$, and Eq.~\ref{eq:pseudogoldstone} is evaluated at the shifted v.e.v. $M_0(b)$.

Explicitly expanding the linear sigma potential with the breaking term about the v.e.v. $M_0(b) = F(b)/\sqrt{2} \identity$, we find that the masses of the sigma and pions are given by
\begin{equation}
M_\sigma^2 = \frac{m_\sigma^2}{2}\left(3 \frac{F^2}{f^2} - 1 \right) \approx m_\sigma^2 + 6\sqrt{2} b/f \quad , \quad M_\pi^2 = \frac{m_\sigma^2}{2}\left(\frac{F^2}{f^2}-1\right) \approx 2\sqrt{2}b/f
\end{equation}
The pion mass agrees with the GMOR expression at $\mathcal{O}(b)$.
The variation in the breaking potential from an axial transformation is
\begin{equation}
\delta_A V_{\text{SB}} = 4 b \theta_i \pi_i
\end{equation}
Then Eq.~\ref{eq:pseudogoldstone} in the $(\sigma,\vec{\pi})$ basis reads.
\begin{equation}
\left( \begin{array}{cccc} 
m_\sigma^2 + 6\sqrt{2}b/f & 0 & 0 & 0 \\ 
0 & 2\sqrt{2}b/f & 0 & 0 \\ 
0 & 0 & 2\sqrt{2}b/f & 0 \\ 
0 & 0 & 0 & 2\sqrt{2}b/f 
\end{array} \right) \left(\begin{array}{c} 0 \\ \sqrt{2} F \theta_1 \\ \sqrt{2}F\theta_2 \\ \sqrt{2}F\theta_3 \end{array} \right) = \left(\begin{array}{c} 0 \\ 4b\theta_1 \\ 4b\theta_2 \\ 4b\theta_3 \end{array} \right)
\end{equation}
We find that the explicit computation of the PNGB masses in the linear sigma model is in agreement with the general PNGB formula Eq.~\ref{eq:pseudogoldstone} up to the order that we have worked $\mathcal{O}(b)$.

In conclusion, we have discussed some general properties of theories with spontaneously broken global chiral symmetries.
Using only the PCAC relation Eq.~\ref{eq:PCACsu2} and the algebraic relationships amongst the chiral Noether currents, one may derive the GMOR relation for the PNGB masses Eq.~\ref{eq:GMOR} as we have shown explicitly.
We have summarized some of the other key results from current algebra studies including the Gell-Mann Okubo mass formula and Weinberg's derivation of the pion scattering amplitudes.
We introduced the $\SU(2)\times\SU(2)$ linear sigma model, and showed that it provides an explicit realization of these general features, including PCAC, the GMOR relation, and Goldstone's theorem.
This discussion has also served as an introduction to the renormalizable linear sigma model for the simpler case of $N_f=2$ flavors.
In Section~\ref{sec:LSM}, we will discuss a generalized linear sigma model for $N_f$ flavors as an effective field theory.
%General EFT remarks.  Keep it short.  ~4 sentences.  Cite Weinberg, phenomenological lagrangians.

\section{Nonlinear Realizations of Chiral Symmetry and Chiral Effective Field Theory \label{sec:XPT}}
%\section{Nonlinear Realizations of Chiral Symmetry Breaking \label{sec:XPT}}
%Can start with the gauge theory and build the theory up that way
In the second half of Section~\ref{sec:currentalgebra}, we studied the $\SU(2)\times\SU(2)$ linear sigma model whose Lagrangian contained only three (four) operators in the spontaneously broken (explicitly and spontaneously broken) case.
Implicit in our discussion was an assumption of weak coupling, $\lambda \ll 1$ or $m_\sigma^2 \ll (4\pi f)^2$, which allowed us to study the theory at tree level and omit loop effects.
In addition, we did not introduce any higher dimensional operators into our linear sigma Lagrangian.  
The linear sigma model is renormalizable, so there is no mathematical obstruction to taking the cutoff arbitrarily large (or even to infinity, removing the cutoff).  
Yet if we apply the linear sigma Lagrangian as a model of the low lying states of a confining gauge theory, there must be an infinite tower of higher dimensional operators included to encapsulate the effects of the heavy states that have been integrated out of the theory.
Including these higher dimensional operators promotes the linear sigma model from a renormalizable quantum field theory to an effective field theory.

In this section, we discuss some general considerations for chiral effective field theories.
We begin by studying a particular strong coupling limit of the linear sigma model in which the $\sigma$ particle is decoupled from the theory.
The resulting theory contains only pions and contains an infinite tower of higher dimensional operators.
It is a particular case of a chiral effective field theory which involves only the (pseudo)-Nambu-Goldstone bosons of the spontaneously broken global chiral symmetry.
We will then discuss the general framework for describing Goldstone bosons at low energy known as chiral perturbation theory.
Crucial to this development will be the geometrical nature of Goldstone bosons, which transform nonlinearly under the chiral symmetry.
We will conclude with some discussion of more general effective field theories.

\subsection{Integrating Out the Sigma}
Consider the $\SU(2)\times\SU(2)$ linear sigma Lagrangian introduced in Eq.~\ref{eq:SU2linear} without any explicit chiral breaking term.
Parametrized in terms of the sigma mass $m_\sigma^2$ and v.e.v. $f$, the Lagrangian is
\begin{equation}
\mathcal{L} = \frac{1}{2}\left\langle \partial_\mu M^\dagger \partial^\mu M \right\rangle + \frac{m_\sigma^2}{4}\left\langle M^\dagger M \right\rangle - \frac{m_\sigma^2}{8f^2}\left\langle M^\dagger M \right\rangle^2
\end{equation}
We would like to answer the question whether it is possible to remove the sigma from the theory as its mass becomes very heavy.
Typically for a particle with mass $m^2$ and coupling $y$, we can remove the particle from the theory by taking $m^2 \rightarrow \infty$ while holding $y$ fixed.
Feynman graphs containing the heavy particle on internal lines will be suppressed by factors of the heavy particle propagator.
However, in the case of the linear sigma model, the coupling also grows as $m_\sigma^2 / f^2$, and one might worry that the limit $m_\sigma^2 \rightarrow \infty$ does not exist.
Nonetheless, a careful study of the perturbation theory shows that the contributions that grow with $m_\sigma^2$ all cancel one another, leaving a finite and weakly coupled perturbation theory.

Taking this as a given, we will analyze the $m_\sigma^2 \rightarrow \infty$ limit at the level of the path integral.
In the path integral, the terms in the action proportional to $m_\sigma^2$ will lead to a rapidly oscillating phase as the $\sigma$ mass is taken large.
Stationary phase analysis dictates that the integral is restricted to the region in which the gradient of the integrand phase (the action) with respect to the integration variables (the fields) is zero.
Thus, the limit $m_\sigma^2 \rightarrow \infty$ with $f^2$ held fixed leads to the constraint on the fields,
\begin{flalign}
&\frac{\partial}{\partial M^\dagger_{ba}} \left(\frac{m_\sigma^2}{4}\left\langle M^\dagger M \right\rangle - \frac{m_\sigma^2}{8f^2}\left\langle M^\dagger M \right\rangle^2  \right) = 0 \nonumber \\
&\rightarrow M_{ab} \left( \left\langle M^\dagger M \right\rangle - f^2 \right) = 0
\end{flalign}
The solution $M(x) = 0$ is unstable for $m_\sigma^2 > 0$.  
Therefore, the constraint becomes
\begin{equation}
\left\langle M^\dagger M \right\rangle = \sigma^2 + \vec{\pi}^2 = f^2 \label{eq:sigmaconstraint}
\end{equation}
This is the same condition for the minimum of the potential, which followed from the fact that in this particular case the entire potential was proportional to the parameter being taken to infinity.
The field becomes frozen at the minimum of the potential, but the theory is not trivial because the minimum of the potential has flat directions -- the directions of the Goldstone modes -- and the field remains free to fluctuate in this moduli space.
We apply the constraint by setting $\sigma = f \sqrt{1 - \vec{\pi}^2/f^2}$.
The entire potential is fixed to a constant and can be pulled outside the path integral, and the only nontrivial term in the action is the kinetic term with the constraint applied.
\begin{equation}
\mathcal{L} = \frac{1}{2}\left\langle \partial_{\mu} M^\dagger \partial^\mu M \right\rangle = \frac{1}{2} \left(\partial \vec{\pi}\right)^2 + \frac{1}{2f^2}\frac{\left(\vec{\pi}\cdot \partial \vec{\pi}\right)^2}{1-\left(\vec{\pi}/f\right)^2} \label{eq:nonlinearsigmamodel}
\end{equation}
This is the \emph{nonlinear sigma model}.  
Expanding the denominator of the second operator leads to an infinite tower of pion interaction terms which encapsulate the effects of the sigma particle that has been removed from the theory.
Notice that the nonlinear constraint Eq.~\ref{eq:sigmaconstraint} did not lead to any higher derivative operators; in the tower of pion operators, there are always only two derivatives.
This is due to the fact that we took the strict $m_\sigma^2 \rightarrow \infty$ limit.  
If we had expanded the linear sigma model for large but finite sigma mass, we would find that there are an infinite tower of higher derivative operators, an expansion in $\partial^2 / m_\sigma^2$.  

\subsection{Nonlinear Realizations of Chiral Symmetry}
The nonlinear sigma model may be derived in a more standard form by choosing a basis for our linear sigma field $M(x)$ that makes the geometrical nature of the Goldstone bosons more explicit. 
In such a basis, the Goldstone bosons fields will transform nonlinearly under the chiral symmetry \cite{PhysRev.166.1568}.
These more complicated transformation laws will reflect the true nature of the Nambu-Goldstone bosons as global coordinates on a manifold as we will show.
Weinberg originally introduced the nonlinear representation of chiral symmetry for the pions of the $\SU(2)\times \SU(2)$ invariant theory \cite{PhysRev.166.1568}, but the construction was soon generalized to the case of a general Lie group $G$ breaking spontaneously to a subgroup $H \subset G$ \cite{Coleman:1969sm,Callan:1969sn}.
We will first briefly review some key results for the general case following the discussion in \cite{Weinberg:1996kr} and then return to our example of the $\SU(2)\times \SU(2)$ linear sigma model.

Consider a theory with fields $\psi$ whose Lagrangian is invariant under transformations 
\begin{equation}
\psi \rightarrow g \circ\psi
\end{equation}
for $g \in G$, where $G$ is a compact Lie group.
For this general discussion, we omit indices and denote the group action by ``$\circ$''.
The theory is spontaneously broken such that the vacuum is invariant under only a subgroup $H \subset G$ of the original symmetries.  
We denote the vacuum here as $\langle \psi \rangle$ (not to be confused with our notation for the trace in the preceding discussion).  
Under an infinitesimal G-transformation, the vacuum transforms as
\begin{equation}
g \circ \langle \psi \rangle \approx \langle \psi \rangle + \delta \langle \psi \rangle + \mathcal{O}(\epsilon^2) \label{eq:infg}
\end{equation}
where $\delta\langle\psi\rangle = \mathcal{O}(\epsilon^1)$ and $\epsilon$ is the small parameter in the infinitesimal group transformation. 
Recall from Eq.~\ref{eq:goldstone} that the Nambu-Goldstone modes are exactly the nonzero components of $\delta\langle\psi\rangle$ expressed in an appropriate basis.
We define a parametrization of our field $\psi$ as a G-transformation acting on the field when all Goldstone modes have been set to zero, which we denote by $\tilde{\psi}$.
\begin{equation}
\psi = \gamma \circ \tilde{\psi} \label{eq:nonlingeneral}
\end{equation}
where $\gamma \in G$.
Since $\delta\langle\psi\rangle$ is a general linear combination of the Goldstone modes, the condition that $\tilde{\psi}$ has the Goldstone modes set to zero may be formulated by defining an appropriate G-invariant inner product $(\psi_1,\psi_2) \in \mathbb{R}$ and insisting that $\tilde{\psi}$ is orthogonal to $\delta\langle\psi\rangle$ with respect to this inner product.
\begin{equation}
(\tilde{\psi},\delta\langle\psi\rangle)  = 0
\end{equation}
For the case of $O(N)$, the inner product is a simple dot product, and for the case of fields transforming linearly in a $(N_f,\bar{N_f})$ representation of  $\SU(N_f)\times\SU(N_f)$, the invariant inner product is $(M_1,M_2) = \text{Tr}( M_1^\dagger M_2 )$.
For an infinitesimal G-transformation Eq.~\ref{eq:infg}, the constraint on $\tilde{\psi}$ becomes
\begin{equation}
 (\tilde{\psi},g \circ \langle\psi\rangle) = (\tilde{\psi},\langle\psi\rangle) + \mathcal{O}(\epsilon^2) \quad \forall g\in G \label{eq:psiconsidition}
\end{equation}
The constraint is of course trivial for any G-transformation under which the vacuum is invariant.
Only the G-transformation under which the vacuum is not invariant, the spontaneously broken transformations, give constraints on $\tilde{\psi}$, and as such the number of independent constraints is equal to the number of Goldstone modes.

There is a redundancy in the parametrization of the fields by Eq.~\ref{eq:nonlingeneral}.  
We may reparametrize the fields as
\begin{equation}
\psi = \gamma \circ \identity \circ \tilde{\psi} = (\gamma \circ h) \circ (h^{-1} \circ \tilde{\psi}) = \gamma' \circ \tilde{\psi}'
\end{equation}
where $h \in H$.  Plugging $\psi'$ into Eq.~\ref{eq:psiconsidition}, we see that the condition is still satisfied.
\begin{flalign}
(h^{-1}\circ \tilde{\psi},g \circ \langle\psi\rangle) &= (h^{-1}\circ \tilde{\psi},\langle\psi\rangle) \nonumber\\
(\tilde{\psi},h\circ g \circ \langle\psi\rangle) &= (\tilde{\psi},h \circ \langle\psi\rangle) \nonumber \\
(\tilde{\psi},g' \circ \langle\psi\rangle) &= (\tilde{\psi},\langle\psi\rangle)
\end{flalign}
In the second line, we have used the G-invariance of the inner product.
In the third line, we have used that that vacuum is invariant under H-transformations, and we have used the closure property of the group to write $h\circ g = g' \in G$.
Since Eq.~\ref{eq:psiconsidition} holds for all $g \in G$, it also holds for $\psi'$.  
Therefore, $\gamma$ is only defined up to right multiplication by an element of $H$.
Elements of G which differ only by right multiplication by an element of H may be considered \emph{equivalent} in the mathematical sense.
G may be partitioned into disjoint equivalence classes which are the right cosets and span the quotient space $G/H$.

The see in more detail how one parametrizes the coset space $G/H$, let us introduce generators $t_i$ for the subgroup H satisfying the subalgebra $[t_i,t_j] = i C_{ijk}t_k$.
The remaining generators of G, which are the generators of the coset space $G/H$ are denoted $x_a$.  
The generators together satisfy the commutation relations,
\begin{flalign}
[t_i,t_j] &= i C_{ijk}t_k \\
[t_i,x_a] &= i C_{iab}x_b \\
[x_a,x_b] &= i C_{abi}t_i + i C_{abc}x_c
\end{flalign}
which is the Cartan decomposition of the group G.
Any finite group element $g \in G$ may be expressed as
\begin{equation}
g = e^{i \xi_a x_a}e^{i \theta_i t_i}
\end{equation}
where $e^{i \theta_i t_i} \in H$.  
Since $\gamma \in G$ is only defined up to right multiplication by a group element of H, we may always standardize the definition of $\gamma$ by acting on the right with an H-transformation that sets all the $\theta_i = 0$ and uniquely characterizes each element of the coset space.
\begin{equation}
\gamma(x) = e^{i \xi_a(x) x_a}
\end{equation}
This is the CCWZ parametrization \cite{Coleman:1969sm,Callan:1969sn} of the fields, which together with Eq.~\ref{eq:nonlingeneral} constitutes a nonlinear realization of the global symmetry and a parametrization of the coset space of Goldstone bosons.
While this parametrization is always possible, there are many other choices for how one may parametrize the coset space in a given theory.

Now let us return to our example of the $\SU(2)\times\SU(2)$ linear sigma model.
Rather than the linear parametrization Eq.~\ref{eq:SU2linear}, we now parametrize the degree of freedom of the matrix field by a nonlinear representation of the chiral symmetry.
We write the heavy (non-Goldstone) degrees of freedom ($\tilde{\psi}$) by setting $\vec{\pi} = 0$ in Eq.~\ref{eq:SU2linear}: $\tilde{\psi} = (\sigma(x)/\sqrt{2}) \identity$.  
One can check that the heavy fields defined this way and the v.e.v. $\langle \psi\rangle \rightarrow M_0 = (f/\sqrt{2}) \identity$ satisfy the condition Eq.~\ref{eq:psiconsidition} under an infinitesimal axial transformation Eq.~\ref{eq:SU2infaxial} (as well as trivially under a $\SU_V(2)$ transformation which preserves the vacuum).
Next we reintroduce the Goldstone fields as a $\SU_L(2)\times\SU_R(2)$ transformation acting on the heavy $\sigma$ degree of freedom.
\begin{equation}
M(x) = (L(x),R(x)) \circ \left(\frac{\sigma(x)}{\sqrt{2}}\identity\right) = L(x)R^\dagger(x) \frac{\sigma(x)}{\sqrt{2}}
\end{equation}
Notice that for this example, because the heavy fields $\tilde{\psi}$ are invariant under H-transformations, there is no ambiguity in how one parametrizes the coset space.
\begin{equation}
(L,R) \circ (T,T) \circ (T^\dagger,T^\dagger) \circ \frac{\sigma}{\sqrt{2}}\identity = (LT,RT)\circ \frac{\sigma}{\sqrt{2}}\identity = (LT)(RT)^\dagger \frac{\sigma}{\sqrt{2}}\identity = L R^\dagger \frac{\sigma}{\sqrt{2}}\identity
\end{equation}
Any reparametrization of the Goldstone fields by right multiplication of an H-transformation ($\SU_V(2)$ transformation) cancels when the heavy fields are H-invariant.
Notice also that the quantity $L(x)R^\dagger(x)$ is unitary, and so in general for this simple case, the Goldstone fields are parametrized by a unitary field which we denote by $\Sigma(x)$.
Under G-transformations, the fields $\Sigma(x),\sigma(x)$ transform as
\begin{equation}
g \circ M(x) = (g_L,g_R) \circ (L,R) \circ \frac{\sigma}{\sqrt{2}}\identity = (g_L L ,g_R R) \circ \frac{\sigma}{\sqrt{2}}\identity = g_L \Sigma(x) g_R^\dagger \frac{\sigma}{\sqrt{2}}\identity
\end{equation}
Thus the transformations in terms of the component fields $\Sigma$ and $\sigma$ are
\begin{equation}
\Sigma(x) \rightarrow L \Sigma(x) R^\dagger \quad , \quad \sigma(x) \rightarrow \sigma(x) \label{eq:su2nonlintransform}
\end{equation}
The transformation of the $\Sigma(x)$ field appears linear, but remember that $\Sigma(x)$ is unitary; the individual components of the matrix are constrained to obey $\Sigma^\dagger \Sigma = \Sigma \Sigma^\dagger = \identity$.  
As a result, the $\SU_L(2)\times\SU_R(2)$ transformations act nonlinearly on $\Sigma$.
We can parametrize the unitary matrix field as
\begin{equation}
\Sigma(x) = e^{i\sqrt{2}\pi_i(x)T_i / F_\pi} \label{eq:su2exponential}
\end{equation}
Then the complete nonlinear basis for the $M(x)$ matrix fields takes the form
\begin{equation}
M(x) = e^{i\sqrt{2}\pi_i(x) T_i / F_\pi} \frac{\sigma(x)}{\sqrt{2}}
\end{equation}
In this form, the benefit on the nonlinear field basis is clear.  
The $\sigma(x)$ field parametrizes the radial distance from the origin in field space and the $\pi_i(x)$ fields are the angular rotations of this radial vector.
We emphasize that the field $M(x)$ still transforms linearly under $\SU_L(2)\times \SU_R(2)$, but we have chosen a complicated field basis to parametrize the degrees of freedom of the matrix field in which the individual components $\sigma,\vec{\pi}$ transform in a complicated nonlinear fashion.

In the nonlinear basis, the action Eq.~\ref{eq:LSMSU2} takes the form
\begin{equation}
\mathcal{L} = \frac{1}{2}\left(\partial\sigma\right)^2 + \frac{\sigma^2}{4}\left\langle \partial_{\mu}\Sigma^\dagger \partial^\mu \Sigma \right\rangle + \frac{m_\sigma^2}{4}\sigma^2 - \frac{m_\sigma^2}{8f^2}\sigma^4
\end{equation}
The Goldstones no longer appear explicitly in the potential; however they are not noninteracting.
Indeed, an infinite tower of $\mathcal{O}(\partial^2)$ interaction terms arise from the ``kinetic operator'' for the $\Sigma$ field when one expands the exponent.
In this basis, it is manifest that the pions couple proportional to their momenta.
The constraint Eq.~\ref{eq:sigmaconstraint} when $m_\sigma^2$ is taken to infinity in this basis simply becomes
\begin{equation}
\left\langle M^\dagger M \right\rangle = \sigma^2 = f^2
\end{equation}
So that $M \rightarrow  \Sigma(x)f/\sqrt{2}$ as $m_\sigma^2 \rightarrow \infty$.  
The Lagrangian that is left after the sigma mass is taken to infinity is
\begin{equation}
\mathcal{L} = \frac{f^2}{4}\left\langle \partial_{\mu} \Sigma^\dagger \partial^\mu \Sigma \right\rangle \label{eq:LOXPT}
\end{equation}
This is the conventional form of the leading order chiral Lagrangian for the three Nambu-Goldstone bosons living in the coset space $\SU_L(2)\times \SU_R(2) / \SU_V(2)$.
It is equivalent via a field redefinition to the nonlinear sigma model Lagrangian that we derived in Eq.~\ref{eq:nonlinearsigmamodel}.
To normalize the pion kinetic term in the conventional way, one should set $F_\pi=f$ in this case.
We have derived the Lagrangian by imposing a nonlinear constraint to remove the $\sigma$ state from an otherwise renormalizable field theory.
The resulting Lagrangian is nonrenormalizable.  
The leading order chiral Lagrangian contains an infinite tower of higher dimensional pion operators once one expands the exponent.
At loop level, new divergences will be generated at each loop order requiring an infinite number of counterterms.

\subsection{Effective Field Theory and the Chiral Lagrangian}

The chiral Lagrangian may be alternatively derived by a ``bottom up'' approach, making no assumptions about the UV physics and only considering the symmetry properties of the Goldstone boson fields.  
This is the approach of effective field theory (EFT), originally referred to as ``phenomenological Lagrangians'' \cite{Weinberg:1978kz}.
Effective field theory starts with a ``folk theorem'' by Weinberg which states \cite{Weinberg:1978kz,Weinberg:1996kw}:
\begin{quote}
If one writes down the most general possible
Lagrangian, including all terms consistent with assumed symmetry principles, and then
calculates matrix elements with this Lagrangian to any given order of perturbation theory,
the result will simply be the most general possible S-matrix consistent with perturbative unitarity,
analyticity, cluster decomposition, and the assumed symmetry properties.
\end{quote}
So, to write down an EFT, or a phenomenological Lagrangian, one only needs to determine the fields present in the theory and the symmetry properties of these fields.
One then writes down the Lagrangian containing all possible operators invariant under the symmetries.

There are some details that are not captured in Weinberg's folk theorem.
Typically, an EFT will be used to describe the lowest lying states of a deeper, perhaps unknown theory existing at higher energy scales.
As such, the EFT is naturally cut off because the low energy effective field theory is blind to poles arising in scattering amplitudes due to heavier states that have been omitted from the theory.
Nonetheless, at energies well below the masses of omitted states, the EFT will provide a good approximate description of the matrix elements corresponding to low energies processes of light states.
An EFT construction requires a \emph{separation of scales} so that there is a sufficient range of energies between the masses of the dynamical degrees of freedom in the EFT and the cutoff imposed by the heavier states that have been omitted.
A prototypical example is the Fermi Theory of weak interactions \cite{Fermi:1934sk} which provides a good description of weak processes up to around the scale of the W boson mass.

A second consideration is that there will typically be an infinite number of operators that one can write down consistent with the symmetries, so one must formulate a measure of the relative importance of the operators known as a \emph{power counting rule}.
In a theory with relevant, marginal, and irrelevant operators, one can argue that the irrelevant operators will be suppressed in the infrared relative to the relevant and marginal operators operators, and therefore an ordering can be assigned based on the natural size of operator coefficients arising from RG running at loop level based on the engineering dimension of the operator (assuming no large anomalous dimensions arise).

Chiral Perturbation Theory may be constructed as an EFT of Goldstone bosons living in the coset space $G/H$ of a spontaneously broken global symmetry $G \rightarrow H$.
Let us again focus on the case of $\SU(2)\times\SU(2) \rightarrow \SU(2)$.
We have shown in Eq.~\ref{eq:su2nonlintransform} that the Goldstone bosons are a parametrized by a unitary matrix field $\Sigma(x)$ transforming as $\Sigma \rightarrow L \Sigma R^\dagger$ under the global symmetry.
The symmetry is nonlinearly realized on the $\pi_i(x)$ components defined in Eq.~\ref{eq:su2exponential} due to the constraint of unitarity, $\Sigma^\dagger \Sigma = \identity$.
Because the Goldstone fields are unitary and there are no other fields on the theory, it is only possible to construct nontrivial scalar operators by taking derivatives of the $\Sigma(x)$ field.  
We also insist that the Lagrangian respects parity $\Sigma(t,\vec{x}) \rightarrow \Sigma^\dagger(t,-\vec{x})$.
Taking only these transformation properties into account and insisting that operators in the Lagrangian be invariant under chiral symmetry and under parity, we may write down an infinite number of operators such as
\begin{equation}
\left\langle \partial_\mu \Sigma^\dagger \partial^\mu \Sigma \right\rangle \;,\; \left\langle \partial_\mu \Sigma^\dagger \partial^\mu \Sigma \right\rangle^2 \;,\; \left\langle \partial_\mu \Sigma^\dagger \partial_\nu \Sigma \right\rangle \left\langle \partial^\mu \Sigma^\dagger \partial^\nu \Sigma \right\rangle \;,\; \left\langle \left( \partial_\mu \Sigma^\dagger \partial^\mu \Sigma \right)^2 \right\rangle \;,\; ...
\end{equation}
The most general Lagrangian that one can write down invariant under chiral symmetry and parity and containing the minimum number of derivatives is Eq.~\ref{eq:LOXPT}.

The infinite number of possible chirally invariant and parity even operators are organized in powers of derivatives (or momenta).
We have discussed that Goldstone bosons couple to other particles and to one another proportional to their momenta \cite{Weinberg:1966kf}.
If the momenta are sufficiently small relative to the cutoff of the EFT, then momentum factors can be considered small parameters, and the EFT can be organized as an expansion in small powers of momenta (derivatives).

This can be made more qualitative by examining a useful formula introduced by Weinberg \cite{Weinberg:1978kz}.
In a theory without fermions, one can derive the following power counting formula that relates the overall power of momenta of a Feynman diagram in perturbation theory, $N_p$, to the number of explicit momentum factors arising for each interaction vertex in the diagram $N_{p,i}$ and the number of loops $L$ \cite{Weinberg:1978kz,Gavela:2016bzc,Scherer:2005ri}.
\begin{equation}
N_p - 2 = \sum_i (N_{p,i} - 2) + 2 L
\end{equation}
The power counting formula dictates that lower loop diagrams formed out of interaction vertices with higher powers of momenta contribute at the same order in the momentum expansion as higher loop diagrams formed out of interaction vertices with lower powers of momenta.
So, it is possible to organize a systematic expansion in small powers of momenta divided by the EFT cutoff, $\partial / \Lambda$.
For example, working only at $\mathcal{O}(p^2)$, only tree level diagrams constructed out of $N_{p,i} = 2$ interaction vertices contribute.
At $\mathcal{O}(p^4)$, tree level diagrams constructed out of a single $N_{p,i} = 4$ vertex plus any number of $N_{p,i} = 2$ vertices will contribute at the same order as one loop diagrams constructed out of any number of $N_{p,i} = 2$ vertices.
Therefore, the number of derivatives that one must include in the Lagrangian in directly related to the loop order to which one wishes to work.
At tree level, one only needs to include the $\mathcal{O}(p^2)$ operator in the chiral Lagrangian.
At one loop order, one must include all the operators up to $\mathcal{O}(p^4)$, and so on.

The chiral Lagrangian as we have presented it so far is an EFT for exactly massless NGBs. 
To include quark mass effects -- that is, sources of explicit chiral symmetry breaking -- one uses the method of \emph{spurion analysis} to organize the symmetry breaking operators.
The spurion is an auxiliary scalar field that mimics the effect of the quark mass matrix in the underlying gauge theory.
In the gauge theory, the quark mass terms appear as
\begin{equation}
\mathcal{L}_{\text{Gauge}} \supset \bar{\psi}_{L,f} \mathcal{M}_{f,f'}\psi_{R,f'} + \bar{\psi}_{R,f} \mathcal{M}_{f,f'}\psi_{L,f'}
\end{equation}
where $\mathcal{M}$ is a Hermitian mass matrix in flavor space.
$\mathcal{M}$ is typically taken to be diagonal as one works in the mass eigenbasis.
For any constant nonzero matrix, $\mathcal{M}$, the chiral symmetry will be explicitly broken.
But notice that if the mass matrix transforms as $\mathcal{M} \rightarrow L \mathcal{M} R^\dagger$ under the chiral symmetry, these mass terms would preserve the chiral symmetry.

As such, one can introduce a scalar field $\chi(x)$ into the chiral Lagrangian transforming as $\chi(x) \rightarrow L \chi(x) R^\dagger$, and build operators that are invariant under the simultaneous transformation of $\Sigma(x)$ and $\chi(x)$. 
One then sets the spurion field equal to a constant matrix, $\chi \rightarrow B \mathcal{M}$, to break the chiral symmetry explicitly.
%One can show by matching correlation functions computed through the gauge theory generating functional and the chiral Lagrangian generating functional that the spurion field is proportional to the quark mass matrix in the underlying gauge theory.
Constructing all possible chirally (and parity) invariant operators in this way uniquely categorizes all possible breaking operators on symmetry grounds.

As with the derivative expansion, we must devise a power counting rule for ordering the infinite number of operators involving the spurion.
Because the spurion is a small breaking of the symmetry, we will naturally have that operators with higher powers of the spurion are higher order in the power counting than operators with lower powers of the spurion.
However, we must also decide how large powers of the spurion are relative to powers of derivatives.
Because the spurion field is non-dynamical, there is no natural power counting assignment based on its engineering dimension; the spurion field does not have a predetermined engineering dimension.

Instead, one infers a power counting rule for the spurion based on the kinematic regime of interest and on the scale present in the underlying theory.
For low momenta pions, one has that $p \approx M_\pi \propto \sqrt{m_q}$ where the the last proportionality follows from the GMOR relation.
So, for low momenta pions -- which is typically what is considered -- one makes the conventional choice that the spurion counts as dimension 2 in the momentum expansion \cite{Pich:1995bw,Scherer:2005ri}.
We emphasize, however, that this is a particular power counting assignment appropriate for a limited range of quark masses and a specific kinematic regime for the pion momenta \cite{Gavela:2016bzc}.
Other power counting assignments for the spurion may be considered depending on the application (c.f. \cite{Fuchs:1991cq}).
In our presentation of the linear sigma EFT framework for nearly conformal gauge theory in Section~\ref{sec:LSM}, the power counting of the spurion will play a crucial role in the formulation of the EFT.

To close our discussion of chiral perturbation theory, we write down some results for the observables $M_\pi$, $F_\pi$ and the scattering length $M_\pi A$ for the general $\SU(N_f)\times \SU(N_f)$ chiral Lagrangian at NLO \cite{Bijnens:2009qm,Bijnens:2011fm}.
\begin{align}
M_\pi^2 &=M^2 \left[ 1 + \frac{ N_F M^2 }{16\pi^2 F^2} \left( 128\pi^2\left( 2L_6^r - L_4^r + \frac{1}{N_F}(2L_8^r - L_5^r)\right) + \frac{1}{N_F^2} \log(M^2/\mu^2)\right)\right]  \label{eq:XPT1}
\\
F_\pi& = F \left[ 1 + \frac{ N_F M^2 }{16\pi^2 F^2} \left( 64\pi^2\left(L_4^r + \frac{1}{N_F}L_5^r\right) -\frac{1}{2} \log(M^2/\mu^2)\right)\right]  \label{eq:XPT2}
\\
M_\pi A &= \frac{-M^2}{16\pi F^2} \left[ 1 + \frac{N_F M^2}{16\pi^2 F^2} \left( -256\pi^2 \left( (1-\frac{2}{N_F})(L_4^r - L_6^r) + \frac{1}{N_F} ( L_0^r + 2L_1^r + 2L_2^r + L_3^r ) \right) \nonumber \right.\right. \\
& \left. \left.- 2\frac{N_F - 1}{N_F^3} + \frac{(2-N_F + 2N_F^2 + N_F^3)}{N_F^3}\log(M^2/\mu^2)  \right) \right] , \label{eq:XPT3}
\end{align}
where $M^2 = 2 B m_q$ as usual. 
We will use these expressions in Section~\ref{sec:EFTFits} when fitting effective field theory expression to the lattice data for $N_f=8$ QCD.

\section{Generalized Linear Sigma EFT \label{sec:LSM}}
In this section we summarize new considerations for the application of the linear sigma model as a low energy description of nearly conformal gauge theories.
This model has been developed in recent works by the present author and collaborators \cite{Gasbarro:2017fmi,Gasbarro:2017ccf,Gasbarro:2018prep}.  
The model is motivated by a body of work from the lattice gauge theory community in studying gauge theories near and inside the conformal window, a subset of which we have reviewed in Section~\ref{sec:latticewindow} and Chapter~\ref{chapter:Lattice}.
One key feature of conformal and nearly conformal gauge theories exposed by these lattice studies is the ostensibly generic appearance of light flavor-singlet scalar states.
The linear sigma model naturally incorporates a singlet scalar whose mass is tunable within the model.
In the data for $N_f = 8$ QCD discussed in Chapter~\ref{chapter:Lattice}, we have also demonstrated how the infrared scale $a\Lambda_{\text{conf}}$ depends on the quark mass.
In the linear sigma model, we will show that the vacuum expectation value (v.e.v.) which sets the scale of the model depends on the explicit chiral symmetry breaking potential at tree level, which may help to model the quark mass dependence of the infrared scales in the gauge theory.
Finally, the linear sigma EFT naturally incorporates a multiplet of flavored scalar states.
A study by the Lattice Higgs Collaboration (latHC) of the two flavors sextet model has reported that the flavored scalars become lighter than the $\rho$ meson close to the chiral limit \cite{Fodor:2016pls}.
If this is true, the inclusion of the flavored scalars in the EFT may extend the radius of convergence of the EFT.
On the other hand, for an application in which the flavored scalars are not lighter than the cutoff, these states may be removed from the linear sigma EFT by an appropriate limit as we will show.

\subsection{Field Content}
The underlying gauge theory has the global symmetry breaking pattern $\SU_L(N_f)\times\SU_R(N_f)\times U_V(1) \rightarrow U_V(N_f)$ after the $U_A(1)$ symmetry is broken at the quantum level by topological effects.
We have discussed in Section~\ref{sec:XPT} that the EFT will be determined by a specification of the global symmetry group, a choice of field content transforming in some representation of the global symmetry group, and by a power counting rule to designate the relative importance of the operators allowed by the symmetries.

For a moment, let us consider the larger $\text{U}(N_f)\times \text{U}(N_f)$ group.
We choose as our starting point for the fields to transform in a (bi)linear representation of $\text{U}(N_f)\times \text{U}(N_f)$.
This is the generalization of the original $\SU(2)\times\SU(2)$ linear sigma model \cite{Schwinger:1957em,GellMann:1960np} to $N_f > 2$, which was originally introduced by Levy \cite{levy1967m} and later studied in many subsequent works, cf. \cite{Schechter:1971qa,Schechter:1971tc,Bardeen:1969ra,Rosenzweig:1979ay}.
As we have demonstrated in Section~\ref{sec:XPT}, the matrix field $M(x)$ must be complex valued in order for the representation to close for $N_f > 2$.  
The linear sigma fields transform in a $(N_f,\bar{N_f})$ representation of $\text{U}(N_f)\times \text{U}(N_f)$,
\begin{equation}
M_a^{\bar{b}} \rightarrow L_a^c M_c^{\bar{d}} (R^\dagger)_{\bar{d}}^{\bar{b}}
\end{equation}
where $L,R \in \text{U}_{L,R}(N_f)$.  
The unbarred subscript (barred superscript) transform via linear action of a matrix in the fundamental (antifundamental) representation of $\text{U}_L(N_f)$ ($\text{U}_R(N_f)$).
We suppress the group indices in the remainder of the discussion where ever possible.

The complex representation has $2 N_f^2$ real degrees of freedom. 
Depending on the parametrization of the matrix field degrees of freedom, the component fields may transform in a variety of ways under the full group as we saw for the case of $N_f=2$ in Section~\ref{sec:XPT}.
However, under the unbroken subgroup $U_V(N_f)$ and under parity, we will always be able to identify $N_f^2 -1$ pseudoscalar pions and $N_f^2 -1$ scalar $a_0$ states, each set transforming irreducibly in adjoint representations of $\SU_V(N_f)$, as well as one pseudoscalar $\eta'$ and one scalar $\sigma$ state, each transforming irreducibly as singlets under $\SU(N_f)$. 
All of these states are mesonic and therefore transform as singlets under the $U_V(1)$ baryon number symmetry.
Thus, the $U_V(1)$ symmetry is trivial in the EFT and we will neglect it going forward.

We may choose to express the $2 N_f^2$ real degrees of freedom of $M(x)$ in a linear basis as follows.
\begin{equation}
M(x) = \frac{\sigma(x) + i \eta'(x)}{\sqrt{N_f}} + (a_i(x) + i \pi_i(x))T_i \label{eq:SUNlinear}
\end{equation}
This basis has the benefit of making the renormalizability of the theory more manifest because it is a linear function of the field components.
However, it also masks the geometrical nature of the Goldstone bosons as coordinates on the coset space.
We may alternatively use a nonlinear basis in which the Goldstones act as a group transformation on the heavy degrees of freedom.
The field with the Goldstone degrees of freedom set to zero is 
\begin{equation}
\left.M(x)\right|_{\pi,\eta \rightarrow 0} \equiv S(x) = \frac{\sigma(x)}{\sqrt{N_f}}\identity + a_i(x) T_i
\end{equation}
Then we reintroduce the Goldstones as a group transformation acting on the heavy degrees of freedom.
\begin{equation}
M(x) = \left(L(x),R(x)\right) \circ S(x) = L(x) S(x) R^\dagger(x)
\end{equation}
In the $\SU(2)$ case, the heavy state ($\sigma(x)\identity$) was invariant under $\SU_V(2)$ transformations (H-transformations) and so it did not matter how we chose to parametrize the Goldstone bosons in the coset space.
Here, $S(x)$ transforms nontrivially under $\SU_V(N_f)$ transformations.
We choose to parametrize the coset such that the Goldstone bosons act entirely as a $\text{U}_L(N_F)$ transformation.
\begin{equation}
M(x) = \left(L(x),R(x)\right)\circ \left(R^\dagger(x),R^\dagger(x)\right) \circ \left(R(x),R(x)\right) \circ S(x) = \left(L(x)R^\dagger(x),\identity\right) \circ S'(x)
\end{equation}
where $S'(x) = R(x)S(x)R^\dagger(x)$.  
Notice that $S(x)$ is simply an expansion for a general Hermitian matrix in a basis of Hermitian generators.
A vector transformation on a Hermitian matrix yields another Hermitian matrix, therefore we can simply reexpand $S'(x) = \sigma'(x)/\sqrt{N_f} \identity + a_i'(x) T_i$.
Having established this, we will drop the primes.
As in the $\SU(2)$ case, $L(x)R^\dagger(x) = \Sigma(x)$ is a general unitary matrix.
Therefore, the nonlinear basis with our convention for the parametrization of the coset space is 
\begin{equation}
M(x) = \Sigma(x) S(x) \label{eq:SUNnonlinear1}
\end{equation}
where
\begin{equation}
\Sigma(x) = \exp\left[i\frac{\sqrt{N_f}}{F} \left( \frac{\eta'(x)}{\sqrt{N_f}}+ \pi_i(x) T_i\right) \right] \label{eq:SUNnonlinear2}
\end{equation}
is a general unitary $N_f \times N_f$ matrix expressed as an exponential of a sum of Hermitian generators, and 
\begin{equation}
S(x) = \frac{\sigma(x)}{\sqrt{N_f}}\identity + a_i(x)T_i \label{eq:SUNnonlinear3}
\end{equation}
is a general Hermitian $N_f \times N_f$ matrix expressed as a sum of Hermitian generators.

As in the $\SU(2)$ case, the nonlinear basis makes manifest many of the properties of the Goldstone bosons such as the fact that they are massless in the absence of explicit chiral symmetry breaking and that they are derivatively coupled to each other and to the heavy scalar states.
We will choose to use the nonlinear basis not only as a matter of convenience for these useful features, but also because of our treatment of the $\eta'(x)$ degree of freedom.
Under the unbroken $\SU_V(N_f)$ subgroup of the chiral symmetry, the various field components transform irreducibly only amongst themselves; that is, they do not mix with each other
In particular, the $\sigma$ transforms as a $\SU_V(N_f)$ singlet scalar, the $\eta'$ transforms as a $\SU_V(N_f)$ singlet pseudoscalar, the $a_i$ transform as $\SU_V(N_f)$ adjoint scalars, and the $\pi_i$ transform as $\SU_V(N_f)$ adjoint pseudoscalars.
However, under the $\SU_A(N_f)$ axial transformations, the components will mix with one another, and the way in which they mix under axial transformation will depend upon the choice of field basis.
In the linear basis of Eq.~\ref{eq:SUNlinear}, the $\eta'(x)$ degree of freedom mixes with the other components under $\SU_A(N_f)$.
But in the nonlinear basis of Eqs.~\ref{eq:SUNnonlinear1}-\ref{eq:SUNnonlinear3}, the $\eta'(x)$ is a singlet under the complete chiral group $\SU_L(N_f)\times \SU_R(N_f)$ and only transforms under $\U_A(1)$.

A study by the latKMI collaboration of the $\eta'$ mass in many flavor QCD suggests that, as in QCD, the $\eta'$ is heavier than the $\rho$ and in fact that its mass increases relative to the $\rho$ mass as the number of flavors is increased \cite{Aoki:2017fnr}.
As such, we will choose to remove it from the EFT by setting this degree of freedom to zero by hand.
In the nonlinear basis Eqs.~\ref{eq:SUNnonlinear1}-\ref{eq:SUNnonlinear3}, the multiplet remains closed under $\SU_L(N_f)\times\SU_R(N_f)$ chiral transformations when the $\eta'$ degree of freedom is set to zero.
But for the reasons that we have just described, in the linear basis of Eq.~\ref{eq:SUNlinear} the multiplet is not closed under chiral transformations when the $\eta'$ degree of freedom is set to zero.

We emphasize that when the entire multiplet ($\sigma$, $a_i$, $\eta'$, $\pi_i$) is retained, any physical prediction is independent of the choice of field basis as it must be due to the invariance of the quantum field theory path integral under changes in the parametrization of the fields.
However, when a component (the $\eta'$) of the multiplet is set to zero, the linear and nonlinear field bases are no longer equivalent because the degree of freedom that has been removed, while locally equivalent, is a different global degree of freedom in the two field bases.
We will demonstrate this in more detail when we discuss nonlinear constraints.

We remark that a recent work by Meurice \cite{Meurice:2017zng} which also considers a linear sigma model as a description of many flavor gauge theories chooses to retain the $\eta'$ degree of freedom and to study the effect of determinant terms which explicitly break the $U_A(1)$ symmetry.
In this work, the take the viewpoint that the anomaly which gives rise to the $\eta'$ mass is a UV effect which need not be encapsulated in the low energy EFT framework.

\subsection{Leading Order Lagrangian}
Let us write down the leading order Lagrangian containing all operators constructed out of $M(x)$ and derivatives and invariant under the $\SU_L(N_f)\times \SU_R(N_f)$ global chiral symmetry and parity.  
We define the leading order Lagrangian to contain all relevant and marginal operators based on the engineering dimension of the operator.
\begin{flalign}
V_0(M) &= \frac{\mu^2}{2}\left\langle M^\dagger M \right\rangle + \frac{\lambda_1}{4}\left\langle M^\dagger M \right\rangle^2 + \frac{N_f \lambda_2}{4} \left\langle \left(M^\dagger M\right)^2\right\rangle  \\
\mathcal{L} &= \frac{1}{2}\left\langle \partial_\mu M \partial^\mu M \right\rangle - V_0(M) - V_{\text{SB}}(M)
\end{flalign}
$V_0$ contains operators that are invariant under the full $\SU_L(N_f)\times\SU_R(N_f)$ chiral symmetry.
$V_{\text{SB}}$ contains the operators that explicitly break the chiral symmetry down to the vector subgroup.
For now, we allow the symmetry breaking potential to be general; we will specify its exact form in the discussion that follows.
The \emph{chiral limit} is the limit in which the symmetry breaking potential is set to zero, which we will show corresponds to vanishing quark masses in the underlying gauge theory.
We will assume that the symmetry is spontaneously broken at the classical level, and therefore we will take $\mu^2 < 0$.
It is convenient to reparametrize the couplings in the following way.
\begin{equation}
V_0(M) = \frac{-m_\sigma^2}{4}\left\langle M^\dagger M \right\rangle + \frac{m_\sigma^2 - m_a^2}{8f^2}\left\langle M^\dagger M \right\rangle^2 + \frac{N_f m_a^2}{8f^2} \left\langle \left(M^\dagger M\right)^2\right\rangle 
\end{equation}
We have exchanged the couplings $\mu^2$, $\lambda_1$, and $\lambda_2$ for the couplings $f^2$, $m_\sigma^2$, and $m_a^2$.  
The condition that the chiral symmetry is spontaneously broken in the chiral limit now corresponds to $m_\sigma^2 > 0$.
We will show presently that $f^2$, $m_\sigma^2$, and $m_a^2$ are the vacuum expectation value of the field, the mass of the flavor singlet scalar $\sigma$, and the mass of the flavored scalars $a_i$, respectively, in the chiral limit.
We will denote the values of these quantities away from the chiral limit by the corresponding capital letters $F^2$, $M_\sigma^2$, and $M_a^2$.

The field takes on a vacuum expectation value which we choose by convention to be oriented along the $\sigma$ direction, $\langle 0 | M_{ab}(x) | 0 \rangle = F/\sqrt{N_f} \delta_{ab}$, with $F$ determined by the extremization condition $\delta V(M)/\delta M(x) = 0$, which reduces to
\begin{equation}
\frac{F^3}{f^2} - F + \frac{2}{m_{\sigma}^2} \left. \frac{\partial V_{\text{SB}}}{\partial \sigma} \right|_{\sigma = F,\; \pi_i = a_i = 0} = 0 \label{eq:minimum}
\end{equation}
After reexpanding the potential around this v.e.v., the tree level expression for the masses of the pions and scalars are given by
\begin{flalign}
M_{\pi}^2 &= \left.\frac{\partial^2 V_{\text{SB}}}{\partial \pi_i^2} \right|_{\sigma = F,\; \pi_i = a_i = 0} \label{eq:LO1} \\
M_{\sigma}^2 &= m_\sigma^2 \left( \frac{3}{2} \frac{F^2}{f^2} - \frac{1}{2} \right) + \left.\frac{\partial^2 V_{\text{SB}}}{\partial \sigma^2}\right|_{\sigma = F,\; \pi_i = a_i = 0} \label{eq:LO2} \\
M_{a}^2 &= m_a^2 \frac{F^2}{f^2} + \frac{m_{\sigma}^2}{2} \left(  \frac{F^2}{f^2} -1 \right) + \left.\frac{\partial^2 V_{\text{SB}}}{\partial a_i^2}\right|_{\sigma = F,\; \pi_i = a_i = 0} \label{eq:LO3}
\end{flalign}
Setting $V_{\text{SB}} = 0$, we confirm that $F^2 = f^2$, $M_\sigma^2 = m_\sigma^2$, $M_a^2 = m_a^2$, and that the pions are exactly massless NGBs.

As in the case of the $\SU(2)$ linear sigma model discussed in Section~\ref{sec:currentalgebra}, the pion decay constant is not a separate scale from the scalar v.e.v.
The axial vector current takes the same form, derived through the usual Noether procedure.
\begin{equation}
A^{\mu}_i = \frac{i}{2}\left\langle (\partial^{\mu} M^\dagger) \{T_i,M\} - \{T_i,M^\dagger\} \partial^\mu M \right\rangle 
\end{equation}
Expanding around the v.e.v. to leading order in the fields, we find
\begin{equation}
A^{\mu}_i = \frac{2 F}{\sqrt{N_f}} \partial^\mu \pi_i(x) + ...
\end{equation}
Finally, we plug into the matrix element Eq.~\ref{eq:Fpi} to find the pion decay constant with our normalization convention.
\begin{equation}
F_\pi = \sqrt{\frac{2}{N_f}} F
\end{equation}
It is an interesting feature of the linear sigma model that the pion decay constant is tied to the scalar v.e.v.
In particular, through Eq.~\ref{eq:minimum} the pion decay constant will depend on the chiral breaking potential $V_{\text{SB}}$ at tree level.
In the lattice results that we have reviewed in Chapter~\ref{chapter:Lattice}, we saw that the chiral breaking operator (the quark mass) had a large effect on the infrared scale of the gauge theory, and that all massive quantities had a dominant behavior determined by the quark mass dependence of the infrared scale.  
Here we see an analogous situation in which the scale of the EFT, $F$, depends on the chiral breaking at tree level.
This effect is not present in the chiral Lagrangian, in which the pion decay constant only depends on the quark mass at loop level.
This feature may make the linear sigma EFT framework better suited for describing the nearly conformal gauge theories discussed in Chapter~\ref{chapter:Lattice}.

\subsection{Nonlinear Constraints and the $\eta'$}
Next let us consider some limiting cases of the leading order linear sigma Lagrangian.
We have discussed that the flavored scalar states in nearly conformal gauge theories corresponding to the $a_i$ degrees of freedom may be lighter than the $\rho$ close to the chiral limit in a nearly conformal gauge theory \cite{Fodor:2016pls}.
However, if the flavored scalar states turn out to be heavy, we may want to remove them from our EFT description.
This can be achieved by taking the limit $m_a^2 \rightarrow \infty$ holding $m_\sigma^2$ and $f^2$ fixed.
As we have discussed when taking the sigma mass to infinity in the $\SU(2)$ linear sigma model in Section~\ref{sec:XPT}, the field becomes constrained such that the terms in the Lagrangian proportional to the coupling that is being taking to infinity are set to their stationary value.
In the case of the scalar mass being taken to infinity, 
\begin{equation}
\frac{\delta}{\delta M^\dagger(x)} \left( N_f \left\langle \left(M^\dagger M\right)^2\right\rangle - \left\langle M^\dagger M \right\rangle^2   \right) = 0
\end{equation}
Performing the functional differentiation and rearranging the expression, one finds that the nonlinear constraint imposed on the fields by taking the flavored scalar mass to infinity is
\begin{equation}
M^\dagger M = \frac{1}{N_f}\left\langle M^\dagger M \right\rangle \identity
\end{equation}
which is independent of the choice of field basis.
In the nonlinear basis Eq.~\ref{eq:SUNnonlinear1}-\ref{eq:SUNnonlinear3}, this constraint is satisfied by simply setting $a_i(x) = 0$.
In the linear basis Eq.~\ref{eq:SUNlinear}, imposing the constraint is much more complicated and the flavored scalar degrees of freedom are fixed to a complicated function of the remaining ($\sigma$,$\pi_i$) degrees of freedom.
Another advantage of working in the nonlinear basis is that it turns out to be more straightforward to impose this constraint.

We can also consider taking the singlet scalar mass to infinity $m_\sigma^2 \rightarrow \infty$ with $f^2$ and $m_a^2$ fixed.
Carrying out a similar analysis, the nonlinear constraint imposed on the fields is
\begin{equation}
\left\langle M^\dagger M \right\rangle = f^2 \identity
\end{equation}
In the nonlinear basis, this constraint becomes $\sigma^2 + a_i^2 = f^2$.
In the linear basis, the constraint is $\sigma^2 + a_i^2 + \pi_i^2 = f^2$.
Again we see that the nonlinear constraint is simpler in the nonlinear basis than the linear basis.

Finally, let us consider a situation in which we have retained the $\eta'$ degree of freedom in Eq.~\ref{eq:SUNnonlinear2}, and now wish to lift its mass by introducing a term into the Lagrangian that explicitly breaks the $U_A(1)$ symmetry.
There is not a unique choice for this term, and various different $U_A(1)$ breaking terms have been considered in the past (c.f. \cite{Rosenzweig:1979ay,Bardeen:1969ra}).
Let us consider a breaking term that is particularly simple for the nonlinear field basis Eq.~\ref{eq:SUNnonlinear1}-\ref{eq:SUNnonlinear3}.
\begin{equation}
V \supset -\frac{1}{8}\left(\frac{F}{N_f}\right)^2 m_{\eta'}^2 \left[\ln\det M - \ln\det M^\dagger\right]^2 \label{eq:SUNopA1}
\end{equation}
It is easy to check that this term explicitly breaks the $U_A(1)$ symmetry.
In the nonlinear basis, $\det M = \exp(i N_f \eta'(x) / F)\det S(x)$, and the $U_A(1)$ breaking term is nothing more than a mass term for the eta prime: $(1/2)m_{\eta'}^2 \eta'(x)^2$.

As we have done for the flavored scalar mass and the sigma mass, let us consider the nonlinear constraint imposed on the field when the coefficient of the $U_A(1)$ breaking operator is taken to infinity, $m_{\eta'}^2 \rightarrow \infty$.
The fields are constrained to satisfy
\begin{equation}
\frac{\delta}{\delta M^\dagger(x)} \left[\ln\det M - \ln\det M^\dagger\right]^2 = 0
\end{equation}
Making use of the identity for the derivative of a determinant
\begin{equation}
\frac{\partial}{\partial M_{ab}} \det M = \det M (M^{-1})_{ba}
\end{equation}
we arrive at the nonlinear constraint
\begin{equation}
\ln\det M - \ln \det M^\dagger = 0
\end{equation}
Unsurprisingly, in the nonlinear basis to this expression reduces to $\eta'(x) = 0$.
So, when we set the $\eta'(x)$ degree of freedom to zero in the above discussion, we can imagine that we had actually retained the $\eta'$ and the $U_A(1)$ symmetry and then had subsequently broken the symmetry by introducing the operator Eq.~\ref{eq:SUNopA1} and had taken its coefficient to infinity.

A different operator which explicitly breaks the $U_A(1)$ symmetry which we may have considered instead is
\begin{equation}
V \supset c_A \left[\det M + \det M^\dagger\right] \label{eq:SUNopA2}
\end{equation}
While this operator has the advantage of being polynomial in the fields, it does not contribute only a simple quadratic term for the $\eta'$ mass.
Rather, it contributes a degree $N_f$ polynomial involving not only the $\eta'$ degree of freedom but also the scalar degrees of freedom.
In the nonlinear basis, the operator takes the form
\begin{equation}
c_A \left[\det M + \det M^\dagger\right] = 2 c_A \cos(N_f \eta'(x)/F) \det S(x)
\end{equation}
The nonlinear constraint that is yielded when ones takes the coefficient $c_A \rightarrow \infty$ is
\begin{equation}
\det M = \det M^\dagger = 0
\end{equation}
which in the nonlinear basis is
\begin{equation}
e^{\pm i N_f \eta'(x) / F}\det S(x) = 0
\end{equation}
which is not simply satisfied by setting $\eta'(x) = 0$.  

Since there is no unique procedure for giving a mass to the $\eta'$ degree of freedom, and because we have demonstrated that there is at least one simple way to remove the $\eta'$ from the theory without affecting the other states, we feel that our prescription of simply setting the $\eta'$ degree of freedom to zero by hand is justifiable.
Continuing forward, we will omit the $\eta'$ degree of freedom as we have already indicated, and we will retain the flavored scalar degrees of freedom bearing in mind the available option of taking their mass heavy relative to the other states.

\subsection{Chiral Breaking and Power Counting}
Next we must specify the chiral symmetry breaking potential $V_{\text{SB}}$ which encapsulates the effect of quark mass terms in the underlying gauge theory.
We have already discussed spurion analysis in Section~\ref{sec:XPT} for introducing quark mass effects into the chiral Lagrangian.
The analysis in the linear sigma framework is carried out in the same way.
We introduce an auxiliary spurion field $\chi(x)$ transforming as $\chi(x) \rightarrow L \chi(x) R^\dagger$ under chiral rotations.
$V_{\text{SB}}$ contains all operators constructed out of $M(x)$ and $\chi(x)$ that are invariant under the simultaneous chiral transformation of the linear sigma fields and the spurion field.
The spurion is set to a constant matrix proportional to the quark mass matrix to break the symmetry.
\begin{equation}
\chi(x) \rightarrow B \mathcal{M}
\end{equation}
B is a low energy constant, and $\mathcal{M}$ is the quark mass matrix which we take to be proportional to the identity for the case of degenerate quarks, $\mathcal{M} = m_q \identity$.  
Mass split systems may be studied in this framework by making a different choice for the mass matrix.

In our discussion of the $\SU(2)$ linear sigma model, we wrote down the simplest chiral breaking operator, $\langle \chi^\dagger M + M^\dagger \chi \rangle$, for the sake of demonstrating the PCAC and GMOR relationships in the context of a simple model.
In our discussion of the chiral Lagrangian, we have discussed how the spurion is incorporated into the power counting and weighted relative to the chiral limit derivative expansion.
The choice that $\chi$ is counted as dimension two in the derivative expansion leads one to the GMOR relation and the Gell-Mann Okubo formula from the chiral Lagrangian.
In the linear sigma EFT framework, we must identify a power counting rule for the spurion field that is appropriate to the underlying theory that the EFT is to describe, nearly conformal gauge theories.
We focus on the kinematic regime in which the momenta are of the same order of magnitude as the particle masses, which fixes $\partial/\Lambda \sim M(x)/\Lambda$.
Derivatives and fields in Lagrangian operators contribute factors of order the particle masses or field v.e.v. to physical observables, so that the expansion parameters have an order of magnitude size
\begin{equation}
\frac{\partial}{\Lambda} \sim \frac{M(x)}{\Lambda} \sim \frac{M_\sigma}{\Lambda}
\end{equation}

We have discussed in Chapter~\ref{chapter:Lattice} that the chiral breaking effects may be large in the underlying gauge theory at the distance from the chiral limit that is accessible to current lattice calculations.
%We focus on the kinematic regime in which the momenta are of the same order of magnitude as the particle masses, which fixes $\partial/\Lambda \sim M(x)/\Lambda$.
A good measure of chiral symmetry breaking in the gauge theory is $m_q B_\pi$ where $B_\pi = \langle 0 | \bar{\psi}\psi | 0 \rangle|_{m_q=0} / f_\pi^2$.
Close to the chiral limit $M_\pi^2 = 2 m_q B_\pi$ as given by the GMOR relation, but further from the chiral limit $m_q B_\pi$ does not correspond to $M_\pi^2$. 
In the linear sigma model, we quantify the relative size of the chiral breaking effects through the quantity $\alpha(m_q)$.
\begin{equation}
\frac{m_q B_\pi}{\Lambda^2} \sim \left( \frac{M_\sigma}{\Lambda}\right)^\alpha \ll 1 \label{eq:oom}
\end{equation}
We emphasize that $\alpha$ is not a low energy constant of the EFT, but rather a derived quantity that is in one to one correspondence with $B_\pi m_q$ and quantifies the size of explicit chiral symmetry breaking effects.
By matching to the underlying gauge theory and choosing an appropriate normalization for the spurion field, one can show that $B = B_\pi$ at leading order.
Utilizing this fact together with Eq.~\ref{eq:oom}, we assign the power counting rule $\alpha$ to the spurion field.
The linear sigma field $M(x)$ and the derivative operator are each assigned a power counting rule of of one.
Then the EFT is constructed out of the small quantities
\begin{equation}
\label{eq:SH}
\frac{\partial}{\Lambda} \sim \frac{M(x)}{\Lambda} \sim \left(\frac{\chi}{\Lambda^2}\right)^{\frac{1}{\alpha}} \ll 1
\end{equation}
where $\Lambda$ is the cutoff of the EFT.
All terms in the Lagrangian taking the schematic form
\begin{equation}
\mathcal{L} \supset \Lambda^4 \left( \frac{\partial}{\Lambda}\right)^{N_p}\left( \frac{M(x)}{\Lambda}\right)^{N_M} \left( \frac{\chi}{\Lambda^{2}}\right)^{N_\chi}.
\label{eq:nda}
\end{equation}
such that the coefficient of an operator has the order of magnitude $\Lambda^{4-N_p-N_M-2N_\chi}$.
Then the power counting dimension of an operator in the Lagrangian is defined to be $D = N_p + N_M + \alpha N_\chi$.

\begin{table}[t]
	\centering
	\renewcommand\arraystretch{1.2}  % Increase the height of each row
	\addtolength{\tabcolsep}{3 pt}   % Increase separation between columns
	\begin{tabular}{ |c|c||c|c|  }
		\hline
		Symbol  &             Operator                                                 &$3/5 < \alpha \le 1$ & $1 < \alpha \le 3$ \\
		\hline
		$\mathcal{O}_1$ & $\vev{\chi^\dag M + M^\dag \chi}$                        & \checkmark           & \checkmark \\
		$\mathcal{O}_2$ & $\vev{M^\dag M}\vev{\chi^\dag M + M^\dag \chi}$          & \checkmark           & X \\
		$\mathcal{O}_3$ & $\vev{(M^\dag M)(\chi^\dag M + M^\dag \chi)}$            & \checkmark           & X \\
		$\mathcal{O}_4$ & $\vev{\chi^\dag M + M^\dag \chi}^2$                      & \checkmark           & X \\
		$\mathcal{O}_5$ & $\vev{\chi^\dag \chi M^\dag M}$                          & \checkmark           & X \\
		$\mathcal{O}_6$ & $\vev{\chi^\dag \chi}\vev{ M^\dag M}$                    & \checkmark           & X \\
		$\mathcal{O}_7$ & $\vev{\chi^\dag M\chi^\dag M + M^\dag \chi M^\dag \chi}$ & \checkmark           & X \\
		$\mathcal{O}_8$ & $\vev{\chi^\dag \chi}\vev{\chi^\dag M + M^\dag \chi}$    & \checkmark           & X \\
		$\mathcal{O}_9$ & $\vev{(\chi^\dag \chi)(\chi^\dag M + M^\dag \chi)}$      & \checkmark           & X \\
		\hline
	\end{tabular}
	\caption{\label{table1}Operator content of leading-order breaking potential in various power counting regimes, with $\alpha$ the power counting rule for the spurion field. }
\end{table}
For simplicity, we focus on the chiral breaking regime $3/5 < \alpha$.
For $\alpha < 3/5$ the number of chiral breaking operators proliferates.
The leading order potential is taken to contain all operators allowed by the symmetries with power counting dimension $D \le 4$.
Table~\ref{table1} catalogs the chiral breaking operators in the leading order potential for this range of power countings specified by $\alpha$.
The most general leading order breaking potential for this range of $\alpha$ may be parametrized as
\begin{equation}
V_{\text{SB}} = - \sum_{i=1}^9 \tilde{c}_i \mathcal{O}_i(x) \label{eq:Vsb}
\end{equation}
where some of the 9 operators in $V_{\text{SB}}$ may not appear at leading order, depending upon the value of $\alpha$.
After $\chi \rightarrow B\mathcal{M}$, there is a redundancy in the low energy constants $B$ and $\tilde{c}_i$.
Physical observables only depend on the product $B \tilde{c}_1$ and not on $\tilde{c}_1$ individually.  
To remove the redundancy, we set $\tilde{c}_1 = f / \sqrt{N_f}$ which also guarantees that $B = B_\pi$ at leading order.

We compute the leading order expressions for the masses and the scalar v.e.v. (Eqs.~\ref{eq:minimum}-\ref{eq:LO3}) 
for the general breaking potential Eq.~(\ref{eq:Vsb}).
We redefine the coefficients to absorb factors for $N_f$ in order to clean up the expressions: 
$c_{2,9}=\sqrt{N_f}\tilde{c}_{2,9}$, $c_3 = \tilde{c}_3 / N_f$, $c_{4,6} = N_f \tilde{c}_{4,6}$, $c_{5,7} = \tilde{c}_{5,7}$, and $c_8 = N_f^{3/2} \tilde{c}_8$.
\begin{flalign}
\frac{F^2}{f^2} = &1 + \frac{2}{m_{\sigma}^2} \left[ 2 B m_q \frac{f}{F} + 6 B m_q (c_2+c_3)F  \right. \cr
&\left. + 2 B^2 m_q^2 (4 c_4 + c_5 + c_6 + 2 c_7) +
2 B^3 m_q^3 \frac{c_8+c_9}{F}  \right]  \label{eq:SB1}\\
M_{\pi}^2 = &2 B m_q \frac{f}{F} + 2 B m_q (c_2 + c_3) F + 8 B^2 m_q^2 (c_4 + c_7) \cr
&+2 B^3 m_q^3 \frac{c_8 + c_9}{F} \label{eq:SB2} \\
M_{\sigma}^2 = &m_{\sigma}^2 \left (\frac{3}{2} \frac{F^2}{f^2} - \frac{1}{2} \right) - 12 B m_q (c_2 + c_3) F  \cr
&- 2 B^2 m_q^2 (4 c_4 + c_5 + c_6 + 2 c_7) \label{eq:SB3} \\
M_a^2 = &m_a^2 \frac{F^2}{f^2} + m_{\sigma}^2 \left(\frac{F^2}{f^2}-1\right) - 4 B m_q (c_2 +3 c_3) F \cr
&- 2 B^2 m_q^2 (c_5 + c_6 + 2 c_7) \label{eq:SB4}
\end{flalign}
Eqs.~(\ref{eq:SB1})-(\ref{eq:SB3}) can be combined to eliminate $F$ and to express $M_{\sigma}^2$ in terms of $M_{\pi}^2$.
\begin{equation}
\label{eq:msigrelation}
3 M_\pi^2 -  M_{\sigma}^2 + m_{\sigma}^2 = 4 B^2 m_q^2 (2 c_4 -c_5 -c_6 + 4 c_7)
\end{equation}
$m_{\sigma}^2$ must be positive in a theory with underlying spontaneous symmetry breaking, so any $\alpha$ regime in which the operators $\mathcal{O}_{4,5,6,7}$ 
are highly suppressed will give rise to the leading order inequality 
\begin{equation}
M_\sigma^2 \ge 3 M_\pi^2.
\label{eq:ineq}
\end{equation}
This inequality will be relaxed slightly at loop level and at higher orders in the EFT expansion, but if the theory is weakly coupled and if the higher dimensional operators are sufficiently suppressed, these effects will be much smaller than $M^2_\sigma$.

We have demonstrated in Chapter~\ref{chapter:Lattice} that lattice computations of nearly conformal gauge theories have found the $\sigma$ to be similar in mass to the pions  over an appreciable range of quark masses, in tension with the inequality Eq.~\ref{eq:ineq}. 
Therefore, the regime of chiral symmetry breaking in the EFT that is appropriate for nearly conformal gauge theories should be one in which the right hand side of Eq.~\ref{eq:msigrelation} is a positive value of order $M^2_\sigma$. 
To determine which power counting regimes are compatible with the lattice data, we use Eq.~\ref{eq:nda} to estimate the sizes of the operator coefficients $c_{4,5,6,7} \sim \mathcal{O}\left(1/\Lambda^{2}\right)$. 
In addition, Eq.~\ref{eq:SH} implies that $B m_q \sim \Lambda^{2-\alpha} M_\sigma^\alpha$. 
The approximate size of the right hand side of Eq.~\ref{eq:msigrelation} is
\begin{equation}
\label{eq:correction}
4 B^2 m_q^2 (2 c_4 -c_5 -c_6 + 4 c_7) \sim M_{\sigma}^2 \left(\frac{M_\sigma}{\Lambda} \right)^{2\alpha - 2}.
\end{equation}
Therefore $\alpha$ must be close to one in order for Eq.~\ref{eq:msigrelation} to appreciably relax the inequality Eq.~\ref{eq:ineq}. 
Values of $\alpha$ that are too large lead to the constraint Eq.~\ref{eq:ineq}, whereas values of $\alpha$ that are too small require the operator coefficients $c_{4,5,6,7}$ to be finely tuned.

Various subtleties arise in the linear sigma construction, and one should carefully work with the form of the model that is most appropriate for the underlying theory to which the EFT is being applied.
This includes the choice of whether or not to retain the flavored scalar degrees of freedom, and the power counting rule for the spurion.
We also remark that for $N_f \leq 4$ determinant operators are leading order in the power counting and may need to be included \cite{Gasbarro:2018prep}.

In conclusion, we have developed a new EFT framework based on the linear sigma multiplet which we believe is a promising versatile model for describing nearly conformal gauge theories.  
The model includes a flavor singlet scalar and a multiplet of flavor-adjoint scalars with parametrically controlled masses, a pion decay constant which varies with the quark mass at tree level, and a generalized power counting for the spurion fields which should help to model the large quark mass effects in lattice calculations of nearly conformal gauge theories at the currently accessible distances from the chiral limit.
We hope that the linear sigma EFT framework will help to answer important outstanding questions about nearly conformal gauge theories relevant for phenomenology such as the chiral limit value of the sigma mass relative to the pion decay constant.
Further details of the linear sigma EFT may be found in the paper currently in preparation \cite{Gasbarro:2018prep}.

In Section~\ref{sec:EFTFits}, we present a first test the linear sigma EFT framework by performing explicit chiral fits to the spectral data for $N_f = 8$ QCD \cite{Appelquist:2016viq,Gasbarro:2017fmi} discussed in Chapter~\ref{chapter:Lattice}.

%\subsection{Naive Dimensional Analysis of the Linear Sigma EFT}

\section{Fits of Chiral Effective Theories to Lattice Data \label{sec:EFTFits}}

In this Section, we present three separate analyses in which EFT expressions are fit to lattice data for $M_\pi$, $M_\sigma$, and $F_\pi$ in $N_f = 8$ QCD computed by the Lattice Strong Dynamics collaboration \cite{Appelquist:2016viq,Gasbarro:2017fmi} and discussed in Chapter~\ref{chapter:Lattice}.
The first two analyses were presented in a recent paper by the present author \cite{Gasbarro:2017ccf}.
The third analysis is a first attempt at fitting the linear sigma EFT framework presented in Section~\ref{sec:LSM} to data.
We will perform fits to the data in both lattice units and nucleon units in accordance with our discussion of scale setting in Chapter~\ref{chapter:Lattice}.
We will use the $\chi^2$/d.o.f. as a measure of the goodness of the chiral fit, but we emphasize that this is only a qualitative measure.  
We do not have a full understanding of the systematic errors on the lattice data, and so we add conservative 3\% errors when performing all fits.
As such, the $\chi^2$/d.o.f. cannot be consider an absolute measure of the goodness of fit in the usual way.

\begin{figure}[t]
	\begin{center}
		\includegraphics[width=0.7\textwidth]{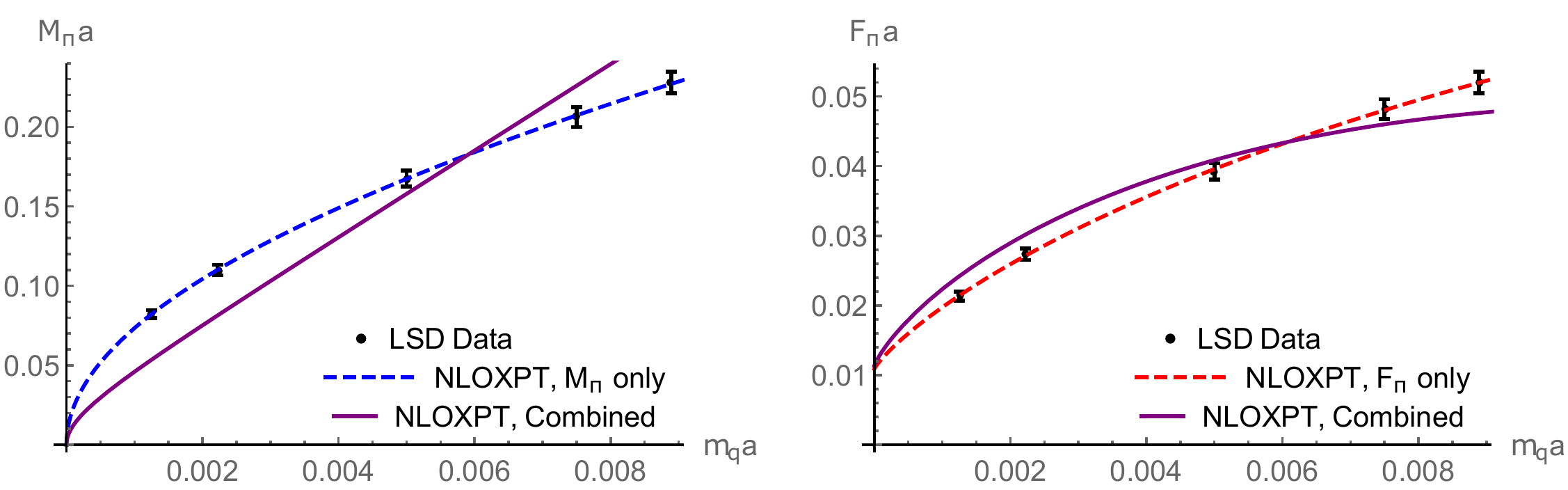}
		\includegraphics[width=0.7\textwidth]{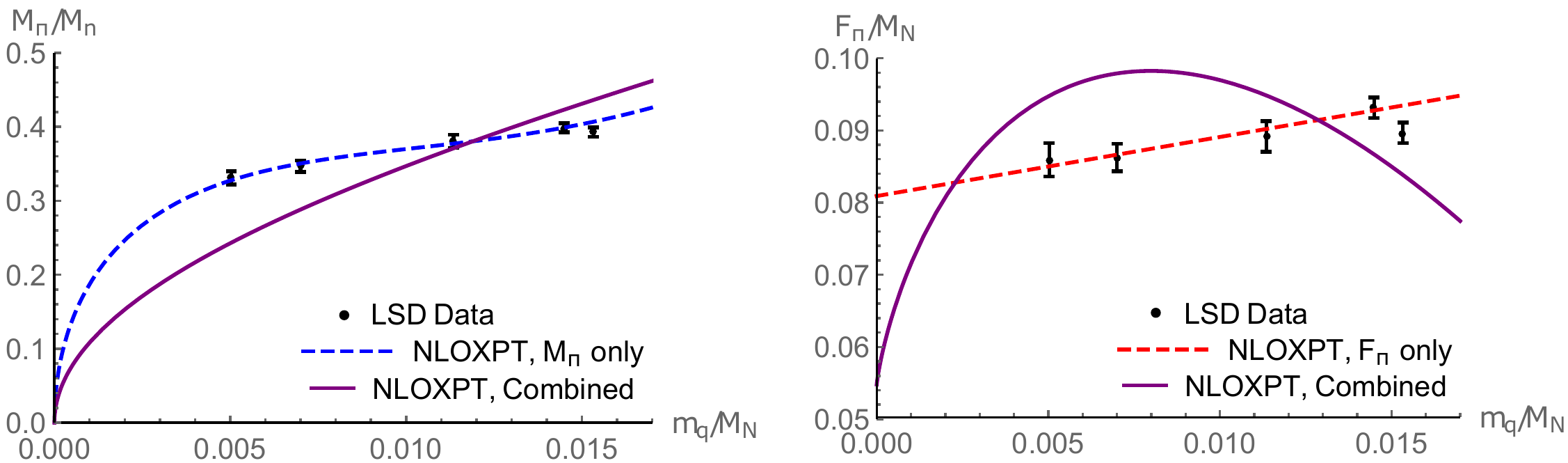}
	\end{center}
	\caption{Fits of $N_f = 8$ LSD data \cite{Appelquist:2016viq,Gasbarro:2017fmi} in lattice units (top) and nucleon units (bottom) to NLO$\chi$PT.  Dashed lines are fits to individual quantities.  The solid purple line is a simultaneous fit of $M_\pi$ and $F_\pi$. Fit lines are drawn for the central values of fit coefficients. Conservative 3\% error bars have been added to lattice data to account for possible systematic errors.  This figure originally appears in Ref.~\cite{Gasbarro:2017fmi}.}
	\label{fig:NLOXPT-fit}
\end{figure}
First we investigate a fit of NLO $\chi$PT.  
Since the chiral Lagrangian does not contain the sigma as a dynamical degree of freedom and therefore makes no prediction about its mass, we fit only to the Goldstone observables $M_\pi$ and $F_\pi$.
The NLO$\chi$PT expressions are given in Eqs.~\ref{eq:XPT1}-\ref{eq:XPT2}.
The results for the fits are presented in Fig.~\ref{fig:NLOXPT-fit}.
Both $M_\pi$ and $F_\pi$ are individually well fit by the NLO$\chi$PT expression in both lattice units and nucleon units.  
These fits are shown by the blue and red dotted lines, respectively, in Fig.~\ref{fig:NLOXPT-fit}.
A simultaneous combined fit to $M_\pi$ and $F_\pi$ was performed for the data both in lattice units and nucleon units.  
The best fit line is shown by the purple curves in Fig.~\ref{fig:NLOXPT-fit}.
The combined fit to $M_\pi$ and $F_\pi$ in lattice units yielded a minimum $\chi^2$/d.o.f. of 29.77.
The combined fit to $M_\pi$ and $F_\pi$ in nucleon units yielded a minimum $\chi^2$/d.o.f. of 7.65.

\begin{figure}[t]
	\begin{center}
		\includegraphics[width=\textwidth]{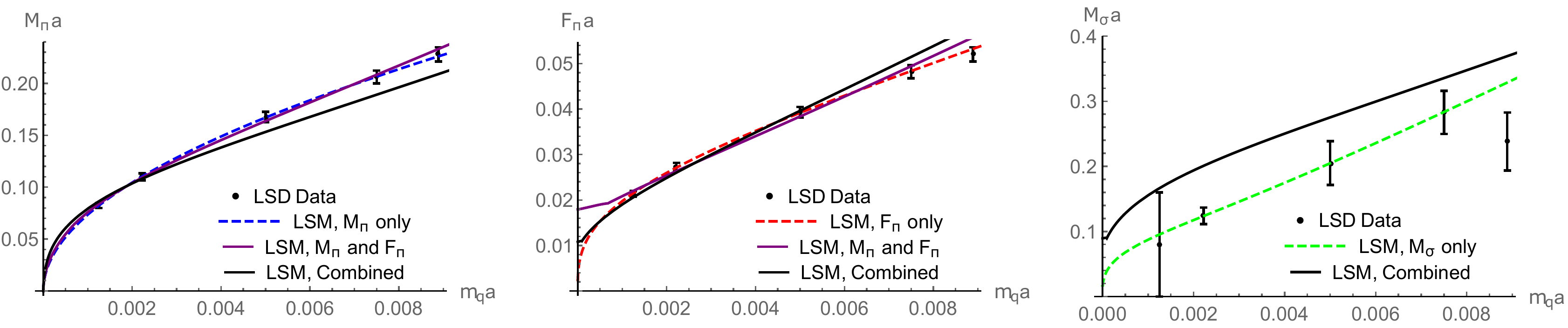}
		\includegraphics[width=\textwidth]{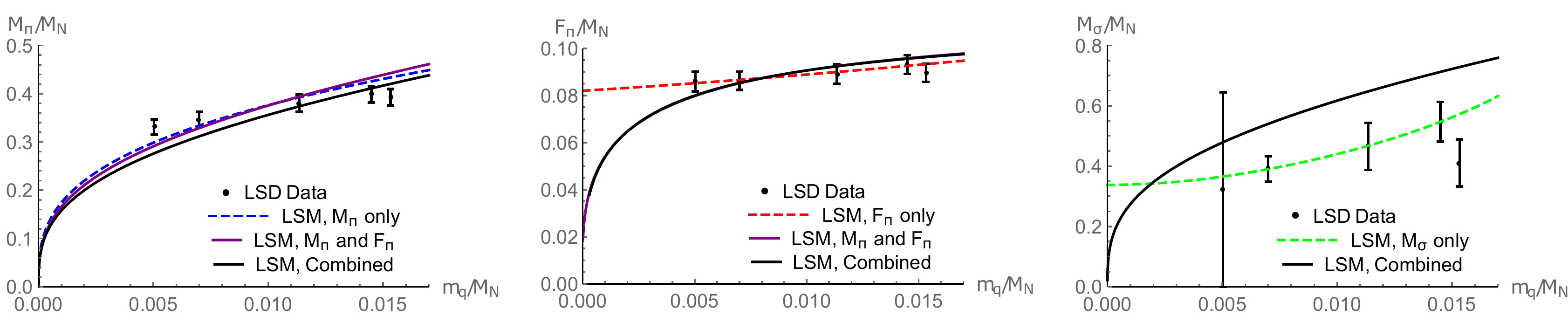}
	\end{center}
	\caption{Fits of $N_f = 8$ LSD data \cite{Appelquist:2016viq,Gasbarro:2017fmi} in lattice units (top) and nucleon units (bottom) to the linear sigma model with $V_{\text{SB}} = - \sum_{i=1}^3 c_i \mathcal{O}_i$.  Dashed lines are fits to individual quantities.  The solid purple line is a simultaneous fit of $M_\pi$ and $F_\pi$, and the solid black line is a simultaneous fit to all three quantities.  Fit lines are drawn for the central values of fit coefficients. Conservative 3\% error bars have been added to lattice data to account for possible systematic errors.  This figure originally appears in Ref.~\cite{Gasbarro:2017fmi}.}
	\label{fig:LSM-fits}
\end{figure}
Next we consider a fit to the linear sigma EFT where the symmetry breaking potential only contains the operators $\mathcal{O}_1$, $\mathcal{O}_2$, $\mathcal{O}_3$ specified in Table.~\ref{table1}.
While this symmetry breaking potential does not correspond to a power counting for a particular value of $\alpha$ as presented in Section~\ref{sec:LSM}, it is interesting to test whether the linear sigma EFT can provide a good fit using only chiral breaking operators that are first order in the spurion field.
The linear sigma EFT expressions are given in Eqs.~\ref{eq:SB1}-\ref{eq:SB3}.
The results for the fits are presented in Fig.~\ref{fig:LSM-fits}.
Each quantity, $M_\pi$, $F_\pi$, and $M_\sigma$ is individually well fit in both lattice units and nucleon units.
The results for the individual fits are shown as dashed blue, red, and green lines respectively.
The combined fit to only $M_\pi$ and $F_\pi$ is shown by the purple curves.
The combined fit to $M_\pi$ and $F_\pi$ in lattice units yielded a minimum $\chi^2$/d.o.f. of 0.6.
The combined fit to $M_\pi$ and $F_\pi$ in nucleon units yielded a minimum $\chi^2$/d.o.f. of 1.9.
We also performed a combined fit to all three quantities, $M_\pi$, $F_\pi$, and $M_\sigma$, shown by the black curves.
The combined fit to all three quantities in lattice units yielded a minimum $\chi^2$/d.o.f. of 6.1.
The combined fit to all three quantities in nucleon units yielded a minimum $\chi^2$/d.o.f. of 3.75.

\begin{figure}[t]
	\begin{center}
		\includegraphics[width=\textwidth]{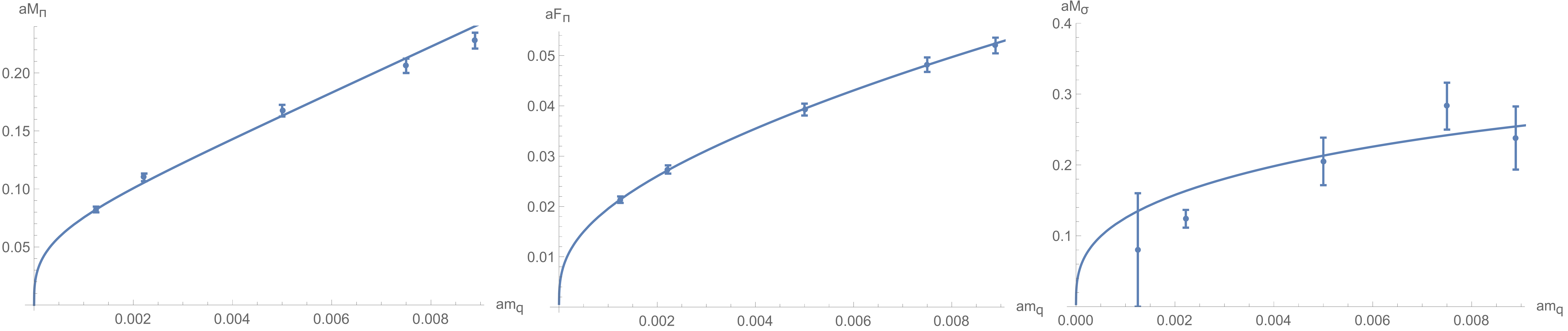}
		\includegraphics[width=\textwidth]{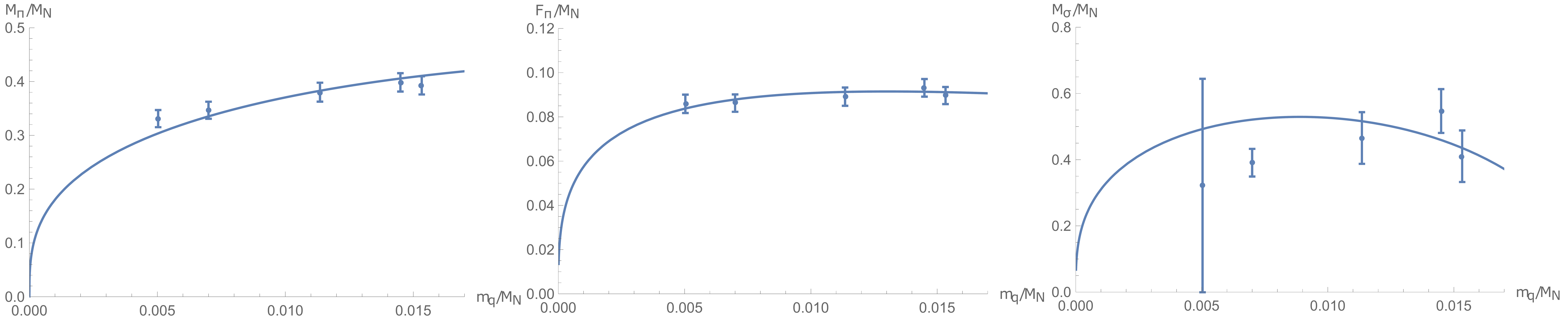}
	\end{center}
	\caption{Fits of $N_f = 8$ LSD data \cite{Appelquist:2016viq,Gasbarro:2017fmi} in lattice units (top) and nucleon units (bottom) to the linear sigma model with $V_{\text{SB}} = - \sum_{i=1}^9 c_i \mathcal{O}_i$.  The solid black line is a simultaneous fit to all three quantities.  Fit lines are drawn for the central values of fit coefficients. Conservative 3\% error bars have been added to lattice data to account for possible systematic errors.}
	\label{fig:LSM4-fits}
\end{figure}
Finally we consider a fit to the linear sigma EFT where the symmetry breaking potential contains the operators $\mathcal{O}_1$-$\mathcal{O}_9$ specified in Table.~\ref{table1}, which corresponds to the power countings with $3/5 < \alpha \leq 1$.
The results for the fits are presented in Fig.~\ref{fig:LSM4-fits}.
We perform a simultaneous fit to $M_\pi$, $F_\pi$, and $M_\sigma$ shown by the black curves.
The combined fit to all three quantities in lattice units yielded a minimum $\chi^2$/d.o.f. of 1.3.
The combined fit to all three quantities in nucleon units yielded a minimum $\chi^2$/d.o.f. of 1.4.

In summary, the leading order linear sigma EFT has provided a substantial improvement over NLO$\chi$PT.  
In fitting the Goldstone observables, $M_\pi$ and $F_\pi$, NLO$\chi$PT was not able to provide a good combined fit.
The linear sigma EFT significantly reduced the $\chi^2$/d.o.f. even with only a restricted set of operators, $\mathcal{O}_1$- $\mathcal{O}_3$.
The full basis of leading order chiral breaking operators $\mathcal{O}_1$- $\mathcal{O}_9$ was able to provide a good combined fit to $M_\pi$, $F_\pi$, and $M_\sigma$, at least qualitatively.
In future works, we would like to extend this analysis to keep track of systematic errors in the lattice data and to extract the errors on the fit coefficients in the chiral fits.
We would also like to fit to an expanded set of observables, including the $\pi\pi$ scattering length studied in Section~\ref{sec:scattering}, in order to further test and constrain the linear sigma EFT.

%%%%%%%%%%
%%%%%%%%%%%
\chapter{New Methods for CFTs on the Lattice \label{chapter:QFE1}}

The preceding sections have demonstrated some of the difficulties that arise when studying conformal (or nearly conformal) systems using traditional lattice methods.
As one approaches a conformal fixed point, correlations grow without bound such that finite volume artifacts become more significant.
While we have not yet discussed conformal correlation functions and critical scaling dimensions, it is generically true that correlation functions at a conformal fixed point take on power law rather than exponential form, which leads to difficulty in extracting the eigenvalues of an individual state.
When studying a conformal fixed point numerically, one is interested in extracting the CFT data -- the scaling dimensions and the OPE coefficients -- rather than extracting masses, but the explicit breaking of dilatation symmetry by the lattice regulator can lead to large lattice artifacts. 
Here we discuss a new formulation of lattice field theory that promises to ameliorate many of the challenges of numerical lattice studies of CFTs.

In developing lattice methodology, it is often useful to first recast the problem into a more agreeable form in the continuum and then to apply a lattice regularization that preserves the important symmetries.
For studying conformal field theories, it is beneficial to make use of \emph{radial quantization} \cite{Fubini:1972mf} in which a Euclidean CFT on $\mathbb{R}^d$ is mapped to the cylinder $\mathbb{R}\times\mathbb{S}^{d-1}$.
\begin{equation}
ds_{\mathbb{R}^d}^2 = dr^2 + r^2 d\Omega_{d-1}^2 = e^{2t} \left( dt^2 + d\Omega_{d-1}^2\right) \rightarrow dt^2 + d\Omega_{d-1}^2 = ds_{\mathbb{R}\times \mathbb{S}^{d-1}}^2
\end{equation}
The coordinates $r$ and $t$ are related simply by an exponential map, $r = e^t$.  
One sees that the two spaces are equivalent up to a Weyl factor $\Omega^2 = e^{2t}$.
When one tunes the theory to the critical region, Weyl factors cancel out of homogeneous ratios of correlation functions such that the conformal field theory is insensitive to these local rescalings of the metric.  
The two spaces $\mathbb{R}^d$ and $\mathbb{R}\times\mathbb{S}^{d-1}$ can be viewed as equivalent in this sense.
Spherical shells at fixed $r$ in $\mathbb{R}^d$ are mapped to spherical cross sections of the cylinder at fixed $t$ in $\mathbb{R}\times\mathbb{S}^{d-1}$, and the dilatation operator becomes an operator which generates translations down the cylinder.  
Correlation functions which exhibit power law behavior in the Euclidean distance at the conformal point are mapped to correlation functions which decay exponentially in $t$ down the cylinder, and the timescale which governs the exponential decay is exactly the scaling dimension of the operator. 
Thus, in this formalism one is poised to extract nonperturbative CFT data from Euclidean correlation functions much in the same way that one would extract hadron masses from Euclidean correlation functions in lattice QCD.

In any numerical implementation of this idea, one will first have to compactify the space in some why such that the volume is finite. 
The simplest choice is to compactify the time direction with periodic boundary conditions such that the geometry becomes $\mathbb{S} \times \mathbb{S}^{d-1}$.  
The infinite volume limit will correspond to taking the circle infinitely long compared to the radius of the sphere; i.e. the aspect ratio goes to infinity.
In two dimensions, the continuum geometry of radial quantization (after compactification) is $\mathbb{S} \times \mathbb{S} = \mathbb{T}^2$.
That is, a 2-torus.  
At this point, one might ask how this is different from the starting point of a canonical lattice construction of flat, two dimensional space with periodic boundary conditions.
The distinction comes in how one takes the infinite volume limit.  
Denote the radii of the the circles as $R_1$ and $R_2$.  
The limit $R_1,R_2 \rightarrow \infty$ with $R_1 / R_2$ fixed corresponds to $\mathbb{S}\times \mathbb{S} \rightarrow \mathbb{R}^2$, while the limit $R_1 \rightarrow \infty$ with $R_2$ fixed corresponds to $\mathbb{S} \times \mathbb{S} \rightarrow \mathbb{R} \times \mathbb{S}$.
Notice that the problem of large finite volume corrections is greatly alleviated in this geometry.  
Roughly speaking, doubling the temporal extent of the cylinder corresponds to squaring the physical volume in the original Euclidean space.
Although, we remark that the map between the two spaces becomes more nontrivial after compactification, and it is only in the infinite volume limit that the two spaces are related by a simple Weyl factor.

Having identified a preferred formulation of the problem in the continuum, we may turn our attention to how this problem may be formulated on the lattice.
It was pointed out by Cardy \cite{Cardy:1984rp} that conformal invariance leads to universal scaling behavior for ``infinite strips,'' which was demonstrated explicitly in a variety of analytical and numerical calculations \cite{Luck:1982nk,Privman:1984zz}.
Cardy subsequently pointed out that for $d > 2$, the infinite strips $\mathbb{R} \times \mathbb{S}^{d-1}$ are curved, which may lead to difficulties in using radial quantization as a numerical (lattice) method for extracting scaling dimensions \cite{Cardy:1985lth}.
More recently, Brower et al.~attempted a lattice study of the 3D Ising conformal fixed point on $\mathbb{R} \times \mathbb{S}^2$ in which the 2-sphere was approximated by a regular icosahedron \cite{Brower:2012vg}.
While the system exhibited critical behavior, the two point correlation function at criticality indicated a breaking of rotational symmetry at the $l=3$ level that persisted in the continuum limit.  
This indicated that the full continuum isometries of the 2-sphere were not being recovered and therefore that the fixed point did not correspond strictly to the 3D Ising CFT on $\mathbb{R}^d$.
A better lattice approximation to the continuum $\mathbb{S}^2$ was needed.
In the last five years, the work has been extended to incorporate smooth approximations to $\mathbb{S}^2$ \cite{Brower:2014daa,Brower:2015zea,Brower:2018szu}.
The more complicated lattices necessary for a smooth approximation to $\mathbb{S}^2$ introduced additional complexity into the problem, including the need for explicit renormalization by quantum counterterms, and for simplicity the authors focused their study on scalar $\phi^4$ theory in two dimensions on the $\mathbb{S}^2$.
  
Recently the authors have achieved a successful implementation of the methodology on $\mathbb{S}^2$ by demonstrating the recovery of the continuum limit and the successful extraction of CFT data corresponding to the minimal $c=1/2$ Ising CFT in two dimensions \cite{Brower:2016moq,Brower:2018szu}
They also argue that the methodology is applicable not only for lattice studies of theories on spheres and cylinders, but more generally for nonperturbative quantum field theory on any smooth Riemannian manifold.  

In this chapter, we review the methodology in it's most general form.
We will discuss the construction of lattice actions for scalar fields \cite{Brower:2016vsl} on simplicial lattices which approximate smooth Riemann manifolds.  
The construction has also been worked out for Dirac-Wilson fermions, and we refer the reader to Ref.~\cite{Brower:2016vsl} for details of the construction.
In Chapter~\ref{chapter:QFE2}, we present examples of the methodology applied to scalar field theories and fermionic theories on $\mathbb{S}^2$ and $\mathbb{R}\times\mathbb{S}^2$.

\section{General Approach to Lattice Regularization of Quantum Field Theory on Curved Riemannian Manifolds}
In this short section, we provide an overview of the general course of action for constructing a lattice regularized quantum field theory in curved space.  
Each step will be discussed in detail in the sections that follow.
For a general quantum field theory consisting of scalar $\phi$, spinor $\psi$, and vector $A$ fields on a Riemannian manifold $\{\mathcal{M},g\}$, the lattice construction consists of four steps:
\begin{enumerate}
\item \textbf{Topology:} The continuum (target) manifold $\mathcal{M}$ is replaced by an infinite sequence of simplicial complexes $\{\mathcal{M}_\sigma\}_{\sigma \in \mathbb{N}}$, each homeomorphic to the target manifold and composed of elementary $d$-simplices.  $\sigma$ denotes the \emph{refinement} of the simplicial complex, which is proportional to the linear size of the system: $\sigma \propto \sqrt[d]{N_V}$ and $N_V$ is the number of vertices (0-simplicies) in the complex. 
\item \textbf{Geometry:} A simplicial metric $g_\sigma$ is constructed at each refinement by assigning lengths $l_{ij}$ to links $\langle ij \rangle$ and extending the metric into the interior of each simplex with piecewise flat volumes, promoting the simplicial complex to a Regge Manifold.  The construction is arranged such that $\lim_{\sigma \rightarrow \infty} (\mathcal{M}_\sigma,g_\sigma) = (\mathcal{M},g)$ corresponds to the continuum limit.
\item \textbf{Hilbert Space:} The field space is truncated at each refinement by introducing a finite element basis for the fields on each simplex.  Alternatively, one may introduce form fields onto the Regge Manifold using Discrete Exterior Calculus.  The truncated space of fields will define the Hilbert space of the quantum theory through the path integral.
\item \textbf{Quantum Correction:} The lattice action is supplemented with explicit counterterms to cancel UV fluctuations sensitive to the nonuniformities in the Regge Manifold near the UV cutoff.
\end{enumerate}
In what follows, we will focus on scalar fields.  
The construction for Dirac-Wilson fermions is detailed in Ref.~\cite{Brower:2016vsl}, and detailed studies of gauge fields and general $k$-form fields are left to future works.

\section{Topology: Simplicial Complexes \label{sec:QFE12}}
The first step is somewhat of a formality, but it is useful both as a conceptual step in the mathematical construction of a discrete manifold and as a practical step when writing code for numerical implementations of discrete manifolds.
For each \emph{refinement}, the target manifold is partitioned into a set of $d$-simplices which cover the manifold yielding a simplicial $d$-complex, and higher refinements are defined such that the number of $d$-simplices in $\mathcal{M}_{\sigma+1}$ is strictly greater than the number of $d$-simplices in $\mathcal{M}_\sigma$.  
$d$-simplices are ``glued together'' at shared faces which are $(d-1)$-simplices, and these $(d-1)$-simplices are in turn glued together at shared $(d-2)$-simplices, iteratively giving a sequence: $\sigma_d \rightarrow \sigma_{d-1} \rightarrow ... \rightarrow \sigma_1 \rightarrow \sigma_0$.
This hierarchy is specified by the boundary operator.
\begin{equation}
\partial \sigma_k (i_0 ... i_k) = \sum_{l=0}^k (-1)^l \sigma_{k-1}(i_0 ... \hat{i_l}...i_k) \label{eq:boundaryop}
\end{equation}
where $\hat{i_l}$ means to exclude this site, and the sign specifies the orientation of the simplex.  
Note that the simplicial complexes that we consider are \emph{homogeneous} because every $k$-simplex for $k<d$ is the face of a $(k+1)$-simplex.
At this point, it is important to recognize that the construction is purely topological in that we have not assigned lengths to links or specified the geometry on the interior of the $d$-simplices  
We have not assigned a metric anywhere.
$\mathcal{M}_\sigma$ is best described as an \emph{abstract simplicial complex}, and a particular \emph{geometrical realization} will be defined in Section~\ref{sec:QFE13} when we introduce lengths and a metric.

An abstract simplicial complex is a purely combinatoric description of the discrete geometry, defined as follows.
A family of nonempty finite sets is an abstract simplicial complex if for every set $X$ which is an element of the family and for every nonempty subset $Y \subseteq X$, Y is also an element of the family. 
The definition implies a simplicial structure in that every subset of vertices of a simplex is itself a simplex of lower dimensionality, and therefore all subsets should be included as part of the simplicial complex.
A set of cardinality $k+1$ in the family is a $k$-simplex, and the elements of the set are vertices of the simplex.
Note that we take these sets to be ordered such that the ordering specifies the orientation of the simplex: even permutations of set elements are equivalent, while odd permutations give the simplex with the opposite orientation.
At the lowest level, the abstract simplicial complex provides a \emph{graph} -- a set of points connected to nearest neighbors by links -- and the graph encodes the topology of the space. 
If one would prefer to work with nonsimplicial complexes, a similar discussion could be carried out for more general abstract cell complexes. 
However, for the discussion that follows we prefer, for simplicity, to restrict the discussion to simplicial complexes.

\begin{figure}[t]
\includegraphics[width=0.75\textwidth]{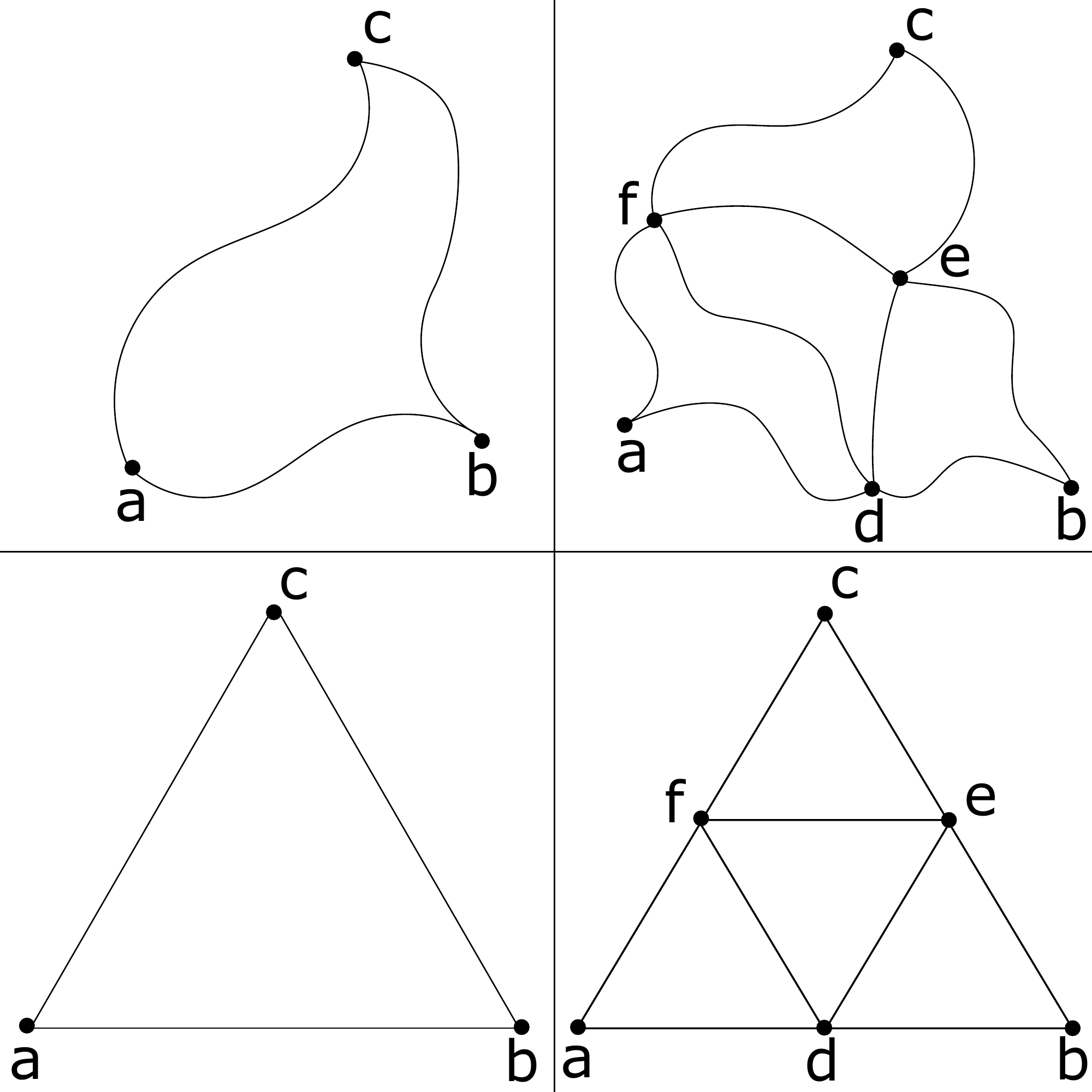}
\centering
\caption{The abstract simplicial complex (top row) and a particular geometrical realization (bottom row) for a simplicial complex consisting of a single 2-simplex (left column) and four 2-simplexes (right column). \label{fig:simplicial_complex_example}}
\end{figure}

For the first refinement $\mathcal{M}_1$ we typically choose our abstract simplicial complex to be isomorphic to the abstract simplicial complex of a regular polyhedron.  
This allows us to choose geometrical realizations that preserves a discrete subgroup of the continuum symmetries of the manifold.
For a simple example, consider the target manifold $\mathcal{M} = \mathbb{R}^2$.  
We may choose our first refinement 
\begin{equation}
\mathcal{M}_1 = \{ \{a,b,c\},\{a,b\},\{b,c\},\{c,a\},\{a\},\{b\},\{c\} \} \label{eq:abstractcomplex}
\end{equation}
which is the abstract simplicial complex of a triangle with vertices $a,b,c$.
When we assign a geometrical realization to the abstract simplicial complex, the first refinement, $(\mathcal{M}_1,g_1)$ typically breaks the continuum isometries to the a discrete lattice subgroup, and higher refinements do not break the symmetries further.
In the example of Eq.~\ref{eq:abstractcomplex}, we may choose the geometrical realization such that $(\mathcal{M}_1,g_1)$ is a flat equilateral triangle, which preserves the discrete subgroup $S_3 \subset O(2)$.
The abstract simplicial complex of Eq.~\ref{eq:abstractcomplex} and its geometrical realization as an equilateral triangle are illustrated in the left column of Fig~\ref{fig:simplicial_complex_example}.
Notice that the abstract simplicial complex itself, being purely topological, does not necessarily preserve any of the symmetries of the target space; it is simply a partitioning. 
However, the partitioning must be chosen judiciously in order to enable the desired symmetries to be preserved when the particular geometrical realization is assigned.

In the examples that we consider, higher refinements are always constructed iteratively from the previous refinement.  
In the example of Eq.~\ref{eq:abstractcomplex}, we may construct the second refinement in an iterative way by introducing the vertices $d$, $e$, and $f$ on the links $ab$, $bc$, and $ca$, respectively, and adding new links $de$, $ef$, and $fd$.  
Note that when we do this $ab$ (for example) is no longer a link in the simplicial complex because it has been broken by the point $d$.  
The resulting abstract simplicial complex is
\begin{flalign}
\mathcal{M}_2 = &\{ \{adf\},\{dbc\},\{fec\},\{fde\},\{ad\},\{db\},\{be\},\{ec\},\{cf\},\{fa\},\{de\},\{ef\},\{fd\},\nonumber \\
&\{a\},\{b\},\{c\},\{d\},\{e\},\{f\} \} \label{eq:abstractcomplex2}
\end{flalign} 
If we choose a particular geometrical realization in which the four 2-simplices in $\mathcal{M}_2$ are flat, equilateral triangles with the same edge length, then the $S_3$ symmetry persists at this refinement.  
The abstract simplicial complex of Eq.~\ref{eq:abstractcomplex2} and its geometrical realization as a regular lattice of equilateral triangles is illustrated in the right column of Fig~\ref{fig:simplicial_complex_example}.
We can continue to refine iteratively in this way and continue to preserve the $S_3$ symmetry at all refinements by choosing our geometrical realization such that all 2-simplices in $\mathcal{M}_\sigma$ are congruent, flat, equilateral triangles. 

There are many schemes for constructing sequences of simplices complexes other than the regular, iterative method described here.
In the classic works on random lattice field theory \cite{Christ:1982zq,Christ:1982ci,Christ:1982ck}, simplicial complexes are constructed by throwing points at random in flat space and connecting the points with links to form simplices via the Delaunay construction \cite{Delaunay}.
Higher refinements are constructed by simply throwing more random points.
In the random lattice construction, one attempts to recover continuum symmetries by summing over many random lattices at each refinement.
In our construction we take a different approach, choosing to preserve the largest possible discrete subgroup of the continuum symmetries and to recover the remainder to the continuum isometries dynamically in the infrared as in conventional lattice field theory.
In the limit of infinite refinement, the simplicial complex contains an infinite number of infinitesimal partitions and is constructed to become equal to the target manifold, $\lim_{\sigma \rightarrow \infty} \mathcal{M}_\sigma = \mathcal{M}$.

At the level of abstract simplicial complexes, it is already possible to introduce form fields onto our discrete manifold. 
This makes sense intuitively because differential forms may be defined on a continuous manifold without specifying a metric tensor.
On our discrete manifold (abstract simplicial complex), a $k$-form field is defined as a map from $k$-simplices to a field $\mathbb{K}$ which is typically either $\mathbb{R}$ or $\mathbb{C}$.
\begin{equation}
\langle w_k,\sigma_k(i_0,...,i_k) \rangle \equiv w_k(i_0,...,i_k) \in \mathbb{K}
\end{equation}
We define this map to be linear on chains (formal sums of simplices) such that $\langle w_k, \sigma_k(i_0,...,i_{k})+\sigma_k(j_0,...,j_{k}) = w_k(i_0,...,i_k)+w_k(j_0,...,j_k)$.
One may arrive at these properties by defining the discrete differential forms to be integrals over simplices of continuously defined interpolating fields, but we instead take these properties as definitions of the discrete differential fields.
The generalized Stokes Theorem
\begin{equation}
\langle dw_k,\sigma_{k+1}\rangle = \langle w_k,\partial\sigma_{k+1}\rangle \label{eq:stokes}
\end{equation}
provides a natural definition of the discrete exterior derivative via the boundary operator Eq.~\ref{eq:boundaryop}.
\begin{equation}
\langle dw_k,\sigma_{k+1}(i_0,...,i_{k+1})\rangle = dw_k(i_0,...,i_{k+1}) = \sum_{l=0}^{k+1} (-1)^l \langle w_k , \sigma_k(i_0,...,\hat{i}_l,...,i_{k+1})\rangle \label{eq:exteriorderivative}
\end{equation}
Note that because the boundary operator is closed, this definition of the discrete exterior derivative is automatically closed.
It is instructive to look at the two simplest examples.  
The discrete exterior derivative of a 0-form field $\phi$ is a 1-form field given by
\begin{equation}
d\phi(ij) = \phi(i) - \phi(j)
\end{equation}
We recover the standard finite difference for the gradient of a scalar field.  The discrete exterior derivative of a 1-form field $U$ is a 2-form field given by
\begin{equation}
dU(ijk) = U(ij) + U(jk) + U(ki)
\end{equation}
We arrive at a plaquette-like object associated with the 2-simplex $\sigma_2(ijk)$.

We emphasize again that we were able to introduce all of these structures without the need for a metric tensor, embedding spaces, or any notion at distances whatsoever.  
In the next section, we will discuss particular geometrical realizations that are well suited for performing real calculations.

\section{Geometry: Regge Calculus \label{sec:QFE13}}

Here we introduce a particular geometrical realization for the abstract simplicial complexes discussed in the previous section.
We will focus on Regge Calculus \cite{Regge:1961px} in which the simplicial complex is taken to be piecewise flat.  
With this choice, all curvature becomes concentrated at the $d-2$ dimensional hinges -- at vertices in two dimensions, at links in three dimensions, etc.
We begin by introducing link lengths on our simplicial complex.  To each 1-simplex, $\sigma_1(ij)$, we assign the length $|\sigma_1(ij)|=l_{ij}$.  
The interiors of the simplices are taken to be piecewise flat volumes.
This lifts the abstract simplicial complex to a Regge representation of a manifold which is continuous but not differentiable.
The metric on the Regge manifold is piecewise constant and is determined entirely by the set of edge lengths in the simplicial complex at each refinement: $g \rightarrow g_\sigma(\{l_{ij}\})$.

\begin{figure}[t]
\includegraphics[width=0.75\textwidth]{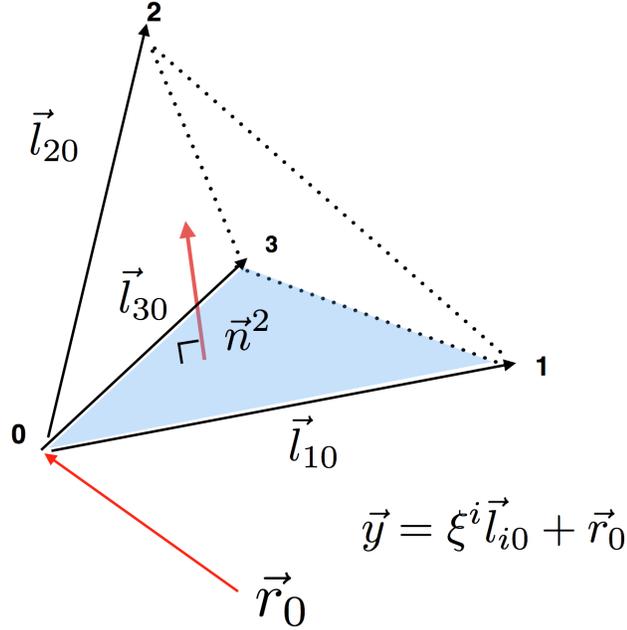}
\centering
\caption{ Visualization of 1-forms $\vec{l}_{i0}$ and dual vectors $\vec{n}^i$ on a 3-simplex. \label{fig:3plex}}
\end{figure}
On each $d$-simplex, we may choose Cartesian coordinates such that the metric tensor is simply the Kronecker delta.  
However, it is complicated to specify the range of the Cartesian coordinates which span the interior of the simplex.
A coordinate system better suited to simplicial geometry is barycentric coordinates.
We introduce the barycentric coordinates $\xi^n$ as follows.  Any point $\vec{y}$ in the interior of the simplex may be parametrized by
\begin{equation}
\vec{y} = \sum_{n=0}^{d} \xi^n \vec{r}_n
\end{equation}
where $\vec{r}_n$ are vectors pointing to the $d+1$ vertices of the simplex and the overcomplete barycentric coordinates always satisfy one constraint $\sum_{n=0}^d \xi^n = 1$.
On the interior of the simplex, the barycentric coordinates satisfy $0 \le \xi^n \le 1$.
Taking $\vec{r}_0$ as an arbitrary origin, we may define the relative vectors $\vec{l}_{i0} = \vec{r}_i - \vec{r}_0$ and use the constraint to eliminate the coordinate $\xi^0$ and remove the degeneracy of the overcomplete basis.
\begin{equation}
\vec{y} = \vec{r}_0 + \sum_{i=1}^d \xi^i \vec{l}_{i0}
\end{equation}
This step has the downside of not treating the $d+1$ vertices on equal footing.
From the measure
\begin{equation}
ds^2 = d\vec{y}\cdot d\vec{y} = \frac{\partial \vec{y}}{\partial \xi^i}\cdot \frac{\partial \vec{y}}{\partial \xi^j}d\xi^i d\xi^j
\end{equation}
we can read off the constant valued metric on each simplex in barycentric coordinates.
\begin{equation}
g_{\sigma,ij} = \frac{\partial \vec{y}}{\partial \xi^i}\cdot \frac{\partial \vec{y}}{\partial \xi^j} = \vec{l}_{i0}\cdot \vec{l}_{j0}
\end{equation}
It is clear from this discussion that the $\vec{l}_{i0}$ are components of a 1-form with basis elements $d\xi^i$.  
The dual tangent vectors $\vec{n}^i$ are given by
\begin{equation}
\vec{\nabla} = \vec{\nabla}\xi^i \frac{\partial}{\partial \xi^i} = \vec{n}^i \frac{\partial}{\partial \xi^i}
\end{equation}
In component notation, the 1-forms and dual tangent vectors resemble a vierbein and dual vierbein, each carrying one orthonormal Cartesian coordinate index $a$ and one barycentric coordinate index $i$.
\begin{equation}
l^a_{i0} = \frac{\partial y^a}{\partial \xi^i} \quad \text{and} \quad n^i_a = \frac{\partial \xi^i}{\partial n^a}
\end{equation}
It is easy to check the duality relationships
\begin{equation}
\sum_a n^i_a l_{j0}^a = \delta^i_j \quad \text{and} \quad \sum_i n^i_a l_{i0}^b = \delta^b_a
\end{equation}
The dual metric tensor is given by
\begin{equation}
g_\sigma^{ij} = \vec{n}^i \cdot \vec{n}^j
\end{equation}
and satisfies $g_\sigma^{ij} g_{\sigma,jk} = \delta^i_k$.
A representative visualization of some of the geometrical structures that we have introduced in the barycentric coordinate description of simplices is shown in Fig.~\ref{fig:3plex}.

At this stage, we have constructed a continuous, piecewise flat (not differentiable) Regge manifold; we have introduced a convenient set of coordinates for doing geometry on simplices; and we have written down the expressions for important geometrical quantities such as the metric tensor and the gradient operator in these coordinates.
We are already equipped with all of the necessary machinery for writing down simple field theories on this Regge manifold.  
For example, the scalar $\phi^4$ action is given simply by
\begin{equation}
S = \frac{1}{2}\sum_{\sigma_d \in \mathcal{M_\sigma}} \int_{\sigma_d} d^d \xi \sqrt{g_\sigma} \left[ g_\sigma^{ij} \frac{\partial}{\partial \xi^i} \phi(\xi) \frac{\partial}{\partial \xi^j}\phi(\xi) + m^2 \phi^2(\xi) + \lambda \phi^4(\xi) \right]  \label{eq:scalaronregge}
\end{equation}
We emphasize that the above action is still an action for a continuum field theory on a continuous Regge manifold.  
We have not yet discretized the problem.
We have only chosen a particular manifold $\mathcal{M}_\sigma$ and coordinate system $\xi^i$ well suited for discretization.
We will discuss the restriction to a finite field space in Section~\ref{sec:QFE14}.

\begin{figure}[t]
\includegraphics[width=0.75\textwidth]{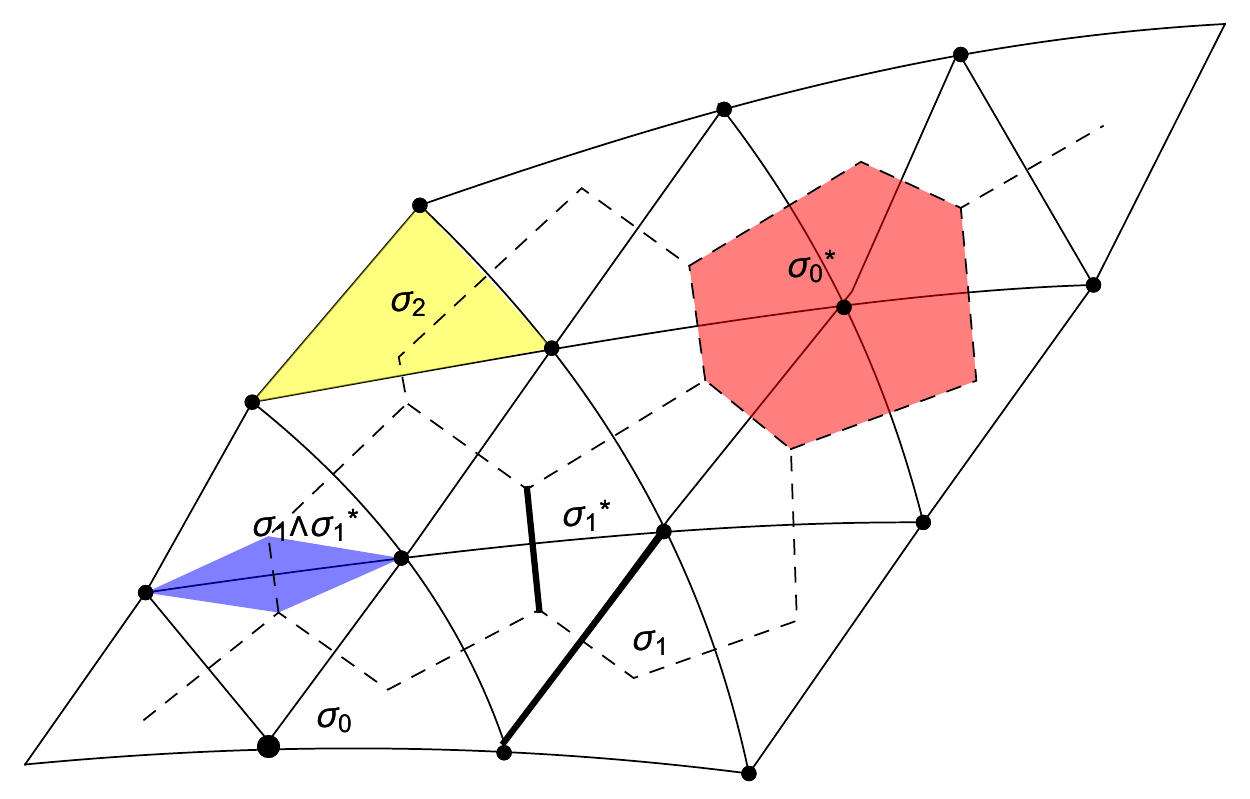}
\centering
\caption{ Regge manifold and Vorono{\"i} dual in two dimensions.  Dotted lines denote links in the dual complex while solid lines denote links in the original simplex.  A representative link $\sigma_1$ and a dual link $\sigma_1^*$ are shaded in black.  A representative 2-simplex $\sigma_2$ is shaded in yellow, and a representative dual 2-cell $\sigma_0^*$ is shaded in red.  A hybrid cell $\sigma_1 \wedge \sigma_1^*$ is shaded in blue. \label{fig:latsimplex}}
\end{figure}
Next we introduce the Vorono{\"i} dual \cite{Voronoi1908} of our Regge lattice which we denote by $\mathcal{M}_\sigma^*$.
The dual lattice is composed of polytopes $\sigma_k^*$ of dimension $d-k$ dual to the original simplices $\sigma_k$.
As in the original lattice, these polytopes form a hierarchy $\sigma_0^* \rightarrow ... \rightarrow \sigma_d^*$ specified by the boundary operator Eq.~\ref{eq:boundaryop}.
One may also think of the duality as a bijective map from simplices to polytopes in the dual lattice and from polytopes to simplices in the original lattice: $*\sigma_k = \sigma_k^*$, $**\sigma_k = *\sigma_k^* = \sigma_k$.
We my also introduce the coboundary operator $*\partial *$ which maps from $k$-simplices to $(k+1)$-simplices or from $k$-cells to $(k+1)$-cells in the dual lattice.
It's action on a $k$-simplex $\sigma_k \in \mathcal{M}_\sigma$ is given by straightforwardly applying Eq.~\ref{eq:boundaryop} and the duality operator.
\begin{equation}
*\partial*\sigma_k = \sum_{\{\sigma_{k+1} | \sigma_k \in \partial \sigma_{k+1}\}} \sigma_{k+1} - \sum_{\{\sigma_{k+1} | -\sigma_k \in \partial \sigma_{k+1}\}} \sigma_{k+1}
\end{equation}

The particular dual complex that we focus on is a circumcenter dual lattice.
It is constructed in an iterative fashion.
One first identifies the circumcenters (midpoints) of links.  
Normals are drawn from the circumcenters of the links into the interior of each 2-simplex.
Inside of each 2-simplex, the three normals intersect at exactly one point which is the circumcenter of the 2-simplex.
This procedure continues iteratively.
Once the circumcenters of all $(k-1)$-simplices are identified, normals are drawn from the circumcenters of the $(k-1)$-simplices into the interior of each $k$-simplex.
The $k+1$ normals intersect at exactly one point inside each $k$-simplex which is the circumcenter.
%Normals are drawn from the circumcenters of each 2-simplex into the interior of each 3-simplex, and again the normals intersect at a single point which is the circumcenter of the 3-simplex.  
The collection of circumcenters and normal lines constructed in this way form the graph of the Vorono{\"i} dual lattice.
Notice that without introducing a metric and a notion of distance on the links, this dual graph could not be constructed.
The notion of geometric duality is closely tied to the metric of the space.  

In the circumcenter construction, dual cells are always orthogonal to the simplex in the Regge lattice to which they correspond.
A consequence of this orthogonality is that hybrid cells $\sigma_k \wedge \sigma_k^*$ formed from a $k$-simplex and it's dual polytope form a proper tiling of the Regge manifold, and the volume formula for the hybrid cell factorizes in a simple way.
\begin{equation}
|\sigma_k \wedge \sigma_k^* | = \frac{k!(d-k)!}{d!}|\sigma_k||\sigma_k^*|
\end{equation}
An example of a two dimension Regge manifold and its Vorono{\"i} dual complex is shown in Fig.~\ref{fig:latsimplex}.

Having introduced the notion of geometric duality, we may also now define the discrete analog of Hodge duality for the discrete differential forms introduced in Section~\ref{sec:QFE12}. 
Recall that for standard differential forms on manifolds, the Hodge star operation is an inner product between the canonical volume form, $\epsilon = \sqrt{g} dx^1\wedge ... \wedge dx^d$, and a $k$-form, $w_k$.  
The metric is a crucial components of Hodge duality.
Accordingly the notion of distance is incorporated into the discrete Hodge star operation by defining it as follows \cite{Hirani:2003:DEC:959640}.
\begin{equation}
\frac{1}{|\sigma_k^*|} *w_k(\sigma_k^*) = \frac{1}{|\sigma_k|} w_k(\sigma_k) \label{eq:hodge}
\end{equation}
The discrete codifferential operator $\delta$ which maps $(k+1)$-forms to $k$ forms may be defined in terms of the discrete exterior derivative Eq.~\ref{eq:exteriorderivative} and the Hodge star operator Eq.~\ref{eq:hodge}.  It's action on a $(k+1)$-form is given by
\begin{equation}
\delta w_{k+1} = (-1)^{dk+1} * d * w_{k+1}
\end{equation}
With these definitions in place, it is possible to write down the discrete Laplace-Beltrami operator.  It's action on a 0-form field $\phi$ is given by
\begin{equation}
-\delta d \phi(i) = \frac{|\sigma_0(i)|}{|\sigma_0^*(i)|} \sum_{j\in \langle ij \rangle} \frac{|\sigma_1^*(ij)|}{|\sigma_1(ij)|} (\phi(i)-\phi(j)) = \frac{1}{\sqrt{g(i)}} \sum_{j\in\langle ij \rangle} V^D_{ij} \frac{\phi(i)-\phi(j)}{l_{ij}^2}
\end{equation}
where $V^D_{ij} = |\sigma_1(ij)\wedge \sigma_1^*(ij)|$.

In this section we have introduced a Regge metric and its Vorono{\"i} dual.  
On the one hand, we are already equipped to study a discretized theory of form fields following the prescription of Discrete Exterior Calculus \cite{Hirani:2003:DEC:959640}.  
On the other hand, the Regge manifold is a continuous space and we may naively write down a field theory on this manifold as in Eq.~\ref{eq:scalaronregge}.
The latter approach consists of an infinite number of field degrees of freedom.
%This leave us with an infinite number of degrees of freedom.
In the next section, we discuss a different approach to constructing a discrete field space, the finite element approach, in which this continuous field space is truncated using a finite element basis.
%Regge Calculus.  Assign lengths.  Flat interiors.  Metric tensor.  Barycentric coordinates and continuous scalar action.  Circumcenter dual.  Hybrid cells.  Hodge duality.  Codifferential.  Laplace Beltrami.  FIGURES!

%Spherical Finite Elements (maybe in next section?)
%
%
\section{Hilbert Space: Finite Elements and Discrete Exterior Calculus \label{sec:QFE14}}

We have discussed how to introduce form fields on our discrete manifold in a natural way by associating numbers with simplices in the complex.
This provides a simple construction of a naturally discrete Hilbert space, but we have distanced ourselves from the continuum description of the field theory by constructing our fields on the discrete manifold.
Here we discuss an alternative construction of discrete Hilbert space which is a truncation of infinite number of degrees of freedom on a continuous Regge manifold.

\subsection{Finite Element Construction for Scalar Field Theory}
Here we discuss the finite element construction for scalar fields (for further reading on finite element methods, we refer the reader to \cite{strang2008analysis}).
We truncate the field basis to a discrete set of degrees of freedom $\phi_n$ living at the sites of the Regge manifold.
The field everywhere in the interior of each simplex is given by an interpolation.
\begin{equation}
\phi_\sigma(x) = \sum_{n=0}^d E^n(x) \phi_n
\end{equation}
This defines our finite element basis. 
The elements obey $E^n(x_m) = \delta^n_m$ such that the interpolating field takes the proper value at the vertices.
We impose the constraint $\sum_{n=0}^d E^n = 1$ in order to properly interpolate a constant function.
Using this constraint to eliminate $E^0$, we have
\begin{equation}
\phi_\sigma(x) = \phi_0 + \sum_{i=1}^d E^i(x) \left(\phi_i - \phi_0\right)
\end{equation}

Linear finite elements are the simplest choice. 
We take the finite elements to be the barycentric coordinates themselves, $E^i(\xi) = \xi^i$.
The gradients of the field with respect to the barycentric coordinates are constant functions.
\begin{equation}
\partial_i \phi_\sigma(\xi) = \phi_i - \phi_0
\end{equation}
Plugging into our expression for the continuum scalar field action on a Regge manifold in barycentric coordinate Eq.~\ref{eq:scalaronregge}, we find for the kinetic term in the action on a single simplex,
\begin{equation}
I_\sigma = \frac{1}{2} \int_{\sigma_a} d^d \xi \sqrt{g} g^{ij} (\phi_i - \phi_0)(\phi_j - \phi_0)
\end{equation}
The integrand is a constant and the integral reduces to a simple Feynman parameter integral.
\begin{equation}
\int_0^1 d\xi^0 ... d \xi^d \delta(\sum_{n=0}^d \xi^n - 1) = \frac{1}{d!}
\end{equation}
so that
\begin{equation}
I_\sigma = \frac{1}{2d!} \sum_{i,j=1}^d \sqrt{g} g^{ij}(\phi_i - \phi_0)(\phi_j - \phi_0) \label{eq:vertexform}
\end{equation}
This form of the scalar kinetic action is somewhat asymmetric looking.  
It is anchored around the zeroth vertex of each simplex due to our arbitrary choice of an origin, but it is nonetheless correct.

A more symmetric form is achieved by evaluating the action for the linear finite element fields in a different way.
Recall that the gradients of the barycentric coordinates are the dual vectors, $\vec{\nabla}\xi^i = \vec{n}^i$.  
The gradient of $\phi_\sigma$ expanded in a linear finite element basis is
\begin{equation}
\vec{\nabla}\left(\sum_{n=0}^d \xi^n \phi_n \right) = \sum_{n=0}^d \vec{n}^n \phi_n
\end{equation}
The duals vectors are constant on each simplex, so plugging into the action the integrand is once again constant and the integration yields a volume factor.
\begin{equation}
I_\sigma = \frac{|\sigma_d|}{2} \sum_{n,m = 0}^{d} \vec{n}^n \cdot \vec{n}^m \phi_i \phi_j = -\frac{\sqrt{g_\sigma}}{2 d!} \sum_{\langle ij \rangle} \vec{n}^i \cdot \vec{n}^j \left(\phi_i - \phi_j\right)^2 \label{eq:edgeform}
\end{equation}
In the second equality we have made use of the identities $|\sigma_d| = \sqrt{g_\sigma} / d!$ and $\sum_{m}{\vec{n}^m} = 0$. 
Eq.~\ref{eq:edgeform} provides an appealing, symmetric expression which treats all edges in the simplex on equal footing.
One can prove the equivalence of Eq.\ref{eq:vertexform} and Eq.~\ref{eq:edgeform} using the identity $\sum_{m}{\vec{n}^m} = 0$ for the dual vectors.

Focusing for a moment on two dimensions, we may compute the dual vectors and the metric explicitly in terms of the edge lengths $l_{12},l_{23},l_{31}$ on each simplex.  
The expression for the action on a single triangle in two dimensions in terms of the edge lengths is
\begin{flalign}
I_\sigma = &\frac{l^2_{31} + l^2_{23} - l^2_{12}}{8 A_{123}} (\phi_1 - \phi_2)^2 + (23) + (31) \nonumber \\
=&\frac{1}{2} A^{(3)}_{12} \frac{(\phi_1 - \phi_2)^2}{l_{ij}^2} + (23) + (31) 
\end{flalign}
where $A_{12}^{(3)}$ is the area of the triangle formed by the vertices $1$, $2$, and the circumcenter $\sigma_2^*(123)$.
Summing the action over the entire simplicial complex yields the simple form
\begin{equation}
\sum_\sigma I_\sigma[\phi] = \frac{1}{2} \sum_{\langle ij \rangle} A^D_{ij} \left(\frac{\phi_i - \phi_j}{l_{ij}} \right)^2 \label{eq:FEMLap2d}
\end{equation}
Each link $\langle ij \rangle$ receives two contributions, one from each triangle that borders it.
The resulting weight is the area of the hybrid dual cell $A^D_{ij} = l_{ij} |\sigma_1^*(ij)|/2 = |\sigma_1(ij) \wedge \sigma_1^*(ij)|$.
We find that in two dimensions, linear finite elements give an equivalent expression to the discrete exterior calculus action.

Let us remark briefly what happens above two dimensions.
The weight of the finite element scalar action for each link is proportional to the inner product of dual vectors, which is proportional to the cosine of the hinge angle opposite the link.
This weight vanishes when a hinge angles becomes right.
On the other hand, the discrete exterior calculus action gives a weight which is proportional to the volume of the hybrid dual cell at the link.
The hybrid cell volume vanishes when a right angle forms at a vertex $k$ sharing a face with the link $\langle ij \rangle$, causing the face circumcenter to coincide with the link circumcenter and the dual volume to vanish.
In two dimensions, the hybrid dual volume vanishes precisely when the dual vectors become orthogonal.
In higher than two dimensions it is possible to have a right angled hinge without the hybrid dual volume vanishing.  
By this logic, it is easy to show by counterexample that the FEM action and the DEC action are not equivalent for $d > 2$.
Therefore, in $d > 2$ the FEM and DEC methods provide alternate formulations for the classical scalar field theory, and the choice of classical action is a matter of preference which should be explored for each particular application.

%FEM basis, linear fem, scalar field theory, site form and link form (figures), FEM vs DEC for scalar fields, 
%FEM for dirac fields, comments on gauging the theory

%\subsection{Finite Element Approach to Dirac Wilson Fermions}
%FEM will also provide a way to introduce Dirac spinor which is an advantage over the exterior calculus approach which does not naturally incorporate spinor fields.
%
%
\section{Interactions and Quantum Corrections \label{sec:QFE15}}
Thus far, we have presented a systematic method for constructing classical field theory on a discrete Riemann manifold.
The geometric space was constructed using simplicial geometry and Regge calculus, and fields have been introduced through either discrete exterior calculus or the finite element method.
Here we make some general remarks about interactions and the renormalizability of the quantum field theory.
It is difficult to make general statements or to provide proofs about renormalizability for these lattice theories on curved spaces.
Instead, we will motivate some general principles and guidelines for renormalizing these types of lattice theories by working in close analogy with existing methods for conventional lattice field theory.

\subsection{Interaction Terms in DEC and FEM}
We have mainly focused on differential operators and kinetic terms in the preceding discussion.  
Interaction terms may be introduced into the Lagrangians following either the prescription of discrete exterior calculus or the finite element method.
In the discrete exterior calculus picture, a potential for scalar (0-form) fields may be introduced by simply summing with the proper metric weight.  For example
\begin{equation}
V_{\text{int, DEC}} \supset \sqrt{g(i)} \lambda_p \phi^p(i)
\end{equation}
where the local metric at a site is the volume of the Vorono{\"i} dual cell.
More generally, interactions between $k$-form fields may be constructed by defining an appropriate inner product between forms \cite{Hirani:2003:DEC:959640}.
We leave this discussion to a future work.

Alternatively, one may construct interactions following the finite element prescription to the letter by plugging the (linear) finite element expansion for the field into the continuum action and performing the integration over each simplex as we have done for the kinetic term.
Strict adherence to the finite element prescription has the advantage that the finite element method constitutes a variational approach.
The spectrum will always converge from above and thus high modes will be suppressed in the propagator which may mitigate lattice artifact due to irregularities in the lattice near the UV cutoff.
However, the FEM prescription also generally leads to point-split expressions in the lattice action for local interaction terms in continuum.
For example, the scalar mass term in two dimensions is given by
\begin{equation}
\int_{\sigma_2(123)} d^2 \xi \sqrt{g_\sigma} \phi^2(\xi) = \frac{A_{123}}{6} (\phi_1^2 + \phi_2^2 + \phi_3^2 + \phi_1\phi_2 + \phi_2\phi_3 + \phi_3\phi_1 )
\end{equation}
These point split interactions complicate the lattice action, and for algorithmic purposes it is often preferable to consider ultralocal interactions.
In our numerical studies presented in Chapter~\ref{chapter:QFE2}, we follow the DEC prescription and introduce ultralocal interactions.
%
%We introduce local interactions into our Lagrangian with the weight $\sqrt{g_\sigma(x)} = |\sigma_0^*(x)|$ equal to the Voronoi dual area at the site.  
%One may instead follow the finite element prescription for the interaction terms, but in general this leads to point split interactions rather than ultralocal interactions, and the difference is an $\mathcal{O}(a^2)$ effect.
%For algorithmic purposes, it is often preferable to consider ultralocal interactions, so we take the former approach in our numerical studies presented in Chapter~\ref{sec:QFE2}.

\subsection{Renormalizability}
We will show by explicit example in Chapter~\ref{chapter:QFE2} that a straightforward application of the DEC/FEM formalism fails to produce a lattice quantum field theory that converges to the continuum field theory in the continuum limit.
Here we present a method for correcting the DEC/FEM action with explicit counterterms, leading to a renormalized lattice field theory with a well defined continuum limit.  
Proofs are difficult for lattice field theory on a general Regge manifold, so we will support our arguments by working closely with the established renormalization theorem for conventional lattice field theory and providing explicit examples to support our claims.

In conventional lattice field theory, one may think about renormalization either perturbatively or nonperturbatively \cite{Luscher:1998pe}. 
In perturbation theory, the lattice is simply a specific type of UV regulator.
It is natural to ask for a renormalizable field theory if one can remove all the divergences that arise in lattice perturbation theory as the lattice spacing $a \rightarrow 0$ with a finite number of counterterms at each order.
%$1/a$ plays the role of a UV regulator.
Reisz \cite{Reisz:1987da,Reisz:1987px,Reisz:1987pw,Reisz:1988kk} introduced a systematic method for keeping tracking divergences in lattice Feynman integrals.
Reisz defines a lattice degree of divergence $\text{Deg}(I)$ for a Feynman integral $I$ -- which is the counterpart of the superficial degree of divergence familiar from continuum field theory -- and proves that any integral with $\text{Deg}(I) < 0$ is finite and given by the naive continuum limit as $a \rightarrow 0$.

\begin{figure}%
    \centering
    \subfloat[1-Loop]{{\includegraphics[width=0.4\textwidth]{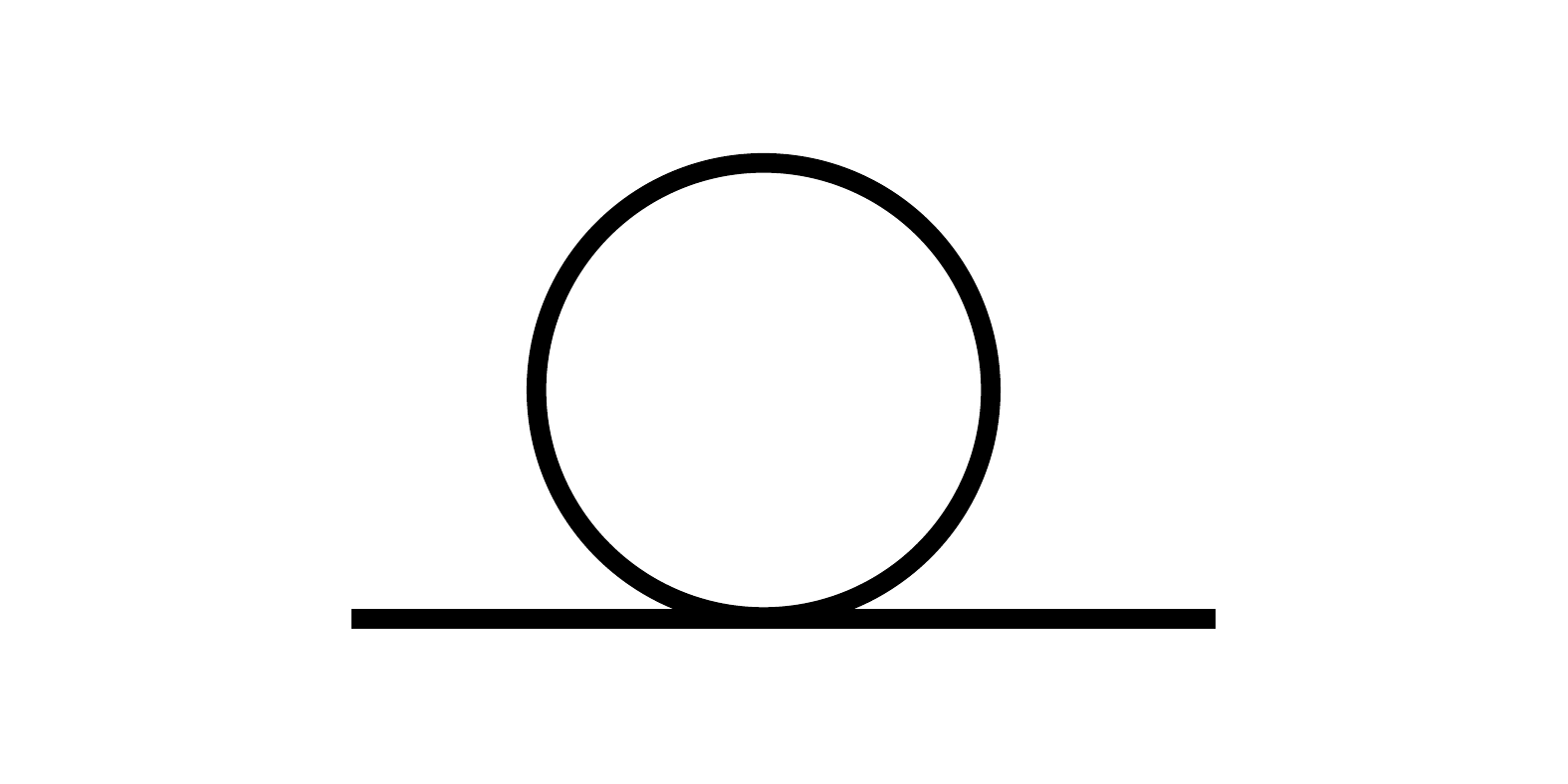} }}%
    \qquad
    \subfloat[2-Loop]{{\includegraphics[width=0.4\textwidth]{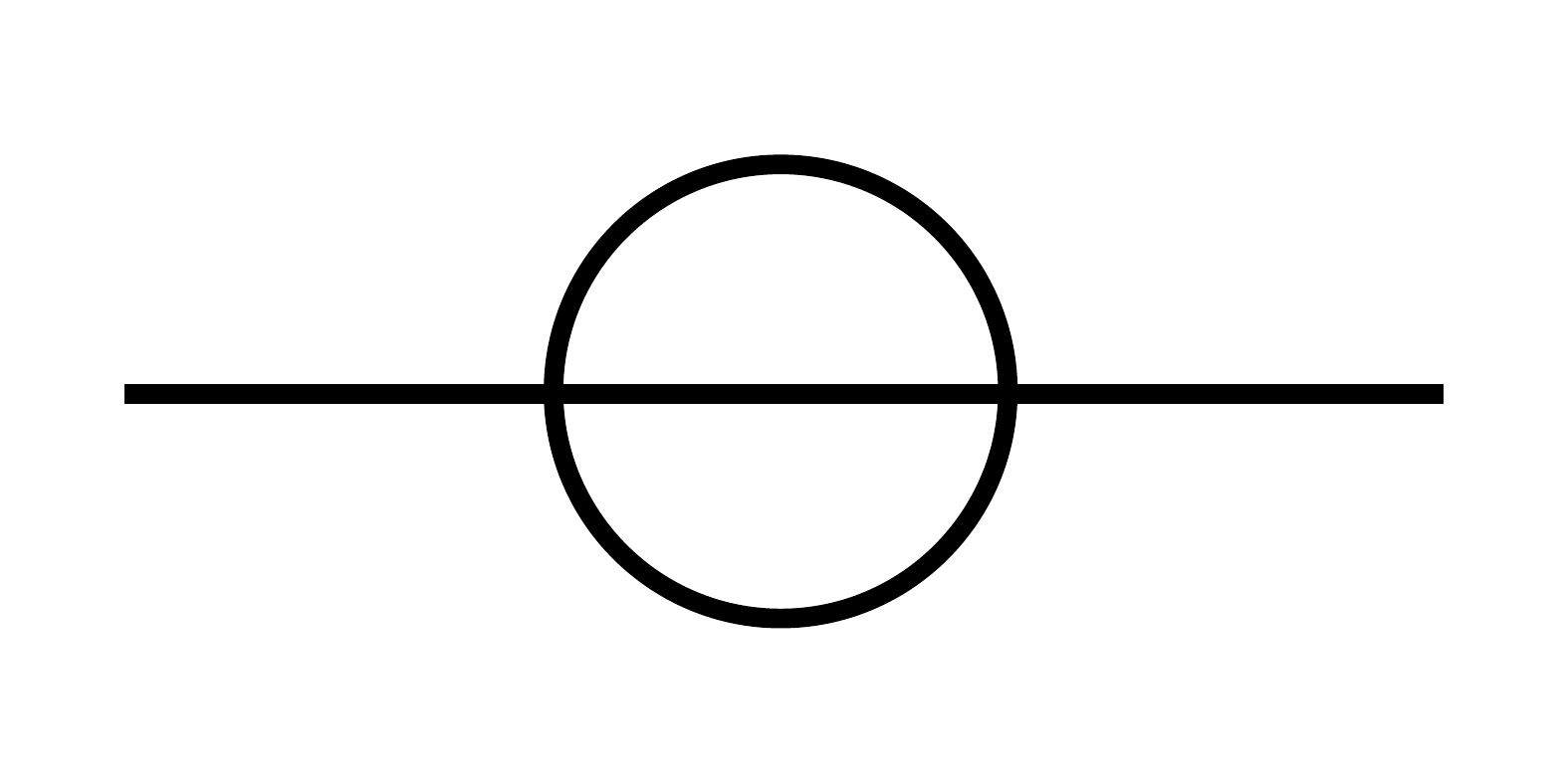} }}%
    \caption{Divergent Feynman diagrams in scalar $\phi^4$ theory in two and three dimensions.}%
    \label{fig:example}%
\end{figure}
Let us consider the example of the 1-loop and 2-loop contributions to the mass in lattice $\phi^4$ theory.  The 1-loop Feynman integral is given by
\begin{equation}
I_1(k,m;a) = \frac{\lambda}{2} \int_{-\pi/a}^{\pi/a} \frac{d^d q}{(2\pi)^d} \frac{1}{\tilde{q}^2 + m^2}
\end{equation}
where $\tilde{q}^2 = (2/a)^2 \sum_\mu \sin(a q_\mu /2)^2$ is the standard momentum factor appearing in the scalar propagator.
Notice that discrete translation invariance plays a crucial role by allowing us to express the Feynman amplitude in momentum space.
For a Regge manifold with no translation symmetry we will have to work in position space.
$\text{Deg}(I_1) = d-2$, so the diagram is divergent in $d=2,3$ dimensions.  
However, the divergence is independent of the momentum $k$ flowing through the diagram -- it is a translationally invariant divergence.
Therefore, we may introduce a single universal mass counterterm at $\mathcal{O}(\lambda)$ to cancel the divergence in perturbation theory.

A slightly more nontrivial example is given by the two loop Feynman integral
\begin{equation}
I_2(k,m;a) = \frac{\lambda^2}{3} \int_{-\pi/a}^{\pi/a} \frac{d^d q}{(2\pi)^2} \frac{d^d q'}{(2\pi)^2} \frac{1}{\tilde{q}^2 + m^2}\frac{1}{\left(\widetilde{q'-q}\right)^2 + m^2}\frac{1}{\left(\widetilde{q'-k}\right)^2 + m^2}
\end{equation}
$\text{Deg}(I_2) = 2d-6$, so the integral is divergent in three dimensions but not two.  
Unlike the 1-loop diagram, the 2-loop integral has a nontrivial dependence on the momentum $k$ flowing through the diagram.  
We can reexpress the diagram trivially as a sum of the $k=0$ piece and the difference $D_2$ between the $k=0$ piece and the $k \neq 0$ piece: $I_2(k,m;a) = I_2(0,m;a) + D_2(k,m;a)$.  
It is straightforward to show that the momentum independent piece, $I_2(0,m;a)$, is divergence while the momentum dependent difference $D_2(k,m;a)$ has a lattice degree of divergence less than zero and therefore has a well defined continuum limit.
The divergent piece is canceled by a universal counterterm, and the momentum dependent piece becomes its naive continuum analogue when the lattice spacing is taken to zero.  
Thus, the theory is renormalized by a finite number of universal (position independent) counterterms, and the renormalized lattice theory gives Lorentz invariant expressions in perturbation theory as the lattice spacing is taken to zero.
In this way, the continuum limit can be shown to exist in perturbation theory to all orders for a renormalizable lattice quantum field theory.

%The true power of the lattice comes as a nonperturbative regulator.  
%The Wilsonian picture of renormalization is a nonperturbative one.  
Nonperturbatively, the continuum limit of a lattice theory is taken by the Wilsonian approach of tuning to a critical point.
Bare couplings are tuned so that the theory lies on the critical surface.  
On the critical surface, the renormalized values of masses and other dimensionful couplings go to zero or infinity as correlations in the theory become infinite (or as large as the box in a finite volume).
Divergences are removed by tuning the bare couplings to the critical surface at incrementally smaller values of the lattice spacing, and one is left with a renormalized theory at the critical point in the continuum limit.

The renormalization of the DEC/FEM action on a Regge manifold with require the introduction of perturbative counterterms in the UV using bare perturbation theory near the Gaussian fixed point followed by a nonperturbative tuning of bare couplings to the critical surface in order to remove divergences and take the continuum limit.
Perturbatively, the UV divergent diagrams are sensitive to the irregularities in the Regge manifold are short distances.
These diagrams will be position dependent due to the nonuniform UV cutoff inherent in the Regge manifold, leading to position dependent quantum corrections to physical amplitudes.
If these contributions are not canceled by explicit counterterms, they prohibit the lattice theory from reaching the continuum limit.

To correct our DEC/FEM action, we introduce explicit counterterms to precisely cancel the position dependent quantum loop corrections.
The method relies on two key assumptions:
\begin{enumerate}
\item Only UV divergent diagrams are position dependent in the continuum limit.
\item The divergence is \emph{universal}: it is position independent and equal to it's continuum value.
\end{enumerate}
The first property ensures that for superrenormalizable quantum field theories, the DEC/FEM action can be corrected by a finite number of counterterms.
The second property ensures that the counterterms are finite functions on the manifold.
In a sense, the renormalization procedure is a position dependent scheme change because it involves only finite corrections to the bare couplings.
However, because these finite corrections are position dependent, they have physical consequences (unlike global scheme changes).
After the perturbative counterterms are added to the DEC/FEM action, the divergences are absorbed by tuning the bare couplings to the critical surface in the usual Wilsonian way.

Working in analogy with our discussion of perturbative renormalization for traditional lattice theory, a Feynman diagram in position space $I(x,m;a)$ may be written as
\begin{equation}
I(x,m;a) = \bar{I}(m;a) + D(x,m;a)
\end{equation}
where
\begin{equation}
\bar{I}(m;a) = \sum_x \sqrt{g_{\sigma}(x)} I(x,m;a) / \sum_x \sqrt{g_\sigma(x)}
\end{equation}
and $D = I-\bar{I}$.
If the Feynman diagram is UV divergent, it should have a position independent divergent piece and a position dependent finite piece (following from our key assumptions), $I(x,m;a) = c(m;a) + f(x,m;a)$ where $\lim_{a\rightarrow 0} c = \infty$ and $\lim_{a \rightarrow 0} f(x,m;a) = f(x,m)$.
The average piece of the diagram becomes 
\begin{equation}
\bar{I}(m;a) = c(m;a) + \bar{f}(m;a) \rightarrow \infty \quad \text{as} \quad a\rightarrow 0
\end{equation}
and the subtracted piece becomes
\begin{equation}
D(x,m;a) = f(x,m;a) - \bar{f}(m;a) \rightarrow f(x,m) - \bar{f}(m) \quad \text{as} \quad  a\rightarrow 0
\end{equation}
The universal divergence can be left alone as it will be removed by tuning to the critical surface.
We see that we need only to introduce a counterterm $\propto -D(x,m;a)$ into our DEC/FEM action to explicitly cancel the position dependent quantum corrections and reach the continuum limit. 
We will demonstrate this procedure with explicit examples in Chapter~\ref{chapter:QFE2}.
%
%\section{Spherical Complexes and Free Fields}
%\section{Classical Construction of Finite Element Method}
%\section{Quantum Corrections to Finite Element Method}
%\section{Beyond Radial Quantization}

%%%%%%%%%%
%%%%%%%%%%%
%\chapter{Interacting Scalar Fields on $\mathbb{S}^2$ and $\mathbb{R}\times\mathbb{S}^2$ \label{sec:QFE2}}
%\chapter{Scalars and Spinors on Spheres and Cylinders: Numerical Results \label{sec:QFE2}}
\chapter{Numerical Studies of Lattice Field Theories on Riemann Manifolds \label{chapter:QFE2}}
In this chapter, we present detailed numerical studies of lattice actions constructed through the quantum finite element formalism of Chapter~\ref{chapter:QFE1}.
%First we will study the free scalar Laplacian and the free Dirac-Wilson operator on a nontrivial Riemann manifold.  
First we will study the free scalar Laplacian on a nontrivial Riemann manifold. 
We will focus on confirming the property of \emph{spectral fidelity} --- the rapid convergence of the classical spectrum to the continuum --- for our lattice operator, which is a crucial prerequisite for any lattice field theory constructed in this manner to converge to the continuum limit at the quantum level.
Once we have established that the free theory rapidly converges to the continuum limit, we will turn our attention to interacting quantum field theory.
In particular, we will study scalar $\phi^4$ theory on $\mathbb{S}^2$ and $\mathbb{R}\times\mathbb{S}^2$.
Each of these actions contains an interacting infrared fixed point \cite{Wilson:1971dc}, corresponding to the 2D and 3D Ising CFT respectively.
The former has been solved both as a CFT by identifying it with the minimal $c=1/2$ CFT \cite{Belavin:1984vu,Polyakov:1984yq} and at finite lattice spacing on a square lattice by Onsager \cite{Onsager:1944}.
As such, it serves as a useful, nontrivial check of the QFE formalism.
The latter is an unsolved problem that has recently been greatly constrained by the numerical conformal bootstrap program \cite{ElShowk:2012ht,El-Showk:2014dwa,Kos:2014bka,Kos:2016ysd}.
We hope that our study of the 3D Ising CFT in radial quantization with the QFE method will serve as evidence that this new lattice methodology is a useful tool for studying conformal systems in a way that is complementary to the conformal bootstrap approach.

\section{Scalar Laplacian on the 2-Sphere}
In this section we study the scalar Laplacian of Eq.~\ref{eq:FEMLap2d} on the specific Riemann manifold $\mathbb{S}^2$.
The 2-sphere is a simple example of a manifold with nonzero curvature, and it is a maximally symmetric space with isometry group $O(3)$ in which the spectral properties of the Laplacian are well known.
This will allow us to test the spectrum of the lattice operator against the known continuum results and to establish the property of \emph{spectral fidelity}.
%These numerical tests serve as evidence of the viability of the quantum finite element formalism presented in Chapter~\ref{sec:QFE1} for providing a nonperturbative method for studying field theory on curved space.

%
\begin{figure}[t]
\centering
\includegraphics[width=0.32\textwidth]{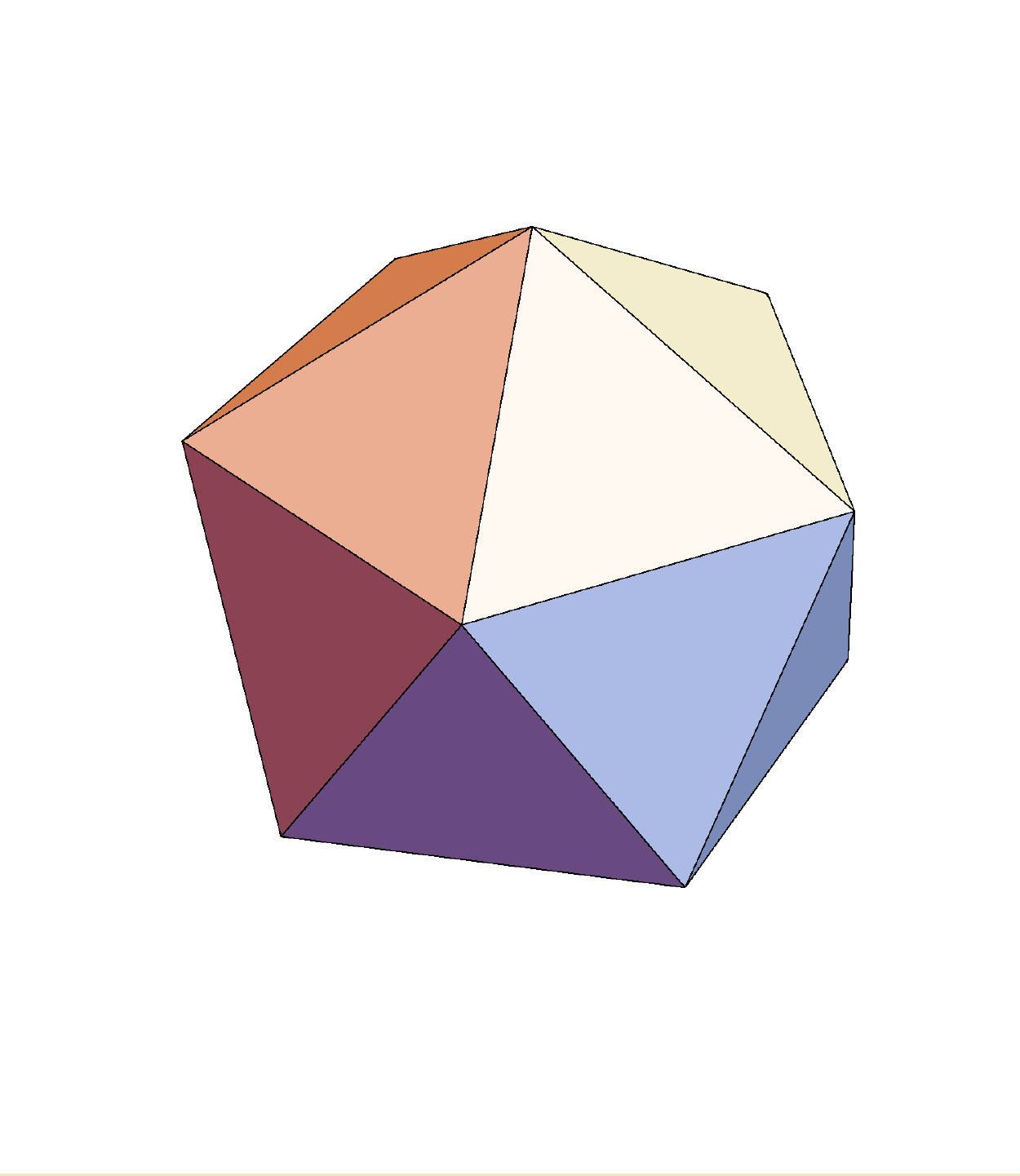}
\includegraphics[width=0.32\textwidth]{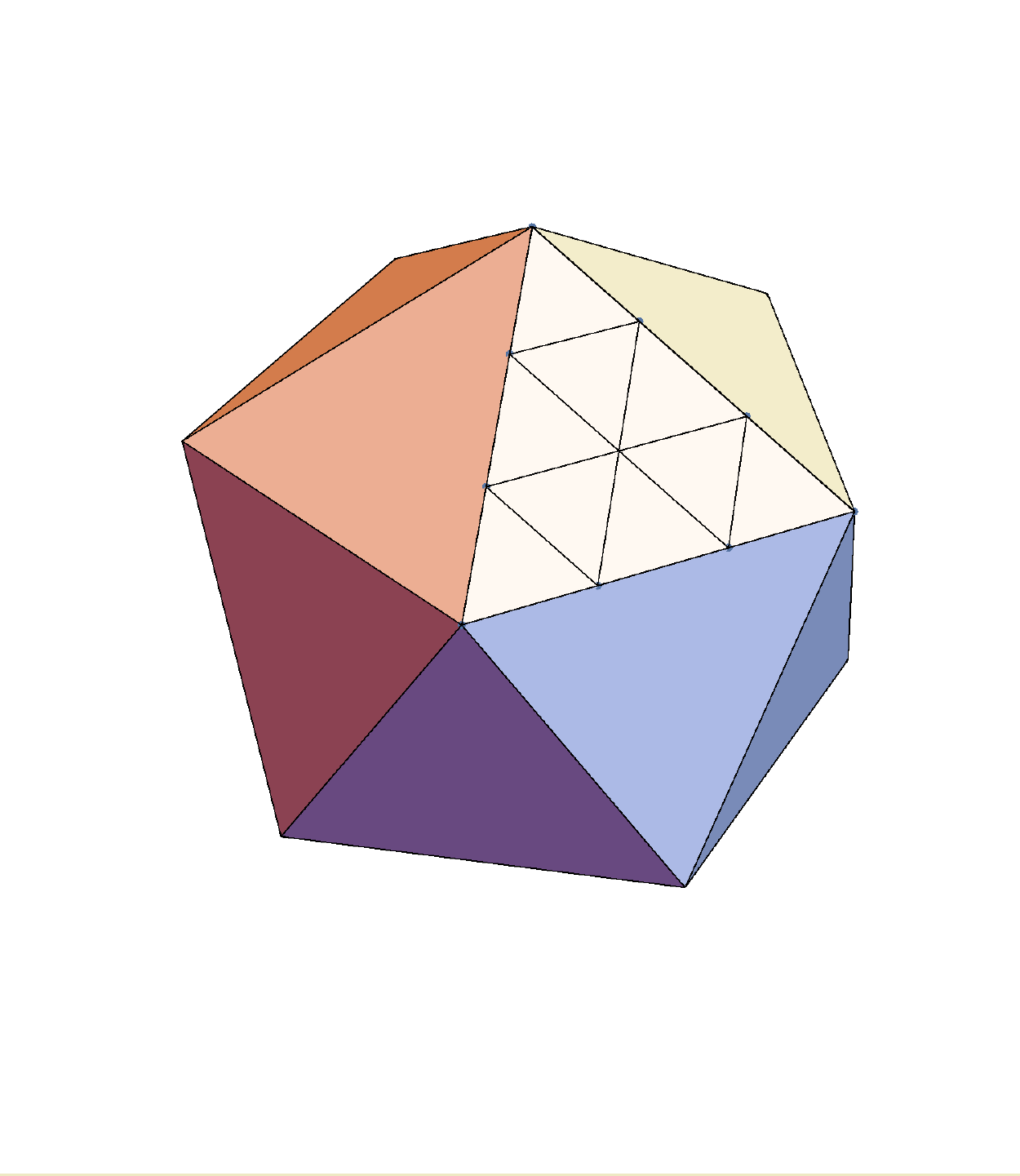}
\includegraphics[width=0.32\textwidth]{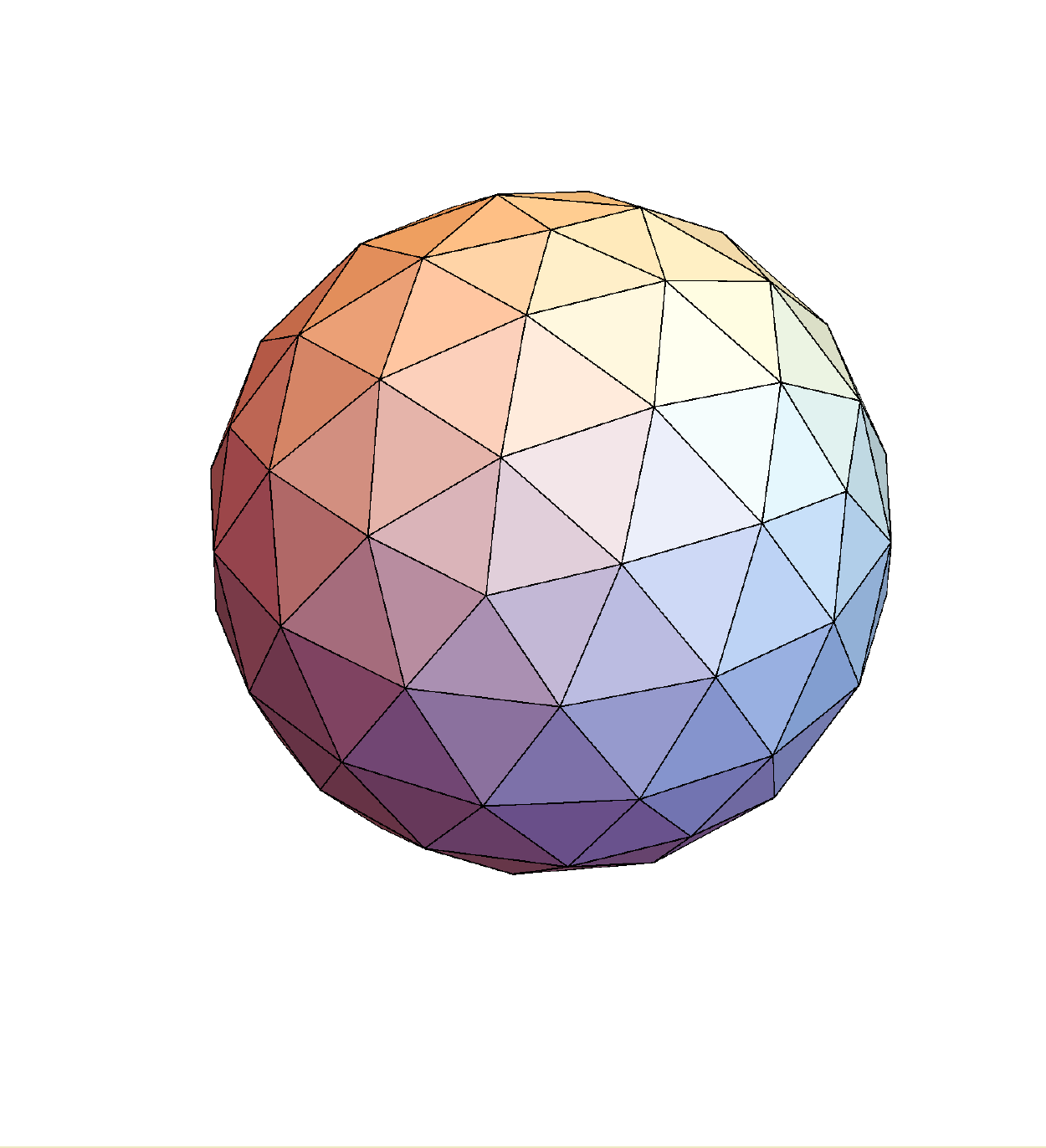}
\vskip -1.0 cm
\caption{\label{fig:icos} The  $L=3$ refinement of
the icosahedron with $V = 2 + 10L^2 = 92$ vertices or sites. The icosahedron on
the left is refined in the middle with $L^2 = 9$ equilateral triangles on 
each face, and  then on the right the new vertices are projected  onto the unit
sphere. The resulting simplicial  complex preserves the icosahedral symmetries. }
\end{figure}
Following the procedure detailed in Chapter~\ref{chapter:QFE1}, we must first construct a sequence of abstract simplicial complexes $\{\mathcal{M}_\sigma \}$ with the topology of $\mathbb{S}^2$.
In our construction, we choose to work with a Regge manifold which preserves the largest possible discrete subgroup of the isometries of the sphere.
The largest discrete subgroup of $O(3)$ is the icosahedral group, so we choose our first refinement $\mathcal{M}_1$ to be the abstract simplicial complex of the icosahedron consisting of 12 vertices, 30 edges, and 20 faces with the appropriate connectivity satisfying $V - E + F = 2$ for a space with the topology of the sphere.
Higher refinements $\mathcal{M}_\sigma$ are constructed by inserting $\sigma-1$ vertices on each icosahedral edge, and connecting them in the fashion of a regular triangular lattice on each on the 20 faces of the icosahedron.
Each icosahedral face is divided into $\sigma^2$ triangular faces.

To promote the abstract simplicial complex to a Regge manifold we must assign lengths to links and choose a geometry for the interior of each simplex.
It is simplest to do this by working in the embedding space $\mathbb{R}^3$.  
A regular icosahedron is circumscribed inside of a unit sphere (the target manifold) in $\mathbb{R}^3$.  
The faces of the icosahedron are divided into a mesh of $\sigma^2$ equilateral triangles.
Next, the vertices of this mesh are projected outwards from the origin onto the circumscribing sphere.
The secant distances (in the embedding space) between the images of these vertices on the sphere are taken to be the link lengths on the Regge manifold.
The interiors of triangles are taken to be flat.
A visualization of this construction for $\sigma = 3$ is shown in Fig.~\ref{fig:icos}.
In what follows we will use $L=\sigma$ to designate the linear size of this particular Regge manifold.

In two dimensions, the DEC and FEM prescriptions for the scalar Laplacian on a Regge manifold coincide.  
For the mass term, we take the DEC approach of defining ultralocal interactions.  
The action for a free scalar field on our Regge manifold of Fig.~\ref{fig:icos} is denoted
\begin{equation}
S_f[\phi] = \frac{1}{2} \phi_i M_{ij} \phi_j = \frac{1}{2} \phi_i \left[ K_{ij} + m_i \delta_{ij} \right] \phi_j  \label{eq:Sfree}
\end{equation}
Where the Laplacian matrix $K$ is an appropriate rewriting of the scalar kinetic action Eq.~\ref{eq:FEMLap2d} in matrix form.
We have defined the position dependent mass to include the metric, $m_i = \sqrt{g_i} m_0^2$.

\begin{figure}[t]
\centering
\includegraphics[width=0.45\textwidth]{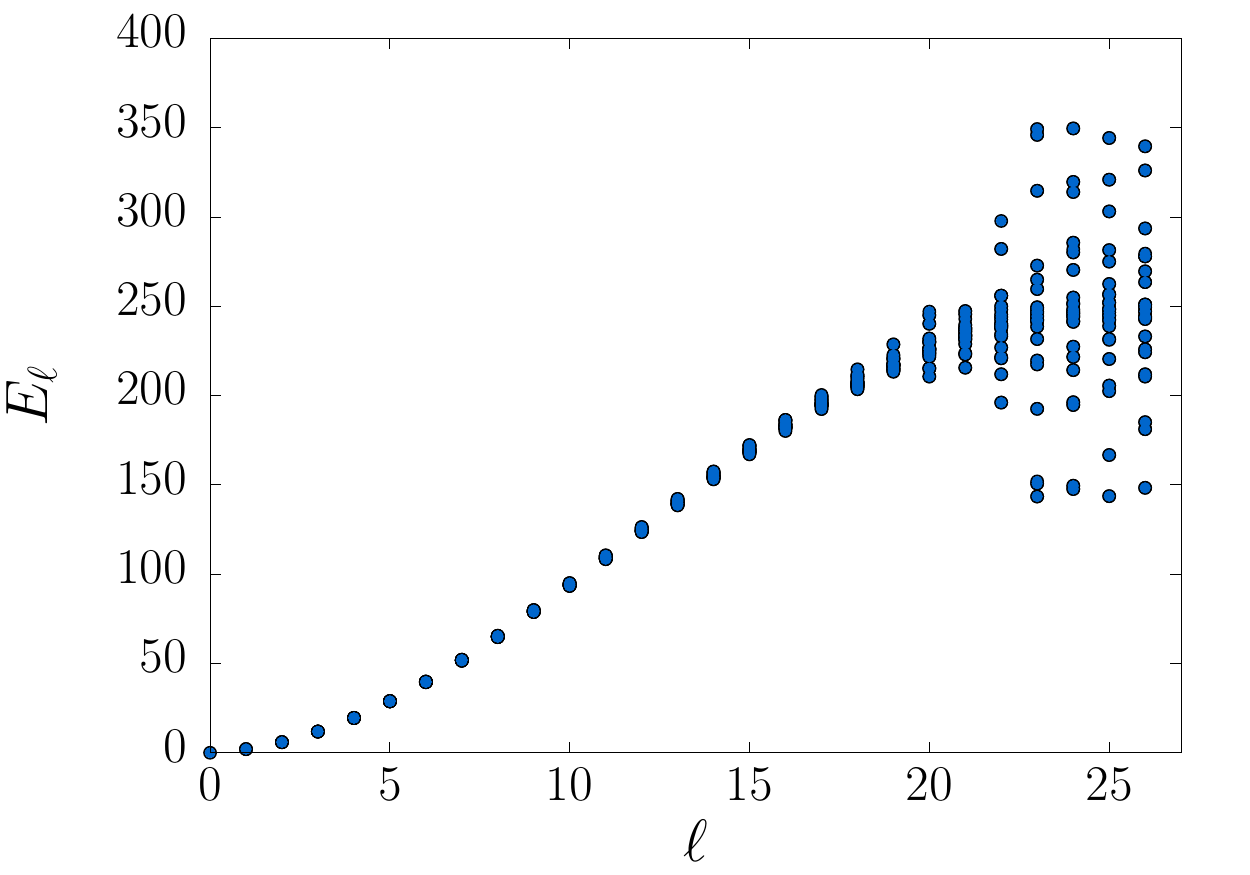}
\includegraphics[width=0.45\textwidth]{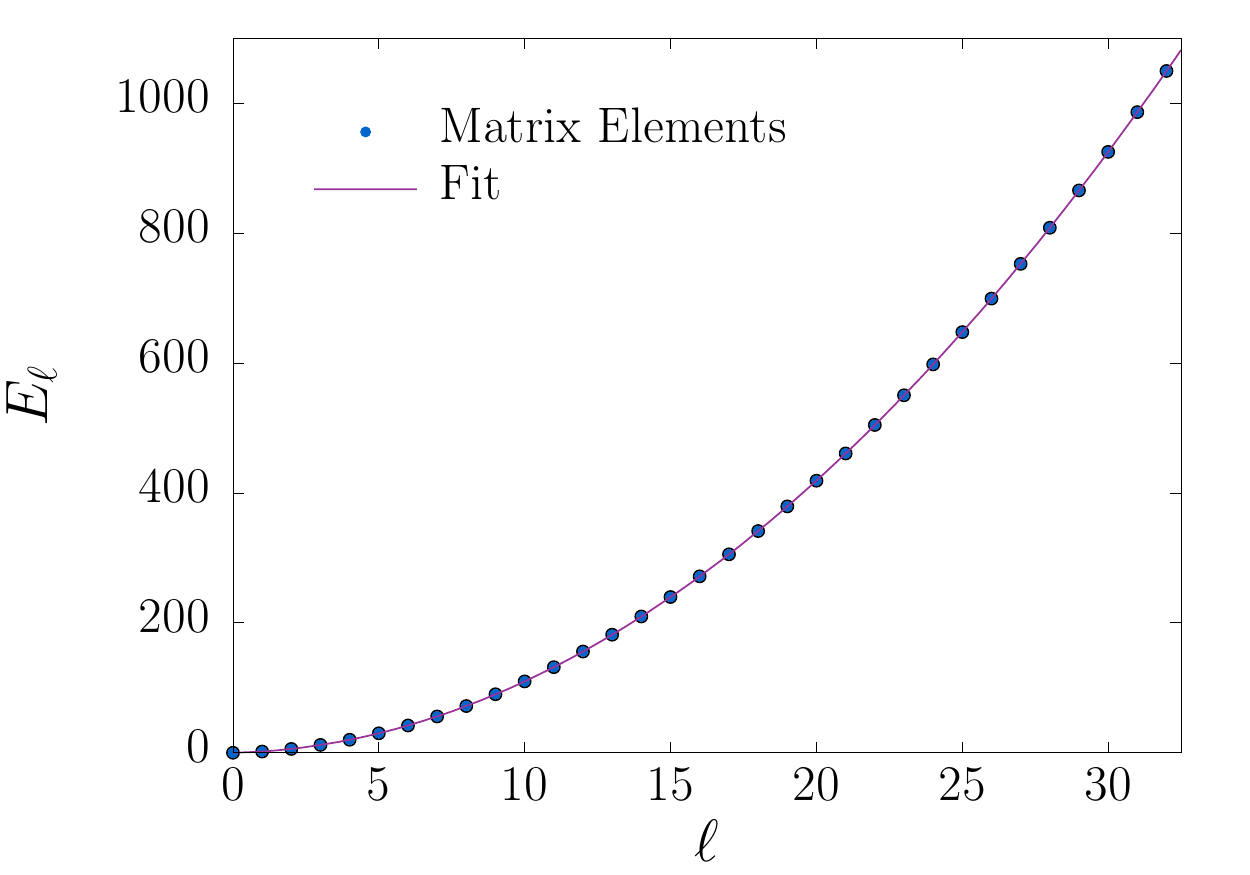}
\caption{\label{fig:FEMdiag} Left: The  $2l+1$  spectral  
 values for  $m \in [-l,l]$ are plotted against $l$ for $L
  = 8$.   Right: The eigenvalues averaged over $m$ compared to the best fit curve $l + 1.00002 \; l^2
  - 1.27455\times 10^{-5} \; l^3 - 5.58246 \times 10^{-6} \; l^4$
  for $L =  128$   and $l \le 32$.  These results originally appear in Refs.~\cite{Brower:2016moq,Brower:2018szu}}
\end{figure}
Our first numerical investigation is to characterize the free theory by studying the spectrum of the operator $M$.
In a curved space, eigenfunctions are orthonormal with respect to the proper metric, and accordingly the metric enters into the eigenvalue equation.
To compute the eigenvectors and eigenvalues of the matrix $M$, we solve the generalized eigenvalue problem for the metric $\sqrt{g_i}$.
\begin{equation}
M_{ij} \phi^{n}_j = E^n \sqrt{g_i} \phi^n_i = \left(E^{n,0} + m_0^2\right) \sqrt{g_i} \phi_i^n
\end{equation}
$E^n$ are the eigenvalues for the eigenfunction $\phi^n$, and $E^{n,0}$ are the eigenvalues at zero mass.
The results are presented in Fig~\ref{fig:FEMdiag}.  
The spectrum of the FEM/DEC Laplacian on the simplicial $\mathbb{S}^2$ rapidly approaches the continuum spectrum $l(l+1)$ with the correct $2l+1$ degeneracy.
One sees that the IR spectrum is close to the continuum result while the spectrum deviates at large $l$ due to lattice artifacts; this is always true of lattice operators.
Each eigenvalue converges to the continuum result like $1/L^2$ as $L\rightarrow \infty$.
The only difference between our FEM/DEC lattice Laplacian and a conventional lattice Laplacian on a square lattice is that we lack the convenient Fourier space techniques to solve for the spectrum analytically, and so we must make do with this numerical confirmation of spectral fidelity.

%
%\section{Wilson-Dirac Operator on the 2-Sphere}
%Dirac Operator Results here.

%
\section{2D Ising CFT from Scalar $\phi^4$ Theory on the 2-Sphere \label{sec:QFES2}}
Our first study of an interacting quantum field theory will be scalar $\phi^4$ theory on the 2-sphere.
At the Wilson Fisher fixed point \cite{Wilson:1971dc}, scalar $\phi^4$ theory on $\mathbb{R}^2$ becomes conformally invariant, and thus the same CFT should exist on any Riemann manifold that is related to $\mathbb{R}^2$ by a Weyl rescaling factor.
A stereographic projection maps between the plane and the Riemann sphere.
The stereographic projection map may be constructed by using the embedding space.
In the embedding space $\mathbb{R}^3$, a sphere is placed with its south pole at the origin of the plane.  
Straight lines in the embedding space are drawn from the north pole of the sphere though a point on the sphere, and each line intersects a unique point on the plane.
This generates a 1:1 mapping between points on the plane and the sphere.  
The point at infinity is mapped to the north pole.
Adopting polar coordinates on the plane $(r,\phi)$ and standard spherical coordinates on the sphere $(\theta,\varphi)$ the stereographic map between the two spaces is
\begin{equation}
r = 2 R \cot(\theta/2) \quad , \quad \phi = \varphi
\end{equation}
where $R$ is the radius of the sphere.  
One finds that the metrics between the two spaces are related by 
\begin{equation}
ds^2_{\mathbb{R}^2} = dr^2 + r^2 d\phi^2 = \Omega^2(\theta,\varphi) \left(d\theta^2 + \sin^2(\theta)d\varphi^2\right) = \Omega^2 ds^2_{\mathbb{S}^2}
\end{equation}
The conformal (Weyl) factor is $\Omega = R\csc^2(\theta/2)$.  
Thus, $\phi^4$ theory on the 2-sphere at the critical Wilson-Fischer fixed point should reproduce the critical behavior of $\phi^4$ theory on the plane, which is the minimal $c=1/2$ CFT.

%In two dimensions, the DEC and FEM prescriptions for the scalar Laplacian on a Regge manifold coincide.  
In introducing the quartic interaction, we continue to follow the DEC approach of defining ultralocal interactions.  
The complete two dimensional scalar action that we study is given by
\begin{equation}
S[\phi] = S_f[\phi] + \lambda_i \phi_i^4 = \frac{1}{2} \phi_i \left[ K_{ij} + m_i \delta_{ij} \right] \phi_j + \lambda_i \phi_i^4 \label{eq:Sfull}
\end{equation}
Where the free scalar action $S_f[\phi]$ was defined in Eq.~\ref{eq:Sfree}.
As with the mass term, we have defined the position dependent quartic coupling to include the metric, $\lambda_i = \sqrt{g_i}\lambda_0$.

A convenient quantity for assessing the critical behavior of the theory is the Binder cumulant \cite{Binder:1981} defined in terms of the magnetization density $m = \sum_i \sqrt{g_i}\phi_i / N$ where $\sum_i \sqrt{g_i} = N$ is normalized to be the number of vertices.
\begin{equation}
U_4 = \frac{3}{2}\left(1 - \frac{\langle m^4 \rangle}{3\langle m^2 \rangle^2}\right) \label{eq:U4}
\end{equation}
In the thermodynamic limit, the binder cumulant is zero in the disordered phase and is one in the ordered phase.  
It should take the form of a Heaviside function when plotted against the relevant coupling with the step occurring at the critical value of the coupling.

\begin{figure}[t]
\centering
\includegraphics[width=0.45\textwidth]{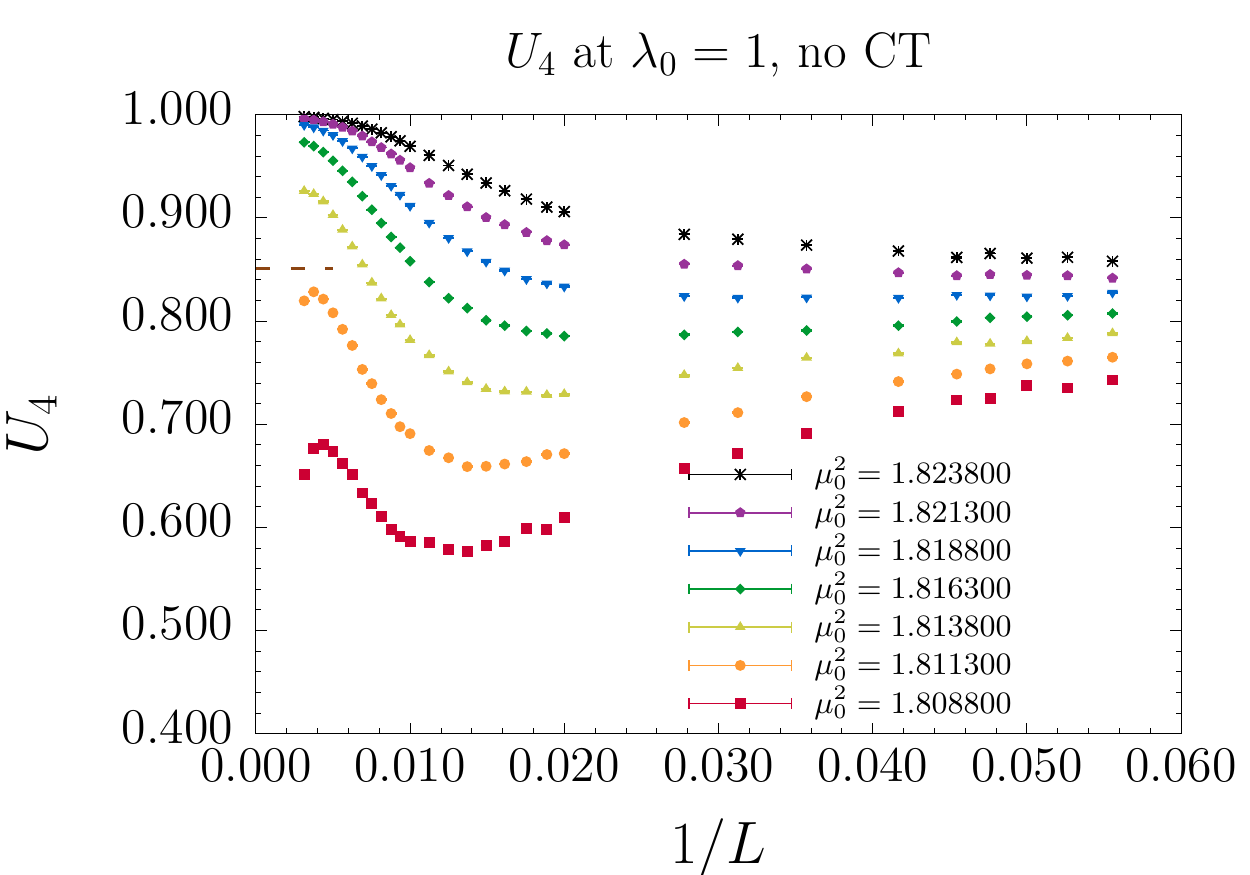}
\includegraphics[width=0.35\textwidth]{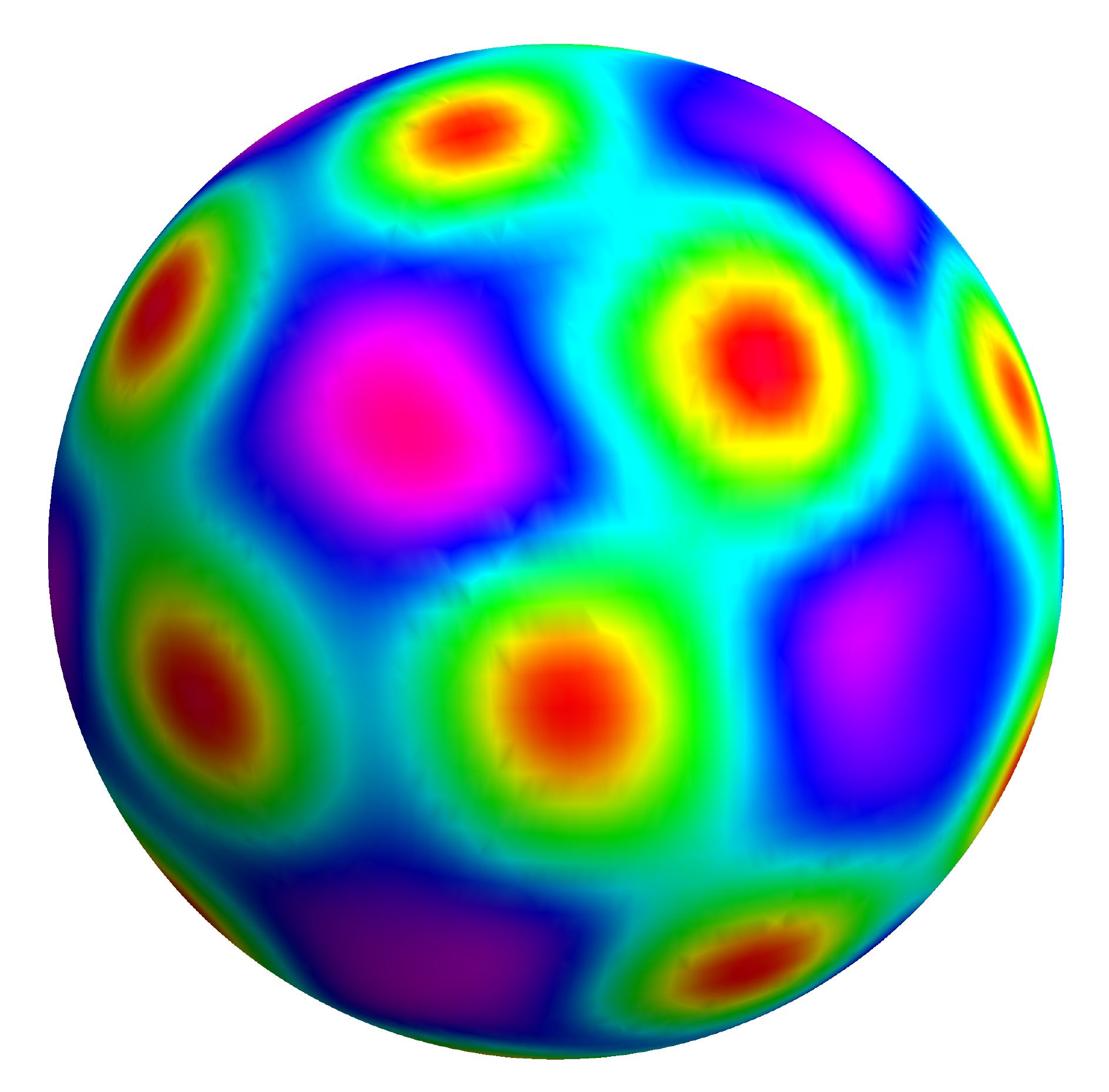}
\caption{\label{fig:noCT2d} Demonstration of the failure of the classical FEM/DEC action for quantum field theory.  Left: the Binder cumulant vs $1/L$ fails to converge to a step function as $L\rightarrow \infty$ indicating an obstruction to criticality.  Right: $\langle \phi(x)^2 \rangle$ plotted as a function on the sphere.  The local susceptibility is sensitive to the variation in the lattice spacing.  This figure originally appears in Ref.~\cite{Brower:2018szu}.}
\end{figure}
The left panel of Fig.~\ref{fig:noCT2d} presents a numerical Monte Carlo study of the Binder cumulant for the FEM/DEC action Eq.~\ref{eq:Sfull}.
For modest lattice sizes $L \lessapprox 64$ the Binder cumulant appears to be approaching a step function as a function of the relevant coupling $\mu_0^2$.
However, at larger $L$, the curves begin to turn around and oscillate signifying an obstruction to reaching the continuum limit.
\emph{The classical FEM/DEC procedure fails to converge to the continuum quantum field theory}.

We can trace the issue back to our construction of the icosahedral Regge manifold in Fig.~\ref{fig:icos}.
The projection from the icosahedron onto the circumscribing sphere caused a distortion in the local lattice spacing.  
The lattice spacing is smaller near the 12 icosahedral vertices and larger near the center of the 20 icosahedron faces.  
While this Regge manifold provided a smooth enough interpolation of the continuum target manifold for the classical spectrum to converge in Fig.~\ref{fig:FEMdiag}, the quantum theory is more demanding because it is sensitive to all length scales.
Quantum loops are affected by the variation in the lattice spacing and contribute different values to the renormalized quark mass at different sites on the lattice. 
We must correct the FEM/DEC action Eq.\ref{eq:Sfull} by introducing explicit geometrical counterterms following the general procedure detailed in Section~\ref{sec:QFE15}.

$\phi^4$ theory is superrenormalizable in two dimensions.  
The only divergent diagram in perturbation theory is the one loop correction to the two point function, which contributes a logarithmic divergence to the mass..
If our key assumptions from Section~\ref{sec:QFE15} are correct, this diagram should contain two terms: a \emph{universal} divergence which is position independent and equal to the value of the continuum divergence, and a position dependent finite piece which must be canceled by an explicit counterterm in our renormalized quantum finite element action.
We can explicitly check these properties by studying the one loop and two loop diagrams in bare lattice perturbation theory.
The amputated 1PI amplitude is given by
\begin{equation}
\Sigma_{ij} = -12 \lambda_i G_{ii} \delta_{ij} + 96 \lambda_i \lambda_j G_{ij}^3 + \mathcal{O}(\lambda_0^3) \label{eq:1PT}
\end{equation}
where we have defined the matrix propagator $G = M^{-1}$.
The diagrams for these two terms are shown in Fig.~\ref{fig:example}. 
When we considered these diagrams on a square lattice in Section~\ref{sec:QFE15}, we were able to write down closed form expressions for them in Fourier space.
Without the convenience of discrete translation invariance and Fourier methods, we are left to study the terms in Eq.~\ref{eq:1PT} by numerically computing the lattice propagator $G$.

The continuum expression for the one loop diagram with a momentum cutoff is 
\begin{equation}
G_{ii} \approx G_{\text{continuum}}(x_i,x_i) = \frac{\sqrt{3}}{2} \int_0^\Lambda \frac{d^2 k}{(2\pi)^2} \frac{1}{k^2 + m^2} = \frac{\sqrt{3}}{8\pi} \ln \left( \frac{\Lambda^2}{m^2} \right)
\end{equation}
The extra factor of $\sqrt{3}/2$ in the density of states is due to the fact that the hexagonal dual areas at a site on an equilateral triangular lattice have volume $a^2 \sqrt{3}/2$ while the square dual cells on a square lattice have area $a^2$, and this factor carries over into momentum space: $(1/N)\sum_i \sqrt{g_i} \rightarrow \int^\Lambda d^2k \rho(k)$ where $\rho$ is the density of states.
We fix the physical mass $m$ so that the dimensionless mass in units of the effective lattice spacing vanishes like $a^2 m^2 = \mathcal{O}(1/N)$ as $L \rightarrow \infty$.  
The UV cutoff is position dependent and given by $\Lambda \rightarrow 1/a_i$.  
Thus the approximate continuum expression for the one loop diagram reads.
\begin{equation}
G_{ii} \approx \frac{\sqrt{3}}{8\pi} \ln \left( N \right) + \frac{\sqrt{3}}{8\pi} \ln \left(\frac{a^2}{a_i^2} \right)
\end{equation}
The first term is position independent and diverges logarithmically as $L \rightarrow \infty$.
The second term approaches a finite value that depends on position as $L \rightarrow \infty$.

\begin{figure}[t]
\centering
\includegraphics[width=0.45\textwidth]{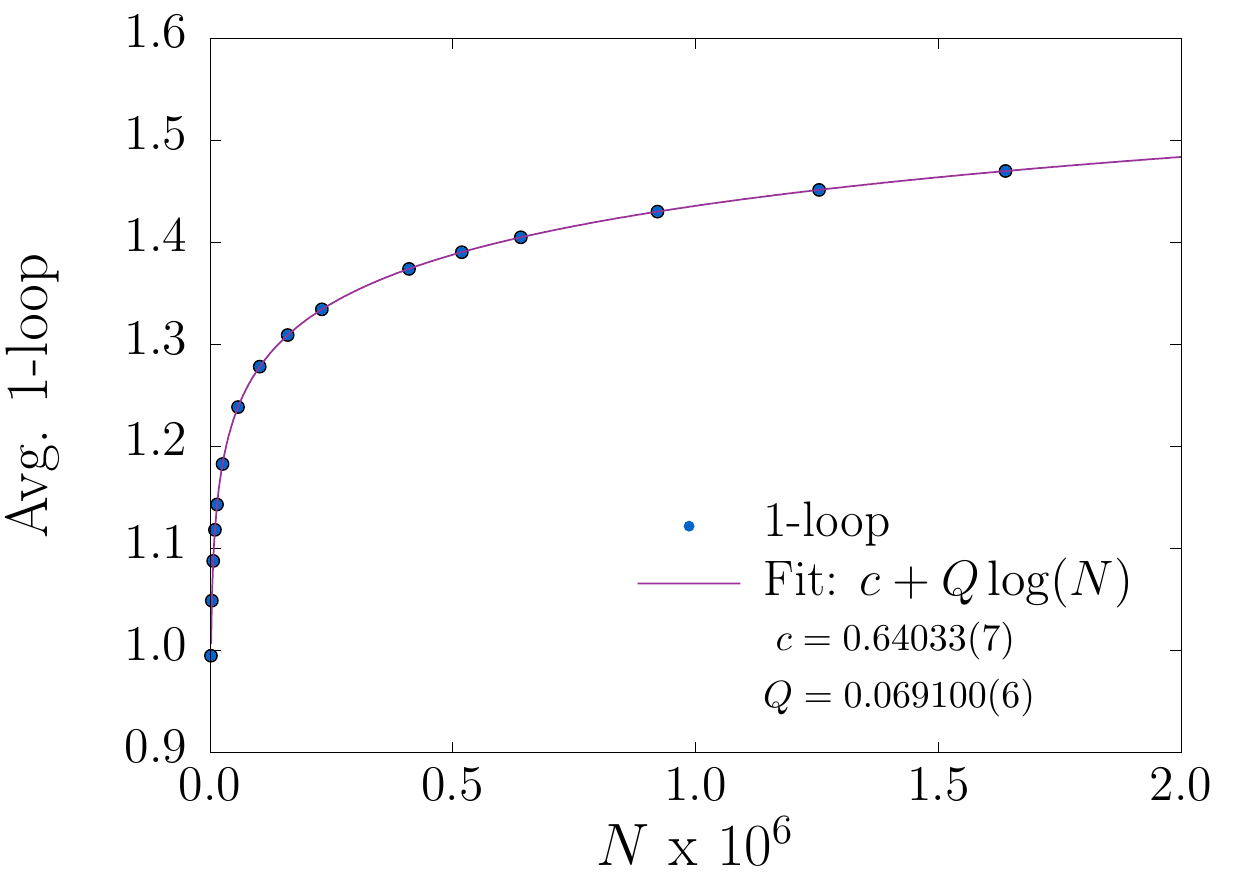}
\includegraphics[width=0.45\textwidth]{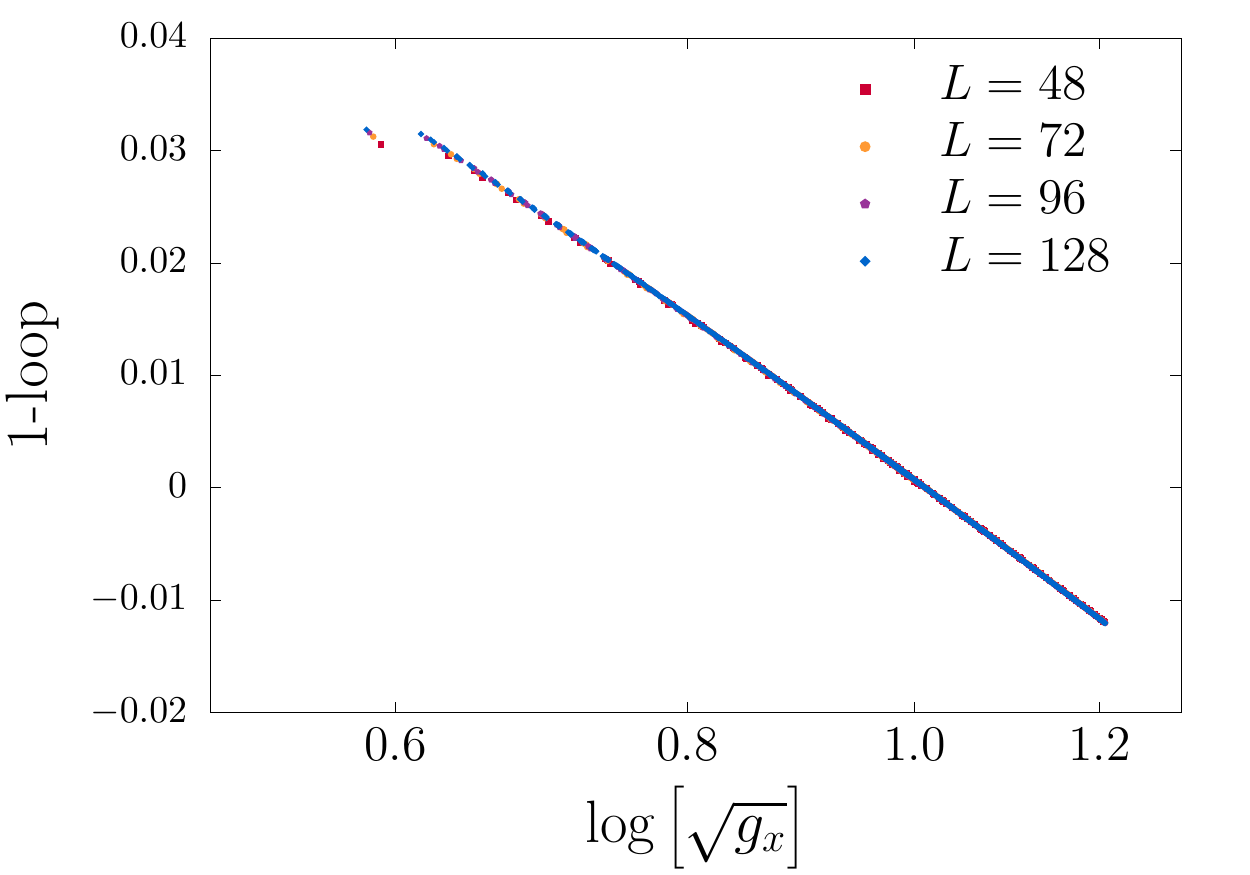}
\caption{\label{fig:OneLoopCT2d} One loop diagram $G_{ii}$ in lattice perturbation theory.  Left: The average value of the one loop diagram $(1/N)\sum_i \sqrt{g_i} G_{ii}$ plotted against $N = 10 L^2 +2$ is shown to fit the approximate continuum expression $(\sqrt{3}/8\pi)\ln N + c \approx 0.068916 \ln N + c$.  Right: The one loop counterterm with the average piece subtracted is shown to approach as finite function as $L \rightarrow \infty$, and this finite function is approximately a linear function of $\ln \sqrt{g_i}$.  This figure originally appears in Ref.~\cite{Brower:2018szu}. }
\end{figure}
We confirm these properties by directly computing the lattice Feynman diagram numerically.  The results are shown in Fig.~\ref{fig:OneLoopCT2d}.  As discussed in Section~\ref{sec:QFE15}, we divide the diagram into an average piece and a subtracted piece (the difference).
\begin{equation}
G_{ii} = \frac{1}{N} \sum_i \sqrt{g_i} G_{ii} + \left( G_{ii} - \frac{1}{N} \sum_i \sqrt{g_i} G_{ii} \right)
\end{equation}
The first term -- the average piece -- is shown on the left panel of Fig.~\ref{fig:OneLoopCT2d}.  
One sees that to high accuracy it behaves (and diverges) as a logarithmic function of $N$ with the coefficient $\sqrt{3}/(8\pi)$ given by the continuum expression.  
%The divergence behaves exactly as it does in the continuum, confirming our assertion that the divergence will be \emph{universal}.
The second term -- the subtracted piece -- is shown plotted against $\ln\left[\sqrt{g_i}\right]$ on the right panel of Fig.~\ref{fig:OneLoopCT2d}.  
The subtracted diagram is not diverging; it approaches a finite function as $L \rightarrow \infty$.  
Furthermore, this finite function is given approximately by the continuum expression, although a deviation can be seen at the smallest values of $\sqrt{g_i}$ which correspond to the twelve special icosahedral vertices.
These results confirm our key assumption that the divergence is position independent and universal in that it matches the continuum expression.

We remark that this is highly analogous to our discussion of Riesz's power counting for the one loop diagram on a square lattice.
There, the $k=0$ piece $I_1(k=0,m;a)$ had degree of divergence zero; it was divergent and had no continuum limit.
But because it was the translationally invariant piece, it could be canceled by a single counterterm in perturbation theory or removed nonperturbatively by tuning a single relevant coupling to the critical surface.
The difference $D(k,m;a) = I_1(k,m;a) - I_1(k=0,m;a)$ was finite and translationally invariant and had a well defined continuum value given by the naive continuum limit -- i.e. it was completely innocuous

In our construction, we will allow the divergent average piece of $G_{ii}$ to be removed nonperturbatively by tuning the bare mass to the critical surface.  
However, the subtracted piece is troublesome in that it contributes a position dependent shift to the renormalized mass.
We must introduce an explicit counterterm to cancel it.  This counterterm appears in the action as
\begin{equation}
S_{\text{CT}} = \sum_i 6 \lambda_i \left( G_{ii} - (1/N)\sum_j \sqrt{g_j} G_{jj} \right) \phi_i^2 \label{eq:CT2d}
\end{equation}
where we have shown the summations explicitly for clarity.
In the resulting renormalized perturbation theory, the position dependent contribution from the one loop diagram is exactly canceled.

\begin{figure}[t]
\centering
\includegraphics[width=0.45\textwidth]{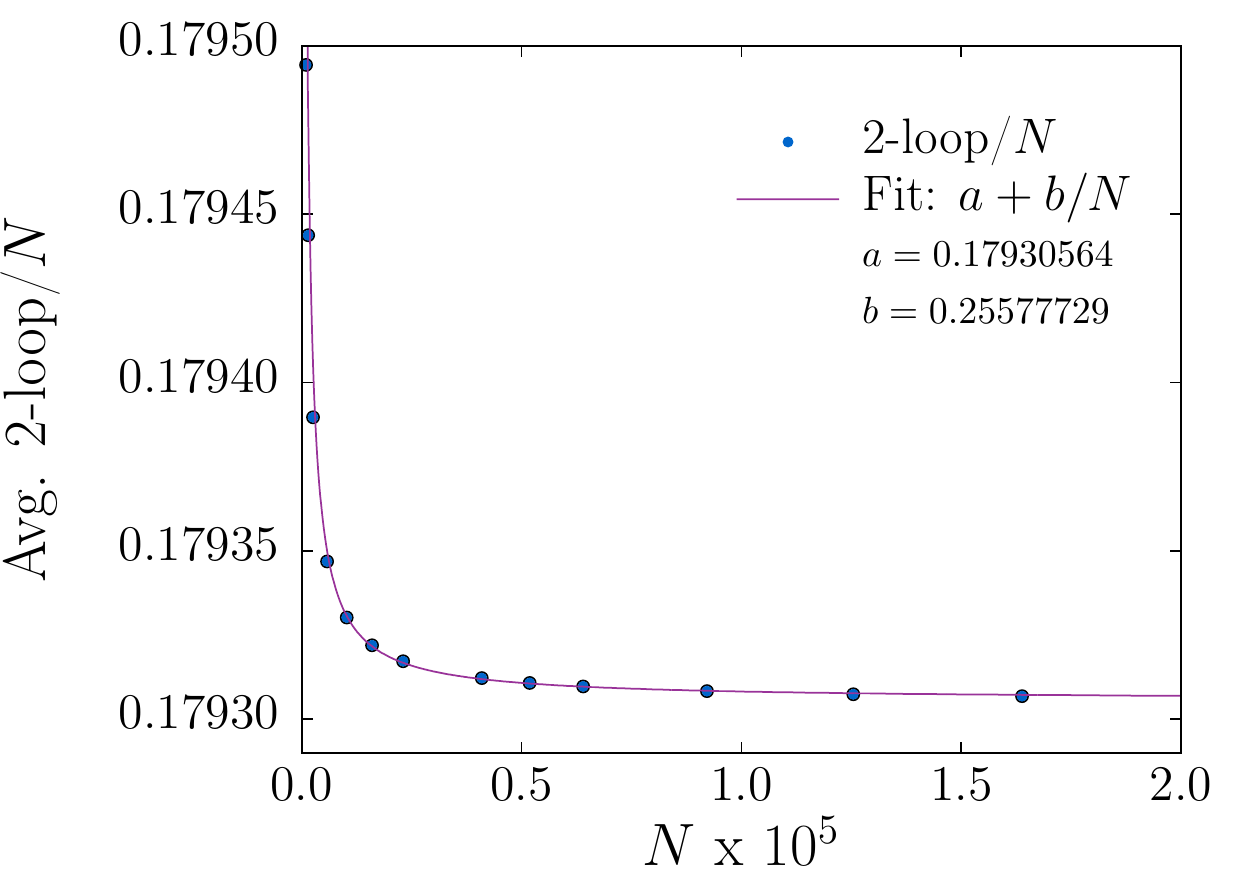}
\includegraphics[width=0.45\textwidth]{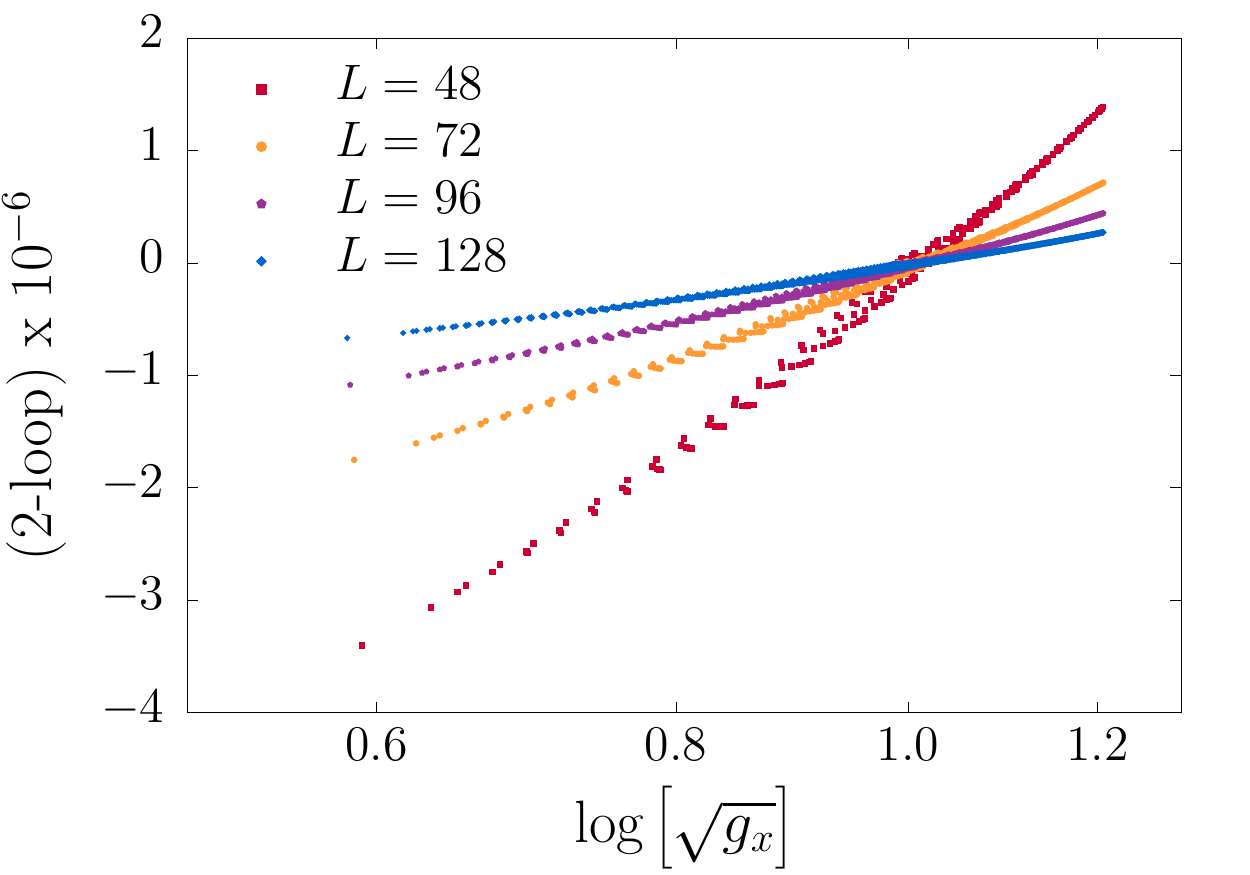}
\caption{\label{fig:TwoLoopCT2d} Local approximation to two loop diagram in two dimensions $\sum_j \sqrt{g_j} G_{ij}^3$.  Left: Average value of diagram plotted against $N$.  Right: Two loop diagram with average piece subtracted plotted against $\sqrt{g_i}$ for various values of $L$.  The right panel of this figure originally appears in Ref.~\cite{Brower:2018szu}.}
\end{figure}
As a check of our other key assumption that UV finite diagrams become position dependent in the continuum limit, we also study the two loop diagram.
Unlike the one loop diagram, the two loop diagram is nonlocal.  
This corresponds to a nontrivial dependence on the momentum $k$ flowing through the diagram when studied in the continuum in momentum space.
In the spirit of dropping higher derivative terms in a momentum expansion, we can replace the two loop diagram by a local approximation.
\begin{equation}
96 \lambda_i \lambda_j G_{ij}^3 \approx 96 \lambda_i (\sum_l \lambda_l G_{il}^3) \delta_{ij}
\end{equation}
We investigate this local approximation to the diagram by again dividing it into the average piece and the subtracted piece.
The results are given in Fig.~\ref{fig:TwoLoopCT2d}.  
Note that we must multiply the dimensionless lattice diagram by $1/N \approx a^2$ to get the correct dimensionful diagram with units of $(\text{mass})^{-2}$.
The left panel shows that the value of the diagram averaged over the lattice approaches a constant value like $\mathcal{O}(1/N)$ as $L \rightarrow \infty$.  
As expected, the diagram contains no divergence. 
In the right panel, we see that the subtracted diagram is rapidly approaching zero.
The diagram has no position dependence as $L \rightarrow \infty$.  
This is an explicit confirmation of our first key assumption that UV finite diagrams are position independent in the continuum limit.
We have no need to introduce an explicit two loop counterterm.

\subsection{Criticality and the Continuum Limit of the QFE Action}
\begin{figure}[t]
\centering
\includegraphics[width=0.65\textwidth]{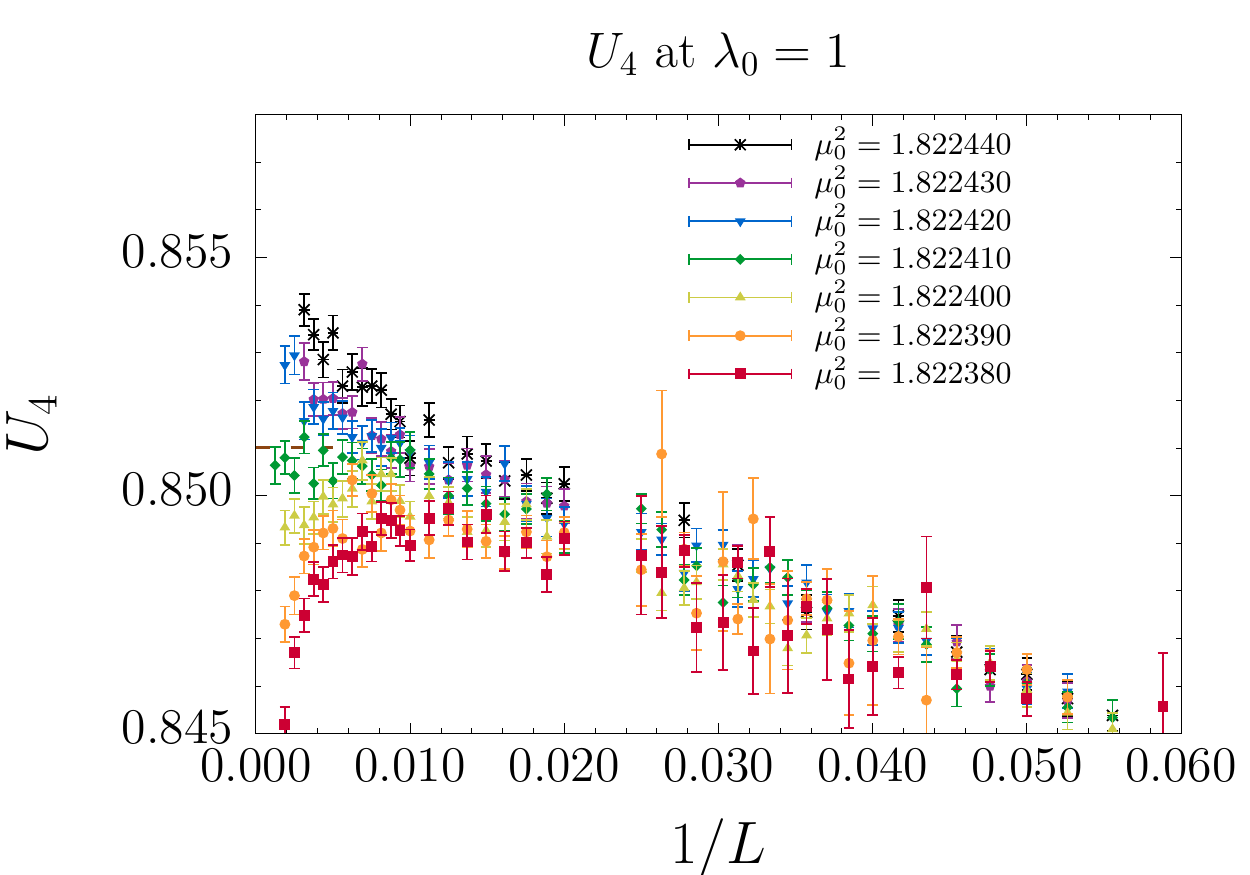}
%\includegraphics[width=0.45\textwidth]{figs/QFE2/mag_U4_vs_musq_2d}
%\caption{\label{fig:magU2d} Binder cumulants for the quantum finite element scalar $\phi^4$ action on $\mathbb{S}^2$.  Left: The fourth binder cumulant $U_4$ vs inverse lattice size $1/L$.  Right: Higher order Binder cumulants vs the relevant coupling $\mu_0^2$ at lattice size $L=200$. }
\caption{\label{fig:magU2d} Binder cumulants for the quantum finite element scalar $\phi^4$ action on $\mathbb{S}^2$.  Left: The fourth binder cumulant $U_4$ vs inverse lattice size $1/L$. This figure originally appears in Ref.~\cite{Brower:2018szu}.}
\end{figure}
We renormalize our finite element action through the addition of the one loop counterterm Eq.~\ref{eq:CT2d}.  
We refer to the resulting action as the \emph{quantum finite element} (QFE) action: $S[\phi] = S_f[\phi] + S_{\text{int}}[\phi] + S_{\text{CT}}[\phi]$.
We could just as well have called it the quantum discrete exterior calculus action, but we choose to use the former term.
To assess the critical behavior of the QFE action, we once again study the fourth Binder cumulant of the magnetization Eq.~\ref{eq:U4}.
%The Binder cumulants have been studied on very large lattice up to $L \approx 800$.
The result is presented in Fig.~\ref{fig:magU2d}.
The Binder cumulant was computed on very large lattices up to $L = 800$ and for very finely spaced relevant couplings $\mu_0^2$ near the critical surface.
There is no visible obstruction to criticality at these large lattice sizes, a marked improvement over the unrenormalized finite element action in which the Binder cumulant showed visible frustration on significantly coarser lattices, near $L=100$.
%We also have studied the higher order Binder cumulants \cite{Mon:1997} which are plotted in the right panel of Fig.~\ref{fig:magU2d} against $\mu_0^2$.  
We interpret Fig.~\ref{fig:magU2d} as strong evidence that the QFE action exhibits genuine critical behavior and that the continuum limit exists as $L \rightarrow \infty$ with no obstruction.

\begin{table}[!b]
	\begin{tabular}{| c || c c c |}
		\hline 
		Observable & QFE + FSS & Analytic Deng and Blote & Analytic Brower et. al. \\
		\hline
		$U_4$ & $0.85020(58)(90)$ & $0.8510061(108)$ & $0.8510207(63)$ \\
		$U_6$ & $0.77193(37)(90)$ & N/A & $0.7731441(213)$ \\
		\hline 
	\end{tabular}
	\caption{ \label{table:2dcumulants} Values for the fourth and sixth Binder cumulants for the Ising CFT on the 2-sphere.  The second column shows results from the Monte Carlo and subsequent Finite Size Scaling (FSS) analysis of the quantum finite element (QFE) action \cite{Brower:2018szu}.  The third and fourth columns show results from numerically integrating the analytic conformal correlators by Deng and Bl\"{o}te \cite{Deng:2003kiw} and Brower et. al. \cite{Brower:2018szu} respectively.}
\end{table}
A more quantitative assessment of the magnetization moments and Binder cumulants may be carried out through a finite size scaling analysis, details of which can be found in the recent work \cite{Brower:2018szu}.
The result for the critical values of the fourth and sixth magnetization Binder cumulants \cite{Mon:1997} from the finite size scaling analysis are found to be $U_{4,cr} = 0.85020(58)(90)$ and $U_{6,cr} = 0.77193(37)(90)$.
We compare these results to the continuum values of the critical Binder cumulants.
In the continuum, magnetization moments are computed from CFT correlators by integrating n-point functions over the manifold.
\begin{equation}
m_n = \langle M^n \rangle = \int_\mathcal{M} \prod_{i=1}^n \frac{d^2 x_i}{V} \sqrt{g(x_i)} \langle \phi(x_1) ... \phi(x_n) \rangle
\end{equation}
As such, the critical values of the magnetization moments and of the Binder cumulants are geometry dependent quantities; under conformal mappings, they are ``covariant'' rather than invariant.
The n-point correlation functions of the 2d Ising CFT may be constructed in principle for any $n$ \cite{Belavin:1984vu,Polyakov:1984yq,Luther:1975wr,Dotsenko:1984nm}.
Explicit expressions have been written down previously for the 4-point and 6-point correlation functions \cite{Burkhardt:1987}.
The two-point correlator integrated over the sphere is a closed form expression \cite{Deng:2003kiw}.
\begin{equation}
m_2 = \int \frac{d\Omega_1}{4\pi}\frac{d\Omega_2}{4\pi} \langle \phi(\Omega_1)\phi(\Omega_2) \rangle = \int_{-1}^{1} \frac{1}{2(2-2\cos \theta_{12})^{1/8}} d\cos\theta_{12} = \frac{2^{11/4}}{7}
\end{equation}
The task of integrating the four- and six-point correlators over the sphere is more nontrivial.
The fourth magnetization moment $m_4$ is an eight dimensional integral which may be reduced to a five dimensional integral using rotational invariance on the sphere.
Deng and Bl\"{o}te \cite{Deng:2003kiw} compute the fourth magnetization moment using 1000 independent Monte Carlo estimates, yielding $m_4^* = 1.19878(2)$ and a prediction for the fourth Binder cumulant $U_4^* = 0.8510061(108)$.
We have recomputed these values to higher accuracy using the \texttt{MonteCarlo} routine of \emph{Mathematica}'s \texttt{NIntegrate[]} function.  
We set \texttt{AccuracyGoal} $\rightarrow$ \texttt{4} yielding a Monte Carlo with standard deviation approximately $10^{-4}$.
We compute 100 Monte Carlo estimates of the integral with these parameters to acquire a sample distribution.
Our result is $m_4^* = 1.1987531(116)$ where the error is the standard error on the mean of the sample distribution.
The corresponding result for the fourth Binder cumulant is $U_4^* = 0.8510207(63)$.
We carry out a similar calculation for the sixth magnetization moment $m_6$.  
The calculation is a twelve dimensional integral which we reduce to a nine dimensional integral using rotational invariance.
Using the same \emph{Mathematica} routine but now setting \texttt{AccuracyGoal}$\rightarrow$\texttt{3} to yield a Monte Carlo distribution with standard deviation approximately $10^{-3}$, generate a sample distribution by computing 50 Monte Carlo samples.
Our result is $m_6^* = 1.632851(253)$.
The corresponding result for the sixth Binder cumulant is $U_6^* = 0.7731441(213)$.
A comparison between the numerical computation of the Binder cumulants and the analytic results can be found in Table~\ref{table:2dcumulants}.
Results agree within errors, providing further confirmation that the QFE action is converging to the correct critical theory.

\begin{figure}[t]
\centering
\includegraphics[width=0.65\textwidth]{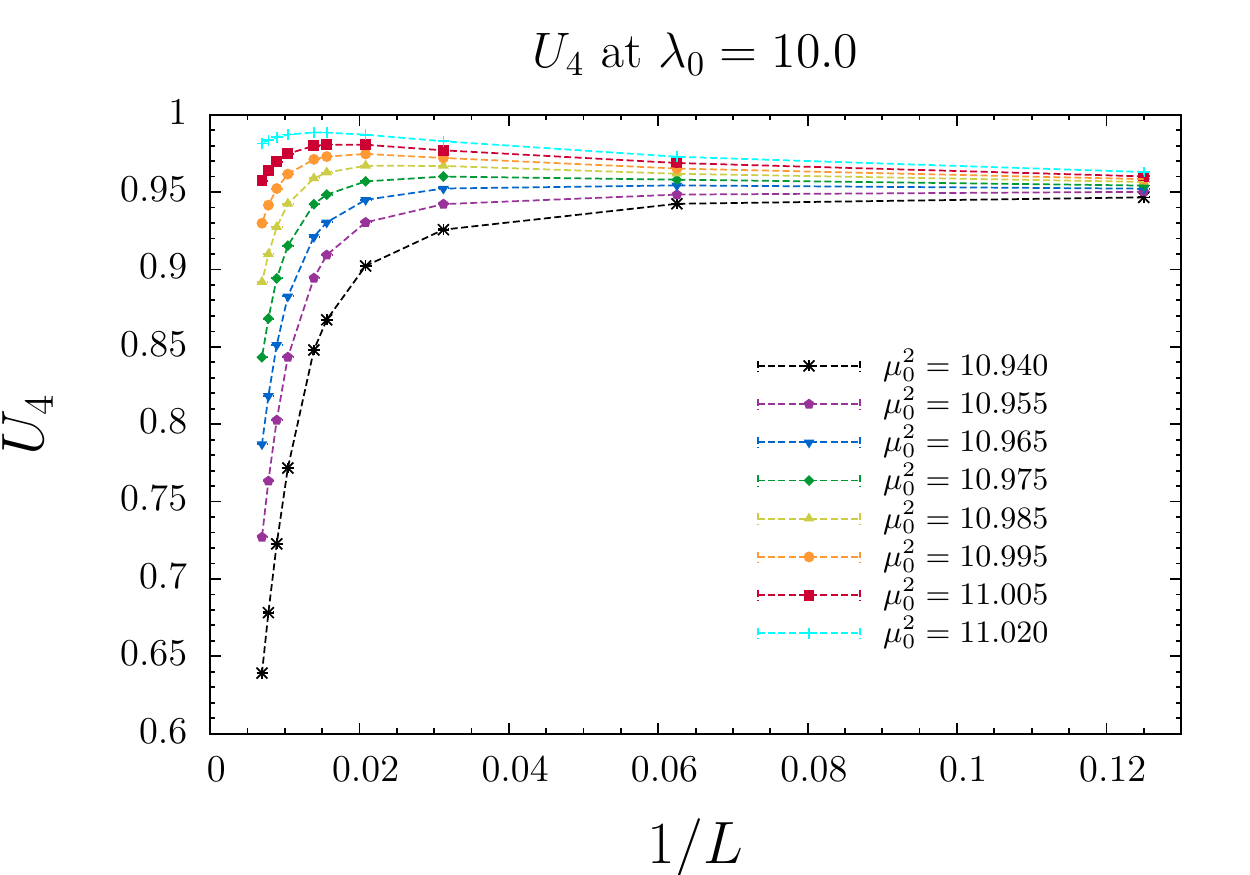}
\caption{\label{fig:magU4_strong_2d} Fourth Binder cumulant in the quantum finite element $\phi^4$ action on the 2-sphere at strong bare coupling.}
\end{figure}
We close this section with an instructive remark on what can happen to the QFE action at larger values of the bare quartic coupling.
So far we have only studied the FEM and QFE actions at a single value of the bare coupling $\lambda_0 = 1.0$.
In Fig.~\ref{fig:magU4_strong_2d} we show the fourth Binder cumulant for the quantum finite element action at a very strong value of the bare coupling $\lambda_0 = 10.0$.  
One notices immediately that the theory does not appear to be reaching the continuum limit, but rather it appears to become frustrated near $L = 100$.
Interestingly, this is roughly the same lattice size at which the uncorrected finite element action become frustrated.
Though, this may be little more than a coincidence.

There are two interpretations of this ostensible problem at strong bare coupling, neither of which undermine the validity of the results that we have presented for weak bare coupling.
The first interpretation is that because the QFE method as we have presented it is based on bare perturbation theory, the renormalization scheme fails for bare couplings outside of the radius of convergence of the perturbation expansion.
In this case, the QFE lattice theory would have a phase diagram with two or more phases.  
At sufficiently weak bare coupling, the theory flows to the continuum limit and the correct critical Ising CFT.
At stronger bare coupling, there is a transition (which could be first or second order) and on the other side of the phase transition line, the theory does not flow to the continuum CFT but flows to a lattice bulk phase in the IR.
This is a familiar phenomenon in lattice QCD; when the bare coupling is too strong the theory does not reach the continuum limit.
In a future work, one could study the phase behavior of the QFE action by running calculations at many different values of $\mu_0^2$ and $\lambda_0$ and mapping out the phase diagram.

On the other hand, it is possible that the QFE theory at strong coupling does eventually reach the continuum limit, but one has to go to very large values of $L$ to see it converge.
If our key assumptions are correct, then any diagram appearing in perturbation theory (besides the one loop diagram) will eventually become position independent as $L \rightarrow \infty$.  
However, diagrams at higher orders in perturbation theory may require one to go to incrementally larger values of $L$ in order for the position dependent contributions arising from these diagrams at finite lattice spacing to be negligible.  
These questions may be addressed in future works by studying the perturbation expansion to higher orders.

Let us emphasize that even if the QFE method with perturbative counterterms only works for sufficiently weak bare couplings in the UV, the theory still flows to strong coupling (to the Wilson Fisher fixed point) in the IR.  
The results for the critical Binder cumulants that we have presented in Fig.~\ref{table:2dcumulants} are features of the strongly coupled fixed point.  
These are nonperturbative results computed to high accuracy with the QFE method.
In the next section, we build on this success by studying the two point and four point correlation functions and extracting nonperturbative CFT data.

\subsection{2- and 4-point Functions}
Here we briefly review results for the two- and four-point correlation functions that have been recently presented in Ref.~\cite{Brower:2018szu}.
The two point function of the primary $\phi(x)$ on $\mathbb{R}^2$ is constrained by conformal symmetry to have the form \cite{Ginsparg:1988ui,DiFrancesco:639405}
\begin{equation}
\langle \phi(\vec{r}_1) \phi(\vec{r}_2) \rangle_{\mathbb{R}^2} = \frac{1}{r^{2\Delta_{\phi}}}
\end{equation}
where $\vec{r} = \vec{r}_1 - \vec{r}_2$ and the scaling dimension of $\phi$ in the $c=1/2$ minimal Ising CFT is $\Delta_{\phi} = 1/8$.
Mapping to the sphere, the expression for the two point function on $\mathbb{S}^2$ is
%\begin{flalign}
%\langle \phi(\theta_1,\phi_1) \phi(\theta_2,\phi_2) \rangle_{\mathbb{S}^2} = &\left[ 4 \left( 
%	\cos^2(\theta_1/2)\sin^2(\theta_2/2) + \sin^2(\theta_1/2)\cos^2(\theta_2/2) \right.\right. \nonumber \\
%	&\left.\left.  - 2 \cos^2(\theta_1/2)\cos^2(\theta_2/2)\cos(\phi) \right)\right]^{-\Delta_{\phi}}
%\end{flalign}
%Rotating coordinates such that $\theta_2 \rightarrow \pi$ which corresponds to $\vec{r}_2 \rightarrow \vec{0}$ in $\mathbb{R}^2$ yields the simple expression
\begin{equation}
g_2(\theta_{12}) \equiv \langle \phi(\hat{n}_1)\phi(\hat{n}_2) \rangle_{\mathbb{S}^2} = \frac{1}{\left(2 - 2 \cos\theta_{12}\right)^{\Delta_{\phi}}}
\end{equation}
where $\theta_{12} = \hat{n}_1 \cdot \hat{n}_2$.
We may express the two point function in angular momentum space by projecting onto Legendre Polynomials.
\begin{equation}
c(l) = \int_{-1}^1 dx P_l(x) \frac{1}{(2-2x)^{1/8}}
\end{equation}

\begin{figure}[t]
	\centering
	\includegraphics[width=0.45\textwidth]{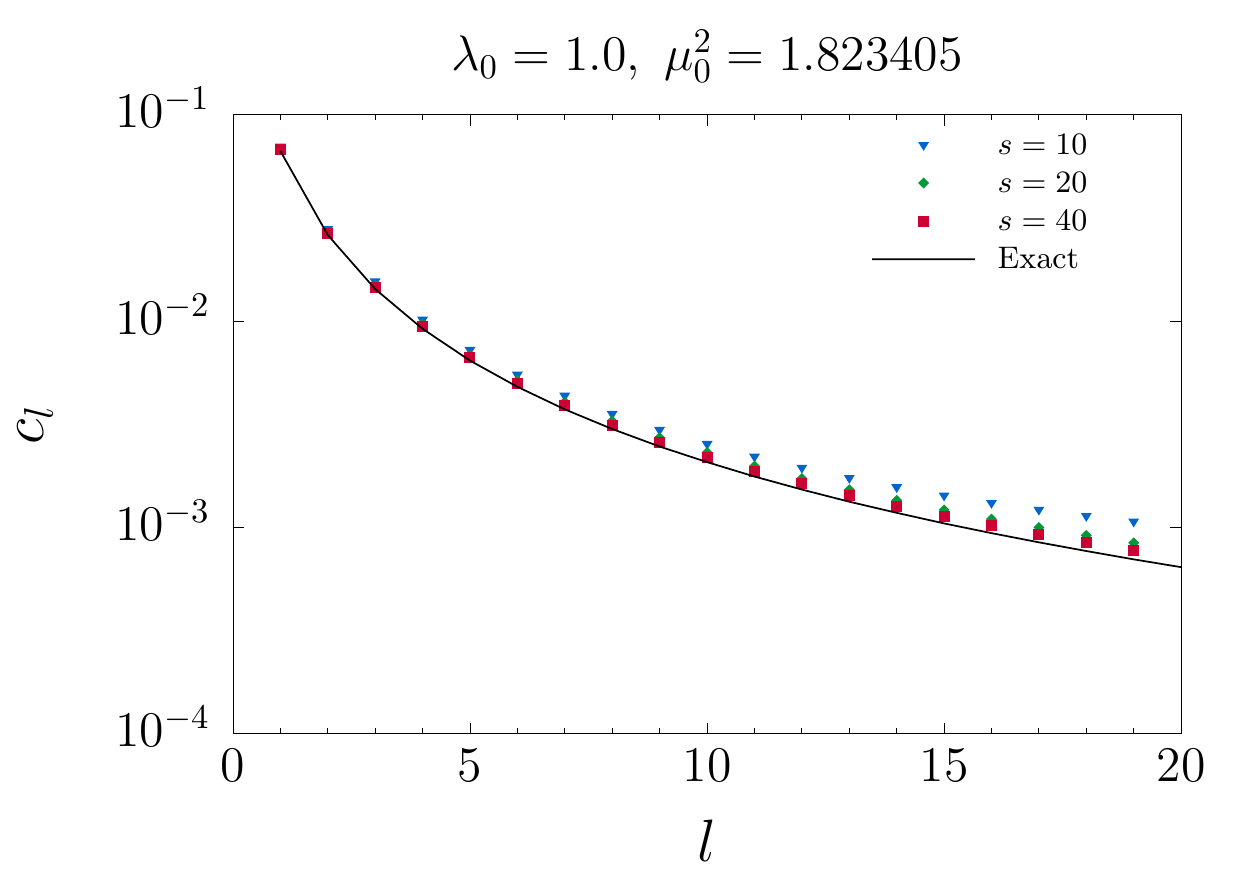}
	\includegraphics[width=0.45\textwidth]{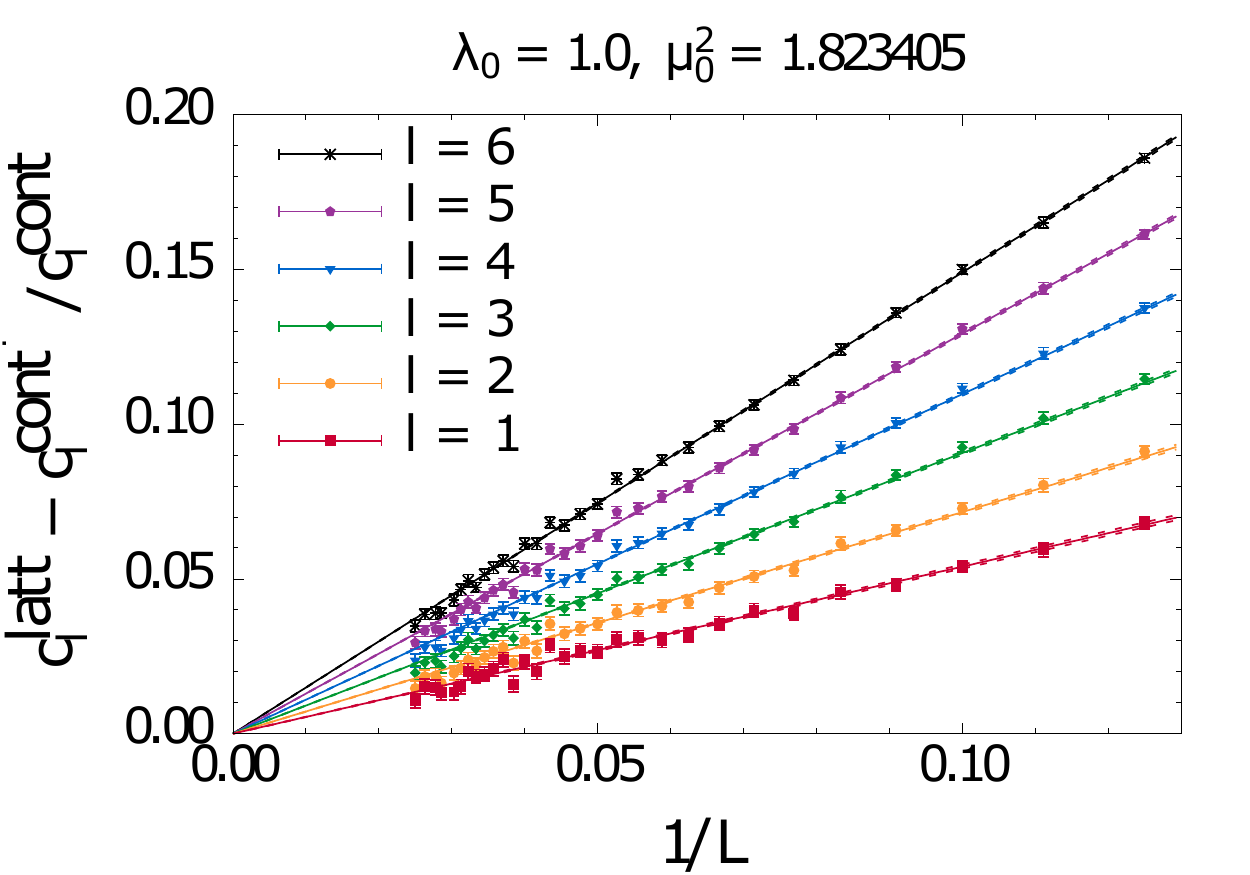}
	\caption{\label{fig:2pt_2d} Legendre Coefficients of two-point correlation function.  Left: Legendre coefficient vs $l$ for several lattice sizes compared to exact continuum expression.  Right: Difference between lattice and continuum Legendre coefficients vs $1/L$ for $l=1,...,6$.  This figure originally appears in Ref.~\cite{Brower:2018szu}.}
\end{figure}
We sample the two point function in our QFE Monte Carlo calculation, binning the values in $\theta_{12}$ in order to project onto Legendre coefficients.
The results are shown in Fig.~\ref{fig:2pt_2d}.
The Legendre coefficients of the lattice two-point function approach the continuum values like $1/L$ as $L\rightarrow \infty$.  
The rapid approach of the two-point function to the exact conformal correlator is strong evidence that the QFE procedure is reaching the correct critical CFT as $L\rightarrow \infty$.

We also study the four-point function by randomly sampling the $\phi$ field on the simplicial complex during the Monte Carlo evolution of the QFE action.
The conformal four point function may be written in a manifestly t-channel symmetric way.
\begin{equation}
G(u,v,\theta_{13},\theta_{24}) = g_2(\theta_{13}) g_2(\theta_{24}) G(u,v)
\end{equation}
where $(u,v)$ are the conformal cross ratios
\begin{equation}
u = \frac{r_{12}^2 r_{34}^2}{r_{13}^2 r_{24}^2} = |z|^2 \quad , \quad v = \frac{r_{14}^2 r_{23}^2}{r_{13}^2 r_{24}^2} = |1-z|^2
\end{equation}
equivalent to the single complex variable $z$.
The explicit form of $G(u,v)$ is known for the minimal $c=1/2$ Ising CFT as a sum of Virasoro blocks \cite{DiFrancesco:639405}.
\begin{equation}
G(u,v) = v^{\Delta}g(u,v) = \frac{1}{2|z|^{1/4}|1-z|^{1/4}} \left[ |1+\sqrt{1-z}|+ |1-\sqrt{1-z}|\right]
\end{equation}

\begin{figure}[t]
	\centering
	\includegraphics[width=0.45\textwidth]{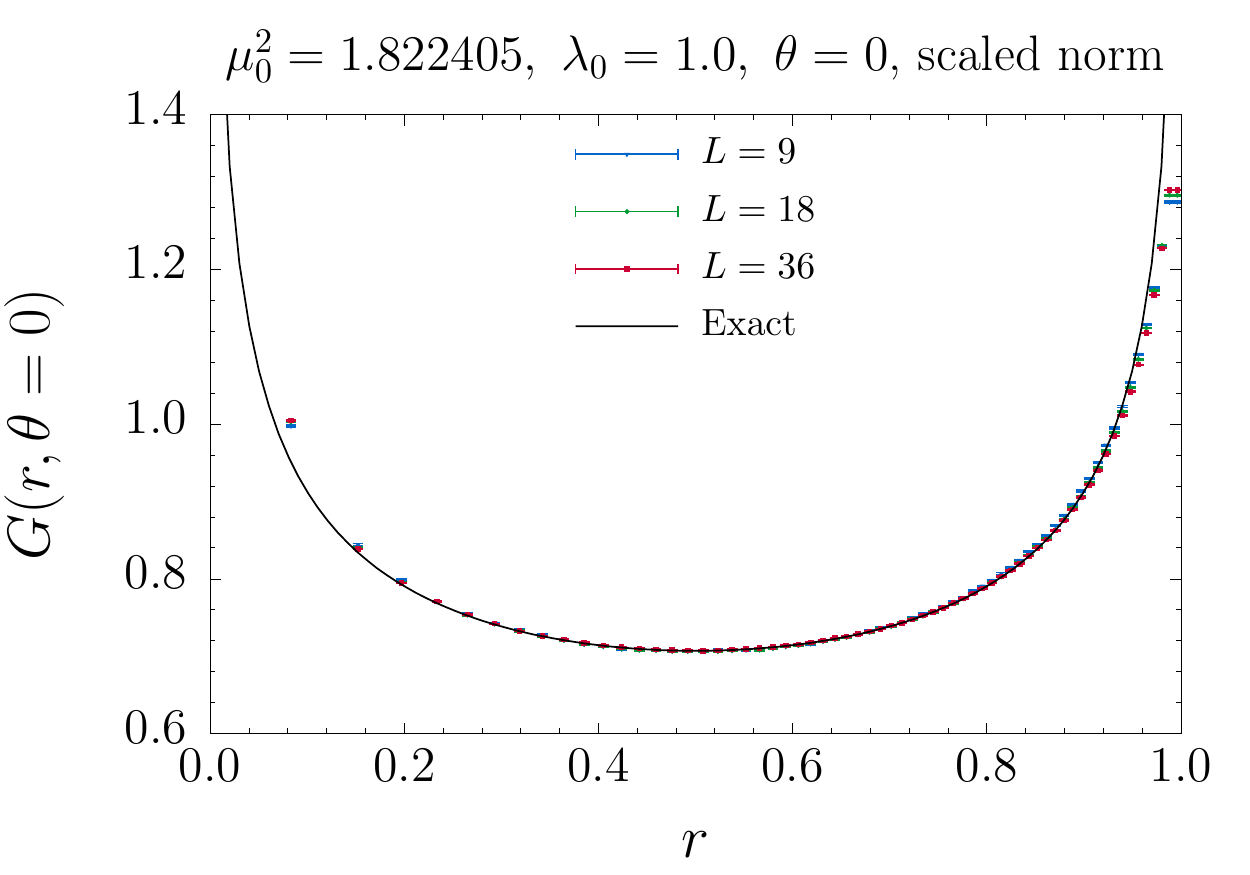}
	\caption{\label{fig:4pt_2d} Four point function at the critical point of the QFE action compared to the exact conformal four point function.  The functions are compared in polar coordinates $z = r e^{i\theta}$ along a fixed radial line, $\theta = 0$.  This figure originally appears in Ref.~\cite{Brower:2018szu}.}
\end{figure}
The exact solution is compared to the numerical computation of the four point function at the critical point using the QFE action in Fig.~\ref{fig:4pt_2d}.
The functions are plotted in polar coordinates $z = r e^{i\theta}$ and compared along a fixed radial line $\theta = 0$. 
The QFE Monte Carlo calculation is in close agreement with the exact solution even at comparatively small refinement, $L = 36$.
The convergence is best near the symmetry point $r = 1/2$.
In a scenario in which the exactly four point function was not known, we would not be in a position to make such a direct comparison.
Instead, one may project the numerically computed four point function onto conformal blocks, which are exactly computable basis functions
By fitting to the conformal block expansion, one may extract the scaling dimensions and O.P.E. coefficients of the CFT.
This procedure has been demonstrated in the recent work Ref.~\cite{Brower:2018szu}.
The conformal dimension $\Delta_{\epsilon}$ and the O.P.E. coefficients $\lambda_{\epsilon}$ and $c$ (the central charge) were found to be in close statistical agreement to the exactly known values.

\section{3D Ising CFT in Radial Quantization}
Next we consider three dimensional scalar $\phi^4$ theory on the manifold $\mathbb{R}\times \mathbb{S}^2$, the geometry of radial quantization.
We have explained at the beginning of Chapter~\ref{chapter:QFE1} that in any number of dimensions the spaces $\mathbb{R}^d$ and $\mathbb{R}\times \mathbb{S}^{d-1}$ are related by a Weyl rescaling factor.
As such, scalar $\phi^4$ theory at the critical point on $\mathbb{R}\times \mathbb{S}^2$ should be equivalent to the critical behavior of the theory on $\mathbb{R}^3$, which is the 3D Ising conformal fixed point.
We have also detailed at the beginning of Chapter~\ref{chapter:QFE1} the anticipated advantages of studying conformal theories in radial quantization on the lattice.
We show here that the QFE $\phi^4$ action on  $\mathbb{R}\times \mathbb{S}^2$ appears to be reaching a critical point which is a major step towards realizing lattice radial quantization in a nontrivial theory.
We leave the complete characterization of this fixed point to be presented in a future work.

The FEM/DEC lattice action is a trivial extension of the lattice action for scalar field theory on $\mathbb{S}^2$; one simply extends the action with a flat time direction with periodic boundary conditions.
\begin{equation}
S[\phi] = S_f[\phi] + \lambda_i \phi_{i,t}^4 = \frac{1}{2}\phi_{i,t}\left[ K_{ij} + m_i \delta_{ij} \right] \phi_{j,t} + \sqrt{g_i}\frac{1}{2}\left(\phi_{i,t} - \phi_{i,t\pm 1 }\right)^2 + \lambda_i \phi_{i,t}^4 \label{eq:free3d}
\end{equation}
The spectrum of the Laplacian operator on $\mathbb{R}\times \mathbb{S}^2$ does not contain any subtleties once the FEM/DEC Laplacian has been analyzed on $\mathbb{S}^2$. 
One may transform to frequency space via a discrete Fourier transform in the periodic time dimension: $t \rightarrow \omega$.
Then, for each frequency mode $\omega_n$, the spectrum is equivalent to the spectrum of the massive Laplacian on $\mathbb{S}^2$ with a shifted mass $m^2 = m_0^2 + \omega_n^2$.

Having understood the classical problem, let us consider the interacting quantum field theory.
As in the two dimensional case, we will have to introduce explicit counterterms in the FEM/DEC Lagrangian in order to cancel position dependent loop contributions to the couplings.
Three dimensional scalar $\phi^4$ theory is also superrenormalizable.  
It generates two divergent diagrams in the perturbative series both contributing to the renormalization of the two point function.
The one loop diagram is linearly divergent, and the two loop diagram is logarithmically divergent.
The computation of the one- and two-loop diagrams in lattice perturbation theory follows exactly the procedure detailed in Section~\ref{sec:QFES2} for the two dimensional theory.
Let us define the quadratic differential operator as a rewriting of the quadratic terms in Eq.~\ref{eq:free3d} into matrix form.
\begin{equation}
S_f[\phi] = \frac{1}{2} \phi_{i,t} M_{i,t;j,t'} \phi_{j,t'}
\end{equation}
Then the propagator is the inverse of this matrix: $G_{i,t;j,t'} \equiv (M^{-1})_{i,t;j,t'}$.

The amputated 1PT amplitude for the two point correlator is given by
\begin{equation}
\Sigma_{i,t;j,t'} = -12 \lambda_i G_{i,t;i,t} \delta_{ij}\delta_{t,t'} + 96 \lambda_i \lambda_j G_{i,t;j,t'}^3 + \mathcal{O}(\lambda_0^3) \label{eq:1PT3d}
\end{equation}
Recall again that by the key assumption of our QFE procedure, UV divergent diagrams should break up into a position independent divergence piece -- which is the universal continuum divergence -- and a position dependent finite piece.

\begin{figure}[t]
	\centering
	\includegraphics[width=0.45\textwidth]{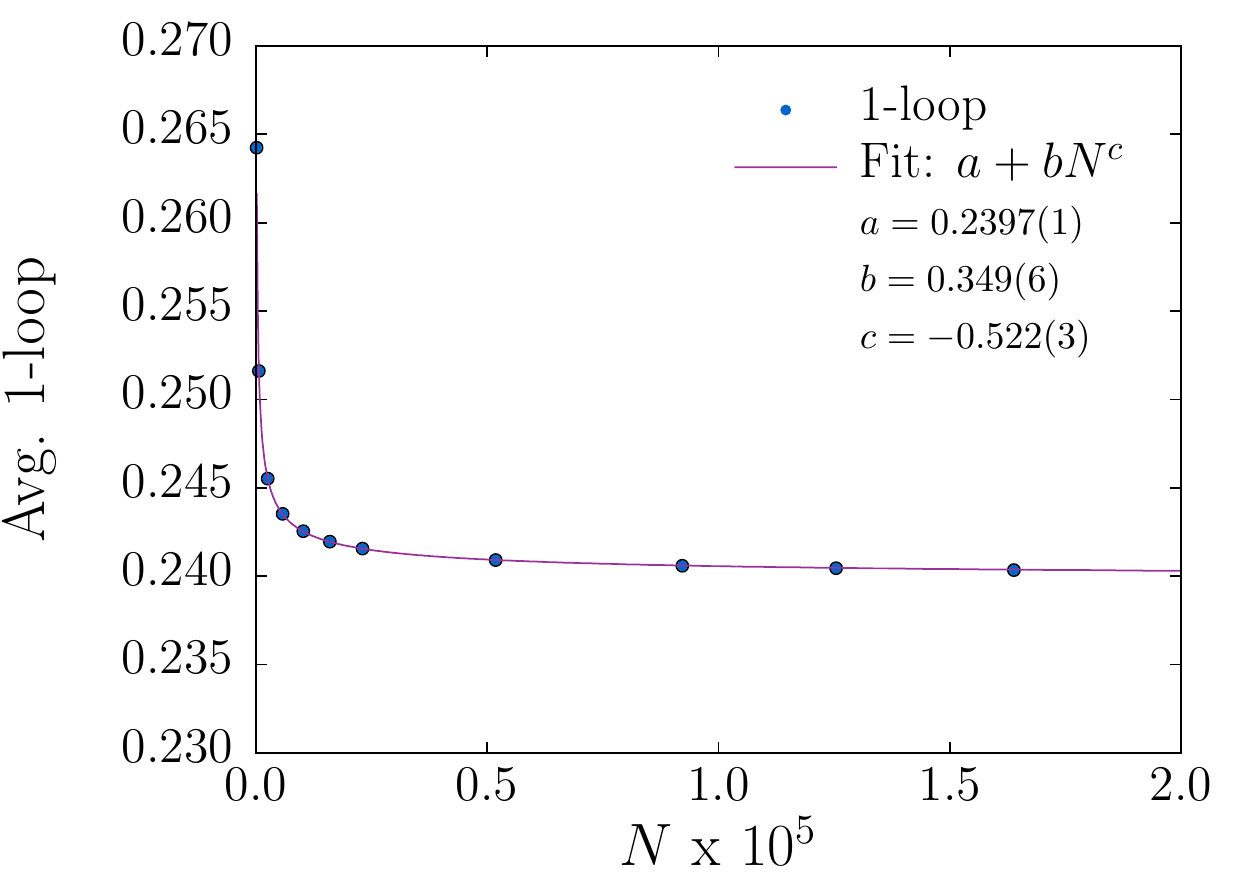}
	\includegraphics[width=0.45\textwidth]{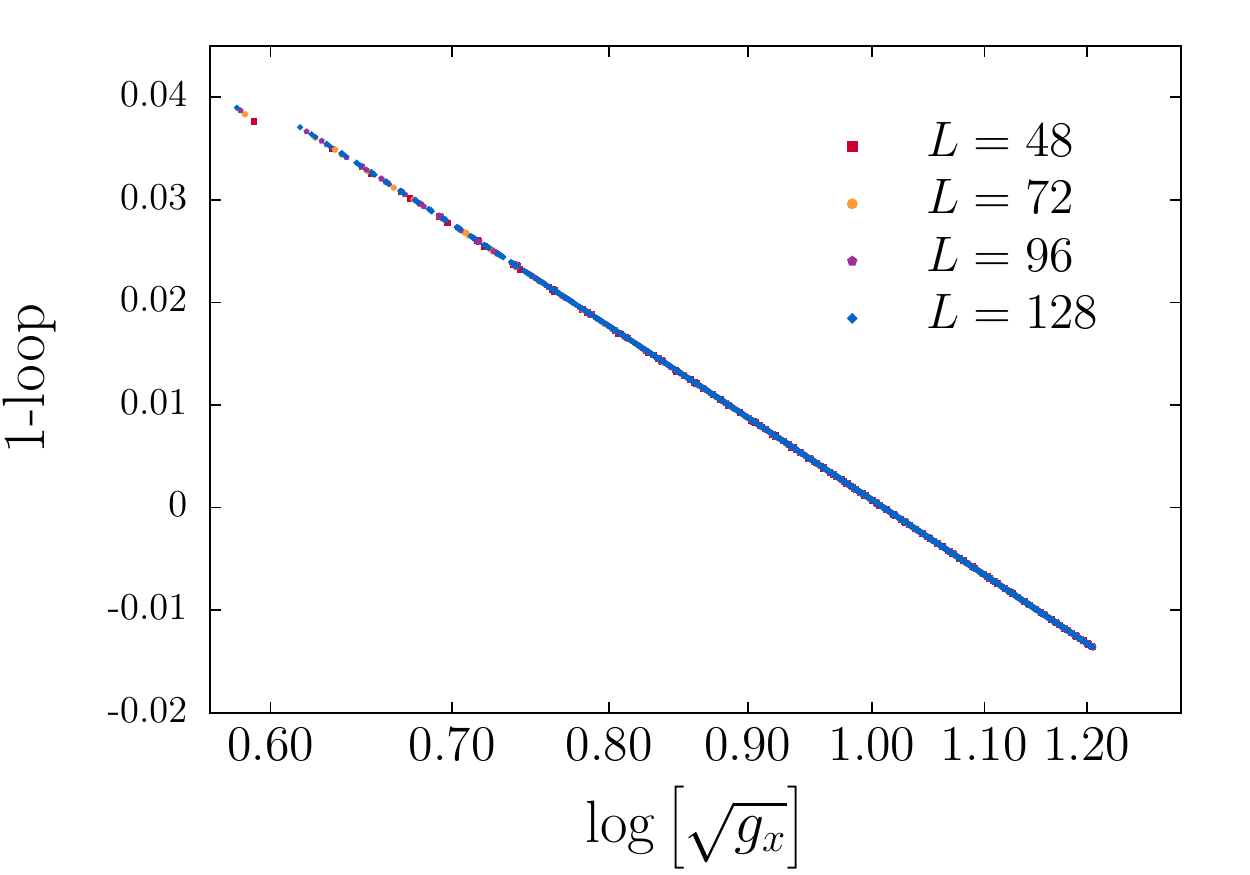}
	\caption{\label{fig:OneLoopCT3d} One loop diagram $G_{i,t;i,t}$ in lattice perturbation theory in three dimensions.  Left: The average value of the one loop diagram $(1/V)\sum_{i,t} \sqrt{g_i} G_{i,t;i,t}$ plotted against $N = 10 L^2 +2$ is shown to converge to a constant as $1/\sqrt{N}$.  Right: The one loop counterterm with the average piece subtracted is shown to approach as finite function as $L \rightarrow \infty$, and this finite function is approximately a linear function of $\ln \sqrt{g_x}$. }
\end{figure}
First we study the one loop diagram.
The average piece of the one loop diagram should contain the linear divergence.
We see in the left panel of Fig.~\ref{fig:OneLoopCT3d} that the average piece of $G_{i,t;i,t}$ approaches a constant as $N \rightarrow \infty$.
This is consistent with the usual intuition that there are no power divergences on the lattice because there are no dimensionful quantities.
To put the one loop diagram in physical units, we multiply by $1/a \approx \sqrt{N} \approx L$.  
Then, since the average piece of the diagram in lattice units approaches a constant as $L \rightarrow \infty$, the average piece in physical units will diverge linearly with $L$ as $L \rightarrow \infty$ as expected.
In the right panel of Fig.~\ref{fig:OneLoopCT3d}, we plot the subtracted one loop diagram as a function of $\ln\left[\sqrt{g_i}\right]$ for various values of $L$.
We see that as in the two dimensional case the subtracted diagram is approaching a finite function that is approximately linear in $\ln\left[\sqrt{g_i}\right]$.

\begin{figure}[t]
	\centering
	\includegraphics[width=0.45\textwidth]{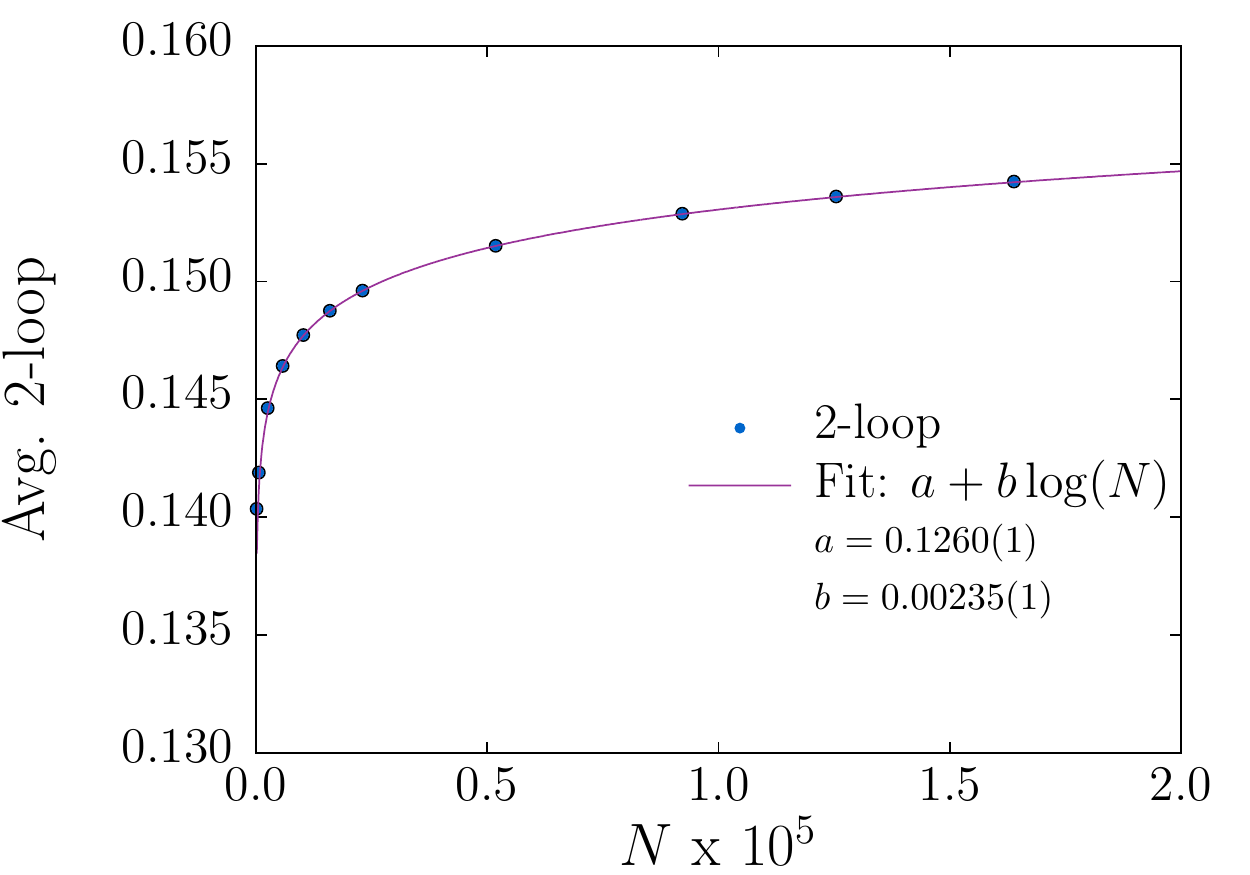}
	\includegraphics[width=0.45\textwidth]{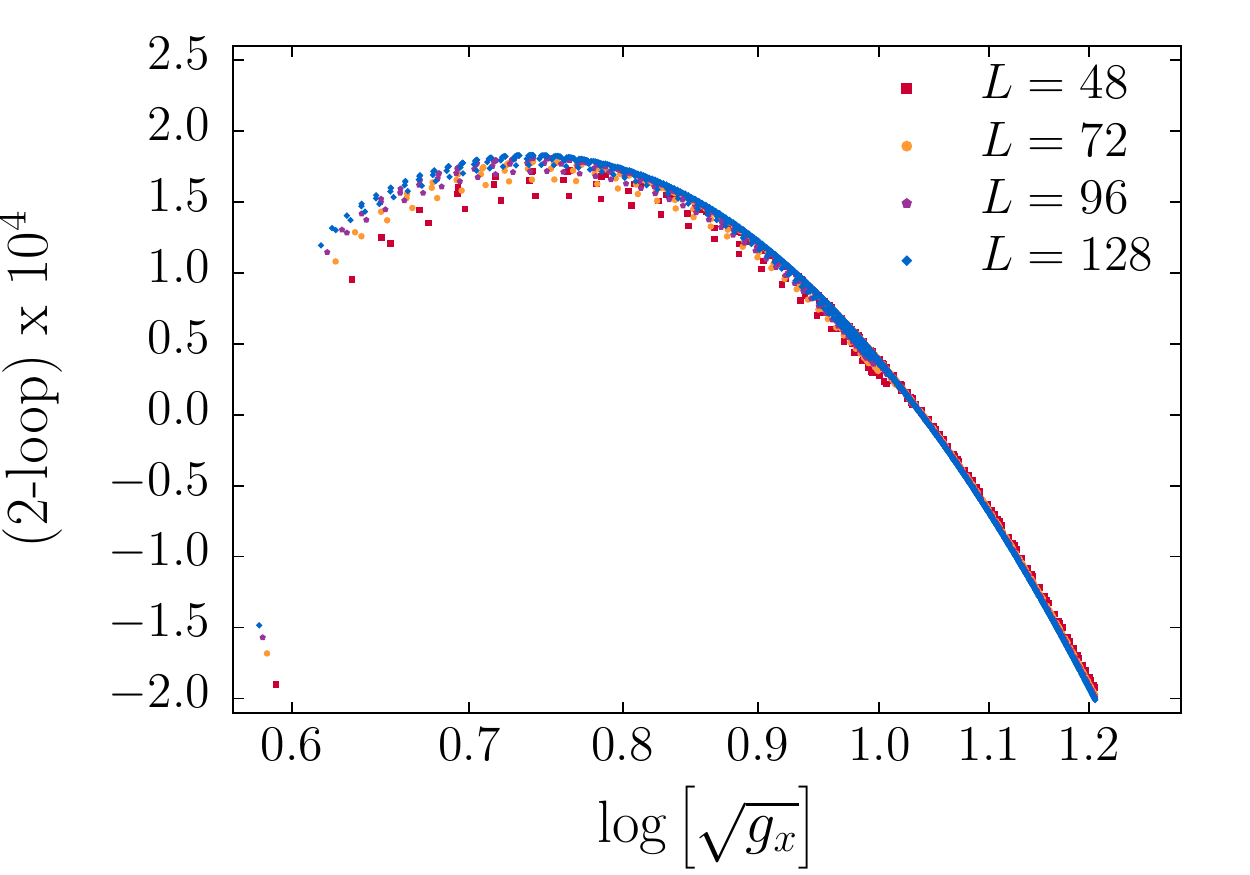}
	\caption{\label{fig:TwoLoopCT3d} Local approximation to two loop diagram in two dimensions $\sum_j \sqrt{g_j} G_{ij}^3$.  Left: Average value of diagram plotted against $N$.  Right: Two loop diagram with average piece subtracted plotted against $\sqrt{g_i}$ for various values of $L$. }
\end{figure}
Next we consider the two loop diagram.
As in Section~\ref{sec:QFES2}, we approximate the point split two loop diagram as an ultralocal function.
\begin{equation}
96 \lambda_i \lambda_j G_{i,t;j,t'}^3 \approx 96 \lambda_i \left(\sum_{j',t''} \lambda_j G_{i,t;j',t''}^3\right) \delta_{ij}\delta_{t,t'}
\end{equation}
This is equivalent to dropping the higher terms in a derivative expansion of the diagram.
Since the 2-loop diagram is log-divergent, the higher derivative components will be convergent and should not add any further position dependent contributions which would require cancellation by an explicit counterterm.
In the left panel of Fig.~\ref{fig:TwoLoopCT3d}, we show the average piece of the local approximation to the two loop diagram as a function of $N$.
We have fixed the temporal extent of the lattice to $L_t = 4 L$.
There is a clear logarithmic divergence which we confirm by an explicit fit to the form $a + b\ln(N)$.
In the right panel of Fig.~\ref{fig:TwoLoopCT3d}, show the subtracted two loop diagram as a function of $\ln\left[\sqrt{g_i}\right]$ for various values of $L$.
We see that it is approaching a smooth function over the sphere.

We introduce two counterterms into our action Eq.~\ref{eq:free3d} to cancel the position dependent contributions shown in the right panels of Fig.~\ref{fig:OneLoopCT3d} and Fig.~\ref{fig:TwoLoopCT3d}.
\begin{flalign}
S_{\text{CT}} = \sum_{i,t} \lambda_i &\left[ 6 \left( G_{i,t;i,t} - \frac{1}{V}\sum_{j,t'} \sqrt{g_j} G_{j,t';j,t'} \right) - \right. \nonumber \\
 &\left.  48 \left( \sum_{j,t'}\lambda_j G_{i,t;j,t'}^3  - \frac{1}{V}\sum_{i',j,t',t''} \sqrt{g_{i'}}\lambda_j G_{i',t';j,t''}^3 \right) \right]  \phi_{i,t}^2 \label{eq:CT3d}
\end{flalign}
we have shown the summations explicitly for clarity, and we have defined $V = L_t N$.
The position dependent contributions from the one- and two-loop diagrams will be exactly canceled

\begin{figure}[t!]
	\centering
	\includegraphics[width=0.65\textwidth]{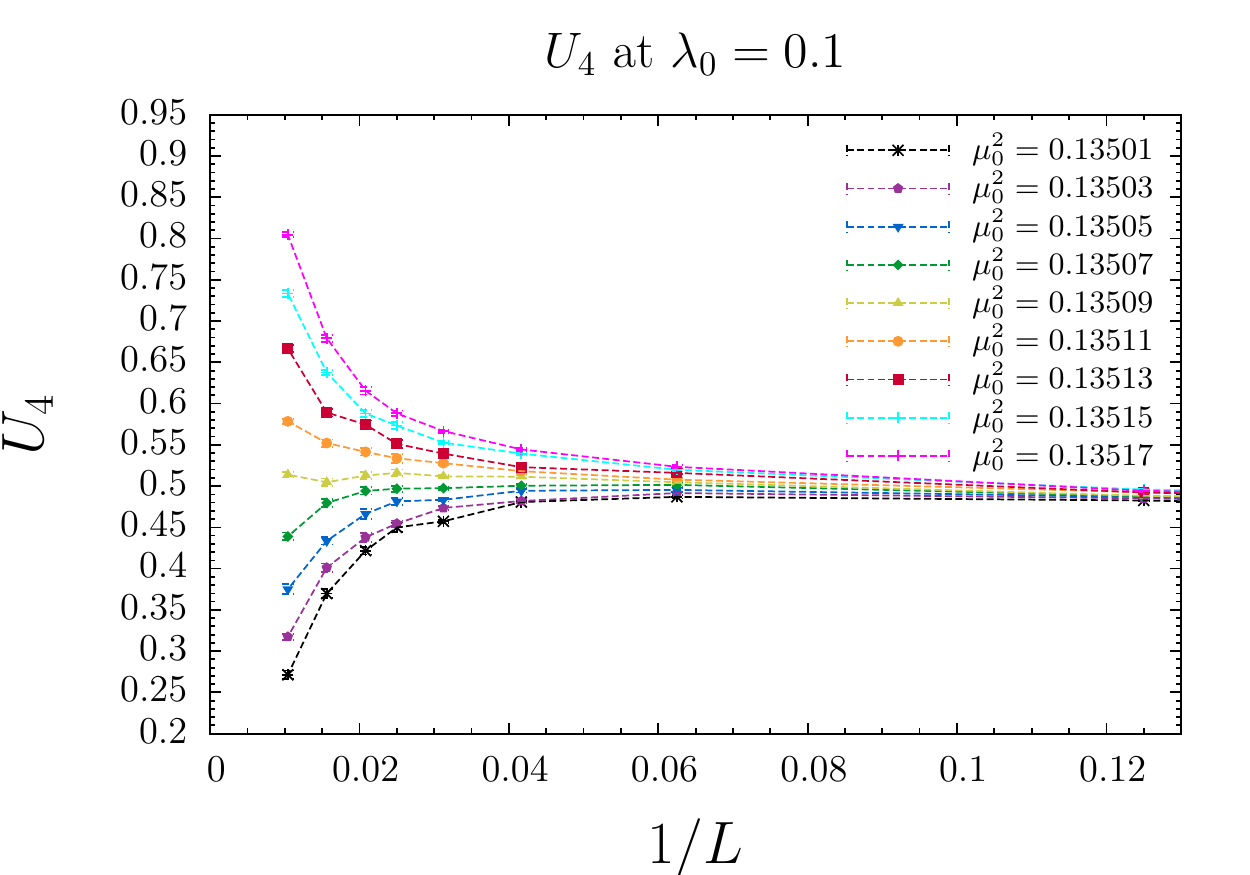}
	\caption{\label{fig:magU43d} Fourth Binder cumulant in the quantum finite element $\phi^4$ action on $\mathbb{R}\times\mathbb{S}^2$ at $\lambda_0 = 0.1$.  The temporal extent of the lattice is fixed to be $L_t = 4 L$.}
\end{figure}
We study the fourth Binder cumulant of the magnetization with a Monte Carlo simulation of our QFE action for the 3D theory.
\begin{equation}
S[\phi] = S_f[\phi] + \lambda_i \phi_{i,t}^4 + S_{\text{CT}}
\end{equation}
The results are shown in Fig.~\ref{fig:magU43d}.  
We have studied the theory up to a refinement of $L = 96$ and see no visible signs of an obstruction to criticality.
As a point of comparison, the theory will become frustrated at $L \approx 64$ when one does not include the counterterms of Eq.~\ref{eq:CT3d}.
We also remark that as in the two dimensional case, when the bare coupling is made stronger the system does not appear to reach criticality.
Because the Monte Carlo calculations are significantly more costly and time consuming for the three dimensional calculation than the calculations in two dimensions, we are not able to study the 3D theory much beyond about $L = 100$.
Our collaboration is working to develop a parallel code so that the critical behavior of the three dimensional theory can be more firmly established by going to larger $L$.
Work is under way to characterize the critical behavior of the ostensible critical surface near $(\lambda_0 = 0.1,\mu_0^2 = 0.13509)$ including finite size scaling analysis of the magnetization moments and an analysis of the two- and four-point functions.
A future work will report on continued progress.

%\subsection{Further 3D Ising Results}

%%%%%%%%%%
%%%%%%%%%%%
%\chapter{D0-Brane Quantum Mechanics}

%%%%%%%%%%
%%%%%%%%%%%
\chapter{Conclusions and Future Work}
In this manuscript, we have reported on a variety of efforts towards characterizing conformal and nearly conformal strongly coupled quantum field theories in two, three, and four dimensions using a combination of traditional and novel techniques.
In four dimensions, we have discussed the interesting interacting conformal fixed points that arise at the conformal window of Yang Mills gauge theories with fermions.
These fixed points are of interest for building models of composite Higgs bosons and other possible applications to beyond the standard model physics.
However, we have demonstrated in Chapter~\ref{chapter:Lattice} some of the difficulties of studying these conformal and nearly conformal gauge theories using traditional lattice methods.
We have proponed two separate novel approaches to making progress on this difficult problem.

The first approach, detailed in Chapters~\ref{chapter:Lattice} and~\ref{chapter:EFT} is to use lattice data to identify the best low energy EFT description of the nearly conformal gauge theories.
This EFT should serve as an aid to guide and compliment ongoing lattice calculations and as a bridge between the numerical studies being carried out on the lattice and phenomenological applications.
We have presented a study of $\pi\pi$ scattering in a nearly conformal gauge theory, and we have demonstrated the tension between the lattice calculation and the prediction from chiral perturbation theory.
In Chapter~\ref{chapter:EFT}, we presented a formulation of a new effective field theory framework based on the linear sigma model and demonstrated that already at leading order it can provide a substantial improvement over NLO chiral perturbation theory in fitting lattice data for a nearly conformal gauge theory.\\

There are numerous future directions for this effort that we hope to pursue in the near future.
One piece of low hanging fruit is to carry out a more comprehensive analysis of the linear sigma EFT as applied to $N_f = 8$ QCD.
This includes incorporating a larger basis of observables from the lattice in the analysis such as the maximal isospin $\pi\pi$ scattering length and the chiral condensate.
We would also like to carry out our analysis in such a way that we can keep track of systematic errors from the lattice and place reliable error bars on the fitted low energy constants.
Continued lattice studies of new observables in nearly conformal gauge theories would help to further constrain the EFT description, including the vector and scalar form factor of the pion and the $I=1$ and $I=0$ $\pi\pi$ scattering channels.

The second novel approach that we have presented for making progress on studying strongly coupled conformal systems is the quantum finite elements method detailed in Chapters~\ref{chapter:QFE1} and~\ref{chapter:QFE2}.
In Chapter~\ref{chapter:QFE1} we have explained the general method for constructing lattice regularizations of superrenormalizable quantum field theories on an arbitrary smooth Riemann manifold.
The general formalism is applicable to field theory in any number of dimensions.
In Chapter~\ref{chapter:QFE2} we have provided explicit numerical computations of interacting scalar field theory on $\mathbb{S}^2$ and $\mathbb{R}\times\mathbb{S}^2$.
In the former study, we have shown a comprehensive numerical analysis characterizing the critical behavior of the theory and showing it to be in close statistical agreement with the exact $c=1/2$ minimal Ising CFT.
In the latter study, we have demonstrated that the theory appears to have been renormalized appropriately at sufficiently weak bare coupling in that the fourth Binder cumulant appears to be showing critical behavior.

The quantum finite element method provides many directions for future research.
In the short term, we will provide a complete characterization of the 3D Ising conformal fixed point in radial quantization.
We hope that the numerical results will be competitive with the numerical conformal bootstrap program.
There are a variety of near future projects that one might consider for study with the quantum finite element radial lattice quantization program, including the O(N) model in 3D, the Gross-Neveu model in 3D, pure Yang Mills gauge theory in various dimensions, scalar QED in 3D, and so on.
Not only are there many interest quantum field theories to study with the existing formalism, but there are many direction for extending the formalism including different fermion discretizations, different renormalization schemes including nonperturbative renormalization, nonperturbative field theory on AdS space, and so on.

Continued study of strongly interacting systems promises to provide deeper insight into the structure of quantum field theory and to unveil new mechanisms that may be realized in physics beyond the standard model.
We hope that the tools and techniques developed in this dissertation will be of use in this important effort.

% Only call appendix once, here.
\appendix

\chapter{Useful Identities for $su(N_f)$ Lie Algebras \label{appendix:lie}}
Lie algebras are defined through the structure constants which determine the the commutation relations between basis elements, or generators, of the algebra.
\begin{equation}
[T_i,T_j] = i f^{ijk} T_k
\end{equation}
Choosing an arbitrary normalization for the Killing metric
\begin{equation}
\langle T_i T_j \rangle = \lambda \delta_{ij}
\end{equation}
the structure constants may be expressed as a trace over three generators.
\begin{equation}
f_{ijk} = \frac{1}{i\lambda} \left\langle [T_i,T_j] T_k \right\rangle
\end{equation}
The $f_{ijk}$ structure constants are totally antisymmetric.
We may define the symmetric structure constants through the anticommutator.
\begin{equation}
\{T_i,T_j\} = \frac{2\lambda}{N_f} \delta_{ij} \identity + d_{ijk} T_k
\end{equation}
Inverting this expression, the symmetric structure constants are given by
\begin{equation}
d_{ijk} = \frac{1}{\lambda} \left\langle \{T_i,T_j\} T_k \right\rangle
\end{equation}
Combining these two definitions for the structure constants, we may express the product of two generators as follows.
\begin{equation}
T_i T_j = \frac{1}{2} \left( [T_i,T_j] + \{T_i,T_j\} \right) = \frac{1}{2} \left(\frac{2\lambda}{N_f} \delta_{ij} \identity + (d_{ijk} + i f_{ijk})T_k \right)
\end{equation}
Defining the complex structure constant $h_{ijk} = d_{ijk} + i f_{ijk}$, 
\begin{equation}
T_i T_j = \frac{\lambda}{N_f} \delta_{ij}\identity + \frac{1}{2}h_{ijk}T_k
\end{equation}

Using this definitions for the structure constants, we may compute the traces of any number of generators in terms of structure constants.
The results for up to four generators are
\begin{flalign}
\langle T_i \rangle &= 0 \\
\langle T_i T_j \rangle &= \lambda \delta_{ij} \\
\langle T_i T_j T_k \rangle &= \frac{\lambda}{2} h_{ijk} \\
\langle T_i T_j T_k T_l \rangle &= \frac{\lambda^2}{N_f} \delta_{ij} \delta_{kl} + \frac{\lambda}{4}h_{ijm}h^{klm}
\end{flalign}

Finally, we have the Jacobi Identities which define the Lie algebra,
\begin{flalign}
[[T_i,T_j],T_k]+[[T_j,T_k],T_i]+[[T_k,T_i],T_j] &= 0 \\
[\{T_i,T_j\},T_k]+[\{T_j,T_k\},T_i]+[\{T_k,T_i\},T_j] &= 0
\end{flalign}
which imply the corresponding Jacobi Identities for the structure constants.
\begin{flalign}
f_{ijm}f_{mjk} + f_{jkm}f_{mil} + f_{kim}f_{mjk} &=0 \\
d_{ijm}f_{mjk} + d_{jkm}f_{mil} + d_{kim}f_{mjk} &=0
\end{flalign}

%\chapter{Perturbative Yang Mills}
%If you need an appendix, it will go here.

%\begin{align}
%a^n + b^n &\ne c^n \\
%n &> 2
%\end{align}

%\chapter{Statistical Analysis Methods}
%A second appendix. Look at you, you over achiever.

% Any chapters such as End Notes go after this.
\backmatter

\bibliography{gasbarro_dissertation}
% for your own sake, use a bibtex file, so all of the numbering of references will be done
% automatically.

\end{document}